%% file: IPHAS_PNe_Main.tex
\pdfoutput=1
\documentclass[usenatbib]{mnras}

\usepackage{mathtools}
\usepackage[load-configurations=astronomy]{siunitx} 
\DeclareSIUnit\parsec{pc}
\DeclareSIUnit\kilo{k}
\DeclareSIUnit\ergs{ergs}
\usepackage[version=3]{mhchem} 
\usepackage{amsmath} 

\usepackage{xcolor}
\newcommand\corrections[1]{\textcolor{red}{#1}}

\usepackage{graphicx}
\usepackage{float}
\usepackage{caption}
\usepackage{subcaption} 
\usepackage{morefloats}

\usepackage{booktabs} 
\usepackage{multirow}
\usepackage{array}
\usepackage{longtable}
\usepackage{tabu}
\usepackage{threeparttable} 
\usepackage{threeparttablex} 
\usepackage{rotating}
\usepackage{xtab}
\usepackage{multicol}
\usepackage{pdflscape}
\newcolumntype{H}{>{\setbox0=\hbox\bgroup}c<{\egroup}@{}} 

\newcommand{\Ha}{H\ensuremath{\alpha}} 
\newcommand{\HB}{H\ensuremath{\beta}} 
\newcommand{\EHaHB}{E(H\ensuremath{\alpha}-H\ensuremath{\beta})} 
\newcommand{\CHa}{C\textsubscript{H\ensuremath{\alpha}}} 
\newcommand{\CHB}{C\textsubscript{H\ensuremath{\beta}}} 
\newcommand{\AV}{A\textsubscript{V}} 
\newcommand{\Av}{A\textsubscript{V}} 
\newcommand{\AvCHB}{A\textsubscript{V}(C\textsubscript{H\ensuremath{\beta}})} 
\newcommand{\AvCHa}{A\textsubscript{V}(C\textsubscript{H\ensuremath{\alpha}})} 
\newcommand{\AvE}{A\textsubscript{V}[E(H\ensuremath{\alpha}-H\ensuremath{\beta})]} 

\title[IPHAS Planetary Nebulae]{H$\alpha$ fluxes and extinction distances 
for planetary nebulae in the IPHAS survey of the Northern Galactic Plane}
\author[T. E. Dharmawardena et al.]
{{\parbox{\textwidth}
{Thavisha E. Dharmawardena,$^{1,2,3,4}$ M. J. Barlow,$^{2}$ J. E. Drew,$^{2}$ A. Seales,$^{2,5}$\\ S. E. Sale,$^6$ D. Jones,$^{7,8}$ A. Mampaso,$^{7,8}$ Q. A. Parker,$^9$ L. Sabin,$^{10}$ R. Wesson$^2$}}
\\ \\
$^{1}$Max Planck Institute for Astronomy, K\"onigstuhl 17, 69117 Heidelberg, Germany.\\
$^{2}$Department of Physics and Astronomy, University College London, Gower Street, London WC1E 6BT, U.K.\\
$^{3}$Academia Sinica Institute of Astronomy and Astrophysics, 11F ASMAB, No.1, Sec. 4, Roosevelt Rd, Taipei 10617, Taiwan.\\
$^{4}$National Central University, No. 300, Zhongda Rd., Zhongli District, Taoyuan City 32001, Taiwan.\\
$^{5}$The Royal Institution of Great Britain,  21 Albemarle Street, London W1S 4BS, U.K.\\
$^6$Rudolf Peierls Centre for Theoretical Physics, Keble Road, Oxford OX1 3NP, U.K. \\
$^7$Instituto de Astrof\'isica de Canarias, E-38205 La Laguna, Tenerife, Spain \\
$^8$Departamento de Astrof\'isica, Universidad de La Laguna, E-38206 La Laguna, Tenerife, Spain \\
$^9$Department of Physics, University of Hong Kong, CYM Physics Building, Hong Kong; \\ The Laboratory for Space Research, University of Hong Kong, Hong Kong \\
$^{10}$Instituto de Astronom\'ia, Universidad Nacional Aut\'onoma de M\'exico, Apdo. Postal 877, C.P. 22860 Ensenada, B.C., M\'exico}

\date{Received:}

\begin{document}

\maketitle

\begin{abstract}
We report {\Ha} filter photometry for 197 northern hemisphere planetary nebulae (PNe) obtained using imaging data from the IPHAS survey. {\Ha}+[N~{\sc ii}] fluxes were measured for 46 confirmed or possible PNe discovered by the IPHAS survey and for 151 previously catalogued PNe that fell within the area of the northern Galactic Plane surveyed by IPHAS. After correcting for [N~{\sc ii}] emission admitted by the IPHAS H$\alpha$ filter, the resulting H$\alpha$ fluxes were combined with published radio free-free fluxes and H$\beta$ fluxes, in order to estimate mean optical extinctions to 143 PNe using ratios involving their integrated Balmer line fluxes and their extinction-free radio fluxes.
Distances to the PNe were then estimated using three different 3D interstellar dust extinction mapping methods, including the IPHAS-based H-MEAD algorithm of Sale (2014). These methods were used to plot dust extinction versus distance relationships for the lines of sight to the PNe; the intercepts with the derived dust optical extinctions allowed distances to the PNe to be inferred.
For 17 of the PNe in our sample reliable Gaia DR2 distances were available and these have been compared with the distances derived using three different extinction mapping algorithms as well as with distances from the nebular radius vs. H$\alpha$ surface brightness relation of Frew et al. (2016). That relation and the H-MEAD extinction mapping algorithm yielded the closest agreement with the Gaia DR2 distances.

\end{abstract}
 

\section{Introduction}

Accurate distance estimation to Galactic planetary nebulae has represented a persistent difficulty over the years, with many different
methods, both direct and statistical, having
been applied - see \citet{Frew2016} for a recent comprehensive summary of the various methods. Although Gaia has now measured parallaxes for a significant number of PN central stars \citep[e.g.][]{Kimeswenger2018,Gonzalez-Santamaria2019}, for many PNe high nebular surface brightnesses or faint central star magnitudes make reliable Gaia parallaxes difficult to determine. For such nebulae alternative distance estimation methods must be sought.

The Galactic PN distance estimation method that will be discussed in this paper is
based on the determination of interstellar dust extinctions and distances to a sufficient number of field stars nearby on the sky to a PN that a plot of the stellar dust extinctions versus distance allows the distance to the
PN to be inferred from its measured dust extinction. Early examples of the use
of this technique for PN distance estimation were published by \citet{Lutz1973}, \citet{Kaler1985} and \citet{Gathier1986}.
In recent years a number of deep optical and infrared imaging surveys have enabled the production of extensive 3D reddening maps of the Galactic plane, allowing distances to be estimated to stars and nebulae by plotting reddening vs. distance for their sight lines \citep[e.g.][]{Marshall2006, Sale2009, Sale2014, Sale2012, Lallement2014, Lallement2019, Green2015, Green2019_Bayestars19}. 

The Isaac Newton Telescope Photometric H-alpha Survey of the northern Galactic plane \citep[IPHAS;][]{Drew2005, Barentsen2014}, in addition to obtaining filter photometry in the Sloan $r'$ and $i'$ bands, has produced a large narrow-band {\Ha} photometric sample, enabling many emission line stars and nebulae to be identified and measured.
Data from IPHAS have been used to identify new examples of young stellar objects \citep{Vink2008}, classical Be stars \cite{Raddi2015}, supernova remnants \citep{Sabin2013} and planetary nebulae \citep{Mampaso2006, Corradi2011, ViironenJuly2009, Sabin2014}. 

From the initial IPHAS data release \citep{Gonzalez-Solares2008} to the second data release \citep{Barentsen2014} the applications of IPHAS data have expanded. \cite{Sale2009}, \cite{Sale2012} and \cite{Sale2014} have demonstrated how the information from
colours using the r$'$, i$'$ and (in particular) H$\alpha$ filters enables spectral types and interstellar dust extinctions to be inferred for objects in the IPHAS survey, enabling 3D extinction mapping to be carried out for sight lines covered by the survey. For example, \cite{Giammanco2011} applied the IPHAS-based extinction mapping technique MEAD \citep[Mapping Extinction Against Distance;][]{Sale2009} to planetary nebulae, using spectroscopic Balmer-line ratios to estimate reddenings and distances for 70 planetary nebulae. 

In the past, dust extinctions to PNe have often been obtained from the ratio of radio free-free fluxes (unaffected by dust extinction) to H$\beta$ filter fluxes. 
In this paper we measure H$\alpha$ fluxes for a sample of 197 PNe in the northern Galactic Plane observed by IPHAS and use them, along with published integrated H$\beta$ fluxes that are available for 37 of the nebulae, to determine interstellar dust extinctions to the nebulae.
In Section 2 of the paper we present H$\alpha$ flux measurements for 197 PNe observed by IPHAS
and correct for the [N~{\sc ii}] 6548, 6583~\AA\ contributions to these fluxes. In Section 3 we 
combine the [N~{\sc ii}]-corrected H$\alpha$ fluxes with published H$\beta$ and radio free-free fluxes in order to
derive interstellar dust extinctions to 143 PNe. In Section 4 we derive extinction distances to these PNe using a number of 3D extinction mapping techniques and for 17 PNe in the sample that are judged to have reliable Gaia DR2 distances we compare these to the extinction mapping distances.

\section{Nebular {\Ha} fluxes for northern planetary nebulae}

\subsection{The IPHAS survey}

The IPHAS survey \citep{Drew2005} was carried out using the Wide Field Camera (WFC) on the 2.5\si{\metre} Isaac Newton telescope located 
in the Observatorio del Roque de los Muchachos on the island of La Palma. It was a digital optical photometric survey of the northern Galactic plane, carried out in three bands, $r', i'$ and {\Ha}. The IPHAS footprint extends over $-5\si{\degree}<b<+5\si{\degree}$ and $30\si{\degree}<l<215\si{\degree}$, forming an area of approximately 1800 square degrees \citep{Drew2005, Barentsen2014}.

The survey obtained images using the WFC's {\Ha} narrow-band filter with a full width at half maximum (FWHM) bandwidth of $95$~\AA\ centred at $6568$~\AA\ and broad-band Sloan $r'$ (bandwidth $1380$~\AA\ centred at $6260$~\AA) and $i'$ (bandwidth $1535$~\AA\ centred at $7670$~\AA) filters. Images were taken sequentially with all three filters at each pointing, with exposure times of $120$s for the {\Ha} filter and mostly $30$s and $10$s for the $r'$ and $i'$ filters, respectively. The WFC detector pixels projected to 0.333 arcsec per pixel on the sky.

Each WFC image is a mosaic from its four CCDs, capturing a sky area of approximately 0.25 square degrees. To fill in the gaps between the CCDs, a set of offset pointings with the same filters and exposure times was obtained after each prime pointing, resulting in 7635 pairs of telescope pointings being used to capture the total IPHAS survey area \citep{Drew2005, Barentsen2014}.

The second IPHAS Data Release (DR2) was made available in 2014 \citep{Barentsen2014}. At this time the survey had achieved $5\sigma$ magnitude limits of $21.2\pm0.5$ in the $r'$ band, $20.0\pm0.3$ in the $i'$ band and $20.3\pm0.03$ in the {\Ha} band and had covered $>$90\% of the 1800 square degree footprint of the IPHAS survey.

\subsection{Planetary nebula source selection}
\label{SelCan}

The IPHAS survey observed a large number of PNe and PN candidates. We have utilised the Hong Kong/AAO/Strasbourg H$\alpha$ database\footnote{\url{http://202.189.117.101:8999/gpne/dbMainPage.php}} \citep[HASH;][]{Parker2016} to select nebulae classified as true, likely or possible PNe from the following sources:

\begin{enumerate}
\item \cite{ViironenJuly2009}: A list of candidate compact PNe from the IPHAS survey, of which we measured 4 that are listed in the HASH database as true or possible PNe. 
\item \cite{Sabin2014}: A catalogue of IPHAS-discovered potential PNe based on the spectroscopic and morphological characteristics of the nebulae \citep[for PN classification criteria see][]{Frew2010}. We measured 42 of these objects that are listed as true, likely or possible PNe by the HASH database, including 5 objects from the bow-shock nebula discovery paper of  \citet{Sabin2010}.
\item  We measured 151 PNe from the Strasbourg-ESO Catalogue \citep{Strasbourg1992} that had been observed by IPHAS and which are listed as true PNe by the HASH database.
\end{enumerate}  


    \label{PNsourceMap}

\subsection{{\Ha} aperture photometry measurements of PNe observed by IPHAS}

The reduced IPHAS {\Ha} images are available for download via the INT WFC Archive's Data Quality Control (DQC) query page, hosted by the Cambridge Astronomy Survey Unit (CASU)\footnote{\url{http://apm3.ast.cam.ac.uk/cgi-bin/wfs/dqc.cgi}}. All sources had multiple observations and all the observations for each source were downloaded, apart from those flagged as having problems. Column 3 of Tables A1 and A2 of Appendix~A lists the estimated angular dimensions of the nebulae. For nebulae less than $\sim$5~arcsec in diameter this was measured as their FWHM in the IPHAS H$\alpha$ filter images, then corrected in quadrature by the FWHM of nearby stars of similar brightness. For larger nebulae, the quoted angular dimensions were 
measured at 10\% of the peak nebular brightnesses in the IPHAS H$\alpha$ filter images.

In order to obtain integrated source counts,
aperture photometry was carried out using the Caltech Aperture Photometry Tool (APT)\footnote{\url{http://www.aperturephotometry.org/aptool/}} and the Starlink Graphical Astronomy and Image Analysis Tool (GAIA)\footnote{\url{http://www.starlink.rl.ac.uk/star/docs/sun214.htx/sun214.html/}}. 
The shape (circular/elliptical) and size of the APT aperture used for each PN was adjusted to best match the shape of the PN, as was the annular sky aperture, and the median sky-subtracted net source counts measured. For a small number of PNe in crowded fields for which there were bright stars within the PN aperture, we aimed to include similarly bright stars in the annular sky aperture in order to remove to first order the contribution from the contaminants. We utilised GAIA for those nebulae whose angular radii were too large to fit within the APT's radius limit of 200 pixels. For a few very extended low surface brightness nebulae, the IPHAS 120s H$\alpha$ exposures might not detect the full extent of their emission.

The source counts per second, obtained by dividing the net source counts by the observation exposure times, were converted to instrumental magnitudes and then to filter magnitudes by adding the zero-point filter magnitudes provided for each WFCAM exposure.

The {\Ha} filter magnitudes were converted to in-filter fluxes using a zero-magnitude flux calibration for the {\Ha} filter of 1.57$\times10^{-7}$ ergs~cm$^{-2}$~s$^{-1}$. This was obtained by multiplying the mean monochromatic flux for Vega in the {\Ha} filter of 1.81$\times10^{-9}$  ergs~cm$^{-2}$~s$^{-1}$~\AA $^{-1}$ \citep{Barentsen2014} by an equivalent width of 84.04~\AA\ for the {\Ha} filter that was obtained by integrating across the transmission profile of the filter as a function of wavelength. At the time of the IPHAS DR2 data release, Vega was adopted to have a magnitude of +0.035 at all optical wavelengths. Between DR2 and the most recent release (Greimel at al., in preparation), the  calibration changes show a 1$\sigma$ scatter of $\pm$0.03 magnitudes. 

In Appendix~A, our {\Ha} filter aperture photometry results are presented in column 5 of Table~A1, for the IPHAS-discovered PNe, while column 5 of Table~A2 presents our photometry for previously catalogued PNe observed by the IPHAS survey. A number of very bright PNe had saturated pixels in their {\Ha} images and were therefore omitted from our sample (e.g. NGC 7027, BD+30$^{\rm o}$3639, Vy~2-2).

\subsection{Correction of the {\Ha} filter fluxes for [N~{\sc ii}] contributions}
\label{NIICorrectionsection}

The [N~{\sc ii}] $\mathrm{\lambda6548.03\ \AA}$ and $\mathrm{\lambda6583.41\ \AA}$ doublet is situated on either side of the {\Ha} $\mathrm{\lambda6562.82\ \AA}$ line. The INT-WFC {\Ha} filter has an effective wavelength of 6568~\AA\ and a transmission FWHM of 95~\AA\ and its transmission at the wavelengths of the [N~{\sc ii}] lines is the same as at {\Ha} to within a few percent.

To correct the {\Ha} filter fluxes for [N~{\sc ii}] line contributions, spectroscopically measured [N~{\sc ii}]/{\Ha} flux ratios for each PN were available from the literature for 39 of the 46 IPHAS-discovered PNe and for all 151 of the previously catalogued PNe. If the published spectroscopic data provided the flux for only one line of the [N~{\sc ii}] doublet, we corrected for the contribution from both lines assuming an intrinsic 6583/6548 flux ratio of 3.0 \citep{storey2000}. Column 9 of Table~A1 and column 6 of Table~A2 list the ([N~{\sc ii}] 6583+6548)/{\Ha} ratios adopted from the literature, and their sources, while the next column lists log~{\Ha} line fluxes {\em after} removing the [N~{\sc ii}] line contributions. For nebulae where the differences between individual flux measurements were less than 0.01~dex, we set the final uncertainty on the mean flux to be 0.01~dex.

Most spectroscopic [N~{\sc ii}]/{\Ha} ratios are based on long-slit spectra. Since observed ratios can sometimes vary significantly with location in extended nebulae, we consider our adopted corrections to be more reliable for those nebulae having angular diameters of less than 5-6~arcsec. 

\begin{figure*}
\centering
\begin{subfigure}[b]{0.495\textwidth}
  \centering
  \includegraphics[width=\textwidth]{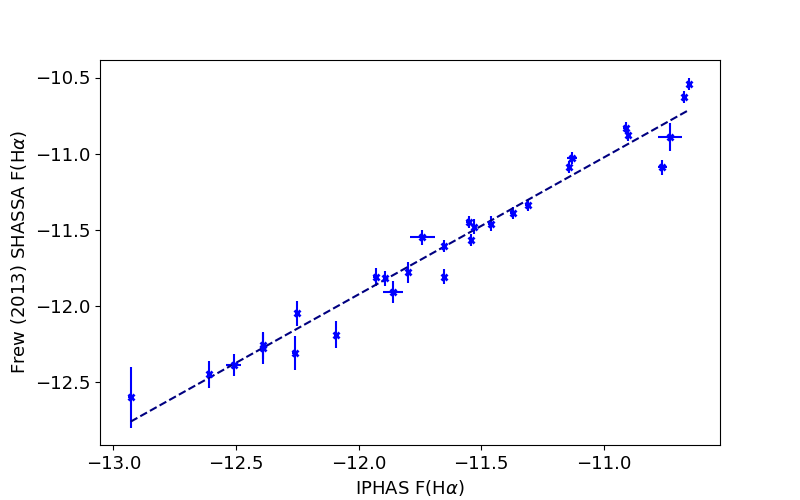}
  \subcaption{}
  \label{fig:SHASSA}
  \end{subfigure}
\begin{subfigure}[b]{0.495\textwidth}
  \centering
  \includegraphics[width=\textwidth]{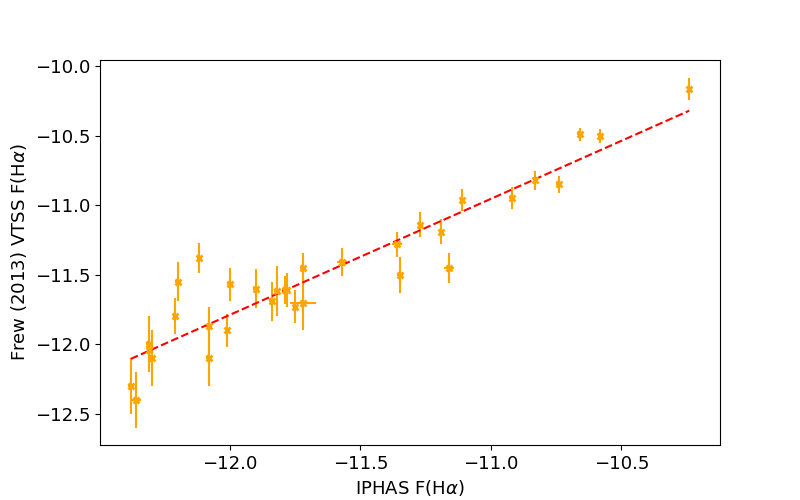}
  \subcaption{}
  \label{fig:VTSS}
  \end{subfigure}
  
  \caption{IPHAS {\Ha} log fluxes (in cgs units) compared to the  
  (a) SHASSA (left) and (b) VTSS (right) H$\alpha$ log fluxes of \citet{Frew2013}, all after correction for [N~{\sc ii}] contributions. The lines of best fit are shown as dashed lines, with slopes of $0.90 \pm 0.03$ and $0.84 \pm 0.06$ for the SHASSA and VTSS comparisons respectively.}
  \label{fig:IPHASHa_FrewHa}
\end{figure*}

Although the [N~{\sc ii}] lines are the major source {\bf of} contamination of the in-filter {\Ha} fluxes, the continuum from stars within the PN measurement aperture is another potential contaminant, although this is corrected to first order via the subtraction of the flux in the sky aperture. The underlying nebular continuum is another potential contributor, although for the case of the narrow-band WFC {\Ha} filter its contribution is estimated to be minor.
Using the {\sc nebcont} routine in the {\sc dipso} package \citep{Howarth2004} then, for a nebular electron density of 10$^3$~cm$^{-3}$ and 10$^4$~K with n(He$^+$)/n(H$^+$) = 0.10, one finds that for a H$\alpha$ flux towards the faint end of the H$\alpha$ flux distribution shown in Figure~\ref{fig:IPHASHa_FrewHa}, i.e. 10$^{-12}$ ergs~cm$^{-2}$~s$^{-1}$,
the in-filter nebular continuum flux in the 95-\AA\ FWHM IPHAS H$\alpha$ filter would contribute 1.9$\times10^{-14}$~ergs~cm$^{-2}$~s$^{-1}$, or 1.9 percent of the H$\alpha$ flux. The in-filter flux that would be contributed by a 17th-magnitude central star would be
less than 1 percent of the same nebular H$\alpha$ flux.

\subsection{Comparison with the {\Ha} fluxes of \citet{Frew2013}}

\cite{Frew2013} presented a catalogue of {\Ha} fluxes for 1258 Galactic PNe obtained using data from the Southern {\Ha} Sky Survey Atlas (SHASSA) \citep{Gaustad2001} and the Virginia Tech Spectral-line Survey (VTSS)\citep{VTech2014}\footnotemark\footnotetext{\url{http://www.phys.vt.edu/~halpha/}}. The VTSS survey used an f/1.2 camera lens with a
58-mm focal length, together with a 512$\times$512 CCD array that projected to 96~arcsec per pixel on the sky. Its integration times were not specified. Its
{\Ha} filter had a FWHM of 17.5~\AA , transmitting mainly the {\Ha} line. The SHASSA survey used an f/1.6 camera lens with a focal length of 52~mm with maximum exposure times of 1200~s onto a 1024$\times$1024 CCD array that projected to 48~arscec per pixel on the sky. Its {\Ha} filter had a FWHM of 32~\AA\ and transmissions of 39\%, 26\% and 78\% at the rest wavelengths of the [N~{\sc ii}] 6548, 6583~\AA\ and {\Ha} lines, respectively \citep{Gaustad2001}. The transmissions of these lines were equal for the 95-\AA\ FWHM {\Ha} filter used by IPHAS. Therefore for the same PN with the same [N~{\sc ii}]/{\Ha} ratio, the [N~{\sc ii}] contribution to the SHASSA {\Ha} filter flux measured by \citet{Frew2013} should be less than that made to the IPHAS WFC {\Ha}-filter flux measured here.

Figure~\ref{fig:IPHASHa_FrewHa}(a) plots our measured IPHAS {\Ha} fluxes against the SHASSA {\Ha} fluxes measured by \citet{Frew2013} for 29 nebulae in common, both after [N~{\sc ii}] corrections have been applied. The SHASSA fluxes are on average fainter than the IPHAS fluxes by 0.08$\pm$0.12 dex, consistent with no net difference within the uncertainty limits. The slope of the line of best fit is 0.90$\pm$0.03. 

A similar comparison of the [N~{\sc ii}]-corrected IPHAS {\Ha} fluxes with the \citet{Frew2013}
VTSS {\Ha} fluxes for 32 nebulae in common showed the VTSS {\Ha} fluxes to be 0.17$\pm0.19$ dex brighter than the IPHAS {\Ha} fluxes, again consistent with no net difference within the uncertainty limits, with the larger dispersion attributable to the larger uncertainties associated with the VTSS fluxes, see Fig.~\ref{fig:IPHASHa_FrewHa}(b).



\section{Nebular Extinctions}

\subsection{Determining dust extinctions to the nebulae}
\label{Extinction Calculations}

Since the radio free-free fluxes of PNe undergo virtually no extinction by interstellar dust, we can estimate absolute extinctions to PNe at optical wavelengths by comparing their measured hydrogen Balmer line fluxes with Balmer line fluxes predicted from their measured radio free-free fluxes using recombination theory. We have therefore used our measured \Ha\ fluxes, as well as \HB\ fluxes from the literature, together with literature radio free-free fluxes, to determine dust extinctions at the wavelengths of the two Balmer lines. 
The logarithmic extinctions at {\HB} ({\CHB}) were calculated using the formulation given by \citet{Milne1975}: 

\begin{equation}
    \centering
    \label{CHB}
    \begin{split}
    C_{H\beta} & = log_{10} \frac{3.28 \times 10^{-9} \times S(5 GHz) \times t^{-0.4}}{ln(9900 \times t^{1.5}) \times [1 + (1-x'')y + 3.7x''y]} \\
    &- log_{10} F(H\beta)
    \end{split}
\end{equation}

where ${S(5~GHz)}$ is the radio flux density at 5~GHz, in Jy, $t$ is the nebular electron temperature in units of $\mathrm{10^{4}\ \si{\kelvin}}$, $y$ is the abundance ratio of He to H nuclei, by number, and $x''$ is the abundance ratio of doubly ionized He to all He atoms. For nebulae where values of $t$, $x''$ and $y$ were not available in the literature, they were assumed to have values of 1, 0 and 0.11, respectively. F(\HB) is the observed integrated \HB~flux, listed in Table~\ref{Fluxsources} from the sources given in Appendix~\ref{Appendix:FluxSources}.

Logarithmic extinctions at {\Ha} ({\CHa}) were also calculated using this formulation, after multiplying the first term in equation (\ref{CHB}) by 2.85 to allow for a theoretical \Ha/\HB\ flux ratio of 2.85 for nebulae at an electron temperature of 10$^4$~K \citep{Storey1995} and replacing the second term by log(\Ha), the [N~{\sc ii}]-corrected IPHAS \Ha\ flux. 

As well as using 5~GHz radio flux densities taken from the literature, we also made use of flux densities measured at frequencies of 1.4~GHz and 30~GHz (listed in Table~\ref{Fluxsources} in Appendix~B). For the {\CHa} and {\CHB} extinction calculations, we scaled the 1.4~GHz and 30~GHz fluxes to 5~GHz by assuming a $\nu^{-0.1}$ optically thin free-free spectrum. 

We also calculated {\EHaHB} reddenings based on the ratio of the [N~{\sc ii}]-corrected IPHAS {\Ha} flux to the published {\HB} filter flux, using equation (\ref{EHaHB}) below to obtain the colour excess between the {\Ha} and {\HB} wave bands: 

\begin{equation}
\centering
\label{EHaHB}
E\left(H\alpha{}-H\beta{}\right)=2.5\ {log}_{10}{\left(\frac{F(H\alpha{})/F(H\beta{})\
}{2.85}\right)}
\end{equation}

Using this color excess we were then able to obtain the absolute visual extinction {\AvE} using equation (\ref{AvEHaHB}) below, which adopts the Galactic Reddening Law of \cite{Howarth1983}, with $R_{V} = A_{V}/E(B-V)$ = 3.10.

\begin{equation}
\centering
\label{AvEHaHB}
A_V\left[E\left(H\alpha{}-H\beta{}\right)\right]=2.669\ \times{}\
E(H\alpha{}-H\beta{}),
\end{equation}

where {\AvE}\ is given in magnitudes.
We also adopted the \cite{Howarth1983} reddening law for R$_{V}$ = 3.1 to calculate visual extinctions from {\CHa} ({\AvCHa}) and {\CHB} ({\AvCHB}), using equations (\ref{AvCHa}) and (\ref{AvCHB}):

\begin{equation}
\centering
\label{AvCHa}
{A_V\left(C_{H\alpha{}}\right)}=\ 3.140\times{}C_{H\alpha{}}
\end{equation}

\begin{equation}
\centering
\label{AvCHB}
{A_V\left(C_{H\beta{}}\right)}=\ 2.135\times{}C_{H\beta{}}
\end{equation}

Table~\ref{Av_EHa-HB} in Appendix~C presents our derived {\EHaHB} values for 37 PNe, together with the corresponding values of {\AV} and {\CHB}. Table~\ref{KnownPNeRadio} in Appendix~D presents the {\AV} values derived for 143 PNe from the various combinations of the three radio frequency fluxes and the {\Ha} and {\HB} fluxes.
In order to calculate the uncertainties on all {\AV} values, standard error propagation was used.
In some cases the literature fluxes did not have any listed uncertainties and in such cases {\AV} values obtained using those fluxes were discarded.
For each PN, using the {\AvCHa} and {\AvCHB} values corresponding to each of the radio frequencies, 1.4 GHz, 5 GHz and 30 GHz, we obtained a weighted mean radio extinction value, {\AV}(Radio), as listed in the penultimate column of Table~\ref{KnownPNeRadio}.

%
%
%

The derived extinctions showing the best inter-agreement (the maximum allowed variation between the {\AV} values selected was $\pm 0.9$ magnitudes)
are shown in bold script in Table~\ref{KnownPNeRadio} and these
were used for obtaining the final weighted averaged {\Av}(Radio) values.
In cases when there were discrepancies between the different radio-based {\AV} values that were significantly larger than the formal flux uncertainties, the {\AV} value corresponding to the highest radio frequency was adopted, on the grounds that the higher frequency free-free radio emission was more likely to be optically thin.

For 143 PNe, a weighted mean of {\AV}(Radio) and {\AvE} was used to generate
a final adopted {\AV} and its uncertainty, which are listed in columns 3 of Table~\ref{KnownPNe_Av_andDists}.

\subsection{Maximum expected extinctions along PN lines of sight}

To provide an upper limit to the extinction along the line of sight to  each PN, we utilised the NASA/IPAC Infrared Archive (IRSA) \textit{Galactic Dust Reddening and
 Extinction}\footnotemark\footnotetext{\url{http://irsa.ipac.caltech.edu/applications/DUST/}}
website. The site uses the data of \cite{Schlegel1998} who created Galactic dust temperature and dust column maps by combining results from the IRAS $\mathrm{100~\mu}$m and COBE/DIRBE surveys. Using these maps and a relation between $\mathrm{100~\mu}$m surface brightness and visual extinction they were able to infer total extinctions along any given line of sight out of the Galaxy. \cite{Schlafly2011} recalibrated the \cite{Schlegel1998} relations ({\Av}(Schlafly) = 0.86{\Av}(Schlegel)) and so we use the \cite{Schlafly2011} values to predict the maximum value of {\Av} expected along a line of sight. These values are presented in Fig.~2 and in Appendix~E.

\section{Extinction distances to Galactic Planetary Nebulae}

The technique that we adopt for determining distances to the northern Galactic PNe in our IPHAS sample is the 3D extinction mapping method, which uses an extinction versus distance relationship derived from field stars nearby on the sky to each PN in order to estimate its distance \citep[see][]{Lutz1973, Gathier1986, Giammanco2011}. 
In this paper we make use of three recent Galactic extinction versus distance mapping tools that are based on independent large stellar photometric databases. In Section 4.1 we
derive extinction distances using the IPHAS-based H-MEAD 3D extinction mapping algorithm \citep{Sale2014}. In Section 4.2 we present distances obtained using the Pan-STARRS~1/2MASS-based {\sc bayestar2019} 3D extinction mapping tool \citep{Green2019_Bayestars19}, while in Section 4.{\bf 3} we make a limited comparison with distances obtained using the Gaia/2MASS-based {\sc stilism} 3D reddening mapping tool \citep{Lallement2019}.
In Section 4.4 we compare distances obtained using each of the above tools with reliable Gaia DR2 distances that are available for 17 of the PNe in our sample.

\subsection{Distances using the H-MEAD 3D extinction mapping algorithm}

This method rests on photometric data from the IPHAS survey in building distance-extinction relationships for A--K stars across the northern Plane.
The source for the extinction-distance relations along the relevant lines of sight is the H-MEAD algorithm (Hierarchical Mapper of Extinction Against Distance) described by \cite{Sale2012} and implemented by \cite{Sale2014}. H-MEAD uses hierarchical Bayesian principles to determine the {\bf required} properties of all the available A--K stars within any specified square pencil beam (voxel) in order to derive a mean relation between increasing extinction and distance.



\begin{figure*}
\centering
\begin{subfigure}[b]{0.49\textwidth}
  \centering
  \includegraphics[width=\textwidth]{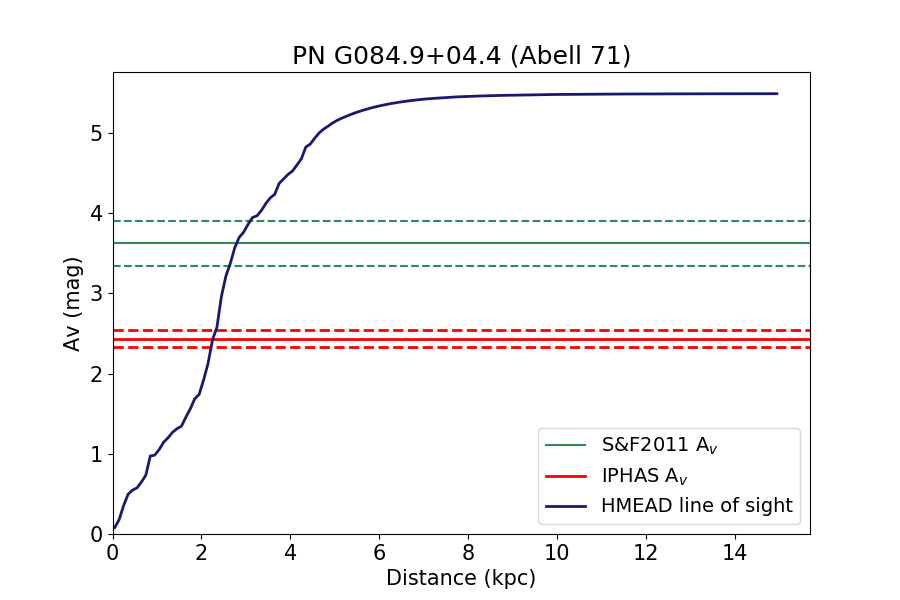}
  \subcaption{}
  \label{fig:Abell47}
  \end{subfigure}
\begin{subfigure}[b]{0.49\textwidth}
  \centering
  \includegraphics[width=\textwidth]{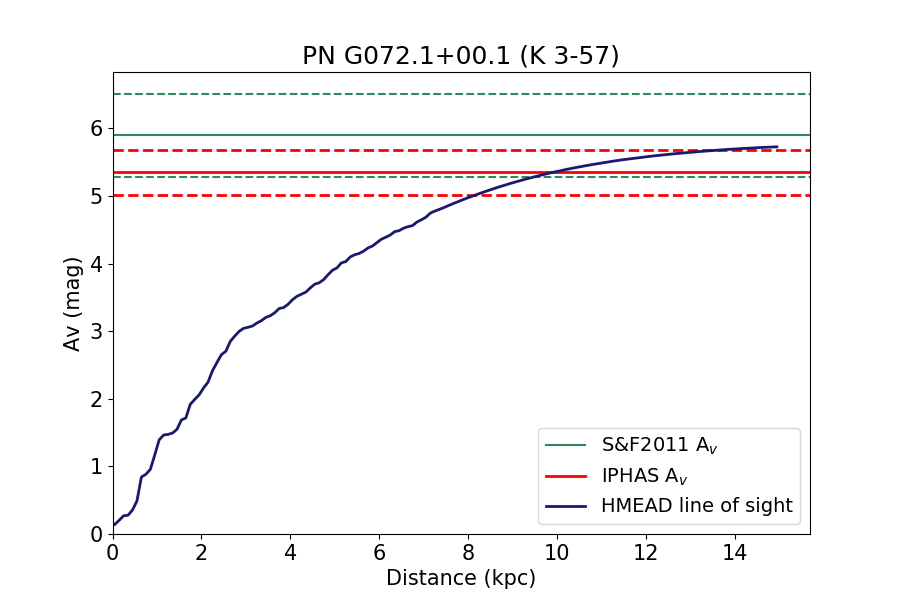}
  \subcaption{}
  \label{fig:K3-6}
  \end{subfigure}
  
\begin{subfigure}[b]{0.49\textwidth}
  \centering
  \includegraphics[width=\textwidth]{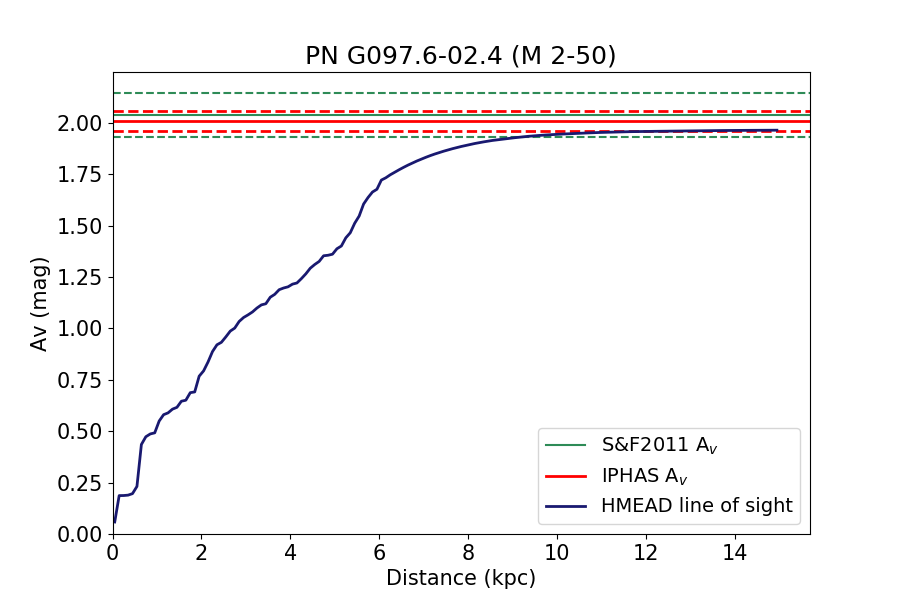}
  \subcaption{}
  \label{fig:HaTr10}
  \end{subfigure}
\begin{subfigure}[b]{0.49\textwidth}
  \centering
  \includegraphics[width=\textwidth]{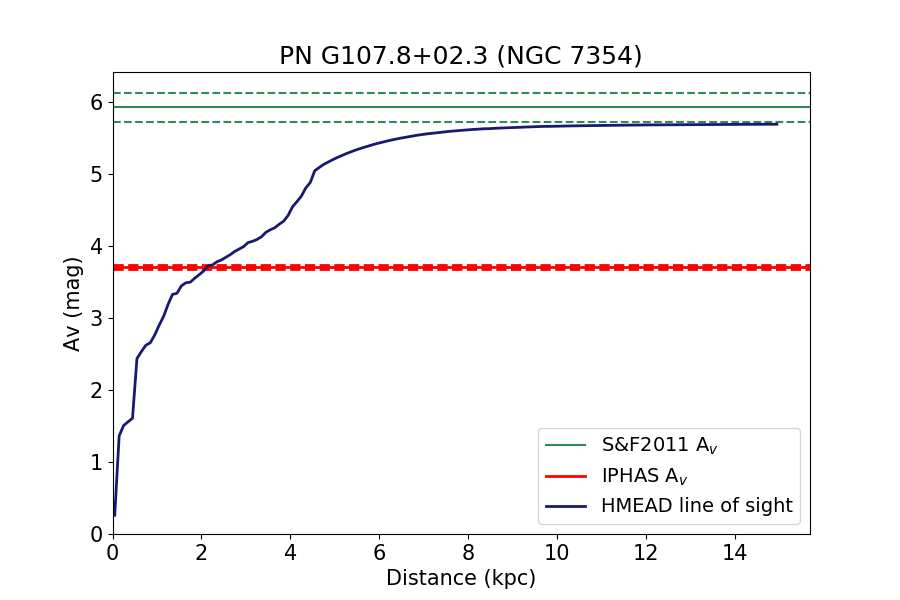}
  \subcaption{}
  \label{fig:PC20}
  \end{subfigure}

\caption{H-MEAD distance-extinction relationships for the sight lines to (a) Abell 71 ($l=85.00, b=+4.49$), (b) K 3-57 ($l=72.20, b=+0.10$), (c) M 2-50 ($l=97.68, b=-2.45$) and (d) NGC 7354 ($l=107.84, b=+2.32$). The solid blue curve in each plot represents the H-MEAD extinction versus distance relation for the sightline to the nebula.
The solid and dashed red horizontal lines correspond to the value of A$_{\rm V}$ derived for the nebula, and its corresponding uncertainties, while the solid and dashed green horizontal lines correspond to the \protect\cite{Schlafly2011} A$_{\rm V}$ limit and its uncertainties.}
  \label{fig:ext_mainBody_All}
\end{figure*}

To give a little more detail, \cite{Sale2014} used the IPHAS $(r'-i')$ and $(r'-H\alpha)$ colour-colour plane to assign (probabilistically) spectral type and luminosity class to every detected star in the voxel of interest. The algorithm rejects any objects that are not plausible A--K stars. This information constrains each accepted star's likely absolute magnitude and monochromatic extinction at 5495 \AA\ ($A_0$, a quantity that is for most purposes nearly identical to the band mean, $A_V$). These data, along with the IPHAS apparent magnitudes, are then used to calculate distances to all the retained stars, which are then dynamically-binned according to
their distances: each bin contains a minimum of eight stars and is at least $100~\si{\parsec}$ in depth. H-MEAD takes into account sources of inaccuracy, scatter and bias such as photometric error, unresolved ISM substructure within the voxel, binarity and photometric magnitude limits \citep[see also][]{Sale2009, Giammanco2011}.

The photometric depth of the IPHAS survey enabled the northern Galactic Plane to be extinction-mapped by \cite{Sale2014} with voxel sizes ranging from $5\times5$~arcmin$^2$ at the highest stellar densities up to, very occasionally, $60\times60$ arcmin$^2$ in the darkest regions. This variation is in response to the requirement of at least 200 stars to work with, within the voxel. The most frequent angular resolution is $10\times10$~arcmin$^2$. The useful distance range is usually from 1 up to 10~kpc, set by the bright and faint limits of the survey photometry, respectively.

The H-MEAD distance-extinction relationship curves from \cite{Sale2014} that we selected as best matching each of the PN sightlines were obtained from the website \url{http://www.iphas.org/data/extinction}.

Once the appropriate distance-extinction relation for the voxel enclosing the coordinates of a PN had been identified and downloaded, it was simply a matter of plotting it and reading off the distance corresponding to the final averaged {\AV} of the PN.  The uncertainty on this distance was
found by obtaining the distances corresponding to the upper and lower limits for {\AV}. We were able to do this for 143 PNe - examples are shown in
Figure~2.
The distance--extinction relationships for all the sightlines analyzed are presented in 
Appendix~E.
The cut-off distance in all of them is 14~kpc.

The distance-extinction curves typically begin with a sharp increase in {\Av} and end as a plateau with very little subsequent increase in {\Av} with distance.  A sharp jump in extinction over a small distance range will be due to the presence of a dense interstellar cloud extinguishing the starlight. There are two reasons for the terminal plateaus commonly seen. The first is that the dust layer is physically thin (100 pc or so above/below the Galactic equatorial plane) and is exited more quickly for sightlines at higher Galactic latitude. 
Lines of sight at high Galactic latitudes tend to run out of after a few $\si{k\parsec}$. This can be seen for the PN Abell~71 ($b=+4.49\si{\degree}$) whose distance-extinction relationship is shown in Figure~2.
Its curve reaches a plateau at a heliocentric distance of $\sim4\ \si{k\parsec}$. The second factor that can be at work, particularly at low Galactic latitudes within the Solar Circle, is high extinction limiting the detected stars to within a short distance -- forcing H-MEAD to rely on the prior in extending the curve further out. In such cases the prior can impose the plateau \citep[see][]{Sale2014}.
The distance-extinction relationship for K~3-57 ($l=72.20\si{\degree}$, $b=00.10\si{\degree}$), shown in Figure~2, is a good example of this.

\input{Tables/KnownPNe_Av_and_HMEAD_Dists}

It is unavoidable that this method of distance estimation does not always deliver a useful result. For some PNe the derived value of {\Av} is above the maximum {\Av} of the plateau region of the relevant extinction-distance curve but below the {\Av}(Schlafly) value for the sightline. Where this happens, it is likely that more heavily reddened field stars along the line of sight were fainter than the IPHAS survey
limits.  In cases of PNe with {\Av} falling in the plateau region of the extinction-distance relationship, the uncertainties on their estimated distances will be much larger, as shown by the relationship for M~2-50 in Figure~2. Seventeen PNe were found to have {\Av} values larger than the maximum {\AV} for their H-MEAD extinction-distance curve as well as lying above the uncertainty limits for the corresponding {\Av}(Schlafly) values. Five of those 17 PNe are located between $l$=184 and $l$=202 degrees. In some of these cases this could be due to
overestimated extinction values due to errors in the  adopted radio fluxes or in the adopted [N~{\sc ii}]/H$\alpha$ correction factors. Another possibility is that there might be significant variations of {\Av} within some of the $10'\times10'$~arcmin voxels used to sample the sight lines. We consider it unlikely that significant internal dust extinction within the PNe themselves is the cause of such discrepancies. There are only a few confirmed cases of PNe with large internal dust extinctions (e.g. NGC 7027, NGC 6302) - typical bright PNe such as NGC 7009 have been found to have relatively small internal dust columns \citep{Walsh2016}.

The extinction-distance relationship for NGC~7354 shown in Figure~2 is an example of a well-behaved curve, where the calculated {\Av} for the PN lies on the sharply increasing portion of the curve, and where the {\Av}(Schlafly) value lies above the {\Av} limit corresponding to the plateau region of the extinction-distance curve.

The measured distances to all 143 PNe are listed in column 4 of Table~1. Table~\ref{KnownPNe_Av_andDists}. For PNe which had calculated {\Av} values greater than the H-MEAD curve's maximum {\Av} limit, they were assigned a minimum distance corresponding to just after the onset of the final plateau in the H-MEAD extinction curve.

\subsection{Distances using the {\sc bayestar2019} 3D extinction mapping algorithm}

\citet{Green2019_Bayestars19} have constructed a 3D dust reddening map for the sky north of declination -30$^{\rm o}$. By making use of Pan-STARRS~1 optical photometry, 2MASS near-infrared photometry and Gaia DR2 parallaxes and applying a hierarchical Bayesian
model ({\sc bayestar2019}), they inferred distances, reddenings and types for 800 million stars in order to create a 3D dust map extending beyond several kpc, which can be accessed at doi:10.7910/DVN/2EJ9TX. We made use of this map to estimate distances for the PNe in our samples, using reddening values corresponding to the extinctions listed in
Table~\ref{KnownPNe_Av_andDists}. The resulting distance estimates, or lower limits, are listed in the final column of the same table. As was the case with H-MEAD distance estimates, where PNe had calculated {\Av} values that were greater than the maximum {\Av} limits of the {\sc bayestar2019} curves, they were assigned a minimum distance corresponding to just after the onset of the final plateau in the extinction curves.

\subsection{Distances using the {\sc stilism} 3D extinction mapping algorithm}

We have also used the {\sc stilism}
3D reddening mapping algorithm of \citet{Lallement2019} (see \url{https://astro.acri-st.fr/gaia\_dev/}),  to estimate reddening distances to PNe in our sample. Because the algorithm makes use of stars having 5$\sigma$ or better Gaia DR2 parallaxes, for most directions this tool does not reach beyond distances of $\sim$3~kpc and so provides only lower limits to the distances to the majority of the PNe in our sample. However,
{\sc stilism} distances, or lower limits, are listed in column 8 of Table~2 for 17 of the PNe in our sample that are judged to have reliable Gaia DR2 distances.

\subsection{Comparison with Gaia DR2 distances}

For a subset of the PNe in our sample that were judged to have reliable central star distances from Gaia DR2 parallaxes, we can compare these reliable astrometric distances with the distances estimated using the different reddening mapping methods.  

The following criteria were used to select the PNe in our sample as having reliable central star parallaxes in the Gaia DR2 archive \citep[][]{Brown2018}: (a) the number of effective Gaia visits, N$_{\rm per}$ $\geq$ 8; (b) a parallax signal-to-noise ratio $> 3.4$; (c) a renormalised unit-weighted error $u/u_0$(G, B$_{\rm P}$ - R$_{\rm P}$) $<$1.4, where  $u = \sqrt{\chi^2/(N-5)}$ and $u_0$(G, B$_{\rm P}$ - R$_{\rm P}$) is the magnitude and colour-dependent reference value.\footnote{See Gaia mission document GAIA-C3-TN-LU-LL-124-01 by L. Lindegren at \url{www.rssd.esa.int/doc_fetch.php?id=3757412}} 17 of the PNe in our sample satisfied these criteria. 
Table~\ref{KnownPNe_GaiaDists} lists Gaia DR2 parallaxes\footnote{As recommended by \cite{Lindegren2018}, the published Gaia-DR2 parallaxes have been increased by 0.03~mas to allow for the global zero-point of -0.03~mas.} in column~3, followed in column~4 by the distances obtained by inverting the parallaxes and in column~5 by Gaia DR2 distances from \citet{Bailer-Jones2018} that are based on a weak distance prior
and which are often adopted for stars with lower signal-to-noise Gaia DR2 parallaxes. For the sample of 17 PNe listed in Table~\ref{KnownPNe_GaiaDists}, the Gaia DR2 parallaxes are of sufficiently high reliability that the \citet{Bailer-Jones2018} distances agree closely with the inverted parallax distances. One of the PNe listed in Table~\ref{KnownPNe_GaiaDists} has a pre-Gaia parallax measurement. \citet{Benedict2009} reported a parallax of 2.47$\pm$0.16 mas for the central star of NGC~6853, versus the Gaia DR2 parallax of 2.658$\pm$0.044 mas.

\input{Tables/KnownPNe_GaiaDist.tex}

\begin{figure*}
    \centering
    \includegraphics[width=\textwidth]{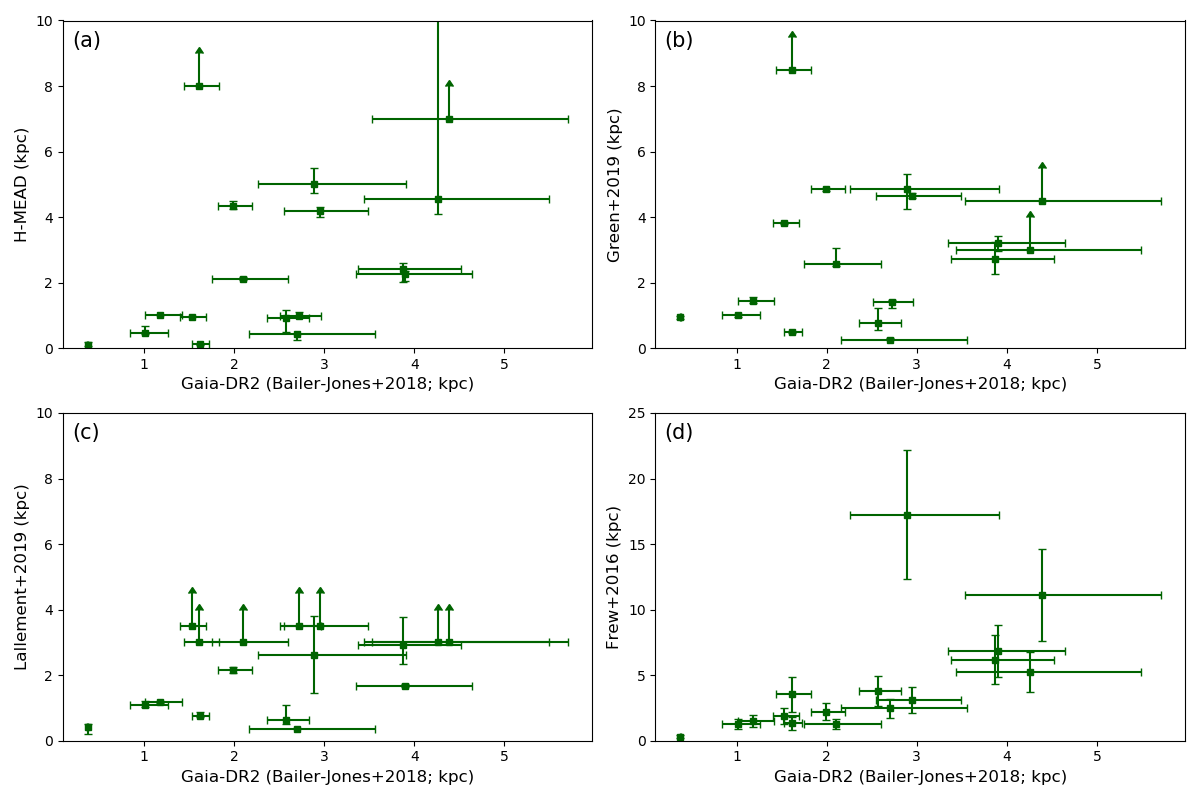}
    \caption{Correlations between Gaia-DR2 distances from \protect\cite{Bailer-Jones2018} and those from the following methods. \textit{Top left}: (a) H-MEAD \protect\citep{Sale2014}, with a Pearson correlation coefficient of r = 0.58 (15 PNe); \textit{top right}: (b) \protect\cite{Green2019_Bayestars19}, with r = 0.35 (14 PNe); \textit{bottom left}: (c) \protect\cite{Lallement2019}, with r = 0.57 (10 PNe); and \textit{bottom left}: (d) \protect\cite{Frew2016}, with r = 0.61 (16 PNe). Upward-pointing arrows indicate lower limits, which were not used in calculating correlation coefficients.
    Note that the y-axis scale for plot (d) differs from those of the other three plots.}
    \label{fig:DistComp}
\end{figure*} 

For comparison with the above Gaia DR2 PN central star distances, column~6 of Table~\ref{KnownPNe_GaiaDists} lists the PN distances obtained using the H-MEAD extinction mapping method of \cite{Sale2014}, while column~7 lists the distances obtained using the {\sc Bayestar2019} extinction mapping method of \cite{Green2019_Bayestars19}.  Column~8 of Table~\ref{KnownPNe_GaiaDists} provides \cite{Lallement2019} {\sc stilism} distances for 10 of the 17 PNe that have reliable Gaia distances. Finally, column~9 of Table~\ref{KnownPNe_GaiaDists} lists PN distances from \cite{Frew2016}, a widely used statistical distance estimator that is based on a H$\alpha$ surface brightness-radius relation for PNe.

Figure~\ref{fig:DistComp} plots the Gaia DR2 distances of \cite{Bailer-Jones2018} against the distance estimates obtained from (a) H-MEAD \citep{Sale2014}; (b) \cite{Green2019_Bayestars19}; (c) \cite{Lallement2019}; and (d) \cite{Frew2016}. The Pearson correlation coefficient (r) and the corresponding probability value (p) for a null correlation have been calculated for each plot (lower limits to distances were not used). In descending order of their correlation with the Gaia DR2 distances are the \cite{Frew2016} distances (r=0.61, p=0.012; 16 PNe), the H-MEAD distances (r=0.58, p=0.023; 15 PNe), the \cite{Lallement2019} distances (r=0.57, p=0.085; for 10 PNe at distances $<$ 3~kpc), with the \cite{Green2019_Bayestars19} distances yielding r=0.35 and a p value of 0.22 for 14 PNe.

\cite{Giammanco2011} used the MEAD reddening-distance algorithm of \cite{Sale2009} to estimate reddening distances for 70 PNe located in IPHAS fields, using reddening estimates that were based mainly on spectroscopic H$\alpha$/H$\beta$ ratios taken from the literature. Only seven of the eighteen PNe in Table~\ref{KnownPNe_GaiaDists} have distance estimates from \cite{Giammanco2011} that are not upper or lower limits, so we have not estimated their correlation coefficient with the Gaia DR2 distances.

\section{Discussion and Conclusions}

Using aperture photometry we have measured in-band IPHAS H$\alpha$ filter fluxes for 46 PNe discovered by IPHAS and for 151 previously catalogued PNe. These in-band fluxes were corrected for [N~{\sc ii}] 6548,6583~\AA\ contributions using published spectroscopic [N~{\sc ii}]/H$\alpha$ ratios for 39 of the IPHAS-discovered PNe and for all 151 of the previously catalogued PNe. Radio free-free flux measurements at 1.4, 5 or 30~GHz were available from the literature for 143 of the previously catalogued PNe and these were combined with the [N~{\sc ii}]-corrected H$\alpha$ fluxes and with previously published H$\beta$ filter fluxes to give a weighted mean extinction estimate for each PN from its radio free-free to optical Balmer line flux ratios.

We have used these derived interstellar extinctions with the IPHAS-based H-MEAD interstellar extinction mapping tool of \cite{Sale2014} in order to estimate distances to each PN from the H-MEAD extinction vs. distance plot for each sight line. For comparison purposes, we have also estimated extinction distances to the PNe using the \cite{Green2019_Bayestars19} extinction mapping tool, which is based on Pan-STARRS~1 and 2MASS photometry and Gaia parallaxes. Gaia DR2 distances that were judged to be reliable were available for 17 of the PNe in our sample and these distances were compared to the H-MEAD and Green et al. extinction distances for this sub-sample, as well as to a smaller number of extinction distances obtained using the \cite{Lallement2019} extinction mapping tool. We also compared the Gaia DR2 distances for this sub-sample with the widely used PN distances from \cite{Frew2016} that are based on a relation between nebular radius and nebular H$\alpha$ surface brightness. For the sub-sample of 17 PNe, the Frew et al. and the H-MEAD distances were found to show similar degrees of correlation with Gaia DR2 distances.

Given that 3D extinction mapping tools are expected to become increasingly sophisticated and to achieve higher angular resolution over time, the weakest link may often be the accuracy of reddening estimates, which need to be significantly improved for many objects. In the case of PNe, reddening values based on observed Balmer line ratios cover too small a wavelength range to be considered ideal. Radio free-free to optical Balmer line ratios cover a much larger wavelength baseline and yield total dust extinctions at the wavelengths of the Balmer lines, but currently many PNe either have no measured radio fluxes or have radio fluxes with relatively low signal to noise ratios. More accurate radio flux measurements, over a range of radio wavelengths, are needed. For the northern PNe with IPHAS H$\alpha$ fluxes
that have been discussed here, the ongoing Very Large Array Sky Survey \citep[VLASS;][]{Lacy2020} will 
help reach this goal, by providing 2-4~GHz radio fluxes at an angular resolution of 2.5~arcsec and with a 1$\sigma$ sensitivity of 0.07~mJy by the completion of the survey.

\section*{Data availability}

Calibrated IPHAS images can be obtained
from \url{http://apm3.ast.cam.ac.uk/cgi-bin/wfs/dqc.cgi}. H-MEAD distance-extinction data for fields within the IPHAS footprint can be obtained from \url{http://www.iphas.org/data/extinction}.

\section*{Acknowledgements}

TD's research has been supported under grants MOST104-2628-M-001-004-MY3 and MOST107-2119-M-001-031-MY3 from the Ministry of Science and Technology of Taiwan, and grant  AS-IA-106-M03 from Academia Sinica. Her work was also partially funded by the Sonderforschungsbereich SFB881 ``The Milky Way System'' of the German Research Foundation (DFG). MB and RW acknowledge support from 
European Research Council grant
SNDUST ERC-2015-AdG-694520. DJ and AM acknowledge support from the State Research Agency (AEI) of the Spanish Ministry of Science, Innovation and Universities (MCIU) and the European
Regional Development Fund (FEDER) under grant AYA2017-83383-P.


\bibliographystyle{mnras}
\bibliography{IPHAS_Bib}

\clearpage
\input{Appendix.tex}

\end{document}

%% file: Tables/KnownPNe_Av_and_HMEAD_Dists.tex

\begin{table*}
\centering 
\caption{Adopted extinctions and distances derived from \protect\cite{Sale2014} H-MEAD and \protect\cite{Green2019_Bayestars19}}  
\label{KnownPNe_Av_andDists} 
\begin{tabular}{lllll}
\hline 
\\
PN G & Name & A$_{\mathrm V}$ (mag) & H-MEAD (kpc) & Green+2019 (kpc)\\ 
\\
\hline

030.8+03.4 & Abell 47 & 8.27 $\pm$ 0.33 & $ > 10.5$ & $ > 5.2 $ \\
031.0+04.1 & K 3-6 & 7.18 $\pm$ 0.17 & $ 5.30 ^{+ 0.50 }_{- 0.26 }$ & $ > 4.5 $ \\
031.3-00.5 & HaTr 10 & 3.26 $\pm$ 0.19 & $ 0.94 ^{+ 0.19 }_{- 0.11 }$ & $ 2.33 ^{+ 0.74 }_{- 0.50 } $ \\
031.7+01.7 & PC 20 & 5.45 $\pm$ 0.10 & $ 4.57 ^{+ 0.33 }_{- 0.33 }$ & $ 12.6 ^{+ 2.31}_{- 4.47 } $ \\
032.0-03.0 & K 3-18 & 3.93	$\pm$ 0.18 & $6.00 ^{+ 0.40 }_{- 0.49 }$ & $>3.42$ \\
032.5-03.2 & K 3-20 & 2.13 $\pm$ 0.19 & $ 2.45 ^{+ 0.77 }_{- 0.56 }$ & $ 3.63 ^{+ 0.50 }_{- 0.59 } $ \\

032.9-02.8 & K 3-19 & 3.40 $\pm$ 0.10 & $ 6.35 ^{+ 0.29 }_{- 0.36 }$ & $ >6.63 $ \\

033.8-02.6 & NGC 6741 & 1.89 $\pm$ 0.03 & $ 2.44 ^{+ 0.07 }_{- 0.06 }$ & $ 3.58 ^{+ 0.03 }_{- 0.03 } $ \\

035.7-05.0 & K 3-26 & 2.12 $\pm$ 0.16 & $ 7.11 ^{+ 6.27 }_{- 0.74 }$ & $ >5.2 $ \\

035.9-01.1 & Sh 2-71 & 1.44 $\pm$ 0.13 & $ 0.12 ^{+ 0.01 }_{- 0.01 }$ & $ 0.49 ^{+ 0.02 }_{- 0.02 } $ \\
036.9-02.6 & HaTr 13 & 2.82 $\pm$ 0.25 & $ 2.58 ^{+ 0.99 }_{- 0.65 }$ & $ 3.90 ^{+ 5.60 }_{- 0.77 } $ \\
036.9-01.1 & HaTr 11 & 5.00 $\pm$ 0.11 & $ 8.35 ^{+ 0.44 }_{- 0.33 }$ & $ > 8.3 $ \\
039.5-02.7 & M 2-47 & 3.08 $\pm$ 0.10 & $ 2.05 ^{+ 0.20 }_{- 0.22 }$ & $ > 3.6 $ \\
039.8+02.1 & K 3-17 & 7.93 $\pm$ 0.09 & $ 5.42 ^{+ 0.30 }_{- 0.27 }$ & $ > 6.3 $ \\
040.3-00.4 & Abell 53 & 4.12 $\pm$ 0.11 & $ 0.87 ^{+ 0.08 }_{- 0.05 }$ & $ 0.86 ^{+ 0.01 }_{- 0.01 } $ \\
040.4-03.1 & K 3-30 & 2.62 $\pm$ 0.11 & $ 1.99 ^{+ 0.23 }_{- 0.49 }$ & $ 3.99 ^{+ 0.74 }_{- 0.08 } $ \\

041.8+04.4 & K 3-15 & 2.41	$\pm$ 0.08 & $> 9.00$ & $>4.83$ \\

043.0-03.0 & M 4-14 & 2.72 $\pm$ 0.06 & $ 4.44 ^{+ 0.44 }_{- 0.43 }$ & $ > 6.2 $ \\
043.1+03.8 & M 1-65 & 1.74 $\pm$ 0.05 & $ 2.25 ^{+ 0.11 }_{- 0.19 }$ & $ 3.22 ^{+ 0.19 }_{- 0.25 } $ \\
043.3+02.2 & PM 1-276 & 3.26 $\pm$ 0.11 & $ 1.99 ^{+ 0.20 }_{- 0.33 }$ & $ 3.03 ^{+ 0.29 }_{- 0.07 } $ \\
045.9-01.9 & K 3-33 & 5.30 $\pm$ 0.05 & $ 3.21 ^{+ 0.09 }_{- 0.18 }$ & $ > 16.9 $ \\
046.3-03.1 & PB 9 & 3.39 $\pm$ 0.06 & $ 6.83 ^{+ 0.18 }_{- 0.12 }$ & $ > 4.3 $ \\
046.4-04.1 & NGC 6803 & 1.24 $\pm$ 0.03 & $ 4.25 ^{+ 0.14 }_{- 0.25 }$ & $ 3.65 ^{+ 0.45 }_{- 0.23 } $ \\
046.8+02.9 & CTSS 4 & 3.03 $\pm$ 0.19 & $ > 9.0 $ & $ > 4.3 $ \\
047.1+04.1 & K 3-21 & 2.77 $\pm$ 0.19 & $ 7.32 ^{+ 6.68 }_{- 0.65 }$ & $ > 4.3 $ \\
048.0-02.3 & PB 10 & 3.59 $\pm$ 0.11 & $ 6.73 ^{+ 0.29 }_{- 0.23 }$ & $ > 8.8 $ \\

048.1+01.1 & K 3-29 & 6.89 $\pm$ 0.05 & $ > 10.0 $ & $ > 9.3 $ \\

048.5+04.2 & K 4-16 & 2.92 $\pm$ 0.06 & $ > 7.0 $ & $ > 4.5 $ \\
048.7+02.3 & K 3-24 & 6.07 $\pm$ 0.19 & $ > 8.0 $ & $ > 2.5 $ \\
048.7+01.9 & Hen 2-429 & 3.50 $\pm$ 0.10 & $ 6.04 ^{+ 0.47 }_{- 0.63 }$ & $ 4.67 ^{+ 1.76 }_{- 0.59 } $ \\
049.4+02.4 & Hen 2-428 & 2.86 $\pm$ 0.09 & $ 2.90 ^{+ 0.36 }_{- 0.30 }$ & $ 3.84 ^{+ 0.05 }_{- 0.09 } $ \\
050.4-01.6 & K 4-28 & 5.15 $\pm$ 0.06 & $ 5.74 ^{+ 0.21 }_{- 0.11 }$ & $ 6.75 ^{+ 0.04 }_{- 0.06 } $ \\
051.0-04.5 & PC 22 & 2.10 $\pm$ 0.15 & $ > 7.0 $ & $ > 5.6 $ \\
051.0+03.0 & Hen 2-430 & 3.56 $\pm$ 0.17 & $ 2.71 ^{+ 0.59 }_{- 0.44 }$ & $ 3.14 ^{+ 2.13 }_{- 0.06 } $ \\
051.3+01.8 & PM 1-295 & 3.56 $\pm$ 0.11 & $ 0.97 ^{+ 0.15 }_{- 0.02 }$ & $ 1.40 ^{+ 0.08 }_{- 0.16 } $ \\
051.9-03.8 & M 1-73 & 1.57 $\pm$ 0.10 & $ 3.46 ^{+ 0.66 }_{- 0.64 }$ & $ 4.45 ^{+ 0.66 }_{- 0.72 } $ \\
052.2-04.0 & M 1-74 & 1.58 $\pm$ 0.12 & $ 6.87 ^{+ 0.63 }_{- 0.87 }$ & $ 7.21 ^{+ 1.81} $ \\
052.5-02.9 & Me 1-1 & 1.15 $\pm$ 0.09 & $ 2.42 ^{+ 0.18 }_{- 0.40 }$ & $ 2.73 ^{+ 0.52 }_{- 0.46 } $ \\
052.9-02.7 & K 3-41 & 1.44 $\pm$ 0.30 & $ 2.74 ^{+ 0.46 }_{- 0.76 }$ & $ 2.74 ^{+ 0.49 }_{- 1.07 } $ \\
052.9+02.7 & K 3-31 & 4.61 $\pm$ 0.08 & $ 4.46 ^{+ 0.21 }_{- 0.23 }$ & $ > 2.5 $ \\
053.2-01.5 & K 3-38 & 4.56 $\pm$ 0.08 & $ 5.88 ^{+ 0.17 }_{- 0.07 }$ & $ 9.74 ^{+ 0.04 }_{- 0.04 } $ \\
053.8-03.0 & Abell 63 & 0.67 $\pm$ 0.20 & $ 0.92 ^{+ 0.25 }_{- 0.41 }$ & $ 0.76 ^{+ 0.47 }_{- 0.19 } $ \\
054.2-03.4 & Necklace & 2.41 $\pm$ 0.19 & $ > 8.0 $ & $ > 4.3 $ \\
054.4-02.5 & M 1-72 & 2.03 $\pm$ 0.32 & $ 6.82 ^{+ 3.12 }_{- 1.74 }$ & $ > 5.5 $ \\
055.1-01.8 & K 3-43 & 3.48 $\pm$ 0.27 & $ 6.65 ^{+ 0.65 }_{- 0.80 }$ & $ > 4.1 $ \\
055.2+02.8 & Hen 2-432 & 3.40 $\pm$ 0.08 & $ 2.50 ^{+ 0.31 }_{- 0.35 }$ & $ > 3.1 $ \\
055.3+02.7 & Hen 1-1 & 3.85 $\pm$ 0.07 & $ 4.15 ^{+ 0.20 }_{- 0.38 }$ & $ > 7.8 $ \\
055.5-00.5 & M 1-71 & 4.39 $\pm$ 0.13 & $ 7.01 ^{+ 0.19 }_{- 0.23 }$ & $ 9.55 ^{+ 0.02 }_{- 0.02 } $ \\
055.6+02.1 & Hen 1-2 & 2.78 $\pm$ 0.07 & $ 0.28 ^{+ 1.11 }_{- 0.02 }$ & $ 2.50 ^{+ 0.23 }_{- 0.20 } $ \\
056.0+02.0 & K 3-35 & 6.63 $\pm$ 0.17 & $ 9.96 ^{+ 4.38 }_{- 1.21 }$ & $ > 10.9 $ \\
056.4-00.9 & K 3-42 & 5.24 $\pm$ 0.07 & $ 8.38 ^{+ 0.20 }_{- 0.19 }$ & $ 10.5 ^{+ 0.03 }_{- 0.03 } $ \\

057.9-01.5 & Hen 2-447 & 4.42 $\pm$ 0.05 & $ 6.77 ^{+ 0.05 }_{- 0.21 }$ & $ >8.0 $ \\

058.9+01.3 & K 3-40 & 3.00 $\pm$ 0.08 & $ 5.43 ^{+ 0.15 }_{- 0.17 }$ & $ 6.64 ^{+ 0.16 }_{- 0.64 } $ \\
059.4+02.3 & K 3-37 & 4.33 $\pm$ 0.07 & $ 7.20 ^{+ 0.25 }_{- 0.21 }$ & $ > 7.5 $ \\
059.0+04.6 & K 3-34 & 2.57 $\pm$ 0.29 & $ > 7.0 $ & $ > 4.7 $ \\

059.9+02.0 & K 3-39 & 4.41 $\pm$ 0.07 & $ 7.76 ^{+ 0.26 }_{- 0.23 }$ & $ > 7.0 $ \\

060.4+01.5 & PM 1-310 & 4.96 $\pm$ 0.27 & $ > 8.0 $ & $ > 5.8 $ \\
060.5-00.3 & K 3-45 & 4.91 $\pm$ 0.19 & $ 5.33 ^{+ 0.56 }_{- 0.65 }$ & $ 10.1 ^{+ 1.22 }_{- 0.45 } $ \\

060.5+01.8 & Hen 2-440 & 3.93 $\pm$ 0.05 & $ 8.47 ^{+ 0.28 }_{- 0.25 }$ & $ 11.1 ^{+ 0.04 }_{- 0.04 } $ \\

060.8-03.6 & NGC 6853 & 0.10 $\pm$ 0.14 & $ 0.11 ^{+ 0.07 }_{- 0.04 }$ & $ 0.96 ^{+ 0.02 }_{- 0.08 } $ \\

\bottomrule 
\end{tabular}
\end{table*}

\begin{table*}
\centering 
\contcaption{Adopted extinctions and distances derived from \protect\cite{Sale2014} H-MEAD and \protect\cite{Green2019_Bayestars19}}
\begin{tabular}{lllll}

\hline 
\\
PN G & Name & A$_{\mathrm V}$ (mag) & H-MEAD (kpc) & Green+2019 (kpc)\\ 
\\
\hline

062.4-00.2 & M 2-48 & 3.45 $\pm$ 0.15 & $ 2.75 ^{+ 0.29 }_{- 0.55 }$ & $ 3.30 ^{+ 0.06 }_{- 0.03 } $ \\
063.8-03.3 & K 3-54 & 4.82 $\pm$ 0.22 & $ 5.30 ^{+ 0.49 }_{- 0.21 }$ & $ > 4.3 $ \\
064.9-02.1 & K 3-53 & 2.91 $\pm$ 1.02 & $ 3.88 ^{+ 2.84 }_{- 2.57 }$ & $ 3.27 ^{+ 11.7 }_{- 2.21 } $ \\
065.1-03.5 & We 1-9 & 3.16 $\pm$ 1.02 & $ 1.34 ^{+ 3.48 }_{- 0.62 }$ & $ 1.60 ^{+ 13.3 }_{- 0.58 } $ \\
065.9+00.5 & NGC 6842 & 2.51 $\pm$ 0.11 & $ 4.34 ^{+ 0.16 }_{- 0.10 }$ & $ 4.85 ^{+ 0.04 }_{- 0.04 } $ \\
066.9+02.2 & K 4-37 & 4.25 $\pm$ 0.19 & $ > 7.0 $ & $ > 8.8 $ \\
067.9-00.2 & K 3-52 & 8.46 $\pm$ 0.18 & $ > 8.0 $ & $ > 7.8 $ \\

068.3-02.7 & Hen 2-459 & 4.61 $\pm$ 0.05 & $ 5.44 ^{+ 0.09 }_{- 0.04 }$ & $ > 5.0 $ \\

068.6+01.1 & Hen 1-4 & 2.36 $\pm$ 0.11 & $ 3.40 ^{+ 0.66 }_{- 0.31 }$ & $ 3.10 ^{+ 0.25 }_{- 0.07 } $ \\
068.7+01.9 & K 4-41 & 3.72 $\pm$ 0.32 & $ > 8.5 $ & $ 11.5 ^{+ 0.19 }_{- 0.20 } $ \\
068.7+03.0 & PC 23 & 3.66 $\pm$ 0.21 & $ > 7.0 $ & $ > 6.7 $ \\
068.8-00.0 & M 1-75 & 4.20 $\pm$ 0.19 & $ 6.20 ^{+ 0.25 }_{- 0.67 }$ & $ 7.75 ^{+ 1.16 }_{- 4.53 } $ \\
069.2+03.8 & K 3-46 & 2.15 $\pm$ 0.26 & $ 4.73 ^{+ 1.49 }_{- 1.80 }$ & $ 5.92 ^{+ 9.03 }_{- 2.50 } $ \\

069.2+02.8 & K 3-49 & 3.37 $\pm$ 0.07 & $ > 8.0 $ & $ > 3.5 $ \\

069.4-02.6 & NGC 6894 & 1.40 $\pm$ 0.14 & $ 1.01 ^{+ 0.01 }_{- 0.01 }$ & $ 1.43 ^{+ 0.14 }_{- 0.07 } $ \\
069.6-03.9 & K 3-58 & 4.11 $\pm$ 0.19 & $ > 7.0 $ & $ > 5.5 $ \\
069.7+00.0 & K 3-55 & 7.05 $\pm$ 0.16 & $ > 8.0 $ & $ > 8.5 $ \\
071.6-02.3 & M 3-35 & 4.61 $\pm$ 0.08 & $ 4.57 ^{+ 0.30 }_{- 0.27 }$ & $ > 5.5 $ \\
072.1+00.1 & K 3-57 & 5.35 $\pm$ 0.33 & $ 9.94 ^{+ 3.88 }_{- 1.75 }$ & $ > 5.1 $ \\
073.0-02.4 & K 3-76 & 4.27 $\pm$ 0.29 & $ 4.30 ^{+ 0.45 }_{- 0.57 }$ & $ > 5.4 $ \\

074.5+02.1 & NGC 6881 & 3.60 $\pm$ 0.03 & $ 5.15 ^{+ 0.16 }_{- 0.05 }$ & $ 5.87 ^{+ 0.05 }_{- 0.05 } $ \\

075.6+04.3 & ARO 342 & 2.26 $\pm$ 0.19 & $ > 7.0 $ & $ > 4.7 $ \\
076.3+01.1 & Abell 69 & 4.20 $\pm$ 0.17 & $ 5.50 ^{+ 0.16 }_{- 0.19 }$ & $ 6.25 ^{+ 0.12 }_{- 0.12 } $ \\
076.4+01.8 & KjPn 3 & 3.42 $\pm$ 0.19 & $ 5.02 ^{+ 0.49 }_{- 0.28 }$ & $ 4.86 ^{+ 0.46 }_{- 0.63 } $ \\
077.5+03.7 & KjPn 1 & 1.50 $\pm$ 0.21 & $ 2.36 ^{+ 0.74 }_{- 0.33 }$ & $ 2.28 ^{+ 0.35 }_{- 0.68 } $ \\
077.7+03.1 & KjPn 2 & 5.51 $\pm$ 0.19 & $ 6.89 ^{+ 2.24 }_{- 0.87 }$ & $ > 4.6 $ \\
078.3-02.7 & K 4-53 & 4.88 $\pm$ 0.10 & $ 9.31 ^{+ 4.69 }_{- 2.02 }$ & $ > 3.9 $ \\
078.9+00.7 & Sd 1 & 5.06 $\pm$ 0.19 & $ 2.84 ^{+ 0.23 }_{- 0.17 }$ & $ 1.07 ^{+ 0.01 }_{- 0.01 } $ \\
084.2+01.0 & K 4-55 & 2.94 $\pm$ 0.18 & $ 0.63 ^{+ 0.10 }_{- 0.01 }$ & $ 1.17 ^{+ 0.33 }_{- 0.38 } $ \\
084.9+04.4 & Abell 71 & 2.44 $\pm$ 0.11 & $ 2.26 ^{+ 0.08 }_{- 0.04 }$ & $ 2.44 ^{+ 0.01 }_{- 0.01 } $ \\
088.7+04.6 & K 3-78 & 4.47 $\pm$ 0.05 & $ 7.70 ^{+ 6.30 }_{- 1.41 }$ & $ > 3.2 $ \\
088.7-01.6 & NGC 7048 & 0.82 $\pm$ 0.04 & $ 2.03 ^{+ 0.05 }_{- 0.06 }$ & $ 1.99 ^{+ 0.34 }_{- 0.75 } $ \\
089.0+00.3 & NGC 7026 & 1.77 $\pm$ 0.05 & $ 2.15 ^{+ 0.11 }_{- 0.11 }$ & $ 2.68 ^{+ 0.01 }_{- 0.01 } $ \\
089.8-00.6 & Sh 1-89 & 3.52 $\pm$ 0.20 & $ 3.39 ^{+ 0.40 }_{- 0.18 }$ & $ 5.99 ^{+ 0.06 }_{- 0.40 } $ \\
091.6-04.8 & K 3-84 & 1.25 $\pm$ 0.24 & $ 6.83 ^{+ 7.17 }_{- 2.01 }$ & $ > 4.3 $ \\
091.6+01.8 & We 1-11 & 4.43 $\pm$ 0.06 & $ 2.16 ^{+ 0.11 }_{- 0.04 }$ & $ 2.07 ^{+ 0.20 }_{- 0.13 } $ \\
093.3-00.9 & K 3-82 & 3.72 $\pm$ 0.14 & $ 4.75 ^{+ 0.15 }_{- 0.13 }$ & $ > 7.8 $ \\
093.3-02.4 & M 1-79 & 0.84 $\pm$ 0.04 & $ 2.31 ^{+ 0.17 }_{- 0.20 }$ & $ 1.33 ^{+ 0.14 }_{- 0.25 } $ \\
094.5-00.8 & K 3-83 & 4.75 $\pm$ 0.04 & $ 14.0 ^{+ 0.47 }_{- 2.5 }$ & $ > 9.1 $ \\
095.1-02.0 & M 2-49 & 3.67 $\pm$ 0.06 & $ 8.05 ^{+ 0.78 }_{- 0.51 }$ & $ > 5.1 $ \\

095.2+00.7 & K 3-62 & 5.35 $\pm$ 0.04 & $ 5.37 ^{+ 0.04 }_{- 0.04 }$ & $ >8.2 $ \\

096.3+02.3 & K 3-61 & 3.78 $\pm$ 0.05 & $ 3.25 ^{+ 0.20 }_{- 0.14 }$ & $ 8.59 ^{+ 2.37 }_{- 0.12 } $ \\
097.6-02.4 & M 2-50 & 2.01 $\pm$ 0.05 & $ > 6.0 $ & $ > 8.0 $ \\
098.1+02.4 & K 3-63 & 2.54 $\pm$ 0.08 & $ 2.84 ^{+ 0.39 }_{- 0.41 }$ & $ 3.70 ^{+ 0.29 }_{- 0.19 } $ \\
098.2+04.9 & K 3-60 & 4.55 $\pm$ 0.05 & $ > 5.0 $ & $ > 2.8 $ \\
102.8-05.0 & Abell 80 & 0.12 $\pm$ 0.06 & $ 1.54 ^{+ 0.43 }_{- 0.45 }$ & $ 0.33 ^{+ 0.05 }_{- 0.03 } $ \\
103.2+00.6 & M 2-51 & 1.79 $\pm$ 0.08 & $ 1.60 ^{+ 0.10 }_{- 0.07 }$ & $ 2.33 ^{+ 0.05 }_{- 0.04 } $ \\
103.7+00.4 & M 2-52 & 2.91 $\pm$ 0.05 & $ 4.19 ^{+ 0.04 }_{- 0.06 }$ & $ 3.85 ^{+ 0.02 }_{- 0.02 } $ \\

104.1+01.0 & Bl 2-1 & 5.31 $\pm$ 0.04 & $ 5.26 ^{+ 0.09 }_{- 0.09 }$ & $ > 4.8 $ \\

104.4-01.6 & M 2-53 & 2.41 $\pm$ 0.05 & $ 2.09 ^{+ 0.07 }_{- 0.05 }$ & $ > 4.8 $ \\
107.4-02.6 & K 3-87 & 4.45 $\pm$ 0.12 & $ > 6.0 $ & $ > 3.7 $ \\
107.4-00.6 & K 4-57 & 3.85 $\pm$ 0.19 & $ > 6.0 $ & $ > 10.7 $ \\
107.7-02.2 & M 1-80 & 1.64 $\pm$ 0.05 & $ 4.33 ^{+ 0.25 }_{- 0.26 }$ & $ 5.09 ^{+ 0.18 }_{- 0.51 } $ \\
107.8+02.3 & NGC 7354 & 3.70 $\pm$ 0.03 & $ 2.12 ^{+ 0.06 }_{- 0.03 }$ & $ 2.56 ^{+ 0.49 }_{- 0.04 } $ \\
112.5-00.1 & KjPn 8 & 1.80 $\pm$ 0.11 & $ 2.36 ^{+ 0.03 }_{- 0.26 }$ & $ 2.88 ^{+ 0.24 }_{- 0.11 } $ \\
112.5+03.7 & K 3-88 & 4.65 $\pm$ 0.06 & $ > 5.0 $ & $ > 3.3 $ \\
119.3+00.3 & BV 5-1 & 2.52 $\pm$ 0.05 & $ 4.56 ^{+ 0.14 }_{- 0.07 }$ & $ > 5.5 $ \\
121.6+00.0 & BV 5-2 & 2.43 $\pm$ 0.06 & $ 1.89 ^{+ 0.15 }_{- 0.13 }$ & $ 3.40 ^{+ 0.03 }_{- 0.04 } $ \\
121.6+03.5 & We 1-1 & 4.05 $\pm$ 0.06 & $ 3.28 ^{+ 0.06 }_{- 0.18 }$ & $ > 2.1 $ \\

\bottomrule 
\end{tabular}
\end{table*} 

\begin{table*}
\centering 
\contcaption{Adopted extinctions and distances derived from \protect\cite{Sale2014} H-MEAD and \protect\cite{Green2019_Bayestars19}}
\begin{tabular}{lllll}

\hline 
\\
PN G & Name & A$_{\mathrm V}$ (mag) & H-MEAD (kpc) & Green+2019 (kpc)\\ 
\\
\hline

122.1-04.9 & Abell 2 & 1.75 $\pm$ 0.05 & $ > 5.0 $ & $ > 2.9 $ \\
126.3+02.9 & K 3-90 & 2.76 $\pm$ 0.05 & $ 3.92 ^{+ 0.08 }_{- 0.07 }$ & $ > 3.9 $ \\
126.6+01.3 & IPHAS PN-1 & 1.47 $\pm$ 1.05 & $ 0.10 ^{+ 2.14 }_{- 0.03 }$ & $ 0.76 ^{+ 2.89 }_{- 0.45 } $ \\
129.5+04.5 & K 3-91 & 3.51 $\pm$ 0.05 & $ > 4.5 $ & $ > 2.3 $ \\
130.2+01.3 & IC 1747 & 2.02 $\pm$ 0.04 & $ 4.20 ^{+ 0.12 }_{- 0.21 }$ & $ 4.64 ^{+ 0.08 }_{- 0.07 } $ \\
130.4+03.1 & K 3-92 & 1.96 $\pm$ 0.19 & $ 4.42 ^{+ 0.39 }_{- 0.90 }$ & $ 4.81 ^{+ 10.1}_{- 1.53 } $ \\
131.5+02.6 & Abell 3 & 0.44 $\pm$ 0.06 & $ 0.43 ^{+ 0.02 }_{- 0.18 }$ & $ 0.26 ^{+ 0.05 }_{- 0.02 } $ \\
132.4+04.7 & K 3-93 & 3.74 $\pm$ 0.06 & $ > 4.5 $ & $ > 2.3 $ \\
136.1+04.9 & Abell 6 & 2.14 $\pm$ 0.19 & $ 0.46 ^{+ 0.22 }_{- 0.03 }$ & $ 1.00 ^{+ 0.05 }_{- 0.05 } $ \\
138.8+02.8 & IC 289 & 2.33 $\pm$ 0.02 & $ 0.96 ^{+ 0.01 }_{- 0.01 }$ & $ 3.82 ^{+ 0.03 }_{- 0.03 } $ \\
142.1+03.4 & K 3-94 & 2.84 $\pm$ 0.15 & $ > 5.0 $ & $ > 3.5 $ \\
147.4-02.3 & M 1-4 & 3.23 $\pm$ 0.06 & $ 2.75 ^{+ 0.07 }_{- 0.09 }$ & $ > 4.0 $ \\
147.8+04.1 & M 2-2 & 2.95 $\pm$ 0.15 & $ 4.55 ^{+ 9.45 }_{- 0.45 }$ & $ > 3.0 $ \\
151.4+00.5 & K 3-64 & 3.28 $\pm$ 0.25 & $ 9.98 ^{+ 4.02 }_{- 3.54 }$ & $ > 10.2 $ \\
173.5+03.2 & Pu 2 & 3.95 $\pm$ 0.17 & $ 3.22 ^{+ 0.25 }_{- 0.26 }$ & $ > 1.5 $ \\
178.3-02.5 & K 3-68 & 2.73 $\pm$ 0.19 & $ 8.88 ^{+ 5.12 }_{- 4.04 }$ & $ > 5.3 $ \\
181.5+00.9 & Pu 1 & 2.55 $\pm$ 0.06 & $ 10.1 ^{+ 3.94 }_{- 2.1 }$ & $ > 4.7 $ \\
184.0-02.1 & M 1-5 & 2.84 $\pm$ 0.02 & $ > 4.5 $ & $ > 5.1 $ \\
184.6+00.6 & K 3-70 & 4.13 $\pm$ 0.12 & $ > 5.0 $ & $ > 5.0 $ \\
184.8+04.4 & K 3-71 & 3.52 $\pm$ 0.27 & $ > 4.0 $ & $ > 2.6 $ \\
194.2+02.5 & J 900 & 1.64 $\pm$ 0.03 & $ > 4.5 $ & $ > 10.2 $ \\
201.7+02.5 & K 4-48 & 3.05 $\pm$ 0.10 & $ > 4.0 $ & $ > 4.4 $ \\
210.3+01.9 & M 1-8 & 2.48 $\pm$ 0.21 & $ > 5.0 $ & $ > 3.5 $ \\
212.0+04.3 & M 1-9 & 1.17 $\pm$ 0.07 & $ 2.90 ^{+ 0.58 }_{- 0.76 }$ & $ > 4.1 $ \\

\bottomrule
\bottomrule 
\end{tabular}
\end{table*}


%% file: Tables/KnownPNe_GaiaDist.tex
\begin{table*}
\centering 
\caption{Comparison to Gaia DR2 distances}
\label{KnownPNe_GaiaDists} 
\begin{tabular}{lllllllll}

\hline 
\\
 & & Gaia DR2 & Gaia DR2 & Gaia DR2 &   & Green & Lallement & Frew \\ 
 
PN G & Name & Parallax & Parallax Distance & Bailer-Jones+ & H-MEAD & +2019 & +2019 & +2016 \\

& & (mas) & (kpc) & (kpc) & (kpc) & (kpc) & (kpc) & (kpc) \\

\hline

035.9-01.1 & Sh 2-71 & 0.588$\pm$0.035 & $ 1.62^{+0.10}_{-0.08}$ & $1.62^{+0.10}_{-0.09}$ & $0.12^{+0.01}_{-0.01}$ & $0.49 ^{+ 0.02 }_{- 0.02 }$ & $0.76^{+0.11}_{-0.05}$ & $1.32\pm0.47$ \\

043.1+03.8 & M 1-65 & 0.224$\pm$0.040 & $ 3.94^{+0.74}_{-0.53}$ & $3.90^{+0.74}_{-0.55}$ & $2.25^{+0.11}_{-0.19}$ &  $ 3.22^{+ 0.19 }_{- 0.25 }$ & $1.68^{+0.06}_{-0.03}$ & $6.85\pm2.01$ \\

048.5+04.2 & K 4-16 & 0.191$\pm$0.053 & $ 4.53^{+1.37}_{-0.86}$ & $4.39^{+1.32}_{-0.86}$ & $> 7.0 $ & $> 4.5 $ & $>3.0$  & $11.1\pm3.5$ \\

051.3+01.8 & PM 1-295 & 0.339$\pm$0.030 & $ 2.71^{+0.24}_{-0.20}$ & $2.72^{+0.24}_{-0.21}$ & $1.40 ^{+ 0.08 }_{- 0.16 }$ & $1.21^{+0.03}_{-0.03}$ & $>3.5$ & --\\

052.5-02.9 & Me 1-1 & 0.226$\pm$0.037 & $ 3.91^{+0.67}_{-0.49}$ & $3.87^{+0.65}_{-0.49}$ & $2.42^{+0.18}_{-0.40}$ & $ 2.73^{+ 0.52 }_{- 0.46 }$ & $2.91^{+0.87}_{-0.57}$ & $6.17\pm1.87 $ \\

053.8-03.0 & Abell 63 & 0.361$\pm$0.034 & $ 2.55^{+0.25}_{-0.20}$ & $2.57^{+0.26}_{-0.21}$ & $0.92^{+0.25}_{-0.41}$ & $ 0.76^{+ 0.47 }_{- 0.19 }$ & $0.63^{+0.47}_{-0.11}$ & $3.79\pm1.12$ \\

060.8-03.6 & NGC 6853 & 2.658$\pm$ 0.044 & $ 0.37^{+0.01}_{-0.01}$ & $0.37^{+0.01}_{-0.01}$ & $0.11^{+0.07}_{-0.04}$ &  $0.96 ^{+ 0.02 }_{- 0.08 }$  & $0.42^{+0.10}_{-0.23}$ & $0.31\pm0.09$ \\

065.9+00.5 & NGC 6842 & 0.476$\pm$0.046 & $ 1.98^{+0.20}_{-0.16}$ & $1.99^{+0.21}_{-0.17}$ & $4.34^{+0.16}_{-0.10}$ & $4.85^{+ 0.04 }_{- 0.04}$ & $2.15^{+0.10}_{-0.09}$  & $2.20\pm0.64$ \\

069.4-02.6 & NGC 6894 & 0.839$\pm$0.131 & $ 1.15^{+0.21}_{-0.15}$ & $1.18^{+0.24}_{-0.17}$ & $1.01^{+0.01}_{-0.01}$ & $ 1.43^{+ 0.14 }_{- 0.07}$  & $1.17^{+0.04}_{-0.02}$ & $1.50\pm0.43$ \\

069.7+00.0 & K 3-55 & 0.600$\pm$0.071 & $ 1.58^{+0.20}_{-0.15}$ & $1.61^{+0.22}_{-0.17}$ & $> 8.0 $ & $> 8.5 $ & $>3.0$  & $3.54\pm1.32$  \\

076.4+01.8 & KjPn 3 & 0.307$\pm$0.090 & $ 2.97^{+1.09}_{-0.63}$ & $2.89^{+1.02}_{-0.63}$ & $5.02^{+0.49}_{-0.28}$ & $ 4.86^{+ 0.46 }_{- 0.63}$ & $2.61^{+1.20}_{-1.15}$  & $17.2\pm4.9$ \\

107.8+02.3 & NGC 7354 & 0.454$\pm$0.085 & $ 2.07^{+0.45}_{-0.31}$ & $2.10^{+0.50}_{-0.35}$ & $2.12^{+0.06}_{-0.03}$ & $ 2.56^{+ 0.49}_{- 0.04}$ & $>3.0$  & $1.26\pm0.37$ \\

130.2+01.3 & IC 1747 & 0.308$\pm$0.051 & $ 2.95^{+0.51}_{-0.38}$ & $2.95^{+0.54}_{-0.40}$ & $4.20^{+0.12}_{-0.21}$ & $ 4.64^{+ 0.08}_{-0.07}$ & $>3.5$  & $3.08\pm1.00$ \\

131.5+02.6 & Abell 3 & 0.343$\pm$0.084 & $ 2.68^{+0.78}_{-0.49}$ & $2.70^{+0.86}_{-0.54}$ & $0.43^{+0.02}_{-0.18}$ & $ 0.26^{+0.05}_{-0.02}$ & $0.37^{+0.01}_{-0.01}$ & $2.47\pm0.73$ \\

136.1+04.9 & Abell 6 & 1.004$\pm$ 0.177 & $ 0.97^{+0.20}_{-0.14}$ & $1.01^{+0.25}_{-0.17}$ & $0.46^{+0.22}_{-0.03}$ & $1.00^{+0.05}_{-0.05}$ & $1.10^{+0.10}_{-0.07}$ & $1.30\pm0.39$ \\

138.8+02.8 & IC 289 & 0.628$\pm$0.060 & $ 1.52^{+0.15}_{-0.13}$ & $1.53^{+0.16}_{-0.13}$ & $0.96^{+0.01}_{-0.01}$ & $3.82^{+0.03}_{-0.03}$ & $>3.5$ & $1.88\pm0.58$ \\

147.8+04.1 & M 2-2 & 0.192$\pm$0.054 & $ 4.54^{+1.46}_{-0.91}$ & $4.26^{+1.23}_{-0.82}$ & $4.55^{+9.45}_{-0.45}$ & $> 3.0 $ & $>3.0$  & $5.22\pm1.51$ \\



\bottomrule
\bottomrule 
\end{tabular}
\end{table*}

%% file: Appendix.tex

\appendix

\section{IPHAS {\Ha} Flux Measurements}

\subsection{IPHAS {\Ha} filter fluxes, [N~{\sc ii}]/{\Ha} ratios and corrected {\Ha} fluxes for IPHAS-discovered PNe}
\label{Appendix:iphasPNe_Fluxes}

Table~A1 presents the IPHAS {\Ha} fluxes measured for the IPHAS-discovered PNe in the sample. Columns (1) and (2) list the PN~G Galactic coordinates and the IPHAS names for the nebulae. 
Column (3) lists the angular dimensions of the nebulae, measured at 10\% of their peak brightnesses in the IPHAS {\Ha} filter images. Column (4) lists the IPHAS run numbers of the images that were used for the nebular aperture photometry, while column (5) lists the measured in-band  {\Ha} fluxes, in cgs units. Columns (6)-(8) respectively list the logs of the individual fluxes, the log of the mean flux and its uncertainty.
Column (9) lists the [N~{\sc ii}]/{\Ha} flux ratios adopted from the literature, when available, with the letter in brackets next to each ratio representing the literature source, listed below. [N~{\sc ii}]-corrected log~{\Ha} fluxes are given in column (10).

The sources for the [N~{\sc ii}]/{\Ha} flux ratios are: (a) \cite{Parker2006}; (b) \cite{ViironenJune2009}; (c) \cite{Acker1991}; (d) \cite{Sabin2014}; (e) \cite{Sabin2008}; (f) \cite{Mampaso2006}.


\subsection{IPHAS {\Ha} filter fluxes, [N~{\sc ii}]/{\Ha} ratios and corrected {\Ha} fluxes for previously catalogued PNe}
\label{Appendix:KnownFluxes}

Table~A2 presents the IPHAS {\Ha} fluxes measured for the previously catalogued PNe. Columns (1) and (2) list PN~G Galactic coordinates and catalogue names for the nebulae. 
Column (3) lists the angular dimensions of the nebulae, measured at 10\% of their peak brightnesses in the IPHAS {\Ha} filter images. Column (4)
lists the IPHAS run numbers of the images that were used for the nebular aperture photometry, while column (5) lists the measured in-band  
{\Ha} fluxes, in cgs units. Column (6) lists the [N~{\sc ii}]/{\Ha} flux ratios adopted from the literature, with the letter in brackets next to each ratio representing the literature source, listed below. [N~{\sc ii}]-corrected log~{\Ha} fluxes are given in column (7), while columns (8) and (9) present the mean log~{\Ha} fluxes (obtained by averaging the linear values of the [N~{\sc ii}]-corrected fluxes), and their corresponding uncertainties, in dex.

The sources for the [N~{\sc ii}]/{\Ha} flux ratios were: (a) \cite{Acker1991}; (b) \cite{Frew2013}; (c) \cite{Rosado1982}; (d) \cite{Kaler1997}; (e) \cite{Milingo2010}; (f) \cite{Guerrero1996}; (g) \cite{Lopez1995}; (h) \cite{Pena2005}; (i) \cite{Aller1979}; (j) \cite{Barker1978}; (k) \cite{Bohigas2001}; (l) \cite{Cantrell2014}; (m) \cite{Henry2010}; (n) 
\cite{Kaler1976}; (o) \cite{Kaler1983a}; (p) \cite{Kwitter2001}; (q) \cite{Pollacco1997}; (r) \cite{Schwarz2006}; (s) \cite{Zhang2005}; (t) \cite{Turatto1990}, (u) \cite{Kaler1983b}.

\input{Tables/iphasPNe_Fluxes}

\input{Tables/KnownPNeNII}

\section{{\HB} and Radio Fluxes for the sample PNe}
\label{Appendix:FluxSources}

The {\HB} fluxes and the radio fluxes at 1.4 GHz, 5 GHz and 30 GHz, together with their uncertainties, are listed in Table~\ref{Fluxsources}. For PNe where fluxes at a given frequency were available from more than one source, we adopted a weighted average. 

The {\HB} fluxes are obtained from the following publications: (a) \citet{Kaler1985}; (b) \citet{Kaler1983b}; (c) \citet{Collins1961}; (d) \citet{Peimbert1971}; (e) \citet{Barker1978}; (f) \citet{Odell1963}; (g) \citet{Osterbrock1960}; (h) \citet{Kaler1983a}; (i) \citet{Weidmann2008}; (j) \citet{Capriotti1960}; (k) \citet{Loup1993}; (l) \citet{Henry2010}; (m) \citet{Kaler1990}; (n) \citet{Carrasco1983}; (o) \citet{Perek1971};  (p) \citet{Kaler1976}.

The references for the radio fluxes are as follows: (a) \citet{Pazderska2009}; (b) \citet{Cahn1992}; (c) \citet{Zijlstra1989}; (d) \citet{Aaquist1990}; (e) \citet{Siodmiak2001}; (f) \citet{Quireza2007} ;(g) \citet{Isaacman1984}; (h) \citet{Cahn1974}; (i) \citet{Turner1984}; (j) \citet{Higgs1971}; (k) \citet{Milne1979}; (l) \citet{Milne1975}; (m) \citet{Condon1998}; (n) \citet{Kerber2003}; (o) \citet{DENISConsortium2003}; (p) \citet{Ofek2011}; (q) \citet{Skiff2014}; (r) \citet{Huggins2005}; (s) \citet{Lumsden2013}; (t) \citet{Stanghellini2008}; (u) \citet{ViironenJuly2009}; (v) \citet{SetiaGunawan2003}; (w) \citet{Kerton2003}; (x) \citet{Karachentsev2001}; (y) \citet{KwokandAaquist1993}; (z) \citet{Irabor2018}.			

\input{Tables/HB_Radio_Fluxes_andReferences}

\section{{\EHaHB}-based {\CHB} and A$_{\rm v}$ values}

Table~\ref{Av_EHa-HB} lists for 38 PNe the values of {\EHaHB}, {\CHB} and {\AvE}, and their uncertainties, derived by comparing the measured {\Ha} fluxes with the literature {\HB} filter fluxes listed in Table~\ref{Fluxsources}.

\input{Tables/KnownPNe_Av_EHa-HB}

\section{{\Av} values from comparisons of Balmer-line and radio fluxes}
\label{Appendix:KnownPNeRadiosec}

Table~\ref{KnownPNeRadio} lists the individual values of {\AV}, and their uncertainties, that were derived from a comparison of the radio fluxes at 1.4 GHz, 5 GHz and 30 GHz (listed in Table~\ref{Fluxsources}) with the {\Ha} fluxes listed in Tables~A1 and A2 and the {\HB} fluxes listed in Table~\ref{Fluxsources}, using the relations given in Section~\ref{Extinction Calculations}. The extinction values used to obtain the weighted averaged {\Av}(Radio) values listed in the penultimate column of Table~\ref{KnownPNeRadio} are indicated in bold script in the table.

\input{Tables/KnownPNeRadio_Avrg}

\section{H-MEAD Extinction versus Distance plots}
\label{Appendix:Distapp}

H-MEAD extinction versus distance plots for the lines of sight to all of the previously catalogued PNe observed by IPHAS, ordered by increasing \mbox{PN G}, are shown in Figures~E1-E19. The solid blue curve in each plot represents the H-MEAD extinction versus distance relation for the sightline to the nebula. 
The solid and dashed red horizontal lines correspond to the value of A$_{\rm V}$ derived for the nebula, and its corresponding uncertainties, while the solid and dashed green horizontal lines correspond to the \cite{Schlafly2011} A$_{\rm V}$ limit and its uncertainties.

\input{App_DistCurves}

%% file: Tables/iphasPNe_Fluxes.tex
\begin{table*}
\centering
 \caption{{\Ha}+[N~{\sc ii}] flux measurements for IPHAS-discovered PNe}
  \label{iphasPNe_Fluxes_Fluxes}
    \begin{tabular}{cccccccccc}

    \toprule
    \toprule

          & \multicolumn{1}{c}{\textbf{Name}} & \multicolumn{1}{c}{\textbf{Angular}} &       & \multicolumn{4}{c}{\textbf{Filter Flux (ergs cm$^{-2}$ s$^{-1}$)}} & \multicolumn{1}{c}{\textbf{[N II]/{\Ha} }} &  \\
    \multicolumn{1}{c}{\textbf{PN G}} & \multicolumn{1}{c}{\textbf{IPHAS}} & \multicolumn{1}{c}{\textbf{ Diam ($''$)}} & \multicolumn{1}{c}{\textbf{Run}} & \multicolumn{1}{c}{\textbf{F}} & \multicolumn{1}{c}{\textbf{LogF}} & \multicolumn{1}{c}{\textbf{Mean}} & \multicolumn{1}{c}{\textbf{unc. }} & \multicolumn{1}{c}{\textbf{Ratio}} & \multicolumn{1}{c}{\textbf{LogF({\Ha})}} \\
    
    (1)   & (2)   & (3)   & (4)   & (5)   & (6)   & (7)   & (8)   & (9) & (10) \\
    
    \hline

          &       &       &       &       &       &       &       &       &  \\

    \multicolumn{4}{l}{\textbf{\protect\cite{ViironenJuly2009}}} &       &       &       &       &       &  \\

    044.4-04.1 & J192717.94+081429.4 & 4.6   & 459511 & 4.73E-14 & -13.33 & -13.33 & 0.01  &       &  \\
          &       &       & 459673 & 4.52E-14 & -13.34 &       &       &       &  \\
    050.0+01.7 & J191701.33+155947.8 & 5.0   & 476362 & 4.38E-14 & -13.36 & -13.36 & 0.01  &       &  \\
          &       &       & 476365 & 4.23E-14 & -13.37 &       &       &       &  \\
          &       &       & 527663 & 4.36E-14 & -13.36 &       &       &       &  \\






    

    
    127.0-01.0 & J012544.68+613611.6 & 5   & 367787 & 2.77E-14 & -13.56 & -13.55 & 0.01  & 0.16(c) & -13.62 \\
          &       &       & 367900 & 2.79E-14 & -13.55 &       &       &       &  \\
          &       &       & 367903 & 2.84E-14 & -13.55 &       &       &       &  \\
    181.4-03.7 & J053440.77+254238.2 & 5   & 372878 & 8.82E-14 & -13.05 & -13.05 & 0.01  & 0.05(b) & -13.08 \\
          &       &       &       &       &       &       &       &       &  \\
    \multicolumn{4}{l}{\textbf{\protect\cite{Sabin2014}}} &       &       &       &       &       &  \\
    029.5-00.2 & XJ184616.3-030625 & 7   & 401522 & 3.87E-14 & -13.41 & -13.40 & 0.01  & 0.71(d) & -13.63 \\
          &       &       & 401525 & 4.03E-14 & -13.39 &       &       &       &  \\
    029.9+03.7 & XJ183249.6-005638 & 7   & 400650 & 2.26E-13 & -12.65 & -12.65 & 0.01  & 0.51(d) & -12.83 \\
          &       &       & 400671 & 2.23E-13 & -12.65 &       &       &       &  \\
    032.3-04.5 & XJ190631.5-023236 & 5   & 407121 & 2.39E-15 & -14.62 & -14.61 & 0.01  & 0.49(d) & -14.78 \\
          &       &       & 407124 & 2.54E-15 & -14.59 &       &       &       &  \\
    038.9-01.3 & XJ190718.1+044056 & 14.5 x 8 & 563823 & 3.34E-14 & -13.48 & -13.48 &       & 0.86(d) & -13.75 \\
    
%
%

    

    
    


    040.5+01.9 & XJ185815.8+073753 & 11  & 407764 & 5.46E-14 & -13.26 & -13.26 & 0.01  & 1.01(d) & -13.56 \\
          &       &       & 407797 & 5.36E-14 & -13.27 &       &       &       &  \\
          &       &       & 407800 & 5.71E-14 & -13.24 &       &       &       &  \\
    040.7+03.4 & XJ185322.1+083018 & 92 x 67 & 406710 & 4.32E-13 & -12.36 & -12.37 & 0.01  &    0.00(d)   &  -12.37 \\
          & (Ear Nebula) &       & 406713 & 4.24E-13 & -12.37 &       &       &       &  \\
    043.3+03.5 & XJ185744.4+105053 & 121 x 86 & 407458 & 3.02E-12 & -11.52 & -11.52 &  & 4.25(d)      & -12.24 \\
    043.6+01.7 & XJ190454.0+101801 & 23 x 24 & 410053 & 6.88E-13 & -12.16 & -12.19 & 0.01  &    1.24(d)   & -12.54 \\
          &       &       & 572482 & 6.31E-13 & -12.20 &       &       &       &  \\
          &       &       & 572485 & 6.16E-13 & -12.21 &       &       &       &  \\
    043.8+02.1 & XJ190338.5+104227 & 641 & 476759 & 8.18E-11 & -10.09 & -10.09 &       &       &  \\
    044.2+01.9 & XJ190518.3+105750 & 7.3   & 572470 & 3.96E-14 & -13.40 & -13.41 & 0.01  & 2.13(d) & -13.91 \\
          &       &       & 572473 & 3.84E-14 & -13.42 &       &       &       &  \\
    045.4+02.6 & J190447.9+121845 & 23 x 12 & 399198 & 3.00E-13 & -12.52 & -12.52 & 0.01  & 1.87(d) & -12.98 \\
          &       &       & 399201 & 3.00E-13 & -12.52 &       &       &       &  \\
    048.1+01.4 & XJ191421.1+140936 & 58 x 49 & 413381 & 4.51E-13 & -12.35 & -12.35 & 0.01  &       &  \\
    048.2+01.9 & J191255.5+143248 & 12.5  & 412044 & 4.90E-14 & -13.31 & -13.34 & 0.02  & 1.0(d) & -13.64 \\
          &       &       & 413381 & 4.22E-14 & -13.37 &       &       &       &  \\
          &       &       & 413384 & 4.55E-14 & -13.34 &       &       &       &  \\
    048.7-00.2 & XJ192152.0+135223 & 18  & 571003 & 1.06E-13 & -12.98 & -12.98 &       & 1.83(d) & -13.43 \\
    048.9+04.3 & XJ190512.4+161347 & 67  & 399234 & 1.40E-12 & -11.85 & -11.86 & 0.01  & 0.28(d)  &  -11.97 \\
          &       &       & 399237 & 1.39E-12 & -11.86 &       &       &       &  \\
    049.2+00.0 & J192153.9+143057 & 25 x 21 & 687803 & 7.97E-14 & -13.10 & -13.10 &       & 6.09(d) & -13.95 \\
    049.3-04.0 & XJ193637.3+123808 & 317 & 459751 & 3.74E-12 & -11.43 & -11.43 &       &       &  \\
    049.5-01.4 & XJ192751.3+140127 & 14.5  & 418221 & 1.07E-13 & -12.97 & -12.98 & 0.01  & 1.26(d) & -13.33 \\
          &       &       & 562419 & 1.02E-13 & -12.99 &       &       &       &  \\
    049.7-00.7 & XJ192543.2+143546 & 281 & 562362 & 1.17E-10 & -9.93 & -9.93 &       &       &  \\
    054.2-03.4 & XJ194359.5+170901 & 33 x 38 & 460367 & 1.55E-12 & -11.81 & -11.81 & 0.01  & 1.26(e) & -12.16 \\
          & (Necklace Nebula) &       &       &       &       &       &       &       &  \\
    057.9-00.7 & XJ194226.0+214521 & 36 x 23 & 460466 & 2.04E-13 & -12.69 & -12.69 & 0.01  & 2.03(d) & -13.17 \\
          &       &       & 460469 & 2.08E-13 & -12.68 &       &       &       &  \\
    059.1+03.5 & J192837.8+245025 & 20 x 12.5 & 414081 & 3.71E-13 & -12.43 & -12.43 &       & 0.81(d) & -12.69 \\
    059.7-01.0 & XJ194727.6+230816 & 43  & 460682 & 6.55E-13 & -12.18 & -12.18 & 0.01  & 1.04(d)   & -12.49 \\
          &       &       & 460685 & 6.67E-13 & -12.18 &       &       &       &  \\
    062.9+01.3 & J194510.7+270930 & 12.5  & 460538 & 5.85E-14 & -13.23 & -13.24 & 0.01  & 2.82(d) & -13.82 \\
          &       &       & 460541 & 5.68E-14 & -13.25 &       &       &       &  \\
    063.3+02.2 & XJ194240.6+275109 & 87  & 460436 & 2.71E-12 & -11.57 & -11.57 &       &  0.39(d)     &  -11.71 \\
    063.5+00.0 & J195126.5+265839 & 9.9   & 461153 & 8.62E-14 & -13.06 & -13.08 & 0.02  & 1.73(d) & -13.52 \\
          &       &       & 571498 & 7.87E-14 & -13.10 &       &       &       &  \\
    065.8+05.1 & XJ193630.3+312810 & 392 & 398786 & 1.29E-11 & -10.89 & -10.89 &       &       &  \\
    067.8+01.8 & XJ195436.4+313326 & 14  & 756015 & 4.20E-13 & -12.38 & -12.36 & 0.02  & 2.60(d) & -12.92 \\
          &       &       & 756018 & 4.54E-13 & -12.34 &       &       &       &  \\
    073.6+02.8 & XJ200522.0+365942 & 12.5  & 570090 & 1.30E-13 & -12.89 & -12.88 & 0.01  & 2.06(d) & -13.37 \\
          &       &       & 570093 & 1.32E-13 & -12.88 &       &       &       &  \\

\bottomrule

\end{tabular}         
\end{table*}

\begin{table*}
\centering
 \contcaption{{\Ha}+[N~{\sc ii}] flux measurements for IPHAS-discovered PNe}
    \begin{tabular}{cccccccccc}

    \toprule

          & \multicolumn{1}{c}{\textbf{Name}} & \multicolumn{1}{c}{\textbf{Angular}} &       & \multicolumn{4}{c}{\textbf{Filter Flux (ergs cm$^{-2}$ s$^{-1}$)}} & \multicolumn{1}{c}{\textbf{[N II]/{\Ha} }} &  \\
    \multicolumn{1}{c}{\textbf{PN G}} & \multicolumn{1}{c}{\textbf{IPHAS}} & \multicolumn{1}{c}{\textbf{ Diam ($''$)}} & \multicolumn{1}{c}{\textbf{Run}} & \multicolumn{1}{c}{\textbf{F}} & \multicolumn{1}{c}{\textbf{LogF}} & \multicolumn{1}{c}{\textbf{Mean}} & \multicolumn{1}{c}{\textbf{unc. }} & \multicolumn{1}{c}{\textbf{Ratio}} & \multicolumn{1}{c}{\textbf{LogF({\Ha})}} \\

    (1)   & (2)   & (3)   & (4)   & (5)   & (6)   & (7)   & (8)   & (9) & (10) \\
    
    \hline

          &       &       &       &       &       &       &       &       &  \\

    077.0+03.6 & J200018.6+365935 & 39  & 474781 & 2.86E-12 & -11.54 & -11.54 & 0.01  & 0.70(d) & -11.77 \\
          &       &       & 474784 & 2.86E-12 & -11.54 &       &       &       &  \\
    114.2+03.7 & J232713.0+650923 & 23  & 538647 & 4.41E-13 & -12.36 & -12.36 & 0.01  & 1.04(d) & -12.67 \\
          &       &       & 538650 & 4.31E-13 & -12.37 &       &       &       &  \\
    114.4+00.0 & XJ233841.2+614146 & 65 x 52 & 541345 & 1.18E-12 & -11.93 & -11.92 & 0.01  & 0.26(d) & -12.02 \\
          &       &       & 541712 & 1.26E-12 & -11.90 &       &       &       &  \\
          &       &       & 541757 & 1.17E-12 & -11.93 &       &       &       &  \\
    114.7-01.2 & J234403.7+603242 & 29  & 527721 & 2.45E-13 & -12.61 & -12.60 & 0.01  & 1.00(d) & -12.90 \\
          &       &       & 527724 & 2.54E-13 & -12.59 &       &       &       &  \\
    126.6+01.3 & XJ012507.9+635652 & 44 x 30 & 367778 & 1.20E-12 & -11.92 & -11.93 & 0.01  & 3.20(f) & -12.55 \\
          & (IPHAS PN-1) &       & 367781 & 1.15E-12 & -11.94 &       &       &       &  \\
          &       &       & 367984 & 1.18E-12 & -11.93 &       &       &       &  \\
          &       &       & 367897 & 1.21E-12 & -11.92 &       &       &       &  \\
    127.6-01.1 & J013109.0+612259 & 30  & 368165 & 1.15E-13 & -12.94 & -12.97 & 0.03  & 0.00(d)  &  -12.97 \\
          &       &       & 368168 & 1.01E-13 & -13.00 &       &       &       &  \\
    132.8+02.0 & J022045.1+631135 & 44 x 32 & 373190 & 9.51E-12 & -11.02 & -11.01 & 0.01  &     0.21(d)  & -11.09 \\
          &       &       & 373193 & 1.00E-11 & -11.00 &       &       &       &  \\
    139.0+03.2 & J031345.6+613710 & 129 & 431413 & 1.87E-12 & -11.73 & -11.73 &       & 0.58(d)  & -11.93 \\
    144.1-00.5 & J033105.4+553852 & 34  & 473618 & 1.05E-13 & -12.98 & -12.89 & 0.09  & 2.21(d) &    -13.41  \\
          &       &       & 473621 & 1.58E-13 & -12.80 &       &       &       &  \\
    150.0-00.3 & J040329.6+520824 & 43 x 108 & 599667 & 1.77E-13 & -12.75 & -12.75 &   &     0.00(d)  & -12.75 \\
    157.1+04.4 & J045627.2+501722 & 108 & 485522 & 3.01E-12 & -11.52 & -11.52 &       &  0.90(d) &    -11.80  \\
    193.5-03.1 & J060328.2+154106 & 66 x 57 & 703212 & 2.75E-13 & -12.56 & -12.56 & & 0.71(d)       & -12.79 \\
    195.4-04.0 & J060416.2+133251 & 40 x 17 & 476270 & 4.40E-13 & -12.36 & -12.35 & 0.01  &     3.53(d)  & -13.01 \\
          &       &       & 476273 & 4.47E-13 & -12.35 &       &       &       &  \\

\bottomrule   
\bottomrule

\end{tabular}         
\end{table*}

%% file: Tables/KnownPNeNII.tex
\begin{table*}
\centering 
 \caption{{[N~{\sc ii}] corrections and final {\Ha} fluxes for catalogued PNe}}
 \label{KnownPNeNII}
    \begin{tabular}{ccccccccc}
    
    \hline
    \hline
    
     \multicolumn{1}{c}{} &       & \multicolumn{1}{c}{\textbf{Angular}} &       & \multicolumn{1}{c}{\textbf{Filter Flux}} & \multicolumn{1}{c}{\textbf{[N II]/{\Ha}}} & \multicolumn{3}{c}{} \\
    \textbf{PN G} & \multicolumn{1}{c}{\textbf{Name}} & \multicolumn{1}{c}{\textbf{Diam. ($''$)}} & \multicolumn{1}{c}{\textbf{Run}} & \multicolumn{1}{c}{\textbf{(ergs cm$^{-2}$ s$^{-1}$)}} & \multicolumn{1}{c}{\textbf{Ratio}} & \multicolumn{1}{c}{\textbf{LogF({\Ha})}} & \multicolumn{1}{c}{\textbf{Mean}} & \multicolumn{1}{c}{\textbf{unc.}} \\

    (1)   & (2)   & (3)   & (4)   & (5)   & (6)   & (7)   & (8)   & (9) \\
    
    \hline
    
      &       &       &       &       &       &       &       &       \\

    030.8+03.4 & Abell 47 & 15 $\times$ 10 & 400722 & 1.95E-13 & 0.60(a) & -12.91 & -12.93 & 0.01 \\
     &       &       & 400725 & 1.84E-13 &       & -12.94 &       &  \\
    031.0+04.1 & K 3-6 & 0.7   & 400677 & 6.51E-13 & 0.20(b) & -12.27 & -12.26 & 0.01 \\
     &       &       & 400680 & 6.71E-13 &       & -12.25 &       &  \\
    031.3-00.5 & HaTr 10 & 21 $\times$ 9.2 & 401663 & 1.47E-12 & 6.50(a) & -12.71 & -12.71 & 0.01 \\
     &       &       & 401666 & 1.44E-12 &       & -12.72 &       &  \\
    031.7+01.7 & PC 20 & 2.0 $\times$ 1.3 & 463947 & 8.59E-13 & 1.10(a) & -12.39 & -12.39 &  \\
    032.0-03.0 & K 3-18 & 1.0   & 568881 & 8.45E-13 & 0.50(a) & -12.25 & -12.25 & 0.01 \\
     &       &       & 568884 & 8.27E-13 &       & -12.26 &       &  \\
    032.5-03.2 & K 3-20 & 4.0   & 568869 & 2.03E-12 & 0.90(a) & -11.97 & -11.97 & 0.01 \\
     &       &       & 568872 & 2.03E-12 &       & -11.97 &       &  \\
    032.9-02.8 & K 3-19 & 0.7   & 406755 & 1.89E-12 & 0.50(a) & -11.90 & -11.86 & 0.04 \\
     &       &       & 464288 & 2.24E-12 &       & -11.83 &       &  \\
    033.8-02.6 & NGC 6741 & 3.0   & 406776 & 5.36E-11 & 1.45(a) & -10.66 & -10.67 & 0.01 \\
     &       &       & 568875 & 5.16E-11 &       & -10.68 &       &  \\
    035.7-05.0 & K 3-26 & 8 $\times$ 5 & 564360 & 1.29E-12 & 0.10(a) & -11.93 & -11.93 & 0.01 \\
     &       &       & 564363 & 1.29E-12 &       & -11.93 &       &  \\
    
    035.9-01.1 & Sh 2-71 & 52 $\times$ 125 & 408156 & 7.23E-11 & 2.50(o,u) & -10.69 & -10.73 & 0.05 \\
     &       &       & 408159 & 5.86E-11 &       & -10.78 &       &  \\

    036.9-01.1 & HaTr 11 & 9.7 $\times$ 10.4 & 478115 & 9.47E-13 & 1.70(a) & -12.46 & -12.46  &   \\
    036.9-02.6 & HaTr 13 & 16.5 $\times$ 13.0 & 410137 & 8.00E-13 & 1.25(a) & -12.45 & -12.43 & 0.01 \\
     &       &       & 410419 & 8.56E-13 &       & -12.42 &       &  \\
    038.7+01.9 & YM 16 & 271 & 407061 & 1.00E-11 & 5.20(a) & -11.79 & -11.75 & 0.04 \\
     &       &       & 407064 & 1.21E-11 &       & -11.71 &       &  \\
    039.5-02.7 & M 2-47 & 1.3   & 750300 & 3.78E-12 & 0.10(a) & -11.46 & -11.46 & 0.01 \\
     &       &       & 750303 & 3.92E-12 &       & -11.45 &       &  \\
    039.8+02.1 & K 3-17 & 4.9 $\times$ 4.3 & 407434 & 1.55E-12 & 0.92(a) & -12.09 & -12.09 & 0.01 \\
     &       &       & 407437 & 1.55E-12 &       & -12.09 &       &  \\
    040.4-03.1 & K 3-30 & 0.7   & 412717 & 1.75E-12 & 0.10(a) & -11.80 & -11.80 &  \\
    041.8+04.4 & K 3-15 & 0.7   & 402091 & 1.30E-12 & 0.27(a) & -11.99 & -12.00 & 0.01 \\
     &       &       & 402094 & 1.25E-12 &       & -12.01 &       &  \\
    043.0-03.0 & M 4-14 & 6.7 $\times$ 24 & 569664 & 4.38E-12 & 2.36(a) & -11.89 & -11.89 & 0.01 \\
     &       &       & 569667 & 4.25E-12 &       & -11.90 &       &  \\
     &       &       & 572512 & 4.38E-12 &       & -11.89 &       &  \\
    043.1+03.8 & M 1-65 & 1.3   & 407458 & 7.34E-12 & 0.50(a) & -11.31 & -11.31 & 0.01 \\
     &       &       & 407503 & 7.38E-12 &       & -11.31 &       &  \\
    043.3+02.2 & PM 1-276 & 14  & 408168 & 1.13E-12 & 0.00(a) & -11.95 & -11.95 & 0.01 \\
     &       &       & 408171 & 1.12E-12 &       & -11.95 &       &  \\
    045.9-01.9 & K 3-33 & 0.6   & 687128 & 2.89E-13 & 0.90(a) & -12.82 & -12.82 & 0.01 \\
     &       &       & 687131 & 2.90E-13 &       & -12.82 &       &  \\
    046.3-03.1 & PB 9  & 4.6   & 418230 & 2.27E-12 & 0.00(a) & -11.64 & -11.65 & 0.01 \\
     &       &       & 418233 & 2.21E-12 &       & -11.66 &       &  \\
    046.4-04.1 & NGC 6803 & 3.0   & 419085 & 3.11E-11 & 0.4(b) & -10.65 & -10.65 & 0.01 \\
     &       &       & 419088 & 3.18E-11 &       & -10.64 &       &  \\
    046.8+02.9 & CTSS 4 & 4.6   & 562802 & 3.68E-13 & 0.71(d,e) & -12.67 & -12.66 & 0.01 \\
     &       &       & 562805 & 3.82E-13 &       & -12.65 &       &  \\
    047.1+04.1 & K 3-21 & 6.9   & 399123 & 7.96E-13 & 2.05(a) & -12.58 & -12.58 & 0.01 \\
     &       &       & 399168 & 8.13E-13 &       & -12.57 &       &  \\
    
    047.1-04.2 & Abell 62 & 164 & 571857 & 2.12E-11 & 0.18(o,u) & -10.75 & -10.76 & 0.02 \\
     &       &       & 571860 & 1.95E-11 &       & -10.78 &       &  \\

    048.0-02.3 & PB 10 & 1.3   & 418503 & 3.53E-12 & 0.20(b) & -11.53 & -11.53 &  \\
    048.1+01.1 & K 3-29 & 0.7 & 562674 & 6.95E-13 & 0.70(b) & -12.39 & -12.39 & 0.01 \\
     &       &       & 562677 & 6.89E-13 &       & -12.39 &       &  \\
    048.5+04.2 & K 4-16 & 1.3   & 399210 & 4.88E-13 & 0.00(b) & -12.31 & -12.31 & 0.01 \\
     &       &       & 399213 & 4.83E-13 &       & -12.32 &       &  \\
    048.7+01.9 & Hen 2-429 & 2.7   & 413423 & 8.09E-12 & 0.90(a) & -11.37 & -11.37 & 0.01 \\
     &       &       & 413426 & 8.06E-12 &       & -11.37 &       &  \\
    048.7+02.3 & K 3-24 & 4.6 $\times$ 1.4 & 412047 & 8.35E-13 & 2.10(a) & -12.57 & -12.56 & 0.01 \\
     &       &       & 412083 & 8.64E-13 &       & -12.55 &       &  \\

    \bottomrule
\end{tabular}
\end{table*}

\begin{table*}
\centering 
 \contcaption{{[N~{\sc ii}] corrections and final {\Ha} fluxes for catalogued PNe}}
 \begin{tabular}{ccccccccc}
    
    \toprule

     \multicolumn{1}{c}{} &       & \multicolumn{1}{c}{\textbf{Angular}} &       & \multicolumn{1}{c}{\textbf{Filter Flux}} & \multicolumn{1}{c}{\textbf{[N II]/{\Ha}}} & \multicolumn{3}{c}{} \\
    \textbf{PN G} & \multicolumn{1}{c}{\textbf{Name}} & \multicolumn{1}{c}{\textbf{Diam. ($''$)}} & \multicolumn{1}{c}{\textbf{Run}} & \multicolumn{1}{c}{\textbf{(ergs cm$^{-2}$ s$^{-1}$)}} & \multicolumn{1}{c}{\textbf{Ratio}} & \multicolumn{1}{c}{\textbf{LogF({\Ha})}} & \multicolumn{1}{c}{\textbf{Mean}} & \multicolumn{1}{c}{\textbf{unc.}} \\

    (1)   & (2)   & (3)   & (4)   & (5)   & (6)   & (7)   & (8)   & (9) \\
    
    \hline
    
      &       &       &       &       &       &       &       &       \\
      
          048.76-01.53 & DeHt 4 & 37 $\times$ 29 & 418158 & 9.08E-13 & 2.00(a) & -12.52 & -12.54 & 0.05 \\
     &       &       & 418161 & 1.04E-12 &       & -12.46 &       &  \\
     &       &       & 687110 & 7.06E-13 &       & -12.63 &       &  \\
    
 049.4+02.4 & Hen 2-428 & 7 $\times$ 22 & 562662 & 4.49E-12 & 0.62(a) & -11.56 & -11.55 & 0.01 \\
     &       &       & 562665 & 4.62E-12 &       & -11.54 &       &  \\
    050.4-01.6 & K 4-28 & 0.1   & 418805 & 1.13E-13 & 0.15(a) & -13.01 & -13.01 &  \\
       
    051.0+03.0 & Hen 2-430 & 0.7   & 413369 & 4.20E-12 & 0.63(b) & -11.59 & -11.57 & 0.02 \\
     &       &       & 570931 & 4.58E-12 &       & -11.55 &       &  \\
    051.0-04.5 & PC 22 & 15 $\times$ 11 & 460170 & 2.07E-12 & 0.30(a) & -11.80 & -11.75 & 0.05 \\
     &       &       & 460173 & 2.63E-12 &       & -11.69 &       &  \\
    051.3+01.8 & PM 1-295 & 1.0   & 528339 & 1.21E-12 & 0.51(t) & -12.09 & -12.08 & 0.01 \\
     &       &       & 528342 & 1.27E-12 &       & -12.07 &       &  \\
    051.9-03.8 & M 1-73 & 2.3   & 460149 & 1.90E-11 & 0.54(b) & -10.91 & -10.91 & 0.01 \\
     &       &       & 460152 & 1.87E-11 &       & -10.92 &       &  \\
    052.2-04.0 & M 1-74 & 1.3   & 460152 & 8.16E-12 & 0.20(a) & -11.17 & -11.13 & 0.02 \\
     &       &       & 570946 & 9.69E-12 &       & -11.09 &       &  \\
     &       &       & 570949 & 8.89E-12 &       & -11.13 &       &  \\
    052.5-02.9 & Me 1-1 & 3.7   & 460074 & 2.51E-11 & 0.94(e) & -10.90 & -10.90 &  \\
    052.9+02.7 & K 3-31 & 0.7   & 571854 & 7.02E-13 & 0.43(a) & -12.31 & -12.31 &  \\
    052.9-02.7 & K 3-41 & 0.7   & 460074 & 5.35E-13 & 0.00(a) & -12.27 & -12.27 &  \\
    053.2-01.5 & K 3-38 & 3.0   & 457828 & 8.22E-13 & 0.00(a) & -12.09 & -12.09 & 0.01 \\
     &       &       & 457831 & 8.16E-13 &       & -12.09 &       &  \\
    053.8-03.0 & Abell 63 & 19  & 562793 & 2.52E-12 & 0.01(q) & -11.61 & -11.62 & 0.01 \\
     &       &       & 687734 & 2.36E-12 &       & -11.63 &       &  \\
    054.4-02.5 & M 1-72 & 1.3   & 625328 & 8.68E-12 & 0.64(a) & -11.28 & -11.28 &  \\
    055.1-01.8 & K 3-43 & 7.6   & 686772 & 2.95E-13 & 0.07(d,e) & -12.56 & -12.46 & 0.08 \\
     &       &       & 921493 & 3.35E-13 &       & -12.50 &       &  \\
     &       &       & 921496 & 5.36E-13 &       & -12.30 &       &  \\
    055.2+02.8 & Hen 2-432 & 0.7   & 413703 & 1.94E-12 & 0.29(a) & -11.82 & -11.79 & 0.02 \\
     &       &       & 413706 & 2.09E-12 &       & -11.79 &       &  \\
     &       &       & 571261 & 2.19E-12 &       & -11.77 &       &  \\
    055.3+02.7 & Hen 1-1 & 3.2 & 413703 & 1.94E-12 & 1.42(a) & -12.10 & -12.08 & 0.01 \\
     &       &       & 413706 & 2.05E-12 &       & -12.07 &       &  \\
    055.5-00.5 & M 1-71 & 1.0   & 570277 & 1.14E-11 & 0.47(b) & -11.11 & -11.11 & 0.01 \\
     &       &       & 570280 & 1.12E-11 &       & -11.12 &       &  \\
    055.6+02.1 & Hen 1-2 & 1.6   & 413754 & 3.19E-12 & 0.83(d,e) & -11.76 & -11.76 & 0.01 \\
     &       &       & 413775 & 3.12E-12 &       & -11.77 &       &  \\
    056.0+02.0 & K 3-35 & 0.7   & 413778 & 2.08E-12 & 4.04(a) & -12.38 & -12.36 & 0.02 \\
     &       &       & 414090 & 2.31E-12 &       & -12.34 &       &  \\
    056.4-00.9 & K 3-42 & 0.7   & 921405 & 2.69E-13 & 0.27(a) & -12.67 & -12.67 & 0.01 \\
     &       &       & 921408 & 2.76E-13 &       & -12.66 &       &  \\
    057.9-01.5 & Hen 2-447 & 0.7   & 460562 & 2.60E-12 & 0.56(b) & -11.78 & -11.78 & 0.01 \\
     &       &       & 460565 & 2.58E-12 &       & -11.78 &       &  \\
    058.9+01.3 & K 3-40 & 2.0   & 420440 & 1.80E-12 & 0.18(a) & -11.82 & -11.81 & 0.01 \\
     &       &       & 420443 & 1.85E-12 &       & -11.81 &       &  \\
    059.0+04.6 & K 3-34 & 8 $\times$ 5 & 413733 & 1.18E-12 & 2.50(a) & -12.47 & -12.45 & 0.01 \\
     &       &       & 571246 & 1.26E-12 &       & -12.44 &       &  \\
     &       &       & 571249 & 1.28E-12 &       & -12.44 &       &  \\
    059.4+02.3 & K 3-37 & 1.3   & 420097 & 5.06E-13 & 0.03(a) & -12.31 & -12.30 & 0.01 \\
     &       &       & 420100 & 5.14E-13 &       & -12.30 &       &  \\
    059.9+02.0 & K 3-39 & 0.7   & 570229 & 6.29E-13 & 0.53(a) & -12.39 & -12.38 & 0.01 \\
     &       &       & 570232 & 6.36E-13 &       & -12.38 &       &  \\
    060.4+01.5 & PM 1-310 & 0.7   & 421140 & 9.73E-14 & 1.04(h) & -13.32 & -13.28 & 0.04 \\
     &       &       & 922041 & 1.18E-13 &       & -13.24 &       &  \\
    060.5+01.8 & Hen 2-440 & 0.7   & 421136 & 2.82E-12 & 0.57(a) & -11.75 & -11.75 & 0.01 \\
     &       &       & 421140 & 2.78E-12 &       & -11.75 &       &  \\
    060.5-00.3 & K 3-45 & 3.9 $\times$ 0.7 & 572264 & 1.68E-13 & 2.02(a) & -13.25 & -13.26 & 0.01 \\
     &       &       & 572267 & 1.63E-13 &       & -13.27 &       &  \\
    060.8-03.6 & NGC 6853 & 331 $\times$ 472 & 571657 & 2.57E-09 & 1.30(a) & -8.95 & -8.95 &  \\

\bottomrule
\end{tabular}
\end{table*}

\begin{table*}
\centering 
 \contcaption{{[N~{\sc ii}] corrections and final {\Ha} fluxes for catalogued PNe}}
 \begin{tabular}{ccccccccc}
    
    \toprule
    
     \multicolumn{1}{c}{} &       & \multicolumn{1}{c}{\textbf{Angular}} &       & \multicolumn{1}{c}{\textbf{Filter Flux}} & \multicolumn{1}{c}{\textbf{[N II]/{\Ha}}} & \multicolumn{3}{c}{} \\
    \textbf{PN G} & \multicolumn{1}{c}{\textbf{Name}} & \multicolumn{1}{c}{\textbf{Diam. ($''$)}} & \multicolumn{1}{c}{\textbf{Run}} & \multicolumn{1}{c}{\textbf{(ergs cm$^{-2}$ s$^{-1}$)}} & \multicolumn{1}{c}{\textbf{Ratio}} & \multicolumn{1}{c}{\textbf{LogF({\Ha})}} & \multicolumn{1}{c}{\textbf{Mean}} & \multicolumn{1}{c}{\textbf{unc.}} \\

    (1)   & (2)   & (3)   & (4)   & (5)   & (6)   & (7)   & (8)   & (9) \\
    
    \hline     
    
      &       &       &       &       &       &       &       &       \\
    062.4-00.2 & M 2-48 & 1.8 $\times$ 1.5 & 461361 & 6.54E-12 & 3.60(a) & -11.85 & -11.85 & 0.01 \\
     &       &       & 461364 & 6.40E-12 &       & -11.86 &       &  \\    
    
063.8-03.3 & K 3-54 & 0.7   & 563008 & 3.38E-13 & 0.84(a) & -12.74 & -12.74 & 0.01 \\
     &       &       & 563011 & 3.38E-13 &       & -12.74 &       &  \\
    064.9-02.1 & K 3-53 & 0.7   & 408279 & 1.16E-12 & 0.19(a) & -12.01 & -12.01 & 0.01 \\
     &       &       & 570256 & 1.16E-12 &       & -12.01 &       &  \\
    065.1-03.5 & We 1-9 & 23 $\times$ 19 & 563071 & 1.29E-12 & 0.89(a) & -12.17 & -12.11 & 0.06 \\
     &       &       & 750375 & 1.70E-12 &       & -12.05 &       &  \\
    065.9+00.5 & NGC 6842 & 52.4  & 474763 & 8.58E-12 & 0.07(a) & -11.09 & -11.09 & 0.01 \\
     &       &       & 572390 & 8.73E-12 &       & -11.09 &       &  \\
    066.9+02.2 & K 4-37 & 1.0   & 399436 & 6.59E-13 & 6.59(a) & -13.06 & -13.05 & 0.01 \\
     &       &       & 399523 & 6.96E-13 &       & -13.04 &       &  \\
    067.9-00.2 & K 3-52 & 0.7   & 408634 & 5.27E-13 & 2.09(a) & -12.77 & -12.77 & 0.01 \\
     &       &       & 475209 & 5.31E-13 &       & -12.76 &       &  \\
    068.3-02.7 & Hen 2-459 & 0.7   & 412200 & 5.85E-12 & 1.00(b,d,e) & -11.53 & -11.55 & 0.01 \\
     &       &       & 412203 & 5.59E-12 &       & -11.55 &       &  \\
     &       &       & 413073 & 5.32E-12 &       & -11.57 &       &  \\
    068.6+01.1 & Hen 1-4 & 22 $\times$ 20 & 407192 & 4.99E-12 & 1.15(l) & -11.63 & -11.64 & 0.01 \\
     &       &       & 407195 & 4.93E-12 &       & -11.64 &       &  \\
    068.7+01.9 & K 4-41 & 1.7   & 686283 & 9.25E-13 & 0.07(a) & -12.06 & -12.06 &  \\
    068.7+03.0 & PC 23 & 0.7   & 399511 & 1.46E-12 & 0.31(a) & -11.95 & -11.95 & 0.01 \\
     &       &       & 399514 & 1.45E-12 &       & -11.96 &       &  \\
    068.8-00.0 & M 1-75 & 17 $\times$ 9 & 410146 & 6.65E-12 & 5.11(d,e,k,l) & -11.96 & -11.97 & 0.01 \\
     &       &       & 410149 & 6.57E-12 &       & -11.97 &       &  \\
    069.2+02.8 & K 3-49 & 0.7   & 399544 & 8.05E-13 & 0.80(a) & -12.35 & -12.35 & 0.01 \\
     &       &       & 570078 & 8.15E-13 &       & -12.34 &       &  \\
    069.2+03.8 & K 3-46 & 26 $\times$ 12 & 399415 & 5.23E-12 & 8.53(a) & -12.26 & -12.26 & 0.01 \\
     &       &       & 399478 & 4.91E-12 &       & -12.29 &       &  \\
     &       &       & 534190 & 5.53E-12 &       & -12.24 &       &  \\

    069.4-02.6 & NGC 6894 & 37  & 530437 & 3.99E-11 & 0.78(o,u) & -10.65 & -10.64 & 0.01 \\
     &       &       & 530440 & 4.19E-11 &       & -10.63 &       &  \\

    069.6-03.9 & K 3-58 & 9 $\times$ 6 & 750399 & 3.24E-12 & 1.38(a) & -11.87 & -11.86 & 0.01 \\
     &       &       & 750426 & 3.38E-12 &       & -11.85 &       &  \\
    069.7+00.0 & K 3-55 & 5.3   & 410760 & 9.79E-13 & 1.09(d,e) & -12.33 & -12.33 & 0.01 \\
     &       &       & 410763 & 9.90E-13 &       & -12.32 &       &  \\
    071.6-02.3 & M 3-35 & 0.7   & 455655 & 5.66E-12 & 0.03(b) & -11.26 & -11.21 & 0.03 \\
     &       &       & 750405 & 6.61E-12 &       & -11.19 &       &  \\
     &       &       & 750408 & 6.84E-12 &       & -11.18 &       &  \\
    072.1+00.1 & K 3-57 & 3.0 $\times$ 4.6 & 412119 & 1.67E-12 & 0.59(a) & -11.98 & -11.97 & 0.01 \\
     &       &       & 412511 & 1.73E-12 &       & -11.96 &       &  \\
    073.0-02.4 & K 3-76 & 1.7   & 459527 & 4.93E-13 & 0.00(a) & -12.31 & -12.32 & 0.02 \\
     &       &       & 572330 & 4.59E-13 &       & -12.34 &       &  \\

    074.5+02.1 & NGC 6881 & 1.2 $\times$ 1.1 & 475983 & 1.04E-11 & 0.70(o,u) & -11.21 & -11.21 & 0.01 \\
    
    075.6+04.3 & ARO 342 & 23  & 471338 & 9.62E-13 & 1.41(l) & -12.40 & -12.41 & 0.01 \\
     &       &       & 569457 & 9.29E-13 &       & -12.41 &       &  \\
     &       &       & 569460 & 8.92E-13 &       & -12.43 &       &  \\
    076.3+01.1 & Abell 69 & 19.1  & 531486 & 1.38E-12 & 5.20(a) & -12.65 & -12.65 & 0.01 \\
     &       &       & 531489 & 1.39E-12 &       & -12.65 &       &  \\
    076.4+01.8 & KjPn 3 & 0.7   & 568992 & 1.65E-13 & 0.00(a) & -12.78 & -12.79 & 0.01 \\
     &       &       & 568995 & 1.63E-13 &       & -12.79 &       &  \\
    077.5+03.7 & KjPn 1 & 1.0   & 472088 & 9.45E-13 & 0.17(a) & -12.09 & -12.07 & 0.03 \\
     &       &       & 472091 & 9.05E-13 &       & -12.11 &       &  \\
     &       &       & 473554 & 1.14E-12 &       & -12.01 &       &  \\
    077.7+03.1 & KjPn 2 & 1.6 $\times$ 1.1 & 569004 & 3.36E-13 & 1.36(a) & -12.85 & -12.85 & 0.01 \\
     &       &       & 569007 & 3.32E-13 &       & -12.85 &       &  \\
    078.3-02.7 & K 4-53 & 9.5 $\times$ 14.5 & 366816 & 1.45E-12 & 0.62(l) & -12.05 & -12.06 & 0.02 \\
     &       &       & 366867 & 1.35E-12 &       & -12.08 &       &  \\
    078.9+00.7 & Sd 1  & 10  & 414455 & 1.12E-12 & 0.86(a) & -12.22 & -12.22 &  \\
    084.2+01.0 & K 4-55 & 7 $\times$ 9 & 367619 & 3.94E-12 & 5.40(f,k,l) & -12.21 & -12.21 &  \\

    084.9+04.4 & Abell 71 & 108 & 369201 & 3.28E-11 & 1.50(o,u) & -10.88 & -10.91 & 0.02 \\
     &       &       & 369834 & 2.96E-11 &       & -10.93 &       &  \\

\bottomrule
\end{tabular}
\end{table*}

\begin{table*}
\centering 
 \contcaption{{[N~{\sc ii}] corrections and final {\Ha} fluxes for catalogued PNe}}
 \begin{tabular}{ccccccccc}
    
    \toprule
    
     \multicolumn{1}{c}{} &       & \multicolumn{1}{c}{\textbf{Angular}} &       & \multicolumn{1}{c}{\textbf{Filter Flux}} & \multicolumn{1}{c}{\textbf{[N II]/{\Ha}}} & \multicolumn{3}{c}{} \\
    \textbf{PN G} & \multicolumn{1}{c}{\textbf{Name}} & \multicolumn{1}{c}{\textbf{Diam. ($''$)}} & \multicolumn{1}{c}{\textbf{Run}} & \multicolumn{1}{c}{\textbf{(ergs cm$^{-2}$ s$^{-1}$)}} & \multicolumn{1}{c}{\textbf{Ratio}} & \multicolumn{1}{c}{\textbf{LogF({\Ha})}} & \multicolumn{1}{c}{\textbf{Mean}} & \multicolumn{1}{c}{\textbf{unc.}} \\

    (1)   & (2)   & (3)   & (4)   & (5)   & (6)   & (7)   & (8)   & (9) \\
    
    \hline 
    
      &       &       &       &       &       &       &       &       \\
     088.7+04.6 & K 3-78 & 1.7   & 369543 & 4.81E-13 & 0.00(a) & -12.32 & -12.33 & 0.01 \\
     &       &       & 370351 & 4.44E-13 &       & -12.35 &       &  \\
     &       &       & 370354 & 4.64E-13 &       & -12.33 &       &  \\     
      
 088.7-01.6 & NGC 7048 & 63  & 419174 & 5.78E-11 & 2.24(l) & -10.75 & -10.74 & 0.01 \\
     &       &       & 419177 & 5.90E-11 &       & -10.74 &       &  \\
    089.0+00.3 & NGC 7026 & 25 $\times$ 8 & 411588 & 9.37E-11 & 0.62(b,l,o) & -10.24 & -10.24 & 0.01 \\
     &       &       & 411591 & 9.28E-11 &       & -10.24 &       &  \\
    089.8-00.6 & Sh 1-89 & 52 $\times$ 9 & 460260 & 3.53E-12 & 3.64(a) & -12.12 & -12.12 &  \\
    091.6+01.8 & We 1-11 & 23 $\times$ 22 & 412607 & 1.26E-12 & 2.00(a) & -12.38 & -12.38 & 0.01 \\
     &       &       & 412610 & 1.23E-12 &       & -12.39 &       &  \\
    091.6-04.8 & K 3-84 & 3.6   & 463449 & 2.96E-12 & 2.05(a) & -12.01 & -12.02 & 0.01 \\
     &       &       & 463452 & 2.92E-12 &       & -12.02 &       &  \\
    093.3-00.9 & K 3-82 & 23  & 530076 & 2.19E-12 & 0.29(a) & -11.77 & -11.72 & 0.05 \\
     &       &       & 756036 & 2.77E-12 &       & -11.67 &       &  \\

    093.3-02.4 & M 1-79 & 24 $\times$ 42 & 756069 & 2.66E-11 & 1.07(o,u) & -10.89 & -10.92 & 0.01 \\
     &       &       & 756072 & 2.53E-11 &       & -10.91 &       &  \\

    094.5-00.8 & K 3-83 & 4.2   & 750456 & 4.85E-13 & 2.82(a) & -12.90 & -12.89 & 0.01 \\
     &       &       & 750459 & 5.01E-13 &       & -12.88 &       &  \\
     &       &       & 756042 & 4.90E-13 &       & -12.89 &       &  \\
    095.1-02.0 & M 2-49 & 0.7   & 419822 & 2.66E-12 & 0.79(d,e) & -11.83 & -11.82 & 0.01 \\
     &       &       & 419825 & 2.82E-12 &       & -11.80 &       &  \\
    095.2+00.7 & K 3-62 & 1.3   & 647763 & 1.81E-12 & 0.26(d,e) & -11.84 & -11.84 & 0.01 \\
     &       &       & 647766 & 1.86E-12 &       & -11.83 &       &  \\
    096.3+02.3 & K 3-61 & 5.6   & 413465 & 9.42E-13 & 0.10(a) & -12.07 & -12.08 & 0.01 \\
     &       &       & 413468 & 8.95E-13 &       & -12.09 &       &  \\
    097.6-02.4 & M 2-50 & 2.0   & 421326 & 1.40E-12 & 0.00(a) & -11.85 & -11.85 & 0.01 \\
     &       &       & 421329 & 1.41E-12 &       & -11.85 &       &  \\

    098.1+02.4 & K 3-63 & 6.0 $\times$ 1.3 & 419524 & 7.04E-13 & 0.12(o,u) & -12.20 & -12.14 & 0.03 \\
     &       &       & 478931 & 8.71E-13 &       & -12.11 &       &  \\
     &       &       & 478934 & 8.54E-13 &       & -12.12 &       &  \\

    098.2+04.9 & K 3-60 & 1.0   & 413124 & 9.43E-13 & 0.19(o,u) & -12.10 & -12.09 & 0.01 \\
     &       &       & 413127 & 9.79E-13 &       & -12.09 &       &  \\

    102.8-05.0 & Abell 80 & 144 $\times$ 106 & 464077 & 1.63E-11 & 1.45(o,u) & -11.18 & -11.16 & 0.02 \\
     &       &       & 464082 & 1.75E-11 &       & -11.15 &       &  \\

    103.2+00.6 & M 2-51 & 47 $\times$ 40 & 537395 & 2.40E-11 & 1.49(o,u) & -11.02 & -11.02 & 0.01 \\

    103.7+00.4 & M 2-52 & 7.3 $\times$ 9.5 & 462315 & 6.42E-12 & 3.36(k,l) & -11.83 & -11.84 & 0.01 \\
     &       &       & 462318 & 6.27E-12 &       & -11.84 &       &  \\
    104.1+01.0 & Bl 2-1 & 0.7   & 462351 & 7.29E-13 & 0.17(a) & -12.21 & -12.20 & 0.01 \\
     &       &       & 462354 & 7.31E-13 &       & -12.20 &       &  \\
     &       &       & 599592 & 7.40E-13 &       & -12.20 &       &  \\
     &       &       & 599595 & 7.34E-13 &       & -12.20 &       &  \\
    104.4-01.6 & M 2-53 & 12 $\times$ 8 & 597700 & 6.12E-12 & 2.12(l) & -11.71 & -11.72 & 0.01 \\
     &       &       & 597703 & 5.85E-12 &       & -11.73 &       &  \\
    107.4-00.6 & K 4-57 & 0.7   & 570770 & 1.26E-13 & 0.06(a) & -12.92 & -12.92 &  \\
    107.4-02.6 & K 3-87 & 2.6   & 418634 & 2.97E-13 & 0.09(a) & -12.56 & -12.52 & 0.04 \\
     &       &       & 570442 & 3.58E-13 &       & -12.48 &       &  \\
    107.7-02.2 & M 1-80 & 5.3   & 418634 & 6.99E-12 & 0.61(a) & -11.36 & -11.35 & 0.01 \\
     &       &       & 418862 & 7.27E-12 &       & -11.35 &       &  \\
    107.8+02.3 & NGC 7354 & 18  & 374924 & 2.81E-11 & 0.27(b) & -10.65 & -10.66 & 0.01 \\
     &       &       & 374972 & 2.79E-11 &       & -10.66 &       &  \\
     &       &       & 375594 & 2.65E-11 &       & -10.68 &       &  \\
    112.5+03.7 & K 3-88 & 7 $\times$ 8 & 464459 & 3.31E-13 & 1.22(a) & -12.83 & -12.81 & 0.01 \\
     &       &       & 464752 & 3.43E-13 &       & -12.81 &       &  \\
     &       &       & 464755 & 3.49E-13 &       & -12.80 &       &  \\
    112.5-00.1 & KjPn 8 & 3.3   & 475765 & 1.49E-12 & 3.80(g) & -12.51 & -12.54 & 0.03 \\
     &       &       & 476488 & 1.26E-12 &       & -12.58 &       &  \\
    119.3+00.3 & BV 1 & 31 $\times$ 11 & 414171 & 7.85E-12 & 4.16(k,l) & -11.82 & -11.82 & 0.01 \\
     &       &       & 414515 & 7.57E-12 &       & -11.83 &       &  \\
     &       &       & 414578 & 7.79E-12 &       & -11.82 &       &  \\
    121.6+03.5 & We 1-1 & 20  & 417705 & 4.74E-13 & 1.07(a) & -12.64 & -12.64 & 0.01 \\
     &       &       & 418383 & 4.68E-13 &       & -12.65 &       &  \\
     &       &       & 418386 & 4.76E-13 &       & -12.64 &       &  \\
    121.6+00.0 & BV 2 & 3.0 $\times$ 8.3 & 418653 & 2.53E-12 & 1.05(l,m) & -11.91 & -11.91 & 0.01 \\
     &       &       & 418656 & 2.50E-12 &       & -11.91 &       &  \\
     &       &       & 922235 & 2.69E-12 &       & -11.88 &       &  \\

\bottomrule
\end{tabular}
\end{table*}

\begin{table*}
\centering 
 \contcaption{{[N~{\sc ii}] corrections and final {\Ha} fluxes for catalogued PNe}}
 \begin{tabular}{ccccccccc}
    
    \toprule
    
     \multicolumn{1}{c}{} &       & \multicolumn{1}{c}{\textbf{Angular}} &       & \multicolumn{1}{c}{\textbf{Filter Flux}} & \multicolumn{1}{c}{\textbf{[N II]/{\Ha}}} & \multicolumn{3}{c}{} \\
    \textbf{PN G} & \multicolumn{1}{c}{\textbf{Name}} & \multicolumn{1}{c}{\textbf{Diam. ($''$)}} & \multicolumn{1}{c}{\textbf{Run}} & \multicolumn{1}{c}{\textbf{(ergs cm$^{-2}$ s$^{-1}$)}} & \multicolumn{1}{c}{\textbf{Ratio}} & \multicolumn{1}{c}{\textbf{LogF({\Ha})}} & \multicolumn{1}{c}{\textbf{Mean}} & \multicolumn{1}{c}{\textbf{unc.}} \\

    (1)   & (2)   & (3)   & (4)   & (5)   & (6)   & (7)   & (8)   & (9) \\
    
    \hline
      &       &       &       &       &       &       &       &       \\
      
           122.1-04.9 & Abell 2 & 27  & 419262 & 2.08E-12 & 0.27(o,u) & -11.79 & -11.79 & 0.01 \\
          &       &       & 419265 & 2.08E-12 &       & -11.79 &       &  \\

    126.3+02.9 & K 3-90 & 6   & 367533 & 1.04E-12 & 0.00(l) & -11.98 & -12.01 & 0.01 \\
     &       &       & 703469 & 9.58E-13 &       & -12.02 &       &  \\
     &       &       & 703472 & 9.38E-13 &       & -12.03 &       &  \\
    128.0-04.1 & Simeiz 22 & 351 $\times$ 48 & 368180 & 7.39E-11 & 5.00(c) & -10.91 & -10.87 & 0.04 \\
     &       &       & 368197 & 8.86E-11 &       & -10.83 &       &  \\
    129.5+04.5 & K 3-91 & 3.6 $\times$ 3.0 & 370422 & 5.63E-13 & 2.09(a) & -12.74 & -12.75 & 0.01 \\
     &       &       & 370425 & 5.48E-13 &       & -12.75 &       &  \\
    130.2+01.3 & IC 1747 & 8.0   & 370158 & 1.87E-11 & 0.089(m) & -10.77 & -10.77 & 0.01 \\
     &       &       & 370416 & 1.80E-11 &       & -10.78 &       &  \\
    130.4+03.1 & K 3-92 & 7.8 $\times$ 3.3 & 364766 & 9.09E-13 & 1.12(k,l) & -12.37 & -12.37 & 0.01 \\
     &       &       & 370490 & 8.80E-13 &       & -12.38 &       &  \\
     &       &       & 370493 & 9.03E-13 &       & -12.37 &       &  \\

    131.5+02.6 & Abell 3 & 54  & 371560 & 4.29E-12 & 0.26(o,u) & -11.47 & -11.48 & 0.01 \\
     &       &       & 372310 & 4.03E-12 &       & -11.50 &       &  \\
     &       &       & 372313 & 4.32E-12 &       & -11.47 &       &  \\

    132.4+04.7 & K 3-93 & 3.0   & 373840 & 4.72E-13 & 1.60(d,e) & -12.74 & -12.74 & 0.01 \\
     &       &       & 373843 & 4.70E-13 &       & -12.74 &       &  \\

    136.1+04.9 & Abell 6 & 166 & 380703 & 2.32E-12 & 0.13(o,u) & -11.69 & -11.62 & 0.07 \\
     &       &       & 380706 & 3.21E-12 &       & -11.55 &       &  \\

    138.8+02.8 & IC 289 & 25  & 431332 & 1.47E-11 & 0.01(b,l) & -10.84 & -10.83 & 0.01 \\
     &       &       & 431335 & 1.49E-11 &       & -10.83 &       &  \\
    142.1+03.4 & K 3-94 & 6.6   & 474115 & 1.71E-12 & 1.72(d,e,k) & -12.20 & -12.21 & 0.01 \\
     &       &       & 474118 & 1.67E-12 &       & -12.21 &       &  \\
    147.4-02.3 & M 1-4 & 2.7   & 376435 & 5.04E-12 & 0.05(b) & -11.32 & -11.36 & 0.02 \\
     &       &       & 428921 & 4.32E-12 &       & -11.39 &       &  \\
     &       &       & 428924 & 4.43E-12 &       & -11.38 &       &  \\

    147.8+04.1 & M 2-2 & 4.3   & 482053 & 6.19E-12 & 0.17(o,u) & -11.28 & -11.27 & 0.01 \\
     &       &       & 482056 & 6.52E-12 &       & -11.25 &       &  \\

    149.0+04.4 & K 4-47 & 8.3 $\times$ 0.9 & 483031 & 2.92E-13 & 2.35(d,e) & -13.06 & -13.07 & 0.01 \\
     &       &       & 483034 & 2.84E-13 &       & -13.07 &       &  \\
    151.4+00.5 & K 3-64 & 9.2   & 479903 & 7.96E-13 & 0.60(a) & -12.30 & -12.32 & 0.01 \\
     &       &       & 479906 & 7.91E-13 &       & -12.31 &       &  \\
     &       &       & 482095 & 7.29E-13 &       & -12.34 &       &  \\
    153.7-01.4 & K 3-65 & 2.3   & 538727 & 3.24E-13 & 1.25(a) & -12.84 & -12.85 & 0.01 \\
     &       &       & 538730 & 3.13E-13 &       & -12.86 &       &  \\
    160.5-00.5 & We 1-2 & 96  & 471195 & 2.36E-12 & 2.57(d,e) & -12.18 & -12.23 & 0.05 \\
     &       &       & 471198 & 1.84E-12 &       & -12.29 &       &  \\
    163.1-00.8 & We 1-3 & 72  & 480495 & 1.10E-11 & 2.94(d,e) & -11.56 & -11.56 &  \\
    173.5+03.2 & Pu 2  & 17  & 597407 & 2.75E-13 & 0.00(a) & -12.56 & -12.56 &  \\
    178.3-02.5 & K 3-68 & 9.3   & 372462 & 6.07E-13 & 0.00(a) & -12.22 & -12.21 & 0.01 \\
     &       &       & 372465 & 6.19E-13 &       & -12.21 &       &  \\
    181.5+00.9 & Pu 1  & 63  & 381282 & 1.79E-12 & 1.01(a) & -12.05 & -12.06 & 0.01 \\
     &       &       & 381285 & 1.69E-12 &       & -12.08 &       &  \\
    184.0-02.1 & M 1-5 & 1.0   & 379317 & 9.58E-12 & 0.45(b,j) & -11.18 & -11.19 & 0.01 \\
     &       &       & 379320 & 9.32E-12 &       & -11.19 &       &  \\
    184.6+00.6 & K 3-70 & 0.7 $\times$ 0.7 & 381765 & 1.06E-12 & 3.60(d,e) & -12.64 & -12.64 & 0.01 \\
     &       &       & 381768 & 1.04E-12 &       & -12.65 &       &  \\
    184.8+04.4 & K 3-71 & 1.7   & 485492 & 3.79E-13 & 0.03(d,e) & -12.44 & -12.44 &  \\

    194.2+02.5 & J 900 & 2.7   & 545436 & 3.08E-11 & 0.18(o,u) & -10.58 & -10.58 & 0.01 \\
     &       &       & 545439 & 3.09E-11 &       & -10.58 &       &  \\

    197.8-03.3 & Abell 14 & 35 $\times$ 18.0 & 477563 & 1.77E-12 & 6.20(a) & -12.61 & -12.61 & 0.01 \\
     &       &       & 478381 & 1.79E-12 &       & -12.60 &       &  \\
    201.7+02.5 & K 4-48 & 1.0   & 603928 & 2.05E-12 & 0.96(a) & -11.98 & -11.97 & 0.01 \\
     &       &       & 603931 & 2.14E-12 &       & -11.96 &       &  \\
    201.9-04.6 & We 1-4 & 12.6 $\times$ 33 & 478856 & 1.81E-12 & 5.35(k) & -12.54 & -12.52 & 0.03 \\
     &       &       & 478859 & 2.07E-12 &       & -12.49 &       &  \\
    
    
    210.0+03.9 & We 2-34 & 222 & 597464 & 8.52E-12 & 4.00(a) & -11.77 & -11.77 &  \\

    210.3+01.9 & M 1-8  & 6.7 $\times$ 2.7 & 599898 & 7.05E-12 & 1.46(a) & -11.54 & -11.54 &  \\
    212.0+04.3 & M 1-9 & 0.7   & 599461 & 9.43E-12 & 0.30(a) & -11.14 & -11.14 &  \\  
    
\bottomrule
\bottomrule

\end{tabular}
\end{table*}

%% file: Tables/HB_Radio_Fluxes_andReferences.tex
\begin{table*}																											
\centering																											
  \caption{{\HB} fluxes and radio fluxes at 1.4, 5 and 30 GHz}																											
  \label{Fluxsources}																											
  \begin{tabular}{llllllllllllll}																											
  																											
\hline																											
& & \multicolumn{3}{l}{logF({\HB})} & \multicolumn{3}{l}{{S$_\nu$(30 GHz)}} & \multicolumn{3}{l}{{S$_\nu$(5 GHz)}} & \multicolumn{3}{l}{{S$_\nu$(1.4 GHz)}}\\																											
& & ergs cm$^{-2}$ &&&&&&&&&&& \\																											
PN G & Name & s$^{-1}$ & unc. & ref. & mJy & unc. & ref. & mJy & unc. & ref. & mJy & unc. & ref. \\ 																											
\hline  																											
    																											
030.8+03.4	&	Abell 47	&	--	&	--	&	--	&	--	&	--	&	--	&	55	&	11	&	(d)	&	7.9	&	0.6	&	(m)	\\
031.0+04.1	&	K 3-6	&	--	&	--	&	--	&	98.6	&	8.3	&	(a)	&	--	&	--	&	--	&	10.9	&	0.6	&	(m)	\\
031.3-00.5	&	HaTr 10	&	--	&	--	&	--	&	--	&	--	&	--	&	--	&	--	&	--	&	5	&	0.5	&	(n)	\\
031.7+01.7	&	PC 20	&	--	&	--	&	--	&	--	&	--	&	--	&	--	&	--	&	--	&	27.4	&	0.9	&	(m)	\\
032.0-03.0	&	K 3-18	&	--	&	--	&	--	&	--	&	--	&	--	&	11	&	1	&	(c)	&	--	&	--	&	--	\\
032.5-03.2	&	K 3-20	&	--	&	--	&	--	&	--	&	--	&	--	&	--	&	--	&	--	&	6.3	&	0.6	&	(o)	\\
032.9-02.8	&	K 3-19	&	--	&	--	&	--	&	--	&	--	&	--	&	22.5 &	0.06	&	(y)	&	12.4	&	0.7	&	(m)	\\
033.8-02.6	&	NGC 6741	&	-11.34	&	0.01	&	(a)	&	--	&	--	&	--	&	197	&	10	&	(g, i)	&	132.8	&	4.8	&	(m)	\\
035.7-05.0	&	K 3-26	&	--	&	--	&	--	&	--	&	--	&	--	&	--	&	--	&	--	&	6.8	&	0.5	&	(m)	\\
035.9-01.1	&	Sh 2-71	&	-11.6	&	0.05	&	(b)	&	--	&	--	&	--	&	66	&	7	&	(m)	&	57.9	&	3.1	&	(m)	\\
036.9-02.6	&	HaTr 13	&	--	&	--	&	--	&	--	&	--	&	--	&	--	&	--	&	--	&	3.6	&	0.5	&	(p)	\\
036.9-01.1	&	HaTr 11	&	--	&	--	&	--	&	--	&	--	&	--	&	--	&	--	&	--	&	17	&	0.7	&	(p)	\\
039.5-02.7	&	M 2-47	&	--	&	--	&	--	&	--	&	--	&	--	&	49	&	6	&	(c, h)	&	39.1	&	1.7	&	(m)	\\
039.8+02.1	&	K 3-17	&	--	&	--	&	--	&	303	&	9	&	(a)	&	398	&	43	&	(c, h)	&	320	&	9.6	&	(m)	\\
040.3-00.4	&	Abell  53	&	-12.61	&	0.05	&	(b)	&	54.9	&	6.5	&	(a)	&	76	&	10	&	(b)	&	34.1	&	1.1	&	(m)	\\
040.4-03.1	&	K 3-30	&	--	&	--	&	--	&	--	&	--	&	--	&	23	&	5	&	(d)	&	12.7	&	0.6	&	(m)	\\
041.8+04.4	&	K 3-15	&	--	&	--	&	--	&	--	&	--	&	--	&	6.5	&	0.1	&	(y)	&	--	&	--	&	--	\\
043.0-03.0	&	M 4-14	&	--	&	--	&	--	&	--	&	--	&	--	&	--	&	--	&	--	&	11.9	&	0.6	&	(m)	\\
043.1+03.8	&	M 1-65	&	--	&	--	&	--	&	--	&	--	&	--	&	23	&	1.2	&	(d, i)	&	21.3	&	0.8	&	(m)	\\
043.3+02.2	&	PM 1-276	&	--	&	--	&	--	&	--	&	--	&	--	&	--	&	--	&	--	&	15.3	&	0.6	&	(q)	\\
045.9-01.9	&	K 3-33	&	--	&	--	&	--	&	--	&	--	&	--	&	17	&	3	&	(d)	&	9.1	&	0.5	&	(m)	\\
046.3-03.1	&	PB 9	&	--	&	--	&	--	&	--	&	--	&	--	&	40	&	8	&	(d)	&	33.6	&	1.1	&	(m)	\\
046.4-04.1	&	NGC 6803	&	-11.18	&	0.01	&	(c, p, d)	&	83.8	&	6.3	&	(a)	&	96.6	&	7.1	&	(g, i, l)	&	69.4	&	2.1	&	(m)	\\
046.8+02.9	&	CTSS 4	&	--	&	--	&	--	&	--	&	--	&	--	&	--	&	--	&	--	&	2.5	&	0.3	&	(n, m)	\\
047.1+04.1	&	K 3-21	&	--	&	--	&	--	&	--	&	--	&	--	&	--	&	--	&	--	&	2.5	&	0.3	&	(r, m)	\\
047.1-04.2	&	Abell 62	&	--	&	--	&	--	&	--	&	--	&	--	&	--	&	--	&	--	&	--	&	--	&	--	\\
048.0-02.3	&	PB 10	&	--	&	--	&	--	&	37.1	&	6.5	&	(a)	&	50	&	10	&	(d)	&	50.7	&	1.6	&	(m)	\\
048.1+01.1	&	K 3-29	&	--	&	--	&	--	&	71.2	&	5.8	&	(a)	&	69.1	&	0.06	&	(y)	&	14.2	&	0.6	&	(m)	\\
048.5+04.2	&	K 4-16	&	--	&	--	&	--	&	--	&	--	&	--	&	3	&	1	&	(d)	&	5.2	&	0.5	&	(m)	\\
048.7+02.3	&	K 3-24	&	--	&	--	&	--	&	--	&	--	&	--	&	26	&	2.6	&	(h)	&	9	&	0.5	&	(m)	\\
048.7+01.9	&	Hen 2-429	&	--	&	--	&	--	&	62.1	&	6.1	&	(a)	&	69	&	6.9	&	(b)	&	59	&	1.8	&	(m)	\\
049.4+02.4	&	Hen 2-428	&	--	&	--	&	--	&	--	&	--	&	--	&	--	&	--	&	--	&	28.2	&	1.3	&	(m)	\\
050.4-01.6	&	K 4-28	&	--	&	--	&	--	&	--	&	--	&	--	&	19	&	4	&	(d)	&	5.4	&	0.5	&	(m)	\\
051.0-04.5	&	PC 22	&	--	&	--	&	--	&	--	&	--	&	--	&	12	&	1.2	&	(b)	&	9.7	&	0.5	&	(m)	\\
051.0+03.0	&	Hen 2-430	&	--	&	--	&	--	&	37.7	&	6.7	&	(a)	&	38.5	&	4	&	(g)	&	18.6	&	0.7	&	(m)	\\
051.3+01.8	&	PM 1-295	&	--	&	--	&	--	&	--	&	--	&	--	&	--	&	--	&	--	&	13.9	&	0.6	&	(q)	\\
051.9-03.8	&	M 1-73	&	--	&	--	&	--	&	36.5	&	4.8	&	(a)	&	43	&	5	&	(c, g)	&	48.1	&	1.5	&	(m)	\\
052.2-04.0	&	M 1-74	&	-11.75	&	0.02	&	(e)	&	--	&	--	&	--	&	31	&	4	&	(c)	&	7.2	&	0.7	&	(m)	\\
052.5-02.9	&	Me 1-1	&	-11.94	&	0.03	&	(f)	&	--	&	--	&	--	&	43	&	5	&	(c)	&	36.3	&	1.2	&	(m)	\\
052.9-02.7	&	K 3-41	&	--	&	--	&	--	&	--	&	--	&	--	&	1.7	&	0.3	&	(d)	&	4.7	&	0.9	&	(m)	\\
052.9+02.7	&	K 3-31	&	--	&	--	&	--	&	--	&	--	&	--	&	39	&	8	&	(d)	&	17.3	&	0.7	&	(m)	\\
053.2-01.5	&	K 3-38	&	--	&	--	&	--	&	--	&	--	&	--	&	25.2	&	2.5	&	(g)	&	29.2	&	1.1	&	(m)	\\
053.8-03.0	&	Abell 63	&	--	&	--	&	--	&	--	&	--	&	--	&	--	&	--	&	--	&	4.9	&	0.5	&	(n)	\\
054.2-03.4	&	Necklace	&	--	&	--	&	--	&	--	&	--	&	--	&	--	&	--	&	--	&	5	&	0.5	&	(m)	\\
054.4-02.5	&	M 1-72	&	--	&	--	&	--	&	--	&	--	&	--	&	26	&	5	&	(d)	&	5.7	&	0.5	&	(m)	\\
055.1-01.8	&	K 3-43	&	--	&	--	&	--	&	--	&	--	&	--	&	--	&	--	&	--	&	5.7	&	0.9	&	(m)	\\
055.2+02.8	&	Hen 2-432	&	--	&	--	&	--	&	--	&	--	&	--	&	33.5	&	5	&	(d, h, g)	&	23.2	&	0.9	&	(m)	\\
055.3+02.7	&	Hen 1-1	&	--	&	--	&	--	&	--	&	--	&	--	&	14	&	1.4	&	(e, c)	&	17.3	&	0.7	&	(m)	\\
055.5-00.5	&	M 1-71	&	--	&	--	&	--	&	178.9	&	6.3	&	(a)	&	203.5	&	13.8	&	(c, z)	&	--	&	--	&	--	\\
055.6+02.1	&	Hen 1-2	&	--	&	--	&	--	&		&		&		&	15	&	1.5	&	(e)	&	16.1	&	1.1	&	(m)	\\
056.0+02.0	&	K 3-35	&	--	&	--	&	--	&	60.1	&	6.4	&	(a)	&	40	&	8	&	(d)	&	14.6	&	0.6	&	(m)	\\
056.4-00.9	&	K 3-42	&	--	&	--	&	--	&		&		&		&	18.6	&	1.7	&	(z)	&	11.6	&	0.8	&	(m)	\\
057.9-01.5	&	Hen 2-447	&	--	&	--	&	--	&	65.5	&	7.1	&	(a)	&	62.3	&	0.06	&	(y)	&	23.6	&	1.2	&	(m)	\\
058.9+01.3	&	K 3-40	&	--	&	--	&	--	&	--	&	--	&	--	&	20	&	4	&	(d)	&	17.2	&	0.7	&	(m)	\\
059.4+02.3	&	K 3-37	&	--	&	--	&	--	&	--	&	--	&	--	&	17	&	3	&	(d)	&	14.7	&	0.6	&	(m)	\\
059.0+04.6	&	K 3-34	&	--	&	--	&	--	&	--	&	--	&	--	&	--	&	--	&	--	&	2.9	&	0.5	&	(r)	\\
059.9+02.0	&	K 3-39	&	--	&	--	&	--	&	--	&	--	&	--	&	11.6	&	0.10	&	(y)	&	3.7	&	0.5	&	(m)	\\
060.4+01.5	&	PM 1-310	&	--	&	--	&	--	&	--	&	--	&	--	&	--	&	--	&	--	&	2.5	&	0.3	&	(s, m)	\\
060.5-00.3	&	K 3-45	&	--	&	--	&	--	&	--	&	--	&	--	&	--	&	--	&	--	&	2.5	&	0.3	&	(r, m)	\\
060.5+01.8	&	Hen 2-440	&	--	&	--	&	--	&	--	&	--	&	--	&	40.6	&	0.06	&	(y)	&	26.9	&	0.9	&	(m)	\\
060.8-03.6	&	NGC 6853	&	-9.46	&	0.06	&	(g)	&	--	&	--	&	--	&	1325	&	19	&	(m)	&	--	&	--	&	--	\\
																											
\bottomrule																											
\end{tabular}																											
\end{table*}%
    																											
\begin{table*}																											
\centering																											
  \contcaption{{\HB} fluxes and radio fluxes at 1.4, 5 and 30 GHz}																											
  \begin{tabular}{llllllllllllll}																											
  																											
\hline									
& & \multicolumn{3}{l}{logF({\HB})} & \multicolumn{3}{l}{{S$_\nu$(30 GHz)}} & \multicolumn{3}{l}{{S$_\nu$(5 GHz)}} & \multicolumn{3}{l}{{S$_\nu$(1.4 GHz)}}\\

& & ergs cm$^{-2}$ &&&&&&&&&&& \\																											
PN G & Name & s$^{-1}$ & unc. & ref. & mJy & unc. & ref. & mJy & unc. & ref. & mJy & unc. & ref. \\ 																											
\hline 																											
																											
062.4-00.2	&	M 2-48	&	--	&	--	&	--	&	--	&	--	&	--	&	19.6	&	1.4	&	(c, z)	&	17	&	0.7	&	(m)	\\
063.8-03.3	&	K 3-54	&	--	&	--	&	--	&	--	&	--	&	--	&	7.2	&	0.9	&	(d, g)	&	5.1	&	0.5	&	(m)	\\
064.9-02.1	&	K 3-53	&	--	&	--	&	--	&	73.8	&	5.3	&	(a)	&	50	&	10	&	(d)	&	10.3	&	0.5	&	(m)	\\
065.1-03.5	&	We 1-9	&	--	&	--	&	--	&	--	&	--	&	--	&	--	&	--	&	--	&	9.9	&	0.5	&	(m)	\\
065.9+00.5	&	NGC 6842	&	-11.73	&	0.03	&	(f)	&	--	&	--	&	--	&	127	&	12.7	&	(j)	&	43.9	&	2.1	&	(m)	\\
066.9+02.2	&	K 4-37	&	--	&	--	&	--	&	--	&	--	&	--	&	--	&	--	&	--	&	2.5	&	0.3	&	(r, m)	\\
067.9-00.2	&	K 3-52	&	--	&	--	&	--	&	90.2	&	9.9	&	(a)	&	65	&	13	&	(d)	&	17.3	&	0.7	&	(m)	\\
068.3-02.7	&	Hen 2-459	&	--	&	--	&	--	&	102.5	&	6.9	&	(a)	&	86.6	&	0.08	&	(y)	&	14	&	0.6	&	(m)	\\
068.6+01.1	&	Hen 1-4	&	--	&	--	&	--	&	--	&	--	&	--	&	--	&	--	&	--	&	16.2	&	0.7	&	(n)	\\
068.7+01.9	&	K 4-41	&	--	&	--	&	--	&	--	&	--	&	--	&	15	&	3	&	(d)	&	11.8	&	0.6	&	(m)	\\
068.7+03.0	&	PC 23	&	--	&	--	&	--	&	--	&	--	&	--	&	18	&	2	&	(c, d)	&	11	&	1.2	&	(m)	\\
068.8-00.0	&	M 1-75	&	--	&	--	&	--	&	--	&	--	&	--	&	26	&	2.6	&	(h)	&	37	&	1.2	&	(m)	\\
069.2+03.8	&	K 3-46	&	--	&	--	&	--	&	--	&	--	&	--	&	--	&	--	&	--	&	3.3	&	0.5	&	(r)	\\
069.2+02.8	&	K 3-49	&	--	&	--	&	--	&	--	&	--	&	--	&	5.9	&	0.05	&	(y)	&	--	&	--	&	--	\\
069.4-02.6	&	NGC 6894	&	-11.41	&	0.05	&	(b, c)	&	--	&	--	&	--	&	61	&	7	&	(c)	&	--	&	--	&	--	\\
069.6-03.9	&	K 3-58	&	--	&	--	&	--	&	--	&	--	&	--	&	31	&	3	&	(c)	&	5.7	&	0.5	&	(m)	\\
069.7+00.0	&	K 3-55	&	--	&	--	&	--	&	76.9	&	7	&	(a)	&	90	&	18	&	(d)	&	87.1	&	2.6	&	(m)	\\
071.6-02.3	&	M 3-35	&	-12.45	&	0.06	&	(e)	&	161.8	&	6.4	&	(a)	&	150	&	28	&	(d, i)	&	29.9	&	1	&	(m)	\\
072.1+00.1	&	K 3-57	&	--	&	--	&	--	&	--	&	--	&	--	&	60	&	12	&	(d)	&	47.3	&	1.5	&	(m)	\\
073.0-02.4	&	K 3-76	&	--	&	--	&	--	&	--	&	--	&	--	&	12	&	2	&	(d)	&	--	&	--	&	--	\\
074.5+02.1	&	NGC 6881	&	-12.26	&	0.03	&	(h)	&	109.1	&	7.4	&	(a)	&	110	&	0.08	&	(y)	&	70.9	&	2.2	&	(m)	\\
075.6+04.3	&	ARO 342	&	--	&	--	&	--	&	--	&	--	&	--	&	--	&	--	&	--	&	2.5	&	0.3	&	(n, m)	\\
076.3+01.1	&	Abell 69	&	-14.1	&	0.2	&	(i)	&	--	&	--	&	--	&	5	&	0.5	&	(t, m)	&	2.5	&	0.3	&	(n, m)	\\
076.4+01.8	&	KjPn 3	&	--	&	--	&	--	&	--	&	--	&	--	&	--	&	--	&	--	&	2.5	&	0.3	&	(n, m)	\\
077.5+03.7	&	KjPn 1	&	--	&	--	&	--	&	--	&	--	&	--	&	2.8	&	0.2	&	(u)	&	--	&	--	&	--	\\
077.7+03.1	&	KjPn 2	&	--	&	--	&	--	&	--	&	--	&	--	&	--	&	--	&	--	&	10	&	1	&	(r)	\\
078.3-02.7	&	K 4-53	&	--	&	--	&	--	&	44.4	&	6.3	&	(a)	&	--	&	--	&	--	&	35.1	&	1.5	&	(m)	\\
078.9+00.7	&	Sd 1	&	--	&	--	&	--	&	--	&	--	&	--	&	27	&	2.7	&	(m)	&	2.5	&	0.3	&	(v, m)	\\
084.2+01.0	&	K 4-55	&	--	&	--	&	--	&	--	&	--	&	--	&	--	&	--	&	--	&	6.6	&	0.6	&	(r)	\\
084.9+04.4	&	Abell 71	&	-11.75	&	0.02	&	(i)	&	--	&	--	&	--	&	83	&	8.3	&	(t, m)	&	5	&	0.5	&	(f, m)	\\
088.7+04.6	&	K 3-78	&	--	&	--	&	--	&	--	&	--	&	--	&	17	&	2	&	(c, d) 	&	15.2	&	0.6	&	(m)	\\
088.7-01.6	&	NGC 7048	&	-11.41	&	0.01	&	(f)	&	--	&	--	&	--	&	37	&	4	&	(c)	&	46.8	&	2.1	&	(m)	\\
089.0+00.3	&	NGC 7026	&	-10.9	&	0.02	&	(a, j)	&	208.2	&	9.5	&	(a)	&	277	&	77	&	(j)	&	267	&	9.5	&	(m)	\\
089.8-00.6	&	Sh 1-89	&	--	&	--	&	--	&	--	&	--	&	--	&	--	&	--	&	--	&	14.7	&	1.6	&	(m)	\\
091.6-04.8	&	K 3-84	&	--	&	--	&	--	&	--	&	--	&	--	&	--	&	--	&	--	&	3	&	0.4	&	(r)	\\
091.6+01.8	&	We 1-11	&	--	&	--	&	--	&	--	&	--	&	--	&	--	&	--	&	--	&	13.3	&	0.6	&	(m)	\\
093.3-00.9	&	K 3-82	&	--	&	--	&	--	&	--	&	--	&	--	&	30	&	3	&	(c)	&	37.3	&	1.6	&	(m)	\\
093.3-02.4	&	M 1-79	&	-11.73	&	0.01	&	(b)	&	--	&	--	&	--	&	19	&	1.9	&	(c)	&	23.6	&	1.4	&	(m)	\\
094.5-00.8	&	K 3-83	&	--	&	--	&	--	&	--	&	--	&	--	&	6.5	&	0.7	&	(c)	&	5.1	&	0.5	&	(m)	\\
095.1-02.0	&	M 2-49	&	--	&	--	&	--	&	--	&	--	&	--	&	74	&	8	&	(d, g)	&	28.3	&	1	&	(m)	\\
095.2+00.7	&	K 3-62	&	--	&	--	&	--	&	101.3	&	7.8	&	(a)	&	118	&	0.10	&	(y)	&	59.9	&	1.9	&	(m)	\\
096.3+02.3	&	K 3-61	&	--	&	--	&	--	&	--	&	--	&	--	&	14	&	1.4	&	(c)	&	16.9	&	0.7	&	(m)	\\
097.6-02.4	&	M 2-50	&	-12.48	&	0.03	&	(e)	&	--	&	--	&	--	&	6.5	&	0.7	&	(c)	&	9.1	&	1.1	&	(m)	\\
098.1+02.4	&	K 3-63	&	-13.97	&	0.06	&	(b)	&	--	&	--	&	--	&	29	&	2.9	&	(c)	&	8.6	&	0.5	&	(m)	\\
098.2+04.9	&	K 3-60	&	-13.17	&	0.18	&	(b)	&	--	&	--	&	--	&	43	&	9	&	(d)	&	28.1	&	0.9	&	(m)	\\
102.8-05.0	&	Abell 80	&	-11.8	&	0.12	&	(b)	&	--	&	--	&	--	&	3	&	0.3	&	(b)	&	9.4	&	2	&	(m)	\\
103.2+00.6	&	M 2-51	&	-11.97	&	0.04	&	(k)	&	--	&	--	&	--	&	41	&	4.1	&	(t, m)	&	28.4	&	1.7	&	(w)	\\
103.7+00.4	&	M 2-52	&	--	&	--	&	--	&	--	&	--	&	--	&	14	&	1.4	&	(c)	&	15.4	&	0.6	&	(m)	\\
104.1+01.0	&	Bl 2-1	&	--	&	--	&	--	&	--	&	--	&	--	&	61.2	&	0.11	&	(y)	&	22	&	0.8	&	(m)	\\
104.4-01.6	&	M 2-53	&	--	&	--	&	--	&	--	&	--	&	--	&	11	&	1.1	&	(c)	&	15.1	&	0.6	&	(m)	\\
107.4-02.6	&	K 3-87	&	--	&	--	&	--	&	--	&	--	&	--	&	4.5	&	0.5	&	(c)	&	11.5	&	1.1	&	(m)	\\
107.4-00.6	&	K 4-57	&	--	&	--	&	--	&	--	&	--	&	--	&	--	&	--	&	--	&	2.5	&	0.3	&	(n, m)	\\
107.7-02.2	&	M 1-80	&	--	&	--	&	--	&	--	&	--	&	--	&	25	&	3	&	(c)	&	18	&	0.7	&	(m)	\\
107.8+02.3	&	NGC 7354	&	-11.58	&	0.01	&	(a, f)	&	439	&	13	&	(a)	&	597	&	76	&	(j)	&	582	&	23	&	(m)	\\
112.5-00.1	&	KjPn 8	&	--	&	--	&	--	&	--	&	--	&	--	&	1	&	0	&	(m)	&	2.5	&	0.3	&	(n, m)	\\
112.5+03.7	&	K 3-88	&	--	&	--	&	--	&	--	&	--	&	--	&	--	&	--	&	--	&	5.8	&	0.9	&	(m)	\\
119.3+00.3	&	BV 5-1	&	-12.7	&	0.03	&	(h)	&	--	&	--	&	--	&	--	&	--	&	--	&	11.7	&	1.2	&	(m)	\\
121.6+00.0	&	BV 5-2	&	--	&	--	&	--	&	--	&	--	&	--	&	--	&	--	&	--	&	9.3	&	1.4	&	(m)	\\
121.6+03.5	&	We 1-1	&	--	&	--	&	--	&	--	&	--	&	--	&	--	&	--	&	--	&	5.6	&	0.5	&	(m)	\\
122.1-04.9	&	Abell  2	&	-12.37	&	0.03	&	(b)	&	--	&	--	&	--	&	2.3	&	0.2	&	(c)	&	9.4	&	1.1	&	(m)	\\
																											
\bottomrule																											
\end{tabular}																											
\end{table*}

\begin{table*}																											
\centering																											
  \contcaption{{\HB} fluxes and radio fluxes at 1.4, 5 and 30 GHz}																											
  \begin{tabular}{llllllllllllll}																											
  																											
\hline																			
& & \multicolumn{3}{l}{logF({\HB})} & \multicolumn{3}{l}{{S$_\nu$(30 GHz)}} & \multicolumn{3}{l}{{S$_\nu$(5 GHz)}} & \multicolumn{3}{l}{{S$_\nu$(1.4 GHz)}}\\

& & ergs cm$^{-2}$ &&&&&&&&&&& \\																											
PN G & Name &  s$^{-1}$ & unc. & ref. & mJy & unc. & ref. & mJy & unc. & ref. & mJy & unc. & ref. \\ 																											
\hline 																											
																											
126.3+02.9	&	K 3-90	&	--	&	--	&	--	&	--	&	--	&	--	&	13.9	&	1.4	&	(c)	&	12.8	&	0.6	&	(m)	\\
126.6+01.3	&	IPHAS PN-1	&	--	&	--	&	--	&	--	&	--	&	--	&	--	&	--	&	--	&	4.2	&	0.5	&	(u)	\\
129.5+04.5	&	K 3-91	&	--	&	--	&	--	&	--	&	--	&	--	&	1.5	&	0.2	&	(c)	&	3.4	&	0.4	&	(m)	\\
130.2+01.3	&	IC 1747	&	-11.49	&	0.03	&	(a, f)	&	67.4	&	3.6	&	(a)	&	128	&	24	&	(j)	&	85.9	&	2.6	&	(m)	\\
130.4+03.1	&	K 3-92	&	--	&	--	&	--	&	--	&	--	&	--	&	2	&	0.2	&	(b)	&	--	&	--	&	--	\\
131.5+02.6	&	Abell 3	&	-12.61	&	0.1	&	(b)	&	--	&	--	&	--	&	2.7	&	0.3	&	(c)	&	7.8	&	1.3	&	(m)	\\
132.4+04.7	&	K 3-93	&	--	&	--	&	--	&	--	&	--	&	--	&	--	&	--	&	--	&	3.5	&	0.6	&	(m)	\\
136.1+04.9	&	Abell 6	&	-12.43	&	0.06	&	(b)	&	--	&	--	&	--	&	12	&	1.2	&	(b)	&	--	&	--	&	--	\\
138.8+02.8	&	IC 289	&	-11.69	&	0.01	&	(f)	&	98.8	&	4	&	(a)	&	212	&	49	&	(j)	&	153.1	&	5.8	&	(m)	\\
142.1+03.4	&	K 3-94	&	--	&	--	&	--	&	--	&	--	&	--	&	5.5	&	0.6	&	(c)	&	7.1	&	0.5	&	(m)	\\
147.4-02.3	&	M 1-4	&	-12.14	&	0.05	&	(h)	&	59.4	&	5.5	&	(a)	&	84.5	&	3.9	&	(d, g)	&	78.2	&	2.4	&	(m)	\\
147.8+04.1	&	M 2-2	&	-12.22	&	0.02	&	(b)	&	--	&	--	&	--	&	54	&	5	&	(c)	&	52.5	&	1.6	&	(m)	\\
149.0+04.4	&	K 4-47	&	--	&	--	&	--	&	--	&	--	&	--	&	--	&	--	&	--	&	--	&	--	&	--	\\
151.4+00.5	&	K 3-64	&	-13.82	&	0.2	&	(l)	&	--	&	--	&	--	&	--	&	--	&	--	&	6.5	&	0.9	&	(n)	\\
153.7-01.4	&	K 3-65	&	--	&	--	&	--	&	--	&	--	&	--	&	--	&	--	&	--	&	--	&	--	&	--	\\
160.5-00.5	&	We 1-2	&	--	&	--	&	--	&	--	&	--	&	--	&	--	&	--	&	--	&	--	&	--	&	--	\\
163.1-00.8	&	We 1-3	&	--	&	--	&	--	&	--	&	--	&	--	&	--	&	--	&	--	&	--	&	--	&	--	\\
173.5+03.2	&	Pu 2	&	--	&	--	&	--	&	--	&	--	&	--	&	--	&	--	&	--	&	6.2	&	0.5	&	(n)	\\
178.3-02.5	&	K 3-68	&	--	&	--	&	--	&	--	&	--	&	--	&	5	&	0.5	&	(c)	&	8.7	&	0.5	&	(m)	\\
181.5+00.9	&	Pu 1	&	--	&	--	&	--	&	--	&	--	&	--	&	7	&	0.1	&	(t, m)	&	2.7	&	0.9	&	(x)	\\
184.0-02.1	&	M 1-5	&	-12.05	&	0.01	&	(e)	&	49.3	&	2.8	&	(a)	&	71	&	9	&	(d, g)	&	42.8	&	1.7	&	(m)	\\
184.6+00.6	&	K 3-70	&	--	&	--	&	--	&	--	&	--	&	--	&	6	&	1	&	(d)	&	4.7	&	0.5	&	(m)	\\
184.8+04.4	&	K 3-71	&	--	&	--	&	--	&	--	&	--	&	--	&	--	&	--	&	--	&	4.9	&	0.5	&	(m)	\\
194.2+02.5	&	J 900	&	-11.32	&	0.02	&	(m)	&	86.1	&	4.8	&	(a)	&	110	&	8	&	(d, g, i)	&	108.4	&	3.3	&	(m)	\\
201.7+02.5	&	K 4-48	&	-12.94	&	0.07	&	(m)	&		&		&		&	--	&	--	&	--	&	12.1	&	0.6	&	(m)	\\
201.9-04.6	&	We 1-4	&	--	&	--	&	--	&	--	&	--	&	--	&	--	&	--	&	--	&	--	&	--	&	--	\\
210.0+03.9	&	We 2-34	&	--	&	--	&	--	&	--	&	--	&	--	&	--	&	--	&	--	&	--	&	--	&	--	\\
210.3+01.9	&	M 1-8	&	-12.37	&	0.02	&	(n)	&	--	&	--	&	--	&	23	&	7	&	(k)	&	16.6	&	0.6	&	(m)	\\
212.0+04.3	&	M 1-9	&	-11.66	&	0.03	&	(o)	&	--	&	--	&	--	&	30	&	5	&	(d)	&	23.9	&	1.1	&	(m)	\\

\bottomrule																											
\end{tabular}																											
\end{table*}

%% file: Tables/KnownPNe_Av_EHa-HB.tex
\begin{table*}
  \centering
  \caption{{\EHaHB}, {\CHB} and {\AvE} values for catalogued PNe}
  \label{Av_EHa-HB}
    \begin{tabular}{cccccccc}
    
    \hline
    \hline
    \\
    \textbf{PN G } & \textbf{Name} & \textbf{{\EHaHB}} & \textbf{unc.} & \textbf{{\CHB}} & \textbf{unc.} & \textbf{{\AvE}} & \textbf{unc.} \\
    \\
    \hline
    033.8-02.6 & NGC 6741 & 0.54  & 0.07  & 0.68  & 0.09  & 1.45  & 0.18 \\
    035.9-01.1 & Sh 2-71 & 1.04  & 0.24  & 1.3   & 0.3   & 2.78  & 0.66 \\
    040.3-00.4 & Abell 53  & 1.27  & 0.17  & 1.59  & 0.21  & 3.4   & 0.46 \\
    046.4-04.1 & NGC 6803 & 0.19  & 0.07  & 0.24  & 0.09  & 0.51  & 0.18 \\
    052.2-04.0 & M 1-74 & 0.42  & 0.09  & 0.52  & 0.12  & 1.11  & 0.25 \\
    052.5-02.9 & Me 1-1 & 1.46  & 0.12  & 1.83  & 0.15  & 3.9   & 0.32 \\
    060.8-03.6 & NGC 6853 & 0.13  & 0.2   & 0.17  & 0.25  & 0.36  & 0.53 \\
    065.9+00.5 & NGC 6842 & 0.46  & 0.12  & 0.57  & 0.15  & 1.23  & 0.32 \\
    069.4-02.6 & NGC 6894 & 0.79  & 0.17  & 0.99  & 0.21  & 2.11  & 0.46 \\
    071.6-02.3 & M 3-35 & 1.97  & 0.2   & 2.46  & 0.25  & 5.25  & 0.55 \\
    074.5+02.1 & NGC 6881 & 1.48  & 0.12  & 1.85  & 0.15  & 3.94  & 0.32 \\
    076.3+01.1 & Abell 69  & 2.48  & 0.62  & 3.1   & 0.78  & 6.63  & 1.66 \\
    084.9+04.4 & Abell 71  & 0.98  & 0.11  & 1.22  & 0.13  & 2.61  & 0.28 \\
    088.7-01.6 & NGC 7048 & 0.53  & 0.07  & 0.66  & 0.09  & 1.41  & 0.18 \\
    089.0+00.3 & NGC 7026 & 0.51  & 0.09  & 0.64  & 0.12  & 1.37  & 0.25 \\
    093.3-02.4 & M 1-79 & 0.93  & 0.07  & 1.17  & 0.09  & 2.49  & 0.18 \\
    097.6-02.4 & M 2-50 & 0.43  & 0.12  & 0.54  & 0.15  & 1.15  & 0.32 \\
    098.1+02.4 & K 3-63 & 0.94  & 0.22  & 1.17  & 0.28  & 2.5   & 0.61 \\
    098.2+04.9 & K 3-60 & 1.56  & 0.55  & 1.95  & 0.69  & 4.15  & 1.48 \\
    102.8-05.0 & Abell 80  & 0.46  & 0.36  & 0.57  & 0.45  & 1.23  & 0.97 \\
    103.2+00.6 & M 2-51 & 1.25  & 0.15  & 1.56  & 0.19  & 3.33  & 0.41 \\
    107.8+02.3 & NGC 7354 & 1.15  & 0.07  & 1.44  & 0.09  & 3.08  & 0.18 \\
    119.3+00.3 & BV 5-1 & 1.05  & 0.12  & 1.32  & 0.15  & 2.81  & 0.32 \\
    122.1-04.9 & Abell 2   & 0.32  & 0.12  & 0.4   & 0.15  & 0.86  & 0.32 \\
    130.2+01.3 & IC 1747 & 0.65  & 0.12  & 0.82  & 0.15  & 1.74  & 0.32 \\
    131.5+02.6 & Abell 3   & 1.7   & 0.3   & 2.12  & 0.38  & 4.53  & 0.82 \\
    136.1+04.9 & Abell 6   & 0.91  & 0.34  & 1.14  & 0.42  & 2.42  & 0.91 \\
    138.8+02.8 & IC 289 & 1.01  & 0.07  & 1.26  & 0.09  & 2.69  & 0.18 \\
    147.4-02.3 & M 1-4 & 0.81  & 0.18  & 1.02  & 0.22  & 2.17  & 0.47 \\
    147.8+04.1 & M 2-2 & 1.25  & 0.09  & 1.56  & 0.12  & 3.34  & 0.25 \\
    151.4+00.5 & K 3-64 & 2.62  & 1.15  & 3.28  & 1.44  & 7.00  & 3.08 \\
    184.0-02.1 & M 1-5 & 1.02  & 0.07  & 1.28  & 0.09  & 2.73  & 0.18 \\
    184.8+04.4 & K 3-71 & 1.82  & 0.3   & 2.28  & 0.38  & 4.87  & 0.82 \\
    194.2+02.5 & J 900 & 0.71  & 0.09  & 0.88  & 0.12  & 1.89  & 0.25 \\
    201.7+02.5 & K 4-48 & 1.29  & 0.22  & 1.61  & 0.28  & 3.43  & 0.6 \\
    210.3+01.9 & M 1-8 & 0.93  & 0.12  & 1.16  & 0.15  & 2.49  & 0.32 \\
    212.0+04.3 & M 1-9 & 0.16  & 0.12  & 0.21  & 0.15  & 0.44  & 0.32 \\
    
    \bottomrule
    \bottomrule
    
    \end{tabular}%

\end{table*}%

%% file: Tables/KnownPNeRadio_Avrg.tex
\begin{landscape}																															
\begin{table}																															
\centering																															
  \caption{Calculated radio {\AvCHa}, {\AvCHB} values and weighted averaged {\AV}(Radio) values for catalogued PNe}																															
  \label{KnownPNeRadio}  																															
    \begin{tabular}{llllllllllllllll}																															
    																															
\hline																															
& & \multicolumn{4}{c}{30 GHz} & \multicolumn{4}{c}{5 GHz} & \multicolumn{4}{c}{1.4 GHz} & Weighted & \\																															
PN G & Name & {\AvCHa} & unc. & {\AvCHB} & unc. & {\AvCHa} & unc. & {\AvCHB} & unc. & {\AvCHa} & unc. & {\AvCHB} & unc. & Averaged & unc. \\																															
& & & & & & & & & & & & & & {\AV}(Radio) &\\																															
\hline																															
																															
030.8+03.4	&	Abell 47	&	--	&	--	&	--	&	--	&	\textbf{8.27}	&	\textbf{0.33}	&	--	&	--	&	5.45	&	0.16	&	--	&	--	&	8.27	&	0.33	\\
031.0+04.1	&	K 3-6	&	\textbf{7.18}	&	\textbf{0.17}	&	--	&	--	&	--	&	--	&	--	&	--	&	3.76	&	0.13	&	--	&	--	&	7.18	&	0.17	\\
031.3-00.5	&	HaTr 10	&	--	&	--	&	--	&	--	&	--	&	--	&	--	&	--	&	\textbf{3.26}	&	\textbf{0.19}	&	--	&	--	&	3.26	&	0.19	\\
031.7+01.7	&	PC 20	&	--	&	--	&	--	&	--	&	--	&	--	&	--	&	--	&	\textbf{5.45}	&	\textbf{0.10}	&	--	&	--	&	5.45	&	0.10	\\
032.0-03.0	&	K 3-18	&	--	&	--	&	--	&	--	&	\textbf{3.93}	&	\textbf{0.18}	&	--	&	--	&	--	&	--	&	--	&	--	&	3.93	&	0.18	\\
032.5-03.2	&	K 3-20	&	--	&	--	&	--	&	--	&	--	&	--	&	--	&	--	&	\textbf{2.13}	&	\textbf{0.19}	&	--	&	--	&	2.13	&	0.19	\\
032.9-02.8	&	K 3-19	&	--	&	--	&	--	&	--	&	\textbf{3.68}	&	\textbf{0.12}	&	--	&	--	&	\textbf{2.69}	&	\textbf{0.19}	&	--	&	--	&	3.40	&	0.10	\\		
033.8-02.6	&	NGC 6741	&	--	&	--	&	--	&	--	&	\textbf{2.56}	&	\textbf{0.12}	&	\textbf{2.21}	&	\textbf{0.07}	&	\textbf{1.85}	&	\textbf{0.03}	&	\textbf{1.72}	&	\textbf{0.05}	&	1.90	&	0.03	\\
035.7-05.0	&	K 3-26	&	--	&	--	&	--	&	--	&	--	&	--	&	--	&	--	&	\textbf{2.12}	&	\textbf{0.16}	&	--	&	--	&	2.12	&	0.16	\\
035.9-01.1	&	Sh 2-71	&	--	&	--	&	--	&	--	&	\textbf{1.33}	&	\textbf{0.29}	&	\textbf{1.80}	&	\textbf{0.21}	&	\textbf{0.98}	&	\textbf{0.22}	&	--	&	--	&	1.39	&	0.13	\\
036.9-02.6	&	HaTr 13	&	--	&	--	&	--	&	--	&	--	&	--	&	--	&	--	&	\textbf{2.82}	&	\textbf{0.25}	&	--	&	--	&	2.82	&	0.25	\\
036.9-01.1	&	HaTr 11	&	--	&	--	&	--	&	--	&	--	&	--	&	--	&	--	&	\textbf{5.00}	&	\textbf{0.11}	&	--	&	--	&	5.00	&	0.11	\\
039.5-02.7	&	M 2-47	&	--	&	--	&	--	&	--	&	\textbf{3.46}	&	\textbf{0.22}	&	--	&	--	&	\textbf{2.98}	&	\textbf{0.11}	&	--	&	--	&	3.08	&	0.10	\\
039.8+02.1	&	K 3-17	&	8.19	&	0.1	&	--	&	--	&	\textbf{8.32}	&	\textbf{0.20}	&	--	&	--	&	\textbf{7.85}	&	\textbf{0.10}	&	--	&	--	&	7.93	&	0.09	\\
040.3-00.4	&	Abell 53	&	\textbf{4.21}	&	\textbf{0.22}	&	\textbf{3.95}	&	\textbf{0.22}	&	\textbf{4.41}	&	\textbf{0.24}	&	\textbf{4.09}	&	\textbf{0.23}	&	3.15	&	0.1	&	--	&	--	&	4.16	&	0.11	\\
040.4-03.1	&	K 3-30	&	--	&	--	&	--	&	--	&	\textbf{3.51}	&	\textbf{0.36}	&	--	&	--	&	\textbf{2.52}	&	\textbf{0.12}	&	--	&	--	&	2.62	&	0.11	\\
041.8+04.4	&	K 3-15	&	--	&	--	&	--	&	--	&	\textbf{2.41}	&	\textbf{0.08}	&	--	&	--	&	--	&	--	&	--	&	--	&	2.41	&	0.08	\\
043.0-03.0	&	M 4-14	&	--	&	--	&	--	&	--	&	--	&	--	&	--	&	--	&	\textbf{2.72}	&	\textbf{0.06}	&	--	&	--	&	2.72	&	0.06	\\
043.1+03.8	&	M 1-65	&	--	&	--	&	--	&	--	&	\textbf{1.97}	&	\textbf{0.13}	&	--	&	--	&	\textbf{1.69}	&	\textbf{0.06}	&	--	&	--	&	1.74	&	0.05	\\
043.3+02.2	&	PM 1-276	&	--	&	--	&	--	&	--	&	--	&	--	&	--	&	--	&	\textbf{3.26}	&	\textbf{0.11}	&	--	&	--	&	3.26	&	0.11	\\
045.9-01.9	&	K 3-33	&	--	&	--	&	--	&	--	&	\textbf{6.29}	&	\textbf{0.30}	&	--	&	--	&	\textbf{5.27}	&	\textbf{0.06}	&	--	&	--	&	5.30	&	0.05	\\
046.3-03.1	&	PB 9	&	--	&	--	&	--	&	--	&	\textbf{3.79}	&	\textbf{0.33}	&	--	&	--	&	\textbf{3.37}	&	\textbf{0.06}	&	--	&	--	&	3.39	&	0.06	\\
046.4-04.1	&	NGC 6803	&	\textbf{1.93}	&	\textbf{0.16}	&	\textbf{1.47}	&	\textbf{0.09}	&	\textbf{1.88}	&	\textbf{0.16}	&	\textbf{1.44}	&	\textbf{0.09}	&	\textbf{1.26}	&	\textbf{0.06}	&	\textbf{1.02}	&	\textbf{0.05}	&	1.26	&	0.03	\\
046.8+02.9	&	CTSS 4	&	--	&	--	&	--	&	--	&	--	&	--	&	--	&	--	&	\textbf{3.03}	&	\textbf{0.19}	&	--	&	--	&	3.03	&	0.19	\\
047.1+04.1	&	K 3-21	&	--	&	--	&	--	&	--	&	--	&	--	&	--	&	--	&	\textbf{2.77}	&	\textbf{0.19}	&	--	&	--	&	2.77	&	0.19	\\
047.1-04.2	&	Abell 62	&	--	&	--	&	--	&	--	&	--	&	--	&	--	&	--	&	--	&	--	&	--	&	--	&	--	&	--	\\
048.0-02.3	&	PB 10	&	\textbf{3.56}	&	\textbf{0.30}	&	--	&	--	&	\textbf{3.73}	&	\textbf{0.33}	&	--	&	--	&	\textbf{3.57}	&	\textbf{0.12}	&	--	&	--	&	3.59	&	0.11	\\
048.1+01.1	&	K 3-29	&	\textbf{7.15}	&	\textbf{0.17}	&	--	&	--	&	\textbf{6.87}	&	\textbf{0.06}	&	--	&	--	&	4.53	&	0.07	&	--	&	--	&	6.89 &	0.05	\\
048.5+04.2	&	K 4-16	&	--	&	--	&	--	&	--	&	\textbf{2.35}	&	\textbf{0.53}	&	--	&	--	&	\textbf{2.92}	&	\textbf{0.06}	&	--	&	--	&	2.92	&	0.06	\\
048.7+02.3	&	K 3-24	&	--	&	--	&	--	&	--	&	\textbf{6.07}	&	\textbf{0.19}	&	--	&	--	&	4.45	&	0.07	&	--	&	--	&	6.07	&	0.19	\\
048.7+01.9	&	Hen 2-429	&	\textbf{3.77}	&	\textbf{0.19}	&	--	&	--	&	\textbf{3.67}	&	\textbf{0.19}	&	--	&	--	&	\textbf{3.28}	&	\textbf{0.14}	&	--	&	--	&	3.50	&	0.10	\\
049.4+02.4	&	Hen 2-428	&	--	&	--	&	--	&	--	&	--	&	--	&	--	&	--	&	\textbf{2.86}	&	\textbf{0.09}	&	--	&	--	&	2.86	&	0.09	\\
050.4-01.6	&	K 4-28	&	--	&	--	&	--	&	--	&	7.04	&	0.35	&	--	&	--	&	\textbf{5.15}	&	\textbf{0.06}	&	--	&	--	&	5.15	&	0.06	\\
051.0-04.5	&	PC 22	&	--	&	--	&	--	&	--	&	\textbf{2.45}	&	\textbf{0.31}	&	--	&	--	&	\textbf{1.98}	&	\textbf{0.18}	&	--	&	--	&	2.10	&	0.15	\\
051.0+03.0	&	Hen 2-430	&	\textbf{3.71}	&	\textbf{0.30}	&	--	&	--	&	\textbf{3.49}	&	\textbf{0.20}	&	--	&	--	&	2.33	&	0.08	&	--	&	--	&	3.56	&	0.17	\\
051.3+01.8	&	PM 1-295	&	--	&	--	&	--	&	--	&	--	&	--	&	--	&	--	&	\textbf{3.56}	&	\textbf{0.11}	&	--	&	--	&	3.56	&	0.11	\\
051.9-03.8	&	M 1-73	&	\textbf{1.60}	&	\textbf{0.24}	&	--	&	--	&	\textbf{1.58}	&	\textbf{0.21}	&	--	&	--	&	\textbf{1.56}	&	\textbf{0.12}	&	--	&	--	&	1.57	&	0.10	\\
052.2-04.0	&	M 1-74	&	--	&	--	&	--	&	--	&	\textbf{1.89}	&	\textbf{0.23}	&	\textbf{1.64}	&	\textbf{0.16}	&	--	&	--	&	--	&	--	&	1.72	&	0.13	\\
052.5-02.9	&	Me 1-1	&	--	&	--	&	--	&	--	&	\textbf{1.23}	&	\textbf{0.21}	&	2.09	&	0.17	&	\textbf{0.83}	&	\textbf{0.10}	&	--	&	--	&	0.91	&	0.09	\\
052.9-02.7	&	K 3-41	&	--	&	--	&	--	&	--	&	\textbf{1.44}	&	\textbf{0.30}	&	--	&	--	&	2.65	&	0.06	&	--	&	--	&	1.44	&	0.30	\\
052.9+02.7	&	K 3-31	&	--	&	--	&	--	&	--	&	\textbf{5.83}	&	\textbf{0.34}	&	--	&	--	&	\textbf{4.55}	&	\textbf{0.08}	&	--	&	--	&	4.61	&	0.08	\\
053.2-01.5	&	K 3-38	&	--	&	--	&	--	&	--	&	\textbf{4.54}	&	\textbf{0.19}	&	--	&	--	&	\textbf{4.56}	&	\textbf{0.09}	&	--	&	--	&	4.56	&	0.08	\\
053.8-03.0	&	Abell 63	&	--	&	--	&	--	&	--	&	--	&	--	&	--	&	--	&	\textbf{0.67}	&	\textbf{0.20}	&	--	&	--	&	0.67	&	0.20	\\
054.2-03.4	&	Necklace	&	--	&	--	&	--	&	--	&	--	&	--	&	--	&	--	&	\textbf{2.41}	&	\textbf{0.19}	&	--	&	--	&	2.41	&	0.19	\\
054.4-02.5	&	M 1-72	&	--	&	--	&	--	&	--	&	\textbf{2.03}	&	\textbf{0.32}	&	--	&	--	&	--	&	--	&	--	&	--	&	2.03	&	0.32	\\
055.1-01.8	&	K 3-43	&	--	&	--	&	--	&	--	&	--	&	--	&	--	&	--	&	\textbf{3.48}	&	\textbf{0.27}	&	--	&	--	&	3.48	&	0.27	\\
																															
\bottomrule																															
\end{tabular}																															
\end{table}																															
\end{landscape}																															
																															
\begin{landscape}																															
\begin{table}																															
\centering																															
  \contcaption{Calculated radio {\AvCHa}, {\AvCHB} values and weighted averaged {\AV}(Radio) values for catalogued PNe}																															
    \begin{tabular}{llllllllllllllll}																															
    																															
\hline																															
& & \multicolumn{4}{c}{30 GHz} & \multicolumn{4}{c}{5 GHz} & \multicolumn{4}{c}{1.4 GHz} & Weighted & \\																															
PN G & Name & {\AvCHa} & unc. & {\AvCHB} & unc. & {\AvCHa} & unc. & {\AvCHB} & unc. & {\AvCHa} & unc. & {\AvCHB} & unc. & Averaged & unc. \\																															
& & & & & & & & & & & & & & {\AV}(Radio) &\\																															
\hline																															
																															
055.2+02.8	&	Hen 2-432	&	--	&	--	&	--	&	--	&	\textbf{4.01}	&	\textbf{0.26}	&	--	&	--	&	\textbf{3.33}	&	\textbf{0.09}	&	--	&	--	&	3.40	&	0.08	\\
055.3+02.7	&	Hen 1- 1	&	--	&	--	&	--	&	--	&	\textbf{3.76}	&	\textbf{0.19}	&	--	&	--	&	\textbf{3.87}	&	\textbf{0.08}	&	--	&	--	&	3.85	&	0.07	\\
055.5-00.5	&	M 1-71	&	--	&	--	&	--	&	--	&	\textbf{4.36}	&	\textbf{0.15}	&	\textbf{4.50}	&	\textbf{0.29}	&	--	&	--	&	--	&	--	&	4.39	&	0.13	\\
055.6+02.1	&	Hen 1-2	&	--	&	--	&	--	&	--	&	\textbf{2.84}	&	\textbf{0.19}	&	--	&	--	&	\textbf{2.77}	&	\textbf{0.08}	&	--	&	--	&	2.78	&	0.07	\\
056.0+02.0	&	K 3-35	&	\textbf{6.85}	&	\textbf{0.20}	&	--	&	--	&	\textbf{6.03}	&	\textbf{0.33}	&	--	&	--	&	4.48	&	0.08	&	--	&	--	&	6.63	&	0.17	\\
056.4-00.9	&	K 3-42	&	--	&	--	&	--	&	--	&	\textbf{5.95}	&	\textbf{0.18}	&	--	&	--	&	\textbf{5.13}	&	\textbf{0.07}	&	--	&	--	&	5.24	&	0.07	\\

057.9-01.5	&	Hen 2-447	&	\textbf{5.13}	&	\textbf{0.20}	&	--	&	--	&	\textbf{4.82}	&	\textbf{0.06}	&	--	&	--	&	\textbf{3.32}	&	\textbf{0.09}	&	--	&	--	&	4.42	&	0.05	\\

058.9+01.3	&	K 3-40	&	--	&	--	&	--	&	--	&	\textbf{3.36}	&	\textbf{0.33}	&	--	&	--	&	\textbf{2.98}	&	\textbf{0.08}	&	--	&	--	&	3.00	&	0.08	\\
059.4+02.3	&	K 3-37	&	--	&	--	&	--	&	--	&	\textbf{4.68}	&	\textbf{0.30}	&	--	&	--	&	\textbf{4.31}	&	\textbf{0.07}	&	--	&	--	&	4.33	&	0.07	\\
059.0+04.6	&	K 3-34	&	--	&	--	&	--	&	--	&	--	&	--	&	--	&	--	&	\textbf{2.57}	&	\textbf{0.29}	&	--	&	--	&	2.57	&	0.29	\\

059.9+02.0	&	K 3-39	&	--	&	--	&	--	&	--	&	\textbf{4.41}	&	\textbf{0.07}	&	--	&	--	&	2.68	&	0.06	&	--	&	--	&	4.41	&	0.07	\\

060.4+01.5	&	PM 1-310	&	--	&	--	&	--	&	--	&	--	&	--	&	--	&	--	&	\textbf{4.96}	&	\textbf{0.27}	&	--	&	--	&	4.96	&	0.27	\\
060.5-00.3	&	K 3-45	&	--	&	--	&	--	&	--	&	--	&	--	&	--	&	--	&	\textbf{4.91}	&	\textbf{0.19}	&	--	&	--	&	4.91	&	0.19	\\

060.5+01.8	&	Hen 2-440	&	--	&	--	&	--	&	--	&	\textbf{4.12}	&	\textbf{0.06}	&	--	&	--	&	\textbf{3.40}	&	\textbf{0.09}	&	--	&	--	&	3.93	&	0.05	\\

060.8-03.6	&	NGC 6853	&	--	&	--	&	--	&	--	&	--	&	--	&	\textbf{0.08}	&	\textbf{0.14}	&	--	&	--	&	--	&	--	&	0.08	&	0.14	\\
062.4-00.2	&	M 2-48	&	--	&	--	&	--	&	--	&	\textbf{3.46}	&	\textbf{0.15}	&	--	&	--	&	\textbf{3.09}	&	\textbf{1.30}	&	--	&	--	&	3.45	&	0.15	\\
063.8-03.3	&	K 3-54	&	--	&	--	&	--	&	--	&	\textbf{4.87}	&	\textbf{0.23}	&	--	&	--	&	\textbf{4.22}	&	\textbf{0.82}	&	--	&	--	&	4.82	&	0.22	\\
064.9-02.1	&	K 3-53	&	--	&	--	&	--	&	--	&	5.23	&	0.33	&	--	&	--	&	\textbf{2.91}	&	\textbf{0.82}	&	--	&	--	&	2.91	&	1.02	\\
065.1-03.5	&	We 1-9	&	--	&	--	&	--	&	--	&	--	&	--	&	--	&	--	&	\textbf{3.16}	&	\textbf{1.02}	&	--	&	--	&	3.16	&	1.02	\\
065.9+00.5	&	NGC 6842	&	--	&	--	&	--	&	--	&	\textbf{3.04}	&	\textbf{0.19}	&	\textbf{2.46}	&	\textbf{0.16}	&	--	&	--	&	--	&	--	&	2.70	&	0.12	\\
066.9+02.2	&	K 4-37	&	--	&	--	&	--	&	--	&	--	&	--	&	--	&	--	&	\textbf{4.25}	&	\textbf{0.19}	&	--	&	--	&	4.25	&	0.19	\\
067.9-00.2	&	K 3-52	&	\textbf{8.65}	&	\textbf{0.21}	&	--	&	--	&	\textbf{7.96}	&	\textbf{0.33}	&	--	&	--	&	5.98	&	1.3	&	--	&	--	&	8.46	&	0.18	\\

068.3-02.7	&	Hen 2-459	&	\textbf{5.02}	&	\textbf{0.15}	&	--	&	--	&	\textbf{4.55}	&	\textbf{0.06}	&	--	&	--	&	1.89	&	1.03	&	--	&	--	&	4.61	&	0.05	\\

068.6+01.1	&	Hen 1-4	&	--	&	--	&	--	&	--	&	--	&	--	&	--	&	--	&	\textbf{2.36}	&	\textbf{0.11}	&	--	&	--	&	2.36	&	0.11	\\
068.7+01.9	&	K 4-41	&	--	&	--	&	--	&	--	&	\textbf{3.76}	&	\textbf{0.33}	&	--	&	--	&	\textbf{3.26}	&	\textbf{1.03}	&	--	&	--	&	3.72	&	0.32	\\
068.7+03.0	&	PC 23	&	--	&	--	&	--	&	--	&	\textbf{3.66}	&	\textbf{0.21}	&	--	&	--	&	--	&	--	&	--	&	--	&	3.66	&	0.21	\\
068.8-00.0	&	M 1-75	&	--	&	--	&	--	&	--	&	\textbf{4.20}	&	\textbf{0.19}	&	--	&	--	&	--	&	--	&	--	&	--	&	4.20	&	0.19	\\
069.2+03.8	&	K 3-46	&	--	&	--	&	--	&	--	&	--	&	--	&	--	&	--	&	\textbf{2.15}	&	\textbf{0.26}	&	--	&	--	&	2.15	&	0.26	\\

069.2+02.8	&	K 3-49	&	--	&	--	&	--	&	--	&	\textbf{3.37}	&	\textbf{0.07}	&	--	&	--	&	--	&	--	&	--	&	--	&	3.37	&	0.07	\\

069.4-02.6	&	NGC 6894	&	--	&	--	&	--	&	--	&	\textbf{1.17}	&	\textbf{0.21}	&	\textbf{1.47}	&	\textbf{0.22}	&	--	&	--	&	--	&	--	&	1.32	&	0.15	\\
069.6-03.9	&	K 3-58	&	--	&	--	&	--	&	--	&	\textbf{4.11}	&	\textbf{0.19}	&	--	&	--	&	--	&	--	&	--	&	--	&	4.11	&	0.19	\\
069.7+00.0	&	K 3-55	&	\textbf{7.06}	&	\textbf{0.18}	&	--	&	--	&	\textbf{7.03}	&	\textbf{0.33}	&	--	&	--	&	--	&	--	&	--	&	--	&	7.05	&	0.16	\\
071.6-02.3	&	M 3-35	&	\textbf{4.54}	&	\textbf{0.12}	&	\textbf{4.76}	&	\textbf{0.14}	&	\textbf{4.19}	&	\textbf{0.32}	&	\textbf{4.53}	&	\textbf{0.31}	&	--	&	--	&	2.92	&	0.16	&	4.60	&	0.08	\\
072.1+00.1	&	K 3-57	&	--	&	--	&	--	&	--	&	\textbf{5.35}	&	\textbf{0.33}	&	--	&	--	&	--	&	--	&	--	&	--	&	5.35	&	0.33	\\
073.0-02.4	&	K 3-76	&	--	&	--	&	--	&	--	&	\textbf{4.27}	&	\textbf{0.29}	&	--	&	--	&	--	&	--	&	--	&	--	&	4.27	&	0.29	\\

074.5+02.1	&	NGC 6881	&	\textbf{3.87}	&	\textbf{0.15}	&	\textbf{3.89}	&	\textbf{0.07}	&	\textbf{3.64}	&	\textbf{0.06}	&	\textbf{3.74}	&	\textbf{0.07}	&	\textbf{2.87}	&	\textbf{0.10}	&	\textbf{3.21}	&	\textbf{0.09}	&	3.59	&	0.03	\\

075.6+04.3	&	ARO 342	&	--	&	--	&	--	&	--	&	--	&	--	&	--	&	--	&	\textbf{2.26}	&	\textbf{0.19}	&	--	&	--	&	2.26	&	0.19	\\
076.3+01.1	&	Abell 69	&	--	&	--	&	--	&	--	&	\textbf{4.11}	&	\textbf{0.19}	&	\textbf{4.91}	&	\textbf{0.61}	&	2.99	&	0.19	&	\textbf{4.15}	&	\textbf{0.61}	&	4.18	&	0.18	\\
076.4+01.8	&	KjPn 3	&	--	&	--	&	--	&	--	&	--	&	--	&	--	&	--	&	\textbf{3.42}	&	\textbf{0.19}	&	--	&	--	&	3.42	&	0.19	\\
077.5+03.7	&	KjPn 1	&	--	&	--	&	--	&	--	&	\textbf{1.50}	&	\textbf{0.21}	&	--	&	--	&	--	&	--	&	--	&	--	&	1.50	&	0.21	\\
077.7+03.1	&	KjPn 2	&	--	&	--	&	--	&	--	&	--	&	--	&	--	&	--	&	\textbf{5.51}	&	\textbf{0.19}	&	--	&	--	&	5.51	&	0.19	\\
078.3-02.7	&	K 4-53	&	\textbf{5.49}	&	\textbf{0.25}	&	--	&	--	&	--	&	--	&	--	&	--	&	\textbf{4.75}	&	\textbf{0.11}	&	--	&	--	&	4.88	&	0.10	\\
078.9+00.7	&	Sd 1	&	--	&	--	&	--	&	--	&	\textbf{5.06}	&	\textbf{0.19}	&	--	&	--	&	1.64	&	0.19	&	--	&	--	&	5.06	&	0.19	\\
084.2+01.0	&	K 4-55	&	--	&	--	&	--	&	--	&	--	&	--	&	--	&	--	&	\textbf{2.94}	&	\textbf{0.18}	&	--	&	--	&	2.94	&	0.18	\\
084.9+04.4	&	Abell 71	&	--	&	--	&	--	&	--	&	\textbf{2.35}	&	\textbf{0.21}	&	\textbf{2.43}	&	\textbf{0.14}	&	--	&	--	&	--	&	--	&	2.41	&	0.11	\\
\bottomrule																															
\end{tabular}																															
\end{table}																															
\end{landscape}																															
																															
\begin{landscape}																															
\begin{table}																															
\centering																															
  \contcaption{Calculated radio {\AvCHa}, {\AvCHB} values and weighted averaged {\AV}(Radio) values for catalogued PNe}																															
        \begin{tabular}{llllllllllllllll}																															
    																															
\hline																															
& & \multicolumn{4}{c}{30 GHz} & \multicolumn{4}{c}{5 GHz} & \multicolumn{4}{c}{1.4 GHz} & Weighted & \\																															
PN G & Name & {\AvCHa} & unc. & {\AvCHB} & unc. & {\AvCHa} & unc. & {\AvCHB} & unc. & {\AvCHa} & unc. & {\AvCHB} & unc. & Averaged & unc. \\																															
& & & & & & & & & & & & & & {\AV}(Radio) &\\																															
\hline																															
																															
088.7+04.6	&	K 3-78	&	--	&	--	&	--	&	--	&	\textbf{4.78}	&	\textbf{0.22}	&	--	&	--	&	\textbf{4.45}	&	\textbf{0.06}	&	--	&	--	&	4.47	&	0.05	\\
088.7-01.6	&	NGC 7048	&	--	&	--	&	--	&	--	&	\textbf{0.55}	&	\textbf{0.20}	&	\textbf{0.82}	&	\textbf{0.12}	&	\textbf{0.70}	&	\textbf{0.06}	&	\textbf{0.92}	&	\textbf{0.06}	&	0.79	&	0.04	\\
089.0+00.3	&	NGC 7026	&	--	&	--	&	--	&	--	&	\textbf{2.08}	&	\textbf{0.45}	&	\textbf{1.85}	&	\textbf{0.31}	&	\textbf{1.85}	&	\textbf{0.07}	&	\textbf{1.70}	&	\textbf{0.08}	&	1.79	&	0.05	\\
089.8-00.6	&	Sh 1-89	&	--	&	--	&	--	&	--	&	--	&	--	&	--	&	--	&	\textbf{3.52}	&	\textbf{0.20}	&	--	&	--	&	3.52	&	0.20	\\
091.6-04.8	&	K 3-84	&	--	&	--	&	--	&	--	&	--	&	--	&	--	&	--	&	\textbf{1.25}	&	\textbf{0.24}	&	--	&	--	&	1.25	&	0.24	\\
091.6+01.8	&	We 1- 11	&	--	&	--	&	--	&	--	&	--	&	--	&	--	&	--	&	\textbf{4.43}	&	\textbf{0.06}	&	--	&	--	&	4.43	&	0.06	\\
093.3-00.9	&	K 3-82	&	--	&	--	&	--	&	--	&	\textbf{3.62}	&	\textbf{0.30}	&	--	&	--	&	\textbf{3.75}	&	\textbf{0.16}	&	--	&	--	&	3.72	&	0.14	\\
093.3-02.4	&	M 1-79	&	--	&	--	&	--	&	--	&	\textbf{0.38}	&	\textbf{0.19}	&	\textbf{1.06}	&	\textbf{0.11}	&	\textbf{0.51}	&	\textbf{0.06}	&	\textbf{1.14}	&	\textbf{0.08}	&	0.76	&	0.04	\\
094.5-00.8	&	K 3-83	&	--	&	--	&	--	&	--	&	\textbf{5.21}	&	\textbf{0.19}	&	--	&	--	&	\textbf{4.71}	&	\textbf{0.06}	&	--	&	--	&	4.75	&	0.05	\\
095.1-02.0	&	M 2-49	&	--	&	--	&	--	&	--	&	5.15	&	0.2	&	--	&	--	&	\textbf{3.67}	&	\textbf{0.06}	&	--	&	--	&	3.67	&	0.06	\\

095.2+00.7	&	K 3-62	&	\textbf{5.89}	&	\textbf{0.16}	&	--	&	--	&	\textbf{5.86}	&	\textbf{0.06}	&	--	&	--	&	\textbf{4.76}	&	\textbf{0.06}	&	4.9	&	0.25	&	5.35	&	0.04	\\

096.3+02.3	&	K 3-61	&	--	&	--	&	--	&	--	&	\textbf{3.70}	&	\textbf{0.19}	&	--	&	--	&	\textbf{3.79}	&	\textbf{0.06}	&	--	&	--	&	3.78	&	0.05	\\
097.6-02.4	&	M 2-50	&	--	&	--	&	--	&	--	&	\textbf{1.83}	&	\textbf{0.19}	&	\textbf{1.62}	&	\textbf{0.16}	&	\textbf{2.12}	&	\textbf{0.06}	&	\textbf{1.81}	&	\textbf{0.18}	&	2.03	&	0.05	\\
098.1+02.4	&	K 3-63	&	--	&	--	&	--	&	--	&	4.38	&	0.23	&	3.78	&	0.23	&	\textbf{2.55}	&	\textbf{0.09}	&	\textbf{2.53}	&	\textbf{0.18}	&	2.55	&	0.08	\\
098.2+04.9	&	K 3-60	&	--	&	--	&	--	&	--	&	\textbf{5.28}	&	\textbf{0.35}	&	\textbf{4.92}	&	\textbf{0.70}	&	\textbf{4.53}	&	\textbf{0.06}	&	\textbf{4.41}	&	\textbf{0.46}	&	4.55	&	0.05	\\
102.8-05.0	&	Abell 80	&	--	&	--	&	--	&	--	&	--	&	--	&	--	&	--	&	\textbf{0.11}	&	\textbf{0.06}	&	\textbf{0.47}	&	\textbf{0.50}	&	0.12	&	0.06	\\
103.2+00.6	&	M 2-51	&	--	&	--	&	--	&	--	&	\textbf{1.83}	&	\textbf{0.20}	&	\textbf{2.31}	&	\textbf{0.18}	&	\textbf{1.15}	&	\textbf{0.15}	&	\textbf{1.85}	&	\textbf{0.14}	&	1.73	&	0.08	\\
103.7+00.4	&	M 2-52	&	--	&	--	&	--	&	--	&	\textbf{2.95}	&	\textbf{0.19}	&	--	&	--	&	\textbf{2.91}	&	\textbf{0.06}	&	--	&	--	&	2.91	&	0.05	\\

104.1+01.0	&	Bl 2-1	&	--	&	--	&	--	&	--	&	\textbf{6.11}	&	\textbf{0.06}	&	--	&	--	&	\textbf{4.54}	&	\textbf{0.06}	&	--	&	--	&	5.31	&	0.04	\\

104.4-01.6	&	M 2-53	&	--	&	--	&	--	&	--	&	\textbf{2.17}	&	\textbf{0.19}	&	--	&	--	&	\textbf{2.43}	&	\textbf{0.06}	&	--	&	--	&	2.41	&	0.05	\\
107.4-02.6	&	K 3-87	&	--	&	--	&	--	&	--	&	\textbf{3.55}	&	\textbf{0.27}	&	--	&	--	&	\textbf{4.67}	&	\textbf{0.13}	&	--	&	--	&	4.45	&	0.12	\\
107.4-00.6	&	K 4-57	&	--	&	--	&	--	&	--	&	--	&	--	&	--	&	--	&	\textbf{3.85}	&	\textbf{0.19}	&	--	&	--	&	3.85	&	0.19	\\
107.7-02.2	&	M 1-80	&	--	&	--	&	--	&	--	&	\textbf{2.22}	&	\textbf{0.22}	&	--	&	--	&	\textbf{1.60}	&	\textbf{0.06}	&	--	&	--	&	1.64	&	0.05	\\
107.8+02.3	&	NGC 7354	&	\textbf{3.95}	&	\textbf{0.10}	&	\textbf{3.67}	&	\textbf{0.03}	&	\textbf{4.12}	&	\textbf{0.23}	&	\textbf{3.79}	&	\textbf{0.14}	&	\textbf{3.92}	&	\textbf{0.09}	&	\textbf{3.65}	&	\textbf{0.06}	&	3.72	&	0.03	\\
112.5-00.1	&	KjPn 8	&	--	&	--	&	--	&	--	&	\textbf{1.58}	&	\textbf{0.12}	&	--	&	--	&	\textbf{2.66}	&	\textbf{0.25}	&	--	&	--	&	1.80	&	0.11	\\
112.5+03.7	&	K 3-88	&	--	&	--	&	--	&	--	&	--	&	--	&	--	&	--	&	\textbf{4.65}	&	\textbf{0.06}	&	--	&	--	&	4.65	&	0.06	\\
119.3+00.3	&	BV 5-1	&	--	&	--	&	--	&	--	&	--	&	--	&	--	&	--	&	\textbf{2.50}	&	\textbf{0.06}	&	\textbf{2.60}	&	\textbf{0.16}	&	2.52	&	0.05	\\
121.6+00.0	&	BV 5-2	&	--	&	--	&	--	&	--	&	--	&	--	&	--	&	--	&	\textbf{2.43}	&	\textbf{0.06}	&	--	&	--	&	2.43	&	0.06	\\
121.6+03.5	&	We 1-1	&	--	&	--	&	--	&	--	&	--	&	--	&	--	&	--	&	\textbf{4.05}	&	\textbf{0.06}	&	--	&	--	&	4.05	&	0.06	\\
122.1-04.9	&	Abell 2	&	--	&	--	&	--	&	--	&	0.05	&	0.19	&	0.31	&	0.16	&	\textbf{1.80}	&	\textbf{0.06}	&	\textbf{1.50}	&	\textbf{0.17}	&	1.77	&	0.05	\\
126.3+02.9	&	K 3-90	&	--	&	--	&	--	&	--	&	\textbf{3.03}	&	\textbf{0.19}	&	--	&	--	&	\textbf{2.74}	&	\textbf{0.06}	&	--	&	--	&	2.76	&	0.05	\\
126.6+01.3	&	IPHAS PN-1	&	--	&	--	&	--	&	--	&	--	&	--	&	--	&	--	&	\textbf{1.47}	&	\textbf{1.05}	&	--	&	--	&	1.47	&	1.05	\\
129.5+04.5	&	K 3-91	&	--	&	--	&	--	&	--	&	\textbf{2.64}	&	\textbf{0.19}	&	--	&	--	&	\textbf{3.58}	&	\textbf{0.06}	&	--	&	--	&	3.51	&	0.05	\\
130.2+01.3	&	IC 1747	&	\textbf{1.88}	&	\textbf{0.06}	&	\textbf{1.83}	&	\textbf{0.09}	&	\textbf{2.60}	&	\textbf{0.32}	&	\textbf{2.32}	&	\textbf{0.24}	&	\textbf{2.42}	&	\textbf{0.09}	&	\textbf{2.20}	&	\textbf{0.24}	&	2.03	&	0.04	\\
130.4+03.1	&	K 3-92	&	--	&	--	&	--	&	--	&	\textbf{1.96}	&	\textbf{0.19}	&	--	&	--	&	--	&	--	&	--	&	--	&	1.96	&	0.19	\\
131.5+02.6	&	Abell 3	&	--	&	--	&	--	&	--	&	--	&	--	&	\textbf{0.85}	&	\textbf{0.32}	&	\textbf{0.40}	&	\textbf{0.07}	&	--	&	--	&	0.41	&	0.06	\\
132.4+04.7	&	K 3-93	&	--	&	--	&	--	&	--	&	--	&	--	&	--	&	--	&	\textbf{3.74}	&	\textbf{0.06}	&	--	&	--	&	3.74	&	0.06	\\
136.1+04.9	&	Abell 6	&	--	&	--	&	--	&	--	&	\textbf{2.03}	&	\textbf{0.36}	&	\textbf{2.16}	&	\textbf{0.23}	&	--	&	--	&	--	&	--	&	2.12	&	0.19	\\
138.8+02.8	&	IC 289	&	\textbf{2.12}	&	\textbf{0.11}	&	\textbf{2.30}	&	\textbf{0.03}	&	\textbf{2.92}	&	\textbf{0.38}	&	\textbf{2.84}	&	\textbf{0.24}	&	--	&	--	&	\textbf{2.42}	&	\textbf{0.06}	&	2.32	&	0.02	\\
142.1+03.4	&	K 3-94	&	--	&	--	&	--	&	--	&	\textbf{2.84}	&	\textbf{0.15}	&	--	&	--	&	--	&	--	&	--	&	--	&	2.84	&	0.15	\\
147.4-02.3	&	M 1- 4	&	\textbf{3.44}	&	\textbf{0.19}	&	\textbf{3.03}	&	\textbf{0.11}	&	\textbf{3.68}	&	\textbf{0.12}	&	\textbf{3.19}	&	\textbf{0.15}	&	--	&	--	&	\textbf{3.00}	&	\textbf{0.14}	&	3.25	&	0.06	\\
147.8+04.1	&	M 2- 2	&	--	&	--	&	--	&	--	&	{2.74}	&	\textbf{0.18}	&	--	&	--	&	--	&	--	&	--	&	--	&	2.74	&	0.18	\\
149.0+04.4	&	K 4-47	&	--	&	--	&	--	&	--	&	--	&	--	&	--	&	--	&	--	&	--	&	--	&	--	&	--	&	--	\\
151.4+00.5	&	K 3-64	&	--	&	--	&	--	&	--	&	--	&	--	&	--	&	--	&	\textbf{3.25}	&	\textbf{0.25}	&	4.45	&	1.3	&	3.25	&	0.25	\\
153.7-01.4	&	K 3-65	&	--	&	--	&	--	&	--	&	--	&	--	&	--	&	--	&	--	&	--	&	--	&	--	&	--	&	--	\\
160.5-00.5	&	We 1-2	&	--	&	--	&	--	&	--	&	--	&	--	&	--	&	--	&	--	&	--	&	--	&	--	&	--	&	--	\\
163.1-00.8	&	We 1-3	&	--	&	--	&	--	&	--	&	--	&	--	&	--	&	--	&	--	&	--	&	--	&	--	&	--	&	--	\\
																															
\bottomrule																															
\end{tabular}																															
\end{table}																															
\end{landscape}																															
																															
\begin{landscape}																															
\begin{table}																															
\centering																															
  \contcaption{Calculated radio {\AvCHa}, {\AvCHB} values and weighted averaged {\AV}(Radio) values for catalogued PNe}																															
    \begin{tabular}{llllllllllllllll}																															
    																															
\hline																															
& & \multicolumn{4}{c}{30 GHz} & \multicolumn{4}{c}{5 GHz} & \multicolumn{4}{c}{1.4 GHz} & Weighted & \\																															
PN G & Name & {\AvCHa} & unc. & {\AvCHB} & unc. & {\AvCHa} & unc. & {\AvCHB} & unc. & {\AvCHa} & unc. & {\AvCHB} & unc. & Averaged & unc. \\																															
& & & & & & & & & & & & & & {\AV}(Radio) &\\																															
\hline 																															
    																															
173.5+03.2	&	Pu 2	&	--	&	--	&	--	&	--	&	--	&	--	&	--	&	--	&	\textbf{3.95}	&	\textbf{0.17}	&	--	&	--	&	3.95	&	0.17	\\
178.3-02.5	&	K 3-68	&	--	&	--	&	--	&	--	&	\textbf{2.73}	&	\textbf{0.19}	&	--	&	--	&	--	&	--	&	--	&	--	&	2.73	&	0.19	\\
181.5+00.9	&	Pu 1	&	--	&	--	&	--	&	--	&	\textbf{2.71}	&	\textbf{0.07}	&	--	&	--	&	\textbf{1.24}	&	\textbf{0.53}	&	--	&	--	&	2.55	&	0.06	\\
184.0-02.1	&	M 1-5	&	\textbf{2.98}	&	\textbf{0.13}	&	\textbf{2.90}	&	\textbf{0.02}	&	3.23	&	0.23	&	3.07	&	0.14	&	--	&	--	&	\textbf{2.48}	&	\textbf{0.06}	&	2.84	&	0.02	\\
184.6+00.6	&	K 3-70	&	--	&	--	&	--	&	--	&	\textbf{4.32}	&	\textbf{0.29}	&	--	&	--	&	--	&	--	&	--	&	--	&	4.32	&	0.29	\\
184.8+04.4	&	K 3-71	&	--	&	--	&	--	&	--	&	--	&	--	&	--	&	--	&	--	&	--	&	\textbf{3.76}	&	\textbf{0.32}	&	3.76	&	0.32	\\
194.2+02.5	&	J 900	&	\textbf{2.09}	&	\textbf{0.13}	&	\textbf{1.62}	&	\textbf{0.05}	&	\textbf{1.50}	&	\textbf{0.13}	&	\textbf{1.68}	&	\textbf{0.11}	&	--	&	--	&	\textbf{1.55}	&	\textbf{0.07}	&	1.63	&	0.03	\\
201.7+02.5	&	K 4-48	&	--	&	--	&	--	&	--	&	--	&	--	&	--	&	--	&	\textbf{2.30}	&	\textbf{0.12}	&	\textbf{3.14}	&	\textbf{0.20}	&	3.04	&	0.10	\\
201.9-04.6	&	We 1- 4	&	--	&	--	&	--	&	--	&	--	&	--	&	--	&	--	&	--	&	--	&	--	&	--	&	--	&	--	\\
210.0+03.9	&	We 2- 34	&	--	&	--	&	--	&	--	&	--	&	--	&	--	&	--	&	--	&	--	&	--	&	--	&	--	&	--	\\
210.3+01.9	&	M 1-8	&	--	&	--	&	--	&	--	&	\textbf{2.47}	&	\textbf{0.49}	&	\textbf{2.48}	&	\textbf{0.36}	&	--	&	--	&	2.06	&	0.1	&	2.47	&	0.29	\\
212.0+04.3	&	M 1-9	&	--	&	--	&	--	&	--	&	\textbf{1.79}	&	\textbf{0.29}	&	\textbf{1.35}	&	\textbf{0.22}	&	\textbf{1.30}	&	\textbf{0.12}	&	\textbf{1.03}	&	\textbf{0.11}	&	1.21	&	0.07	\\

    \bottomrule																															
\end{tabular}																															
\end{table}																															
\end{landscape}

%% file: App_DistCurves.tex
\begin{figure*}
\centering
\begin{subfigure}[b]{0.45\textwidth}
  \centering
  \includegraphics[width=\textwidth]{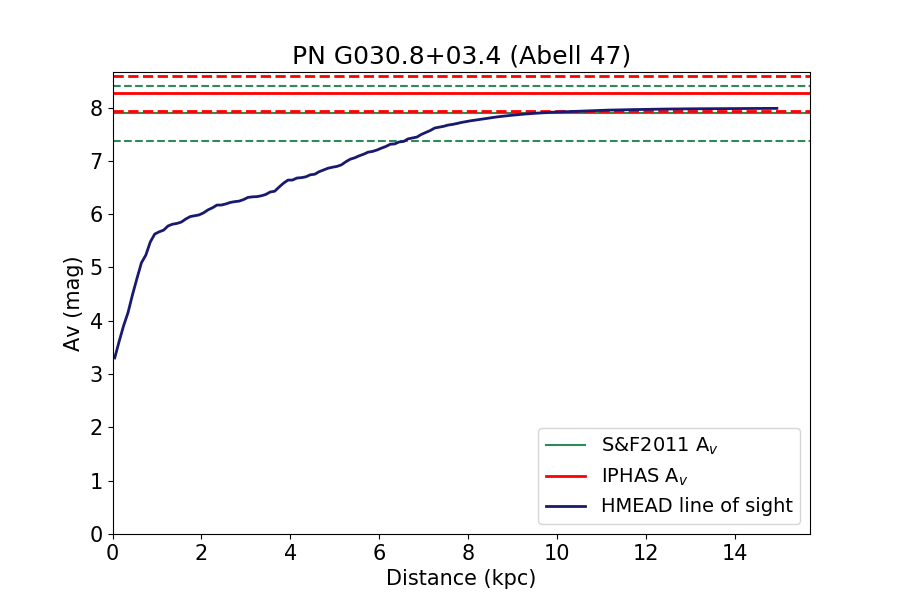}
  \subcaption{}
  \label{fig:Abell47}
  \end{subfigure}
\begin{subfigure}[b]{0.45\textwidth}
  \centering
  \includegraphics[width=\textwidth]{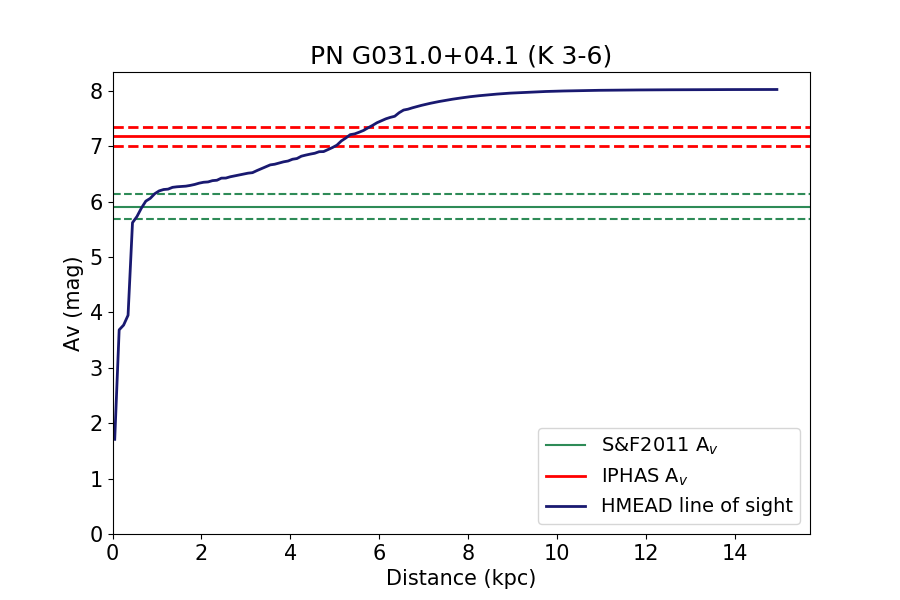}
  \subcaption{}
  \label{fig:K3-6}
  \end{subfigure}
  
\begin{subfigure}[b]{0.45\textwidth}
  \centering
  \includegraphics[width=\textwidth]{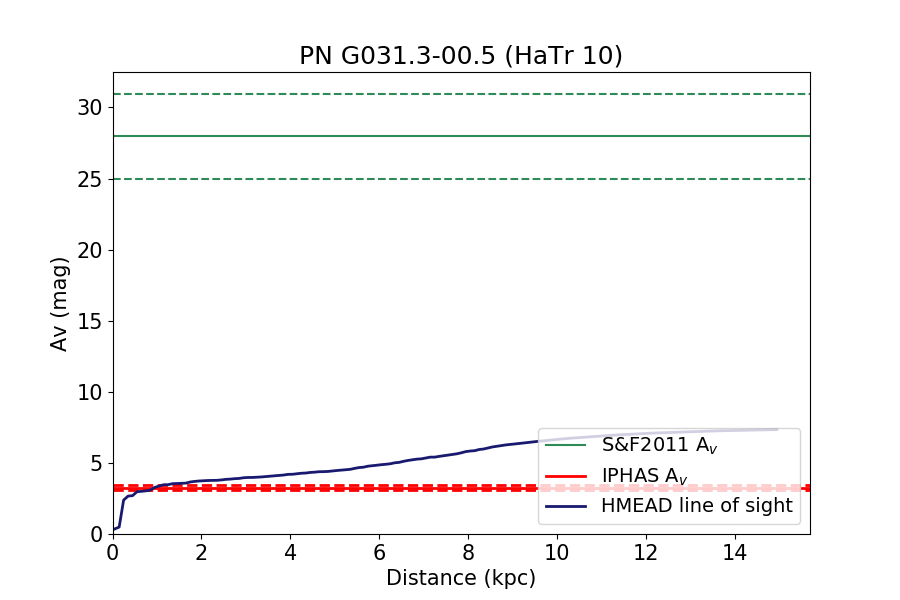}
  \subcaption{}
  \label{fig:HaTr10}
  \end{subfigure}
\begin{subfigure}[b]{0.45\textwidth}
  \centering
  \includegraphics[width=\textwidth]{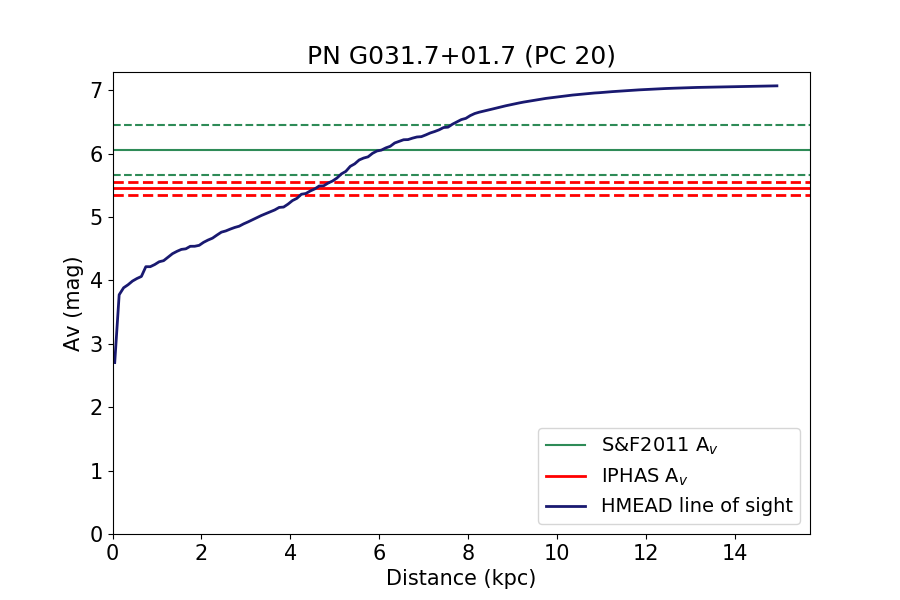}
  \subcaption{}
  \label{fig:PC20}
  \end{subfigure}
  
\begin{subfigure}[b]{0.45\textwidth}
  \centering
  \includegraphics[width=\textwidth]{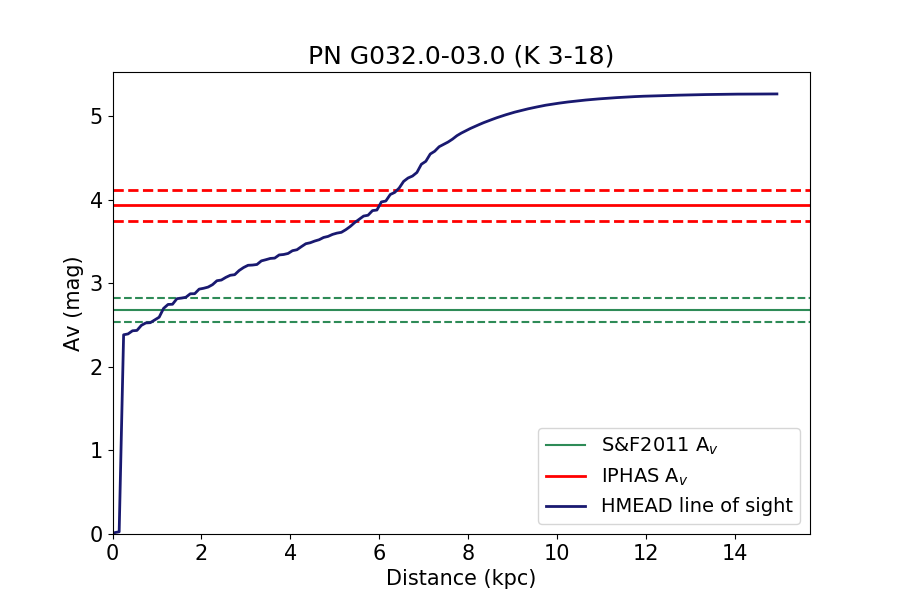}
 \subcaption{}
  \label{fig:K3-18}
  \end{subfigure}
\begin{subfigure}[b]{0.45\textwidth}
  \centering
  \includegraphics[width=\textwidth]{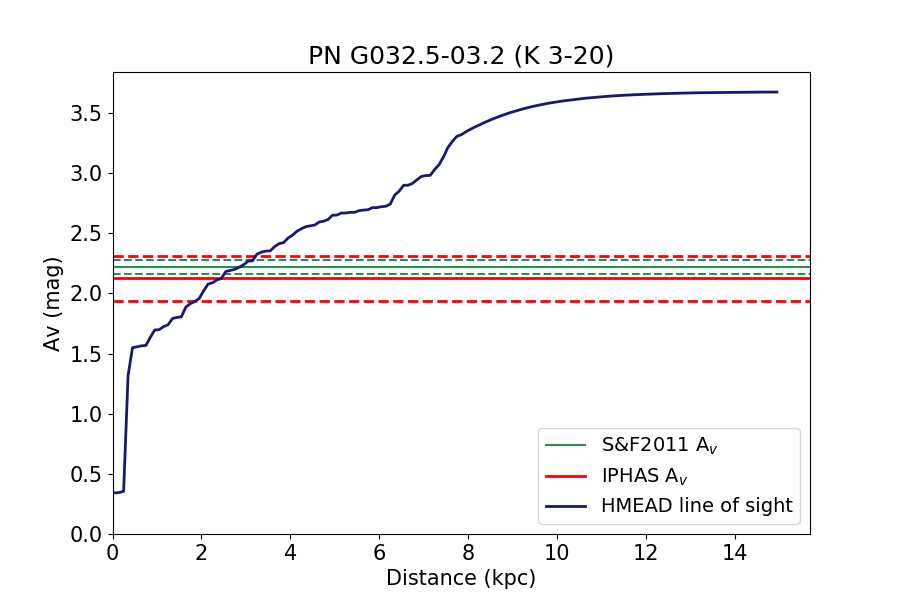}
  \subcaption{}
  \label{fig:K3-20}
  \end{subfigure}
  
\begin{subfigure}[b]{0.45\textwidth}
  \centering
  \includegraphics[width=\textwidth]{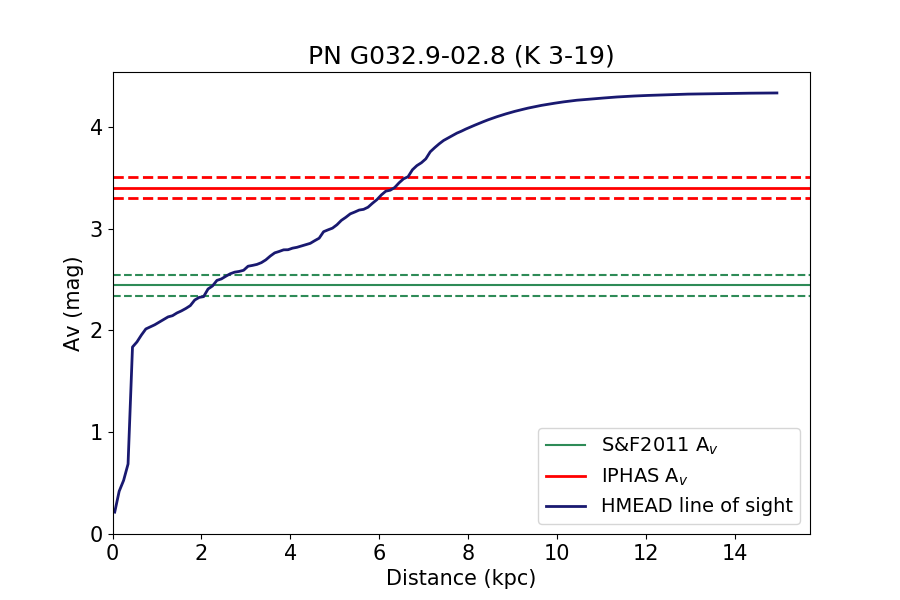}
 \subcaption{}
  \label{fig:K3-19}
  \end{subfigure}
\begin{subfigure}[b]{0.45\textwidth}
  \centering
  \includegraphics[width=\textwidth]{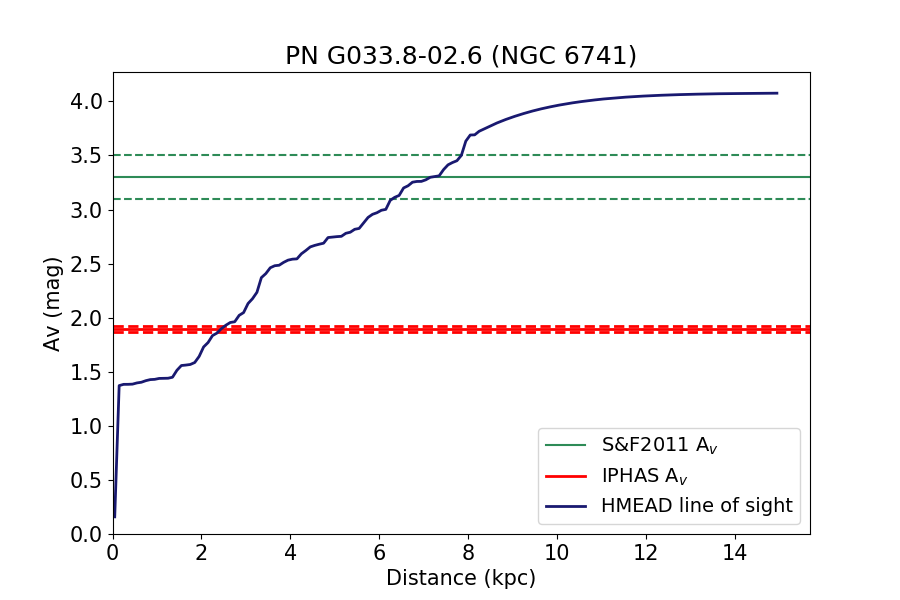}
  \subcaption{}
  \label{fig:NGC6741}
  \end{subfigure}
  \caption{H-MEAD Extinction vs. Distance plots for Abell 47, K 3-6, HaTr 10, PC 20, K 3-18, K 3-20, K 3-19 and NGC 6741.}
  \label{fig:ext_mainBody_All}
\end{figure*}

\begin{figure*}
\centering
\begin{subfigure}[b]{0.45\textwidth}
  \centering
  \includegraphics[width=\textwidth]{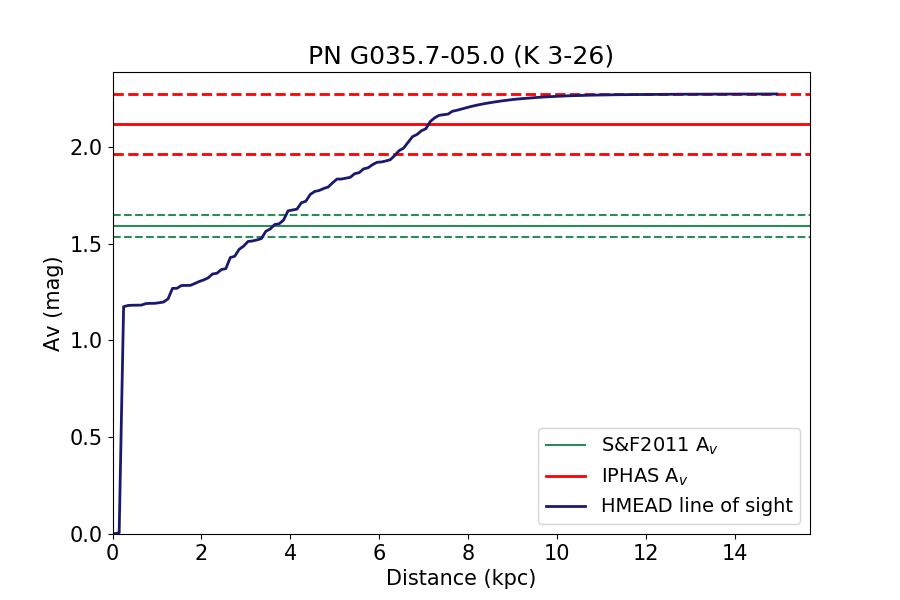}
  \subcaption{}
  \label{fig:K3-26}
  \end{subfigure}
\begin{subfigure}[b]{0.45\textwidth}
  \centering
  \includegraphics[width=\textwidth]{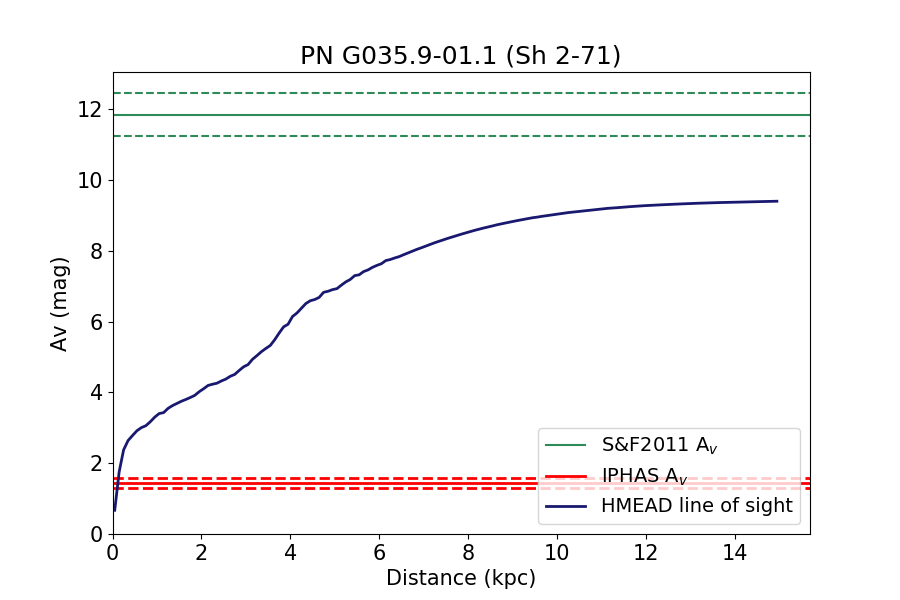}
  \subcaption{}
  \label{fig:Sh2-71}
  \end{subfigure}
  
\begin{subfigure}[b]{0.45\textwidth}
  \centering
  \includegraphics[width=\textwidth]{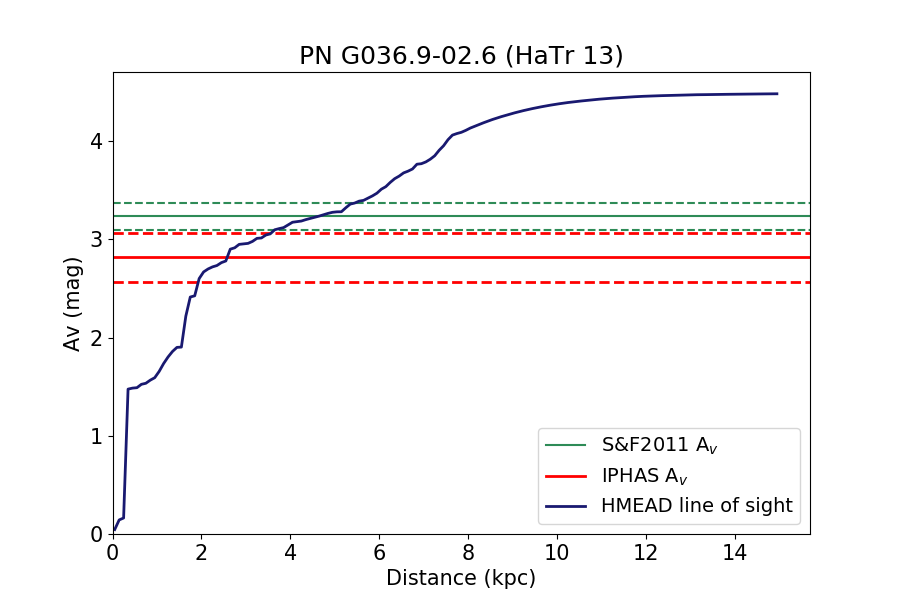}
  \subcaption{}
  \label{fig:HaTr13}
  \end{subfigure}
\begin{subfigure}[b]{0.45\textwidth}
  \centering
  \includegraphics[width=\textwidth]{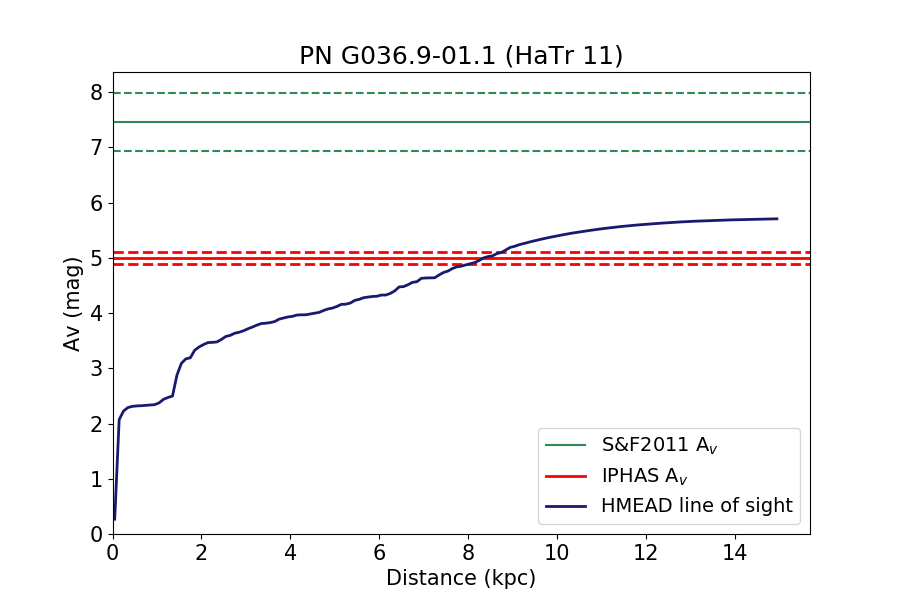}
  \subcaption{}
  \label{fig:HaTr11}
  \end{subfigure}

\begin{subfigure}[b]{0.45\textwidth}
  \centering
  \includegraphics[width=\textwidth]{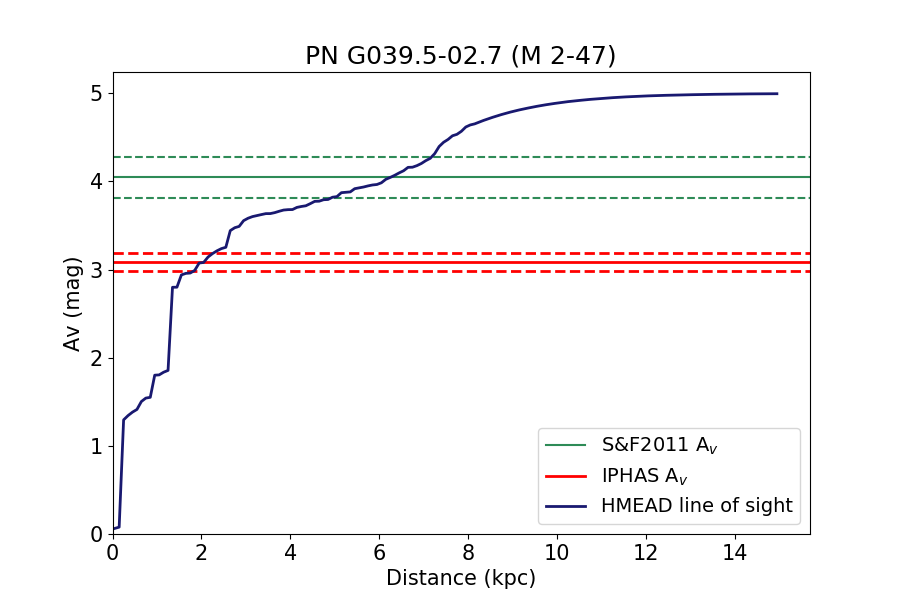}
 \subcaption{}
  \label{fig:M2-47}
  \end{subfigure}
\begin{subfigure}[b]{0.45\textwidth}
  \centering
  \includegraphics[width=\textwidth]{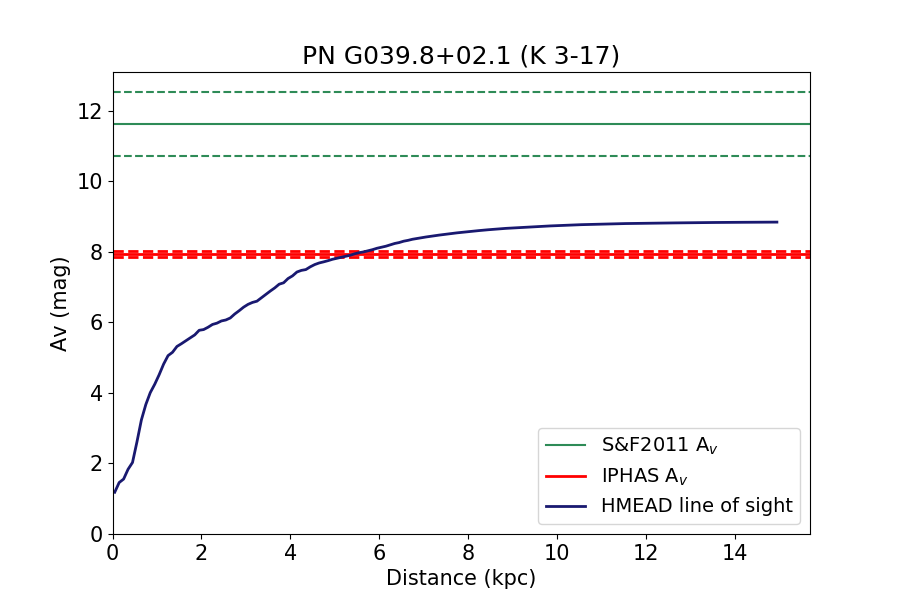}
  \subcaption{}
  \label{fig:K3-17}
  \end{subfigure}
  
\begin{subfigure}[b]{0.45\textwidth}
  \centering
  \includegraphics[width=\textwidth]{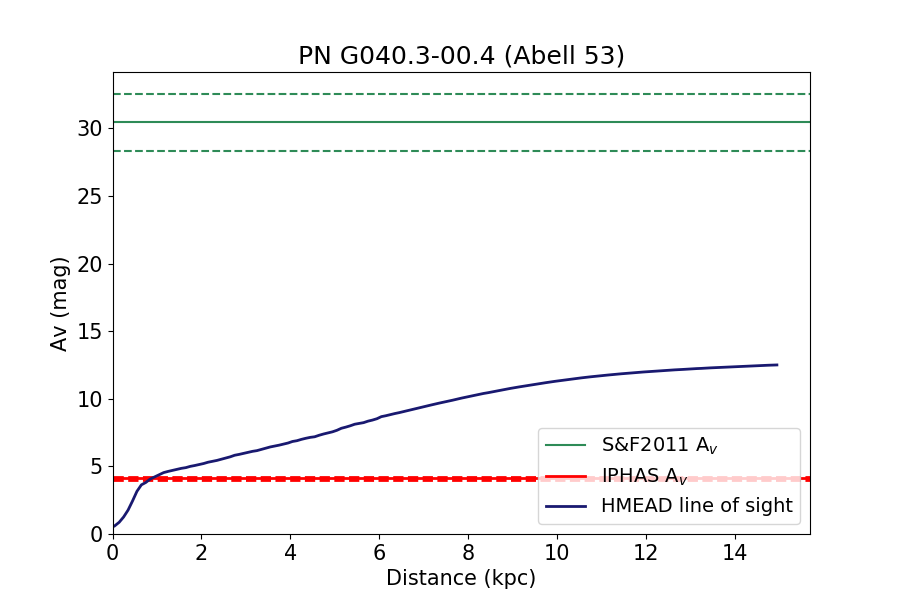}
 \subcaption{}
  \label{fig:Abell53}
  \end{subfigure}
\begin{subfigure}[b]{0.45\textwidth}
  \centering
  \includegraphics[width=\textwidth]{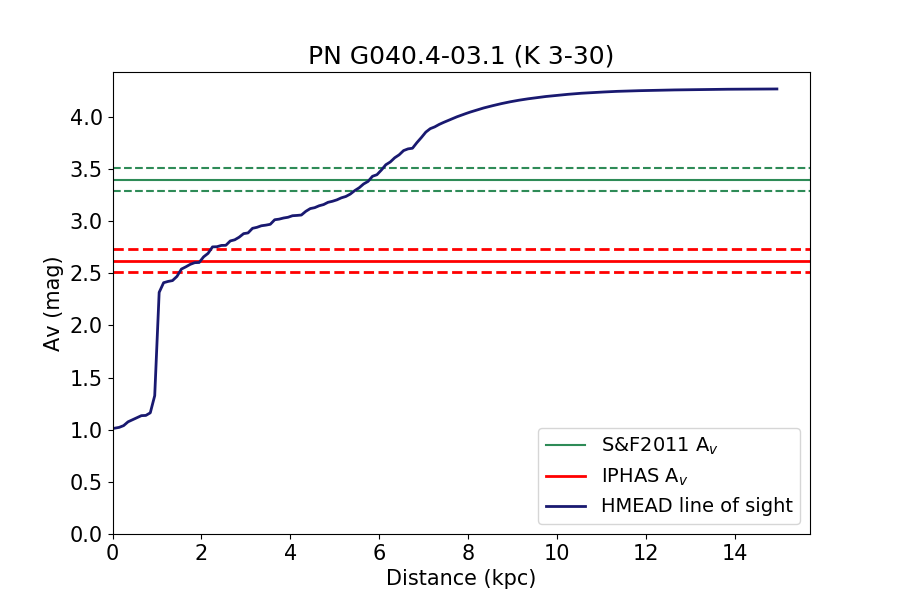}
  \subcaption{}
  \label{fig:K3-30}
  \end{subfigure}
  \caption{H-MEAD Extinction vs. Distance plots for K 3-26, Sh 2-71, 
HaTr 13, HaTr 11, M 2-47, K 3-17, Abell 53 and K 3-30.}
  \label{fig:extCurves_App_1}
\end{figure*}

\begin{figure*}
\centering
\begin{subfigure}[b]{0.45\textwidth}
  \centering
  \includegraphics[width=\textwidth]{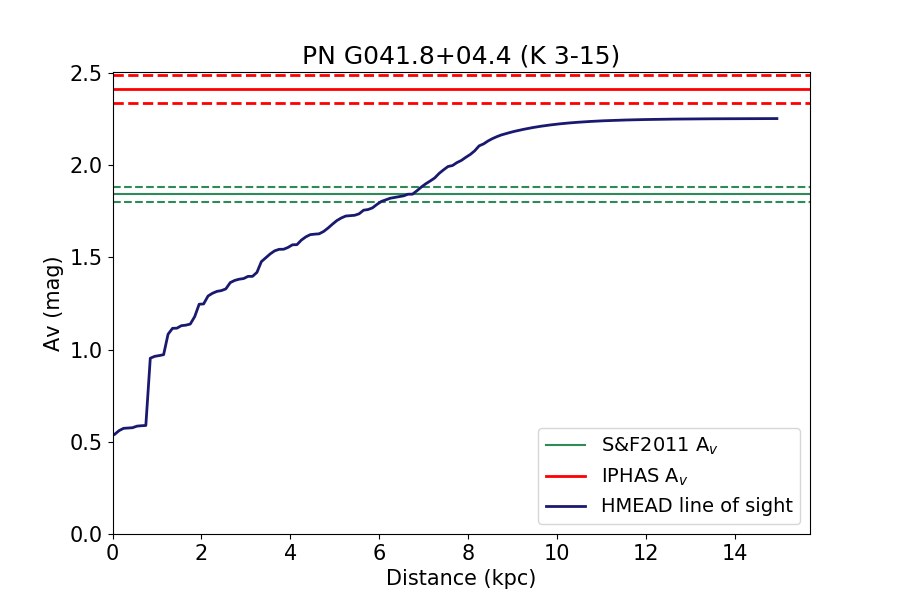}
  \subcaption{}
  \label{fig:K3-15}
  \end{subfigure}
\begin{subfigure}[b]{0.45\textwidth}
  \centering
  \includegraphics[width=\textwidth]{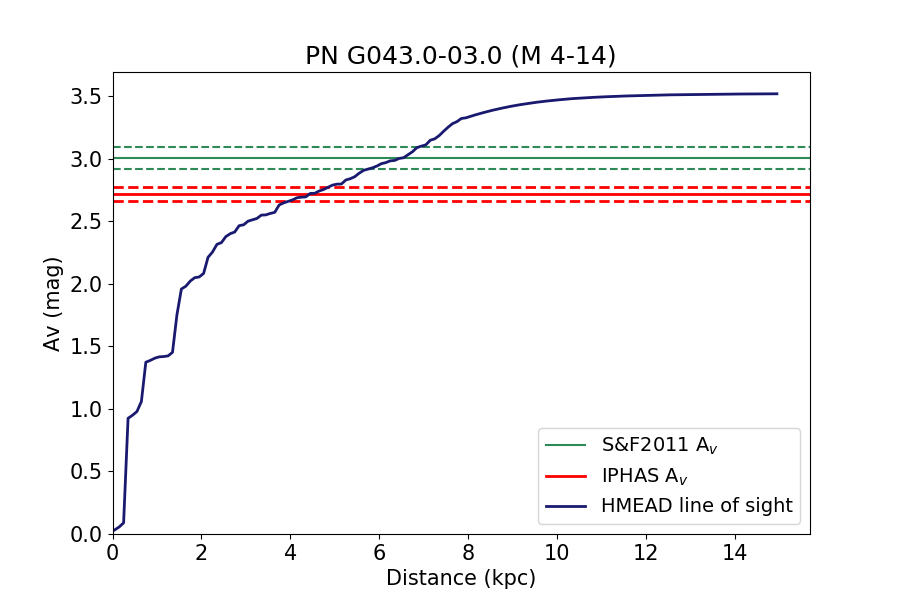}
  \subcaption{}
  \label{fig:M4-14}
  \end{subfigure}
  
\begin{subfigure}[b]{0.45\textwidth}
  \centering
  \includegraphics[width=\textwidth]{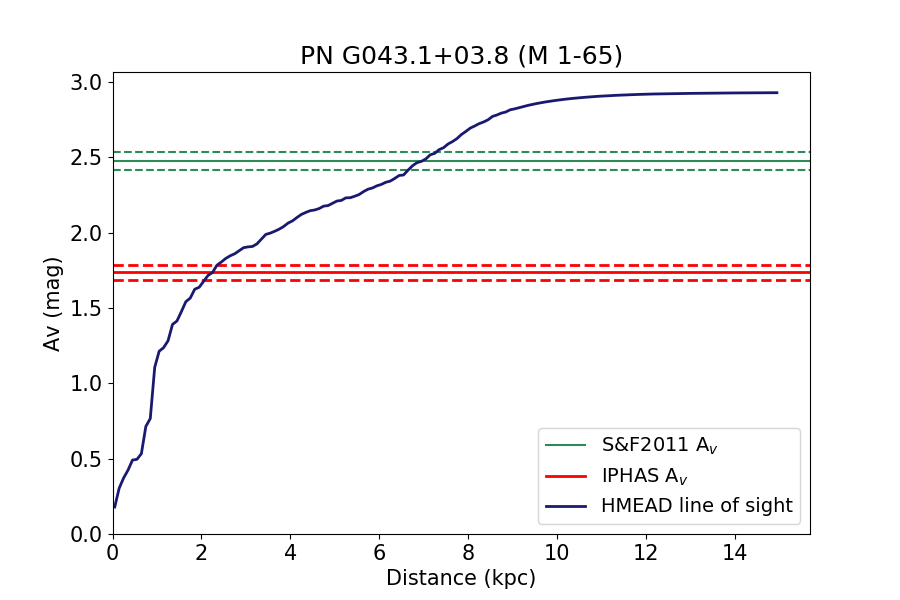}
  \subcaption{}
  \label{fig:M1-65}
  \end{subfigure}
\begin{subfigure}[b]{0.45\textwidth}
  \centering
  \includegraphics[width=\textwidth]{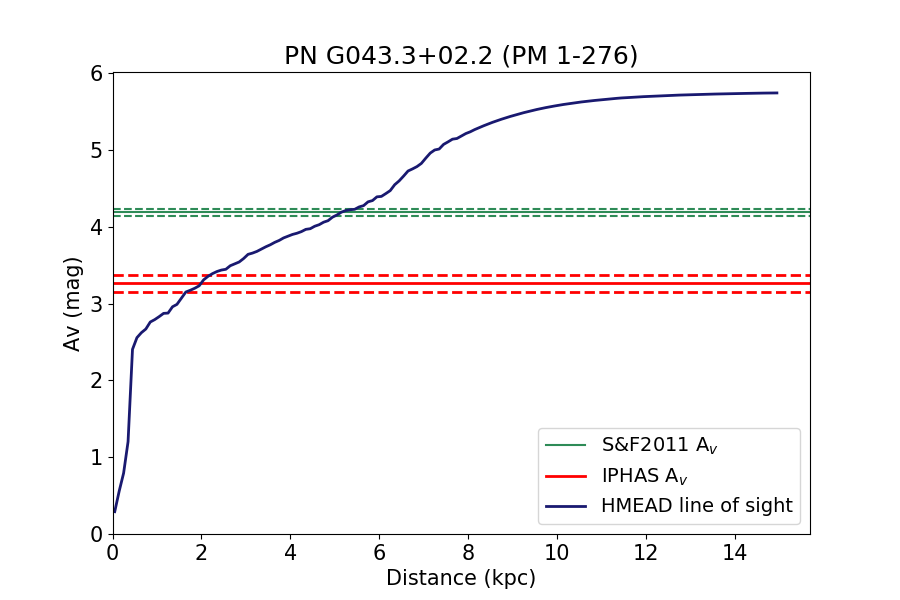}
  \subcaption{}
  \label{fig:PM1-276}
  \end{subfigure}

\begin{subfigure}[b]{0.45\textwidth}
  \centering
  \includegraphics[width=\textwidth]{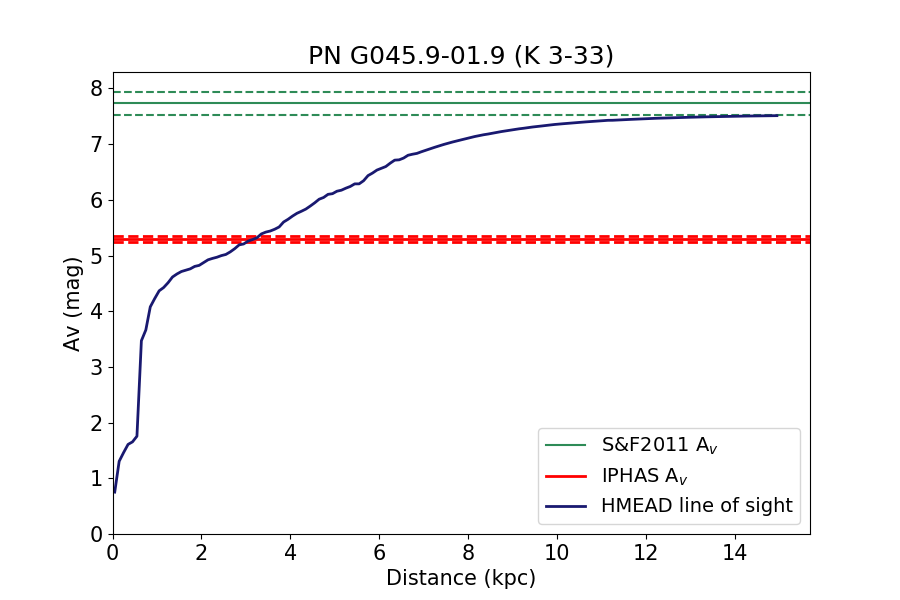}
  \subcaption{}
  \label{fig:K3-33}
  \end{subfigure}
\begin{subfigure}[b]{0.45\textwidth}
  \centering
  \includegraphics[width=\textwidth]{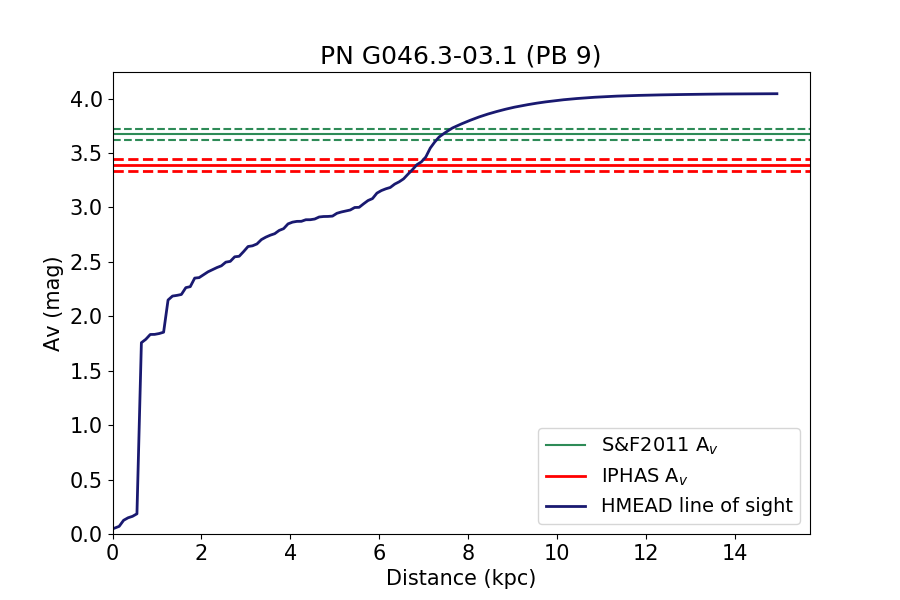}
 \subcaption{}
  \label{fig:PB9}
  \end{subfigure}
  
\begin{subfigure}[b]{0.45\textwidth}
  \centering
  \includegraphics[width=\textwidth]{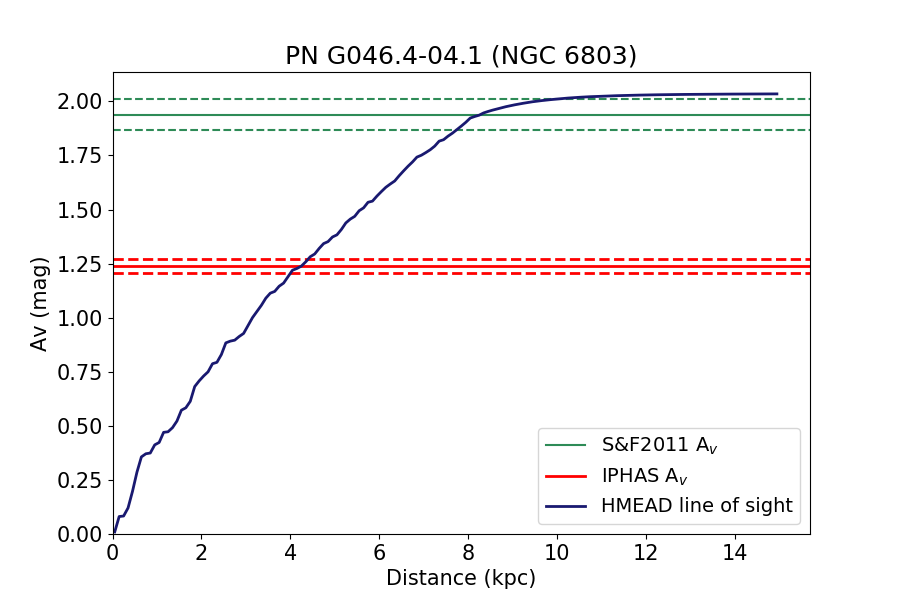}
  \subcaption{}
  \label{fig:NGC6803}
  \end{subfigure}
 \begin{subfigure}[b]{0.45\textwidth}
  \centering
  \includegraphics[width=\textwidth]{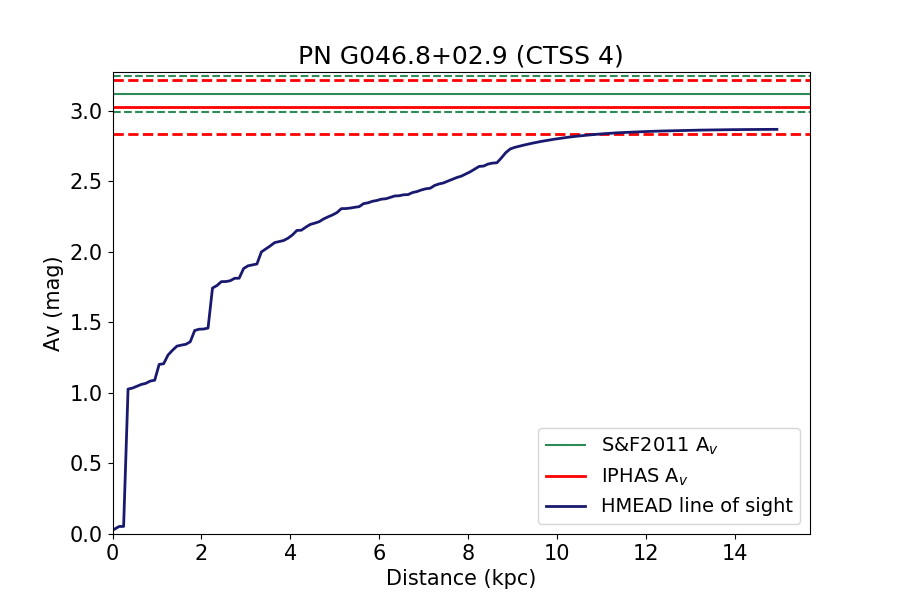}
  \subcaption{}
  \label{fig:CTSS 4}
  \end{subfigure}
  
  \caption{H-MEAD Extinction vs. Distance plots for K 3-15, M 4-14, M 1-65, PM 1-276, K 3-33, PB 9 NGC 6803 and CTSS 4.}
  \label{fig:extCurves_App_2}
\end{figure*}

\begin{figure*}
\centering

\begin{subfigure}[b]{0.45\textwidth}
  \centering
  \includegraphics[width=\textwidth]{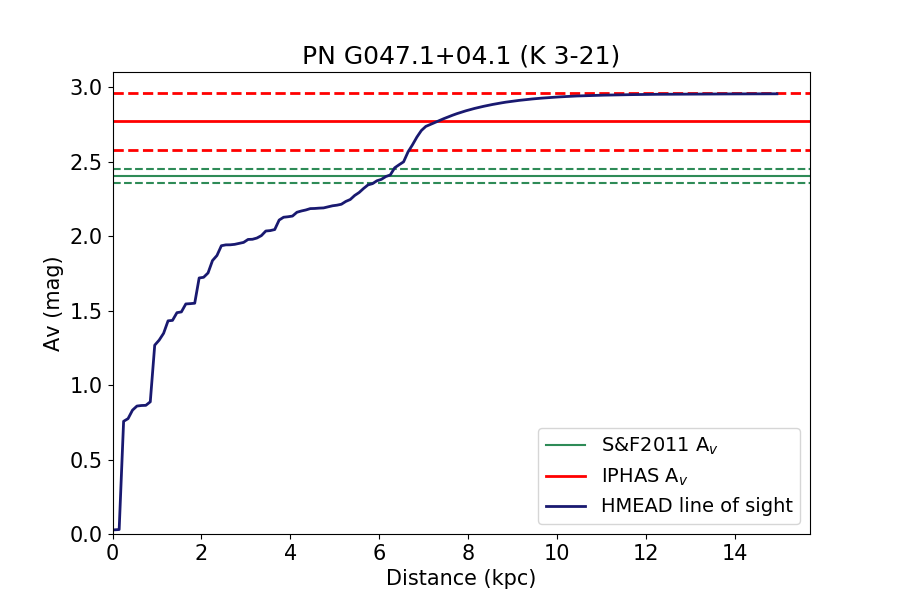}
  \subcaption{}
  \label{fig:K3-21}
  \end{subfigure}
\begin{subfigure}[b]{0.45\textwidth}
  \centering
  \includegraphics[width=\textwidth]{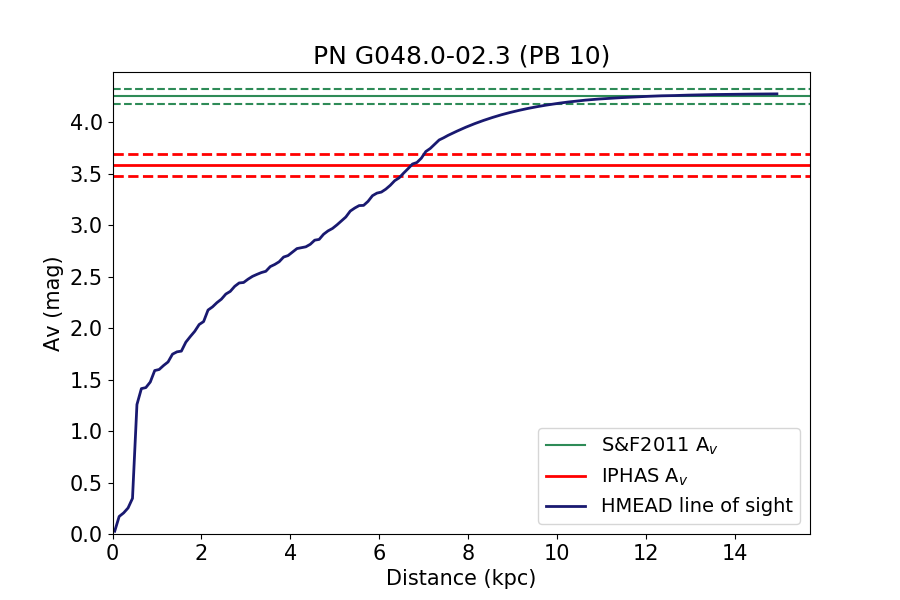}
  \subcaption{}
  \label{fig:PB10}
  \end{subfigure}
  
\begin{subfigure}[b]{0.45\textwidth}
  \centering
  \includegraphics[width=\textwidth]{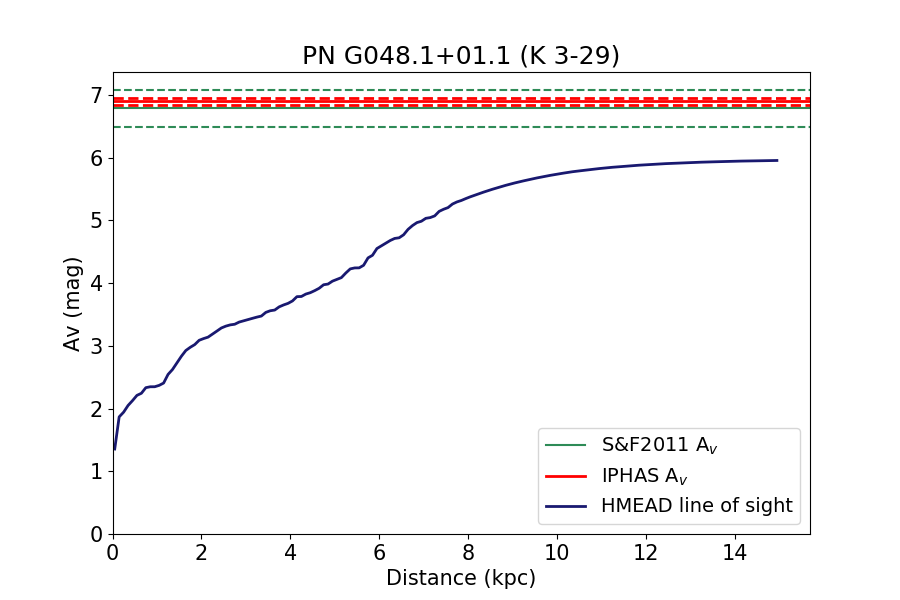}
  \subcaption{}
  \label{fig:K3-29}
  \end{subfigure}
\begin{subfigure}[b]{0.45\textwidth}
  \centering
  \includegraphics[width=\textwidth]{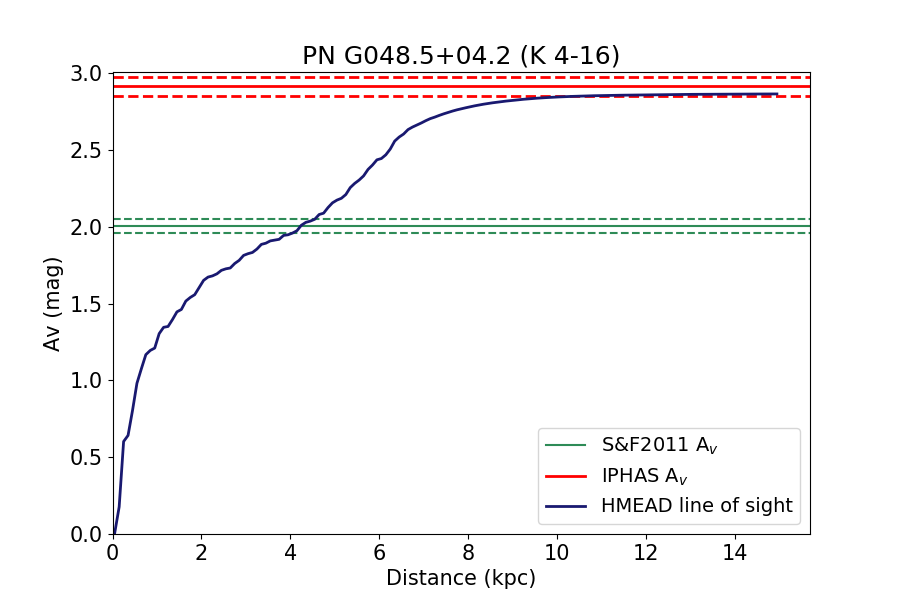}
 \subcaption{}
  \label{fig:K4-16}
  \end{subfigure}
  
\begin{subfigure}[b]{0.45\textwidth}
  \centering
  \includegraphics[width=\textwidth]{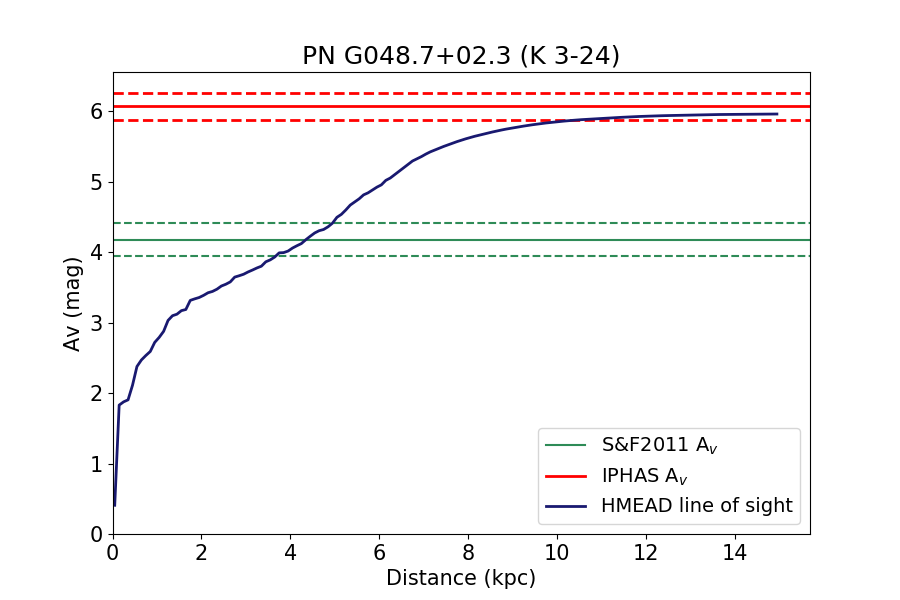}
  \subcaption{}
  \label{fig:K3-24}
  \end{subfigure}
\begin{subfigure}[b]{0.45\textwidth}
  \centering
  \includegraphics[width=\textwidth]{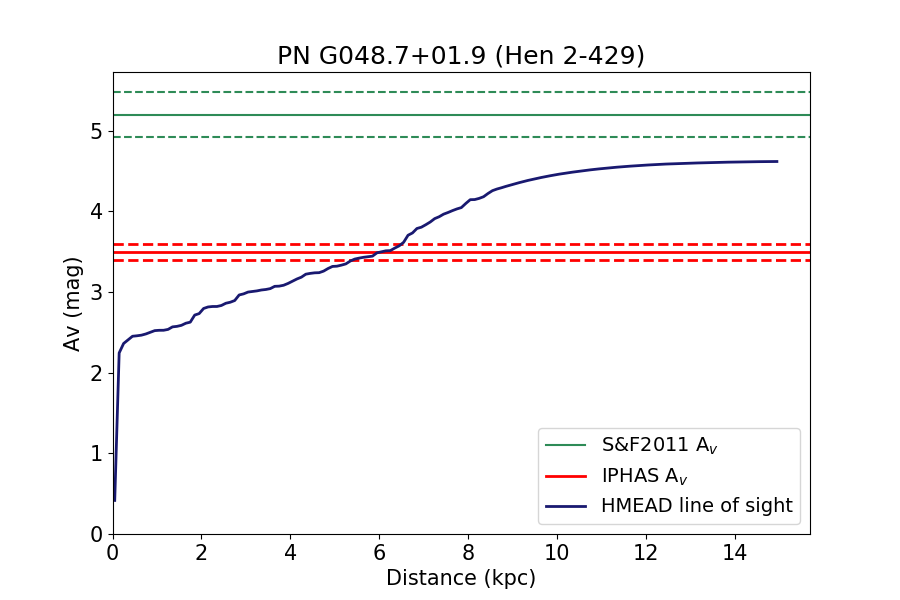}
 \subcaption{}
  \label{fig:Hen2-429}
  \end{subfigure}
  
\begin{subfigure}[b]{0.45\textwidth}
  \centering
  \includegraphics[width=\textwidth]{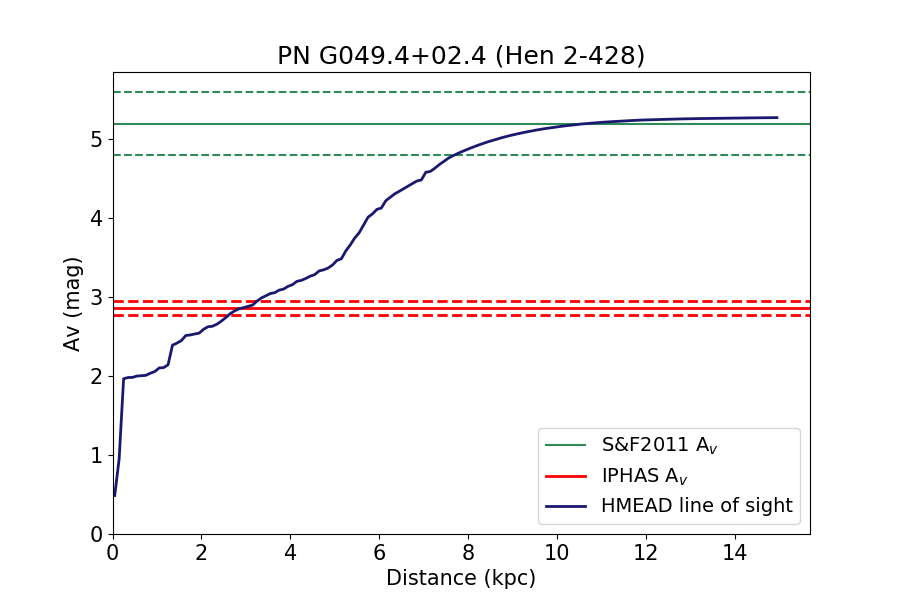}
  \subcaption{}
  \label{fig:Hen2-428}
  \end{subfigure}
 \begin{subfigure}[b]{0.45\textwidth}
  \centering
  \includegraphics[width=\textwidth]{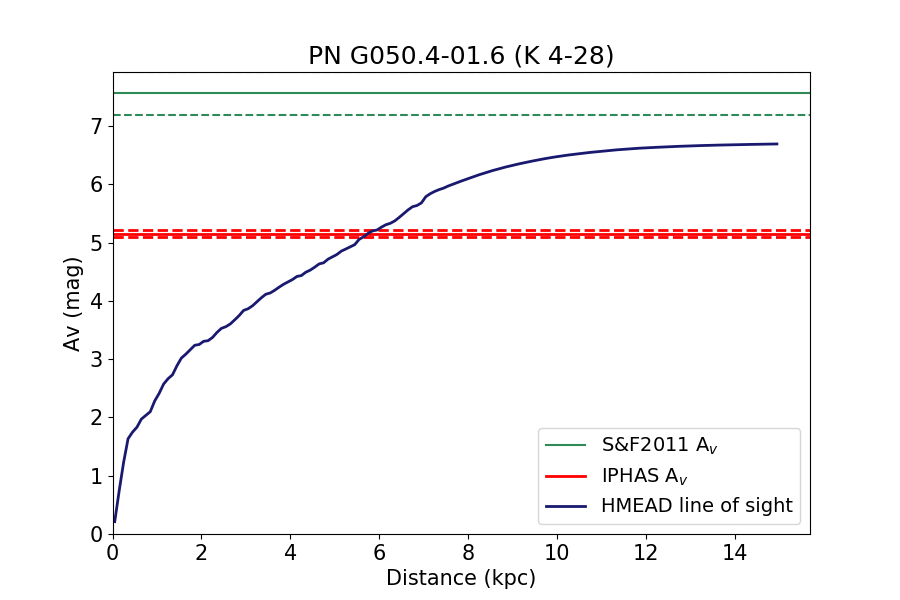}
  \subcaption{}
  \label{fig:K4-28}
  \end{subfigure} 
  
  \caption{H-MEAD Extinction vs. Distance plots for K 3-21, PB 10, K 3-29, K 4-16, K 3-24, Hen 2-429, Hen 2-428 and K 4-28.}
  \label{fig:extCurves_App_3}
\end{figure*}

\begin{figure*}
\centering

\begin{subfigure}[b]{0.45\textwidth}
  \centering
  \includegraphics[width=\textwidth]{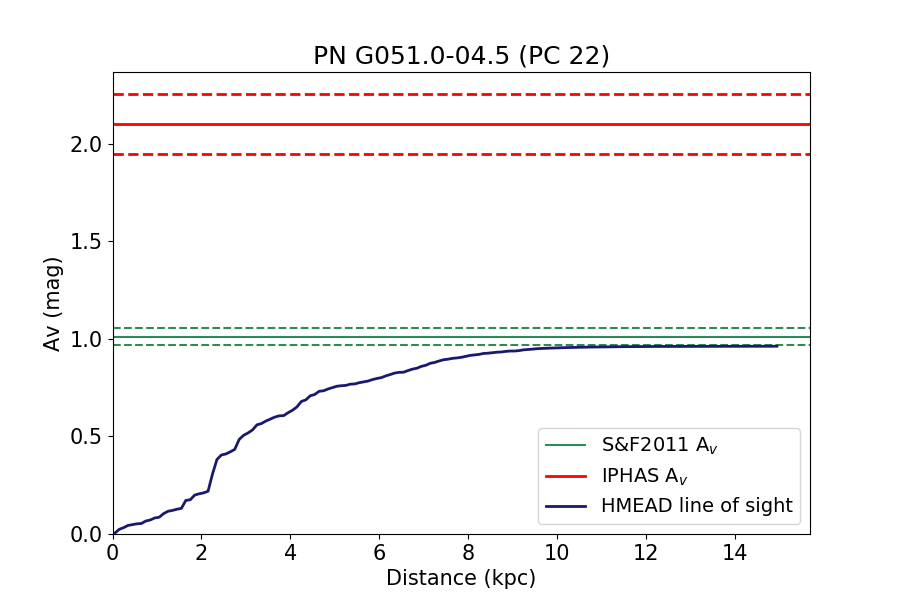}
  \subcaption{}
  \label{fig:PC22}
  \end{subfigure}
\begin{subfigure}[b]{0.45\textwidth}
  \centering
  \includegraphics[width=\textwidth]{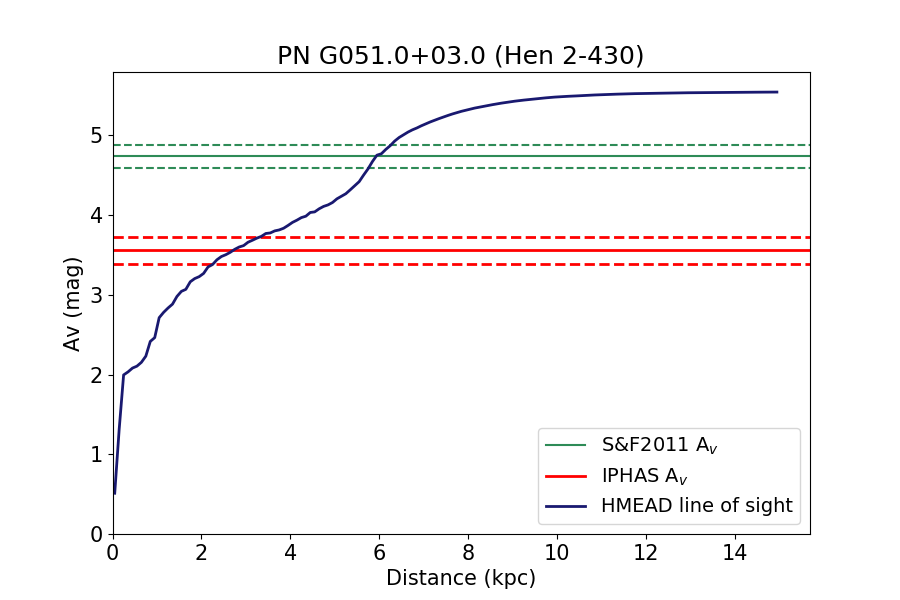}
  \subcaption{}
  \label{fig:Hen2-430}
  \end{subfigure}
  
\begin{subfigure}[b]{0.45\textwidth}
  \centering
  \includegraphics[width=\textwidth]{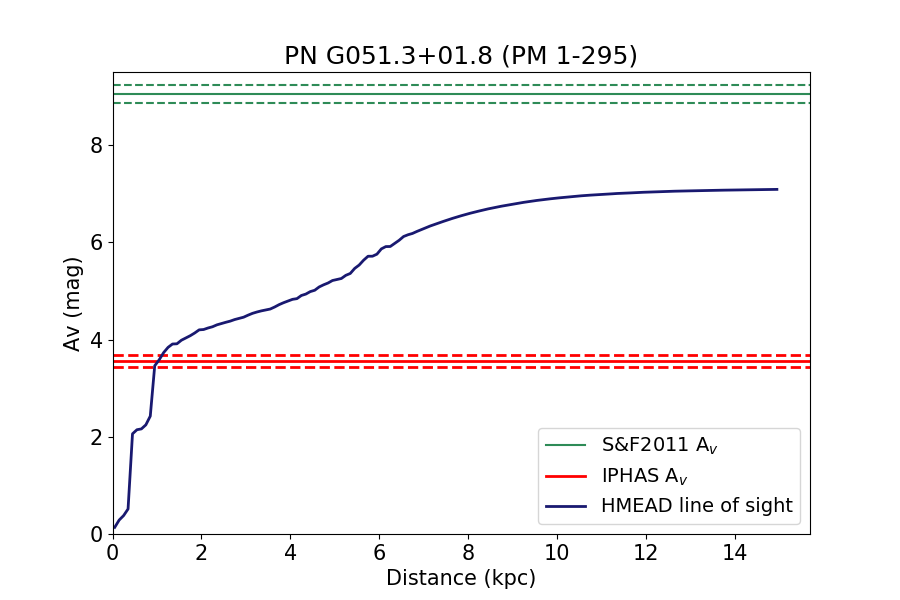}
  \subcaption{}
  \label{fig:PM1-295}
  \end{subfigure}
\begin{subfigure}[b]{0.45\textwidth}
  \centering
  \includegraphics[width=\textwidth]{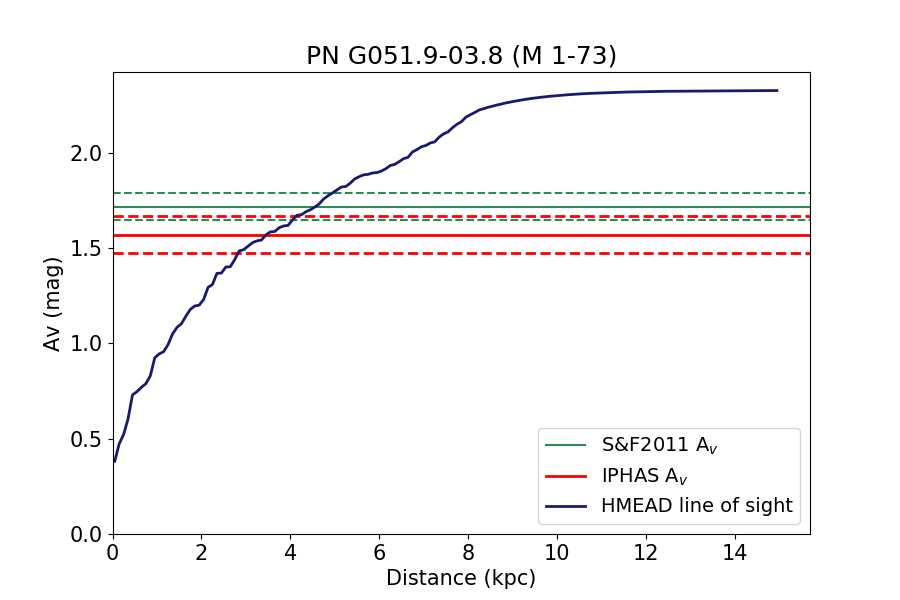}
 \subcaption{}
  \label{fig:M1-73}
  \end{subfigure}
  
\begin{subfigure}[b]{0.45\textwidth}
  \centering
  \includegraphics[width=\textwidth]{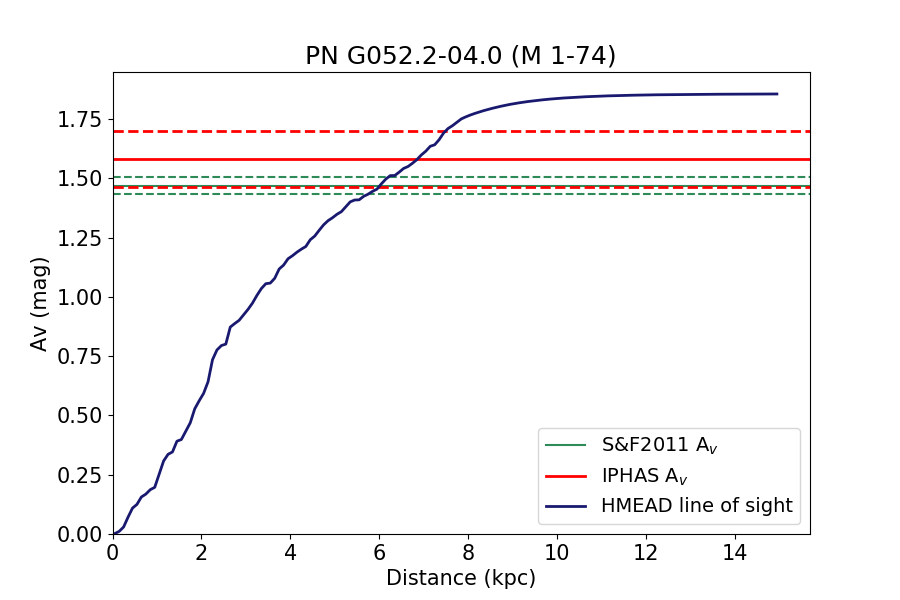}
  \subcaption{}
  \label{fig:M1-74}
  \end{subfigure}
\begin{subfigure}[b]{0.45\textwidth}
  \centering
  \includegraphics[width=\textwidth]{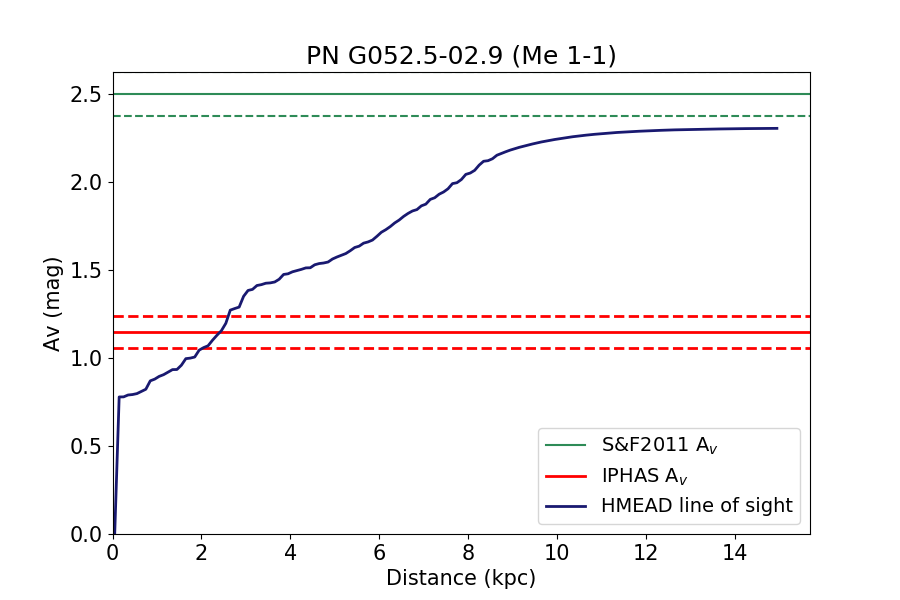}
 \subcaption{}
  \label{fig:Me1-1}
  \end{subfigure}
  
\begin{subfigure}[b]{0.45\textwidth}
  \centering
  \includegraphics[width=\textwidth]{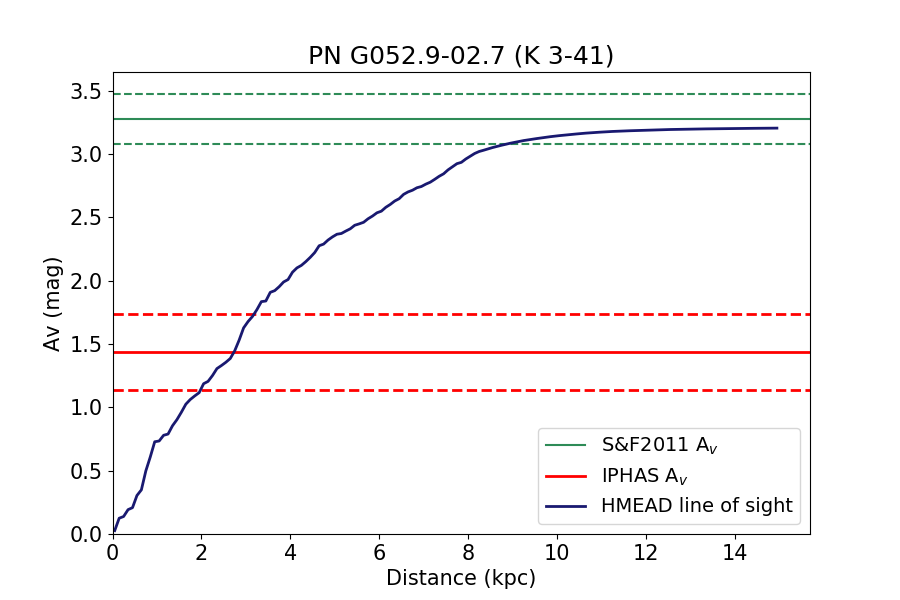}
  \subcaption{}
  \label{fig:K3-41}
  \end{subfigure}
 \begin{subfigure}[b]{0.45\textwidth}
  \centering
  \includegraphics[width=\textwidth]{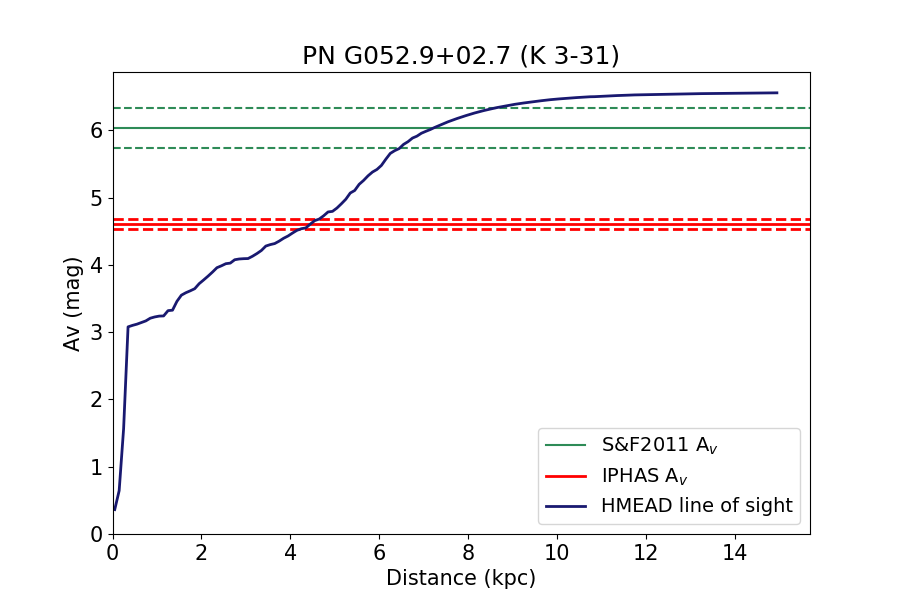}
  \subcaption{}
  \label{fig:K3-31}
  \end{subfigure}

  \caption{H-MEAD Extinction vs. Distance plots for PNe PC 22, Hen 2-430, PM 1-295, M 1-73, M 1-74, Me 1-1, K 3-41 and K 3-31.}
  \label{fig:extCurves_App_4}
\end{figure*}

\begin{figure*}
\centering

\begin{subfigure}[b]{0.45\textwidth}
  \centering
  \includegraphics[width=\textwidth]{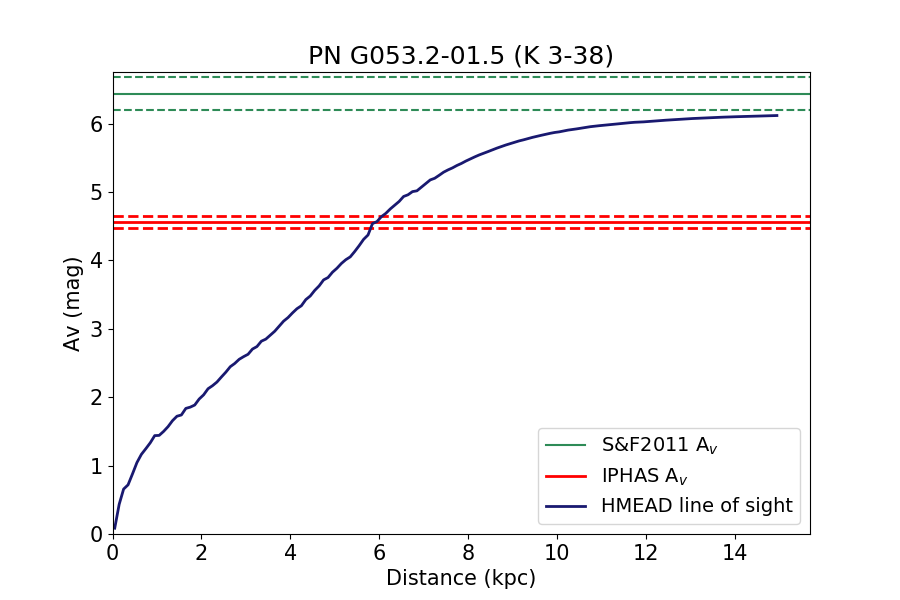}
  \subcaption{}
  \label{fig:K3-38}
  \end{subfigure}
\begin{subfigure}[b]{0.45\textwidth}
  \centering
  \includegraphics[width=\textwidth]{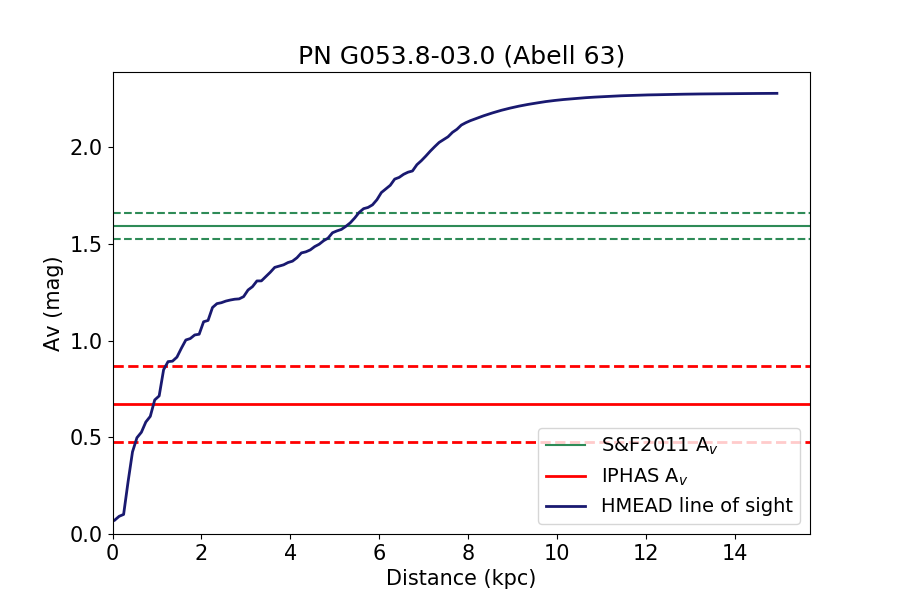}
  \subcaption{}
  \label{fig:Abell63}
  \end{subfigure}
  
\begin{subfigure}[b]{0.45\textwidth}
  \centering
  \includegraphics[width=\textwidth]{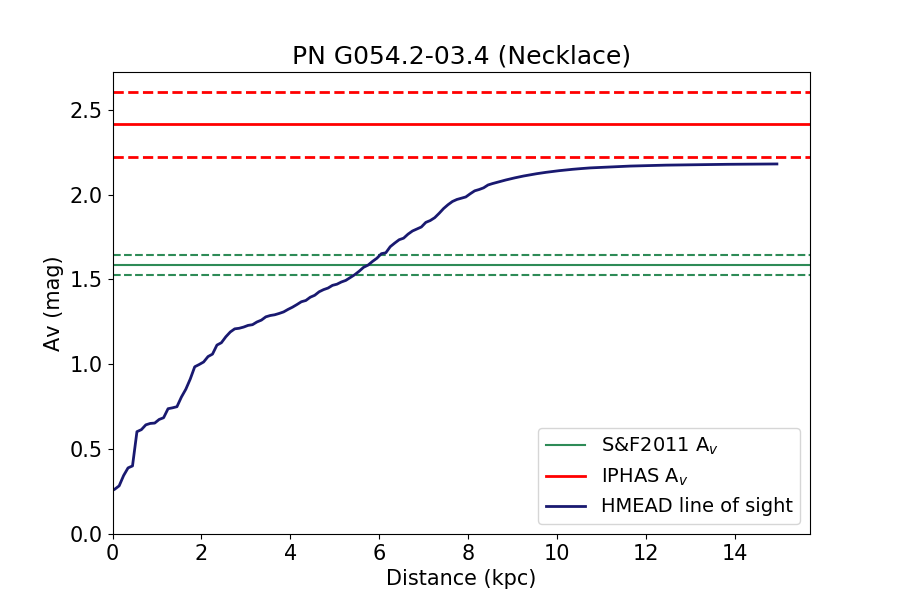}
  \subcaption{}
  \label{fig:Necklace}
  \end{subfigure}
\begin{subfigure}[b]{0.45\textwidth}
  \centering
  \includegraphics[width=\textwidth]{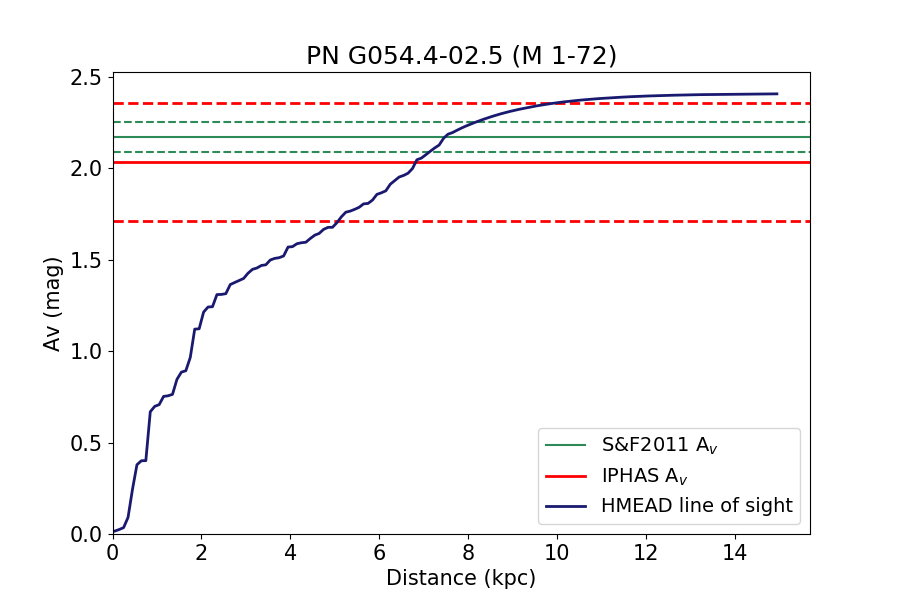}
 \subcaption{}
  \label{fig:M1-72}
  \end{subfigure}
  
\begin{subfigure}[b]{0.45\textwidth}
  \centering
  \includegraphics[width=\textwidth]{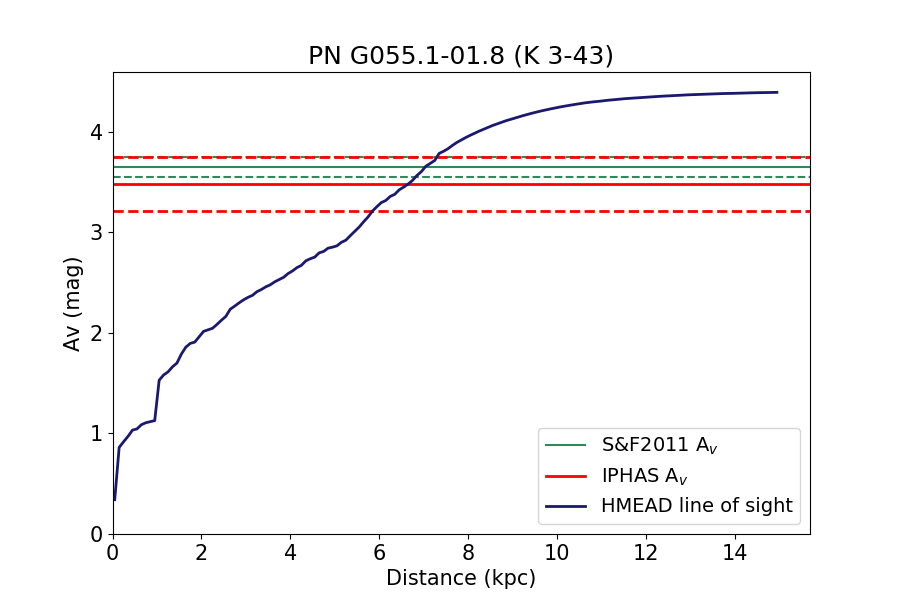}
  \subcaption{}
  \label{fig:K3-43}
  \end{subfigure}
\begin{subfigure}[b]{0.45\textwidth}
  \centering
  \includegraphics[width=\textwidth]{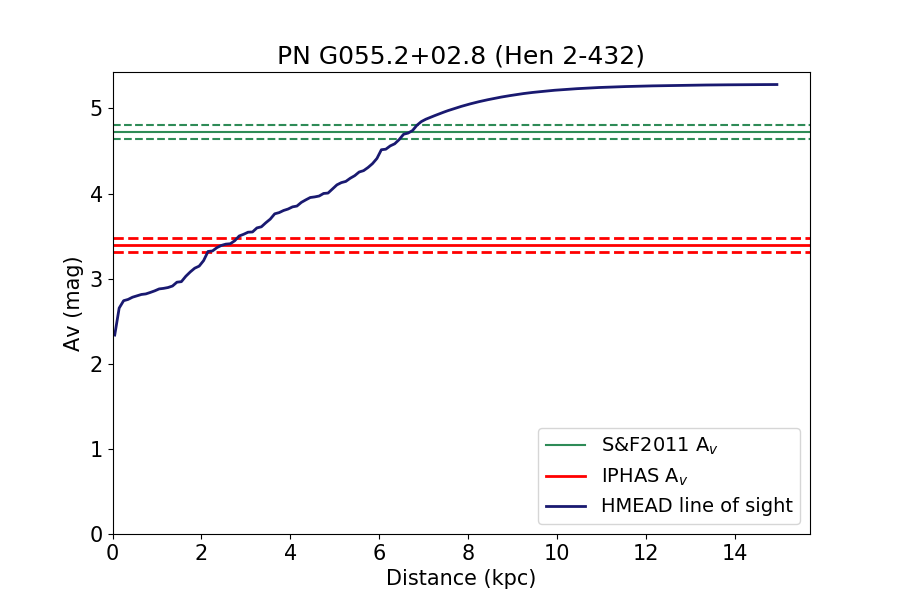}
 \subcaption{}
  \label{fig:Hen2-432}
  \end{subfigure}
  
\begin{subfigure}[b]{0.45\textwidth}
  \centering
  \includegraphics[width=\textwidth]{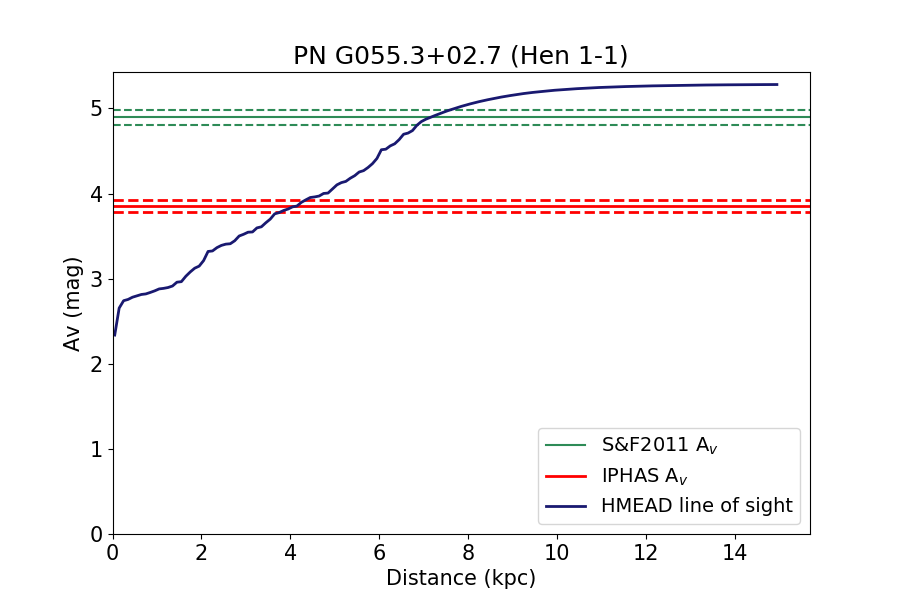}
  \subcaption{}
  \label{fig:Hen1-1}
  \end{subfigure}
 \begin{subfigure}[b]{0.45\textwidth}
  \centering
  \includegraphics[width=\textwidth]{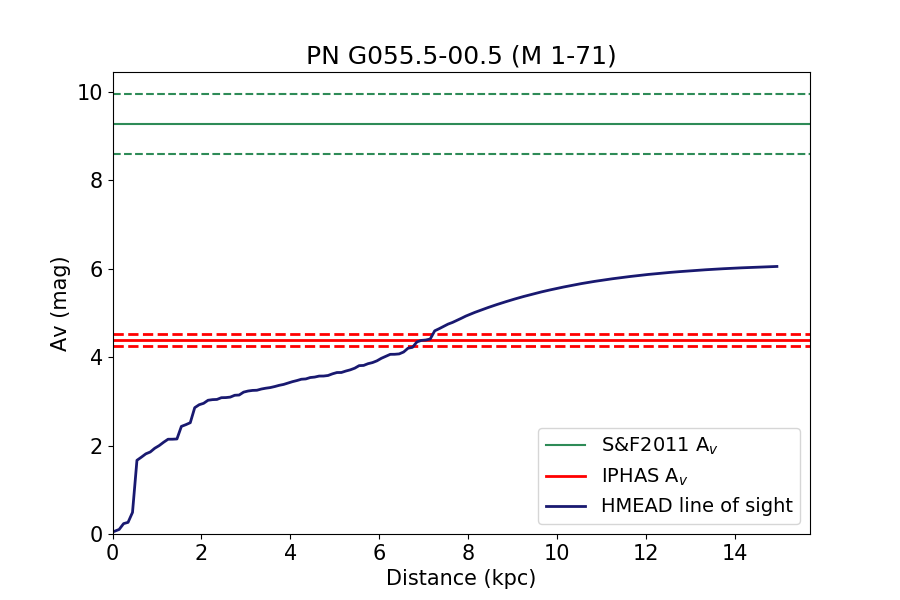}
  \subcaption{}
  \label{fig:M1-71}
  \end{subfigure} 
  
  \caption{H-MEAD Extinction vs. Distance plots for K 3-38, Abell 63, Necklace nebula, M1-72, K 3-43, Hen2-432, Hen 1-1 and M 1-71.}
  \label{fig:extCurves_App_5}
\end{figure*}

\begin{figure*}
\centering

\begin{subfigure}[b]{0.45\textwidth}
  \centering
  \includegraphics[width=\textwidth]{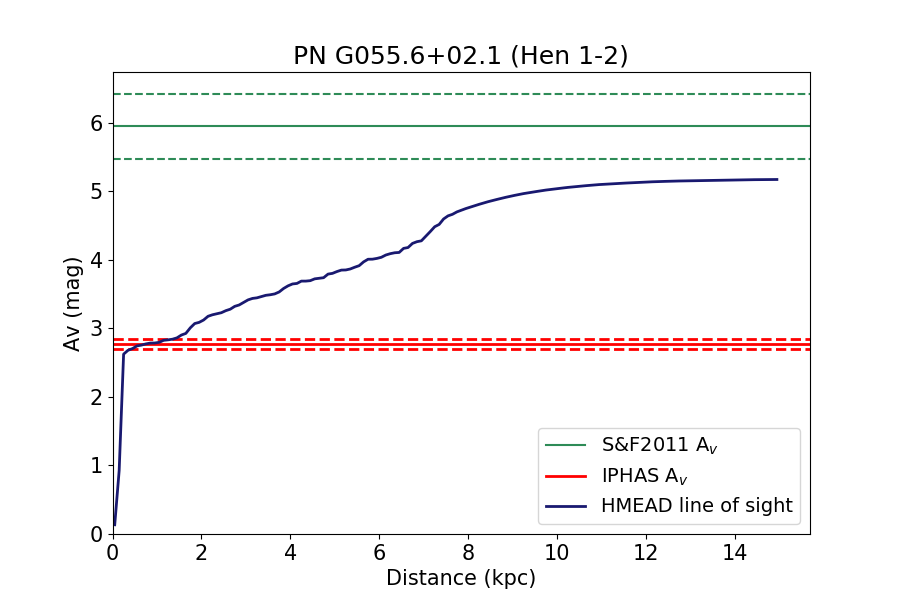}
  \subcaption{}
  \label{fig:Hen1-2}
  \end{subfigure}
\begin{subfigure}[b]{0.45\textwidth}
  \centering
  \includegraphics[width=\textwidth]{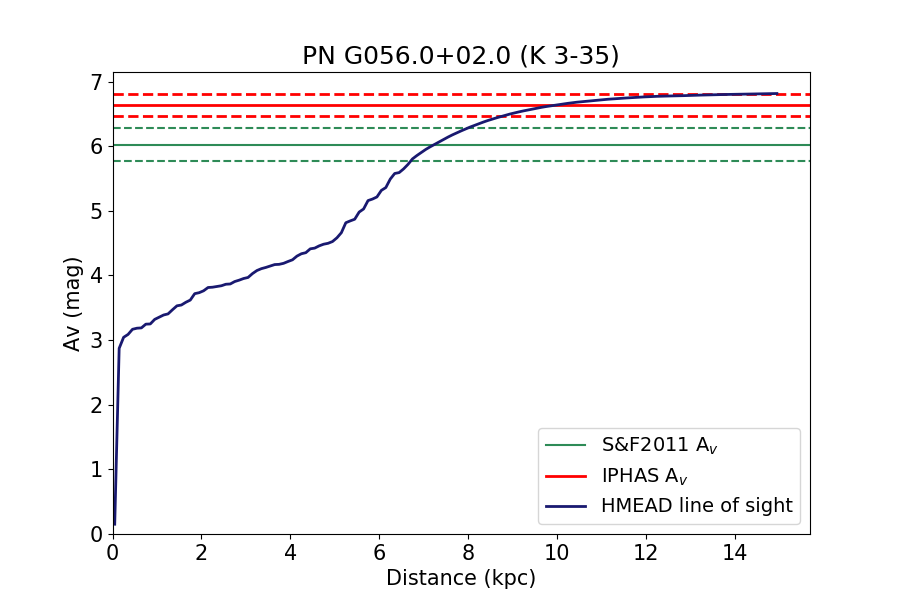}
  \subcaption{}
  \label{fig:K3-35}
  \end{subfigure}
  
\begin{subfigure}[b]{0.45\textwidth}
  \centering
  \includegraphics[width=\textwidth]{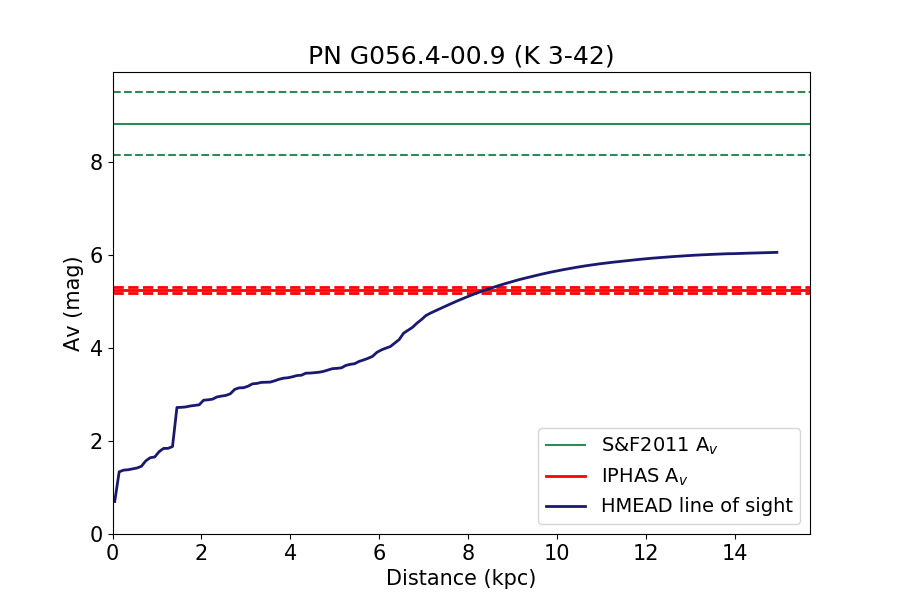}
  \subcaption{}
  \label{fig:K3-42}
  \end{subfigure}
\begin{subfigure}[b]{0.45\textwidth}
  \centering
  \includegraphics[width=\textwidth]{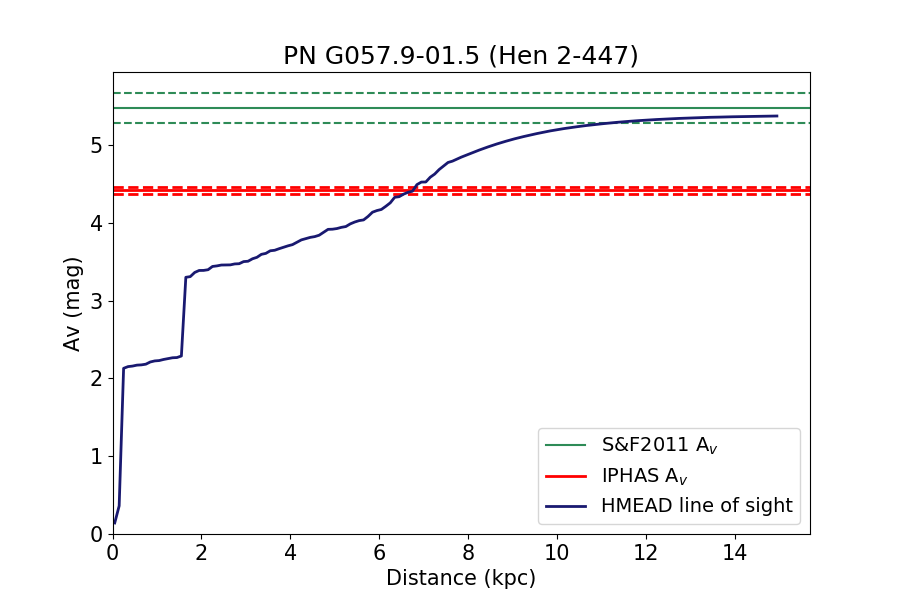}
 \subcaption{}
  \label{fig:Hen2-447}
  \end{subfigure}
  
\begin{subfigure}[b]{0.45\textwidth}
  \centering
  \includegraphics[width=\textwidth]{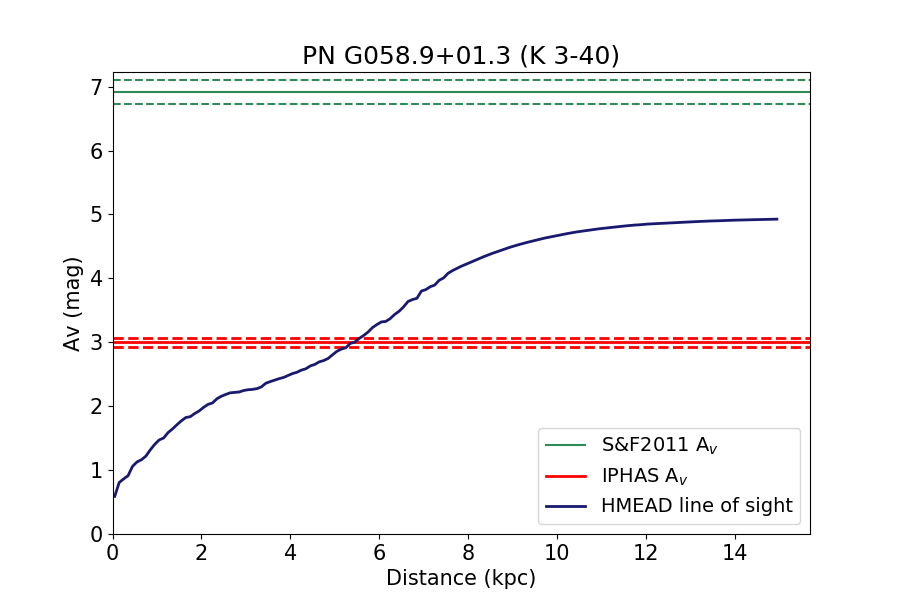}
  \subcaption{}
  \label{fig:K3-40}
  \end{subfigure}
\begin{subfigure}[b]{0.45\textwidth}
  \centering
  \includegraphics[width=\textwidth]{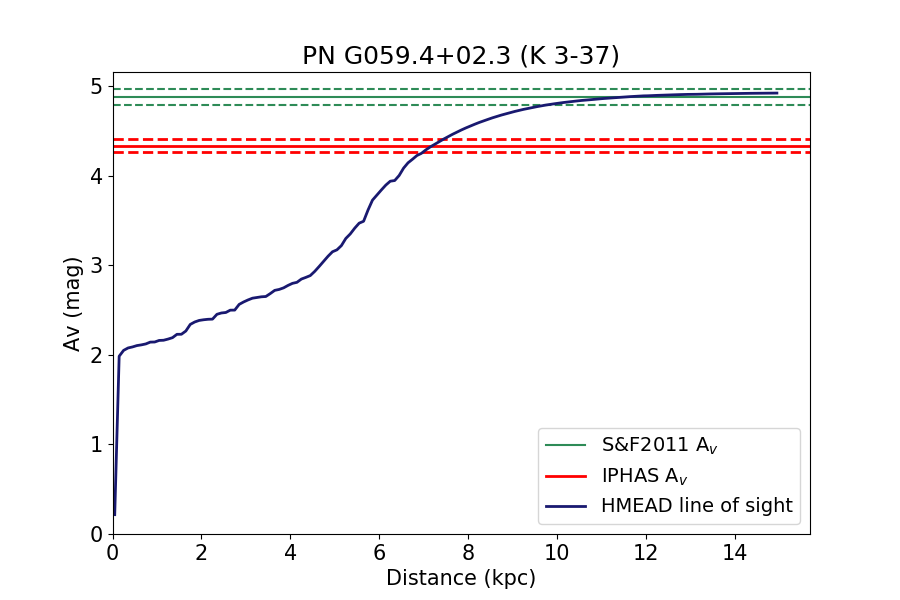}
 \subcaption{}
  \label{fig:K3-37}
  \end{subfigure}
  
\begin{subfigure}[b]{0.45\textwidth}
  \centering
  \includegraphics[width=\textwidth]{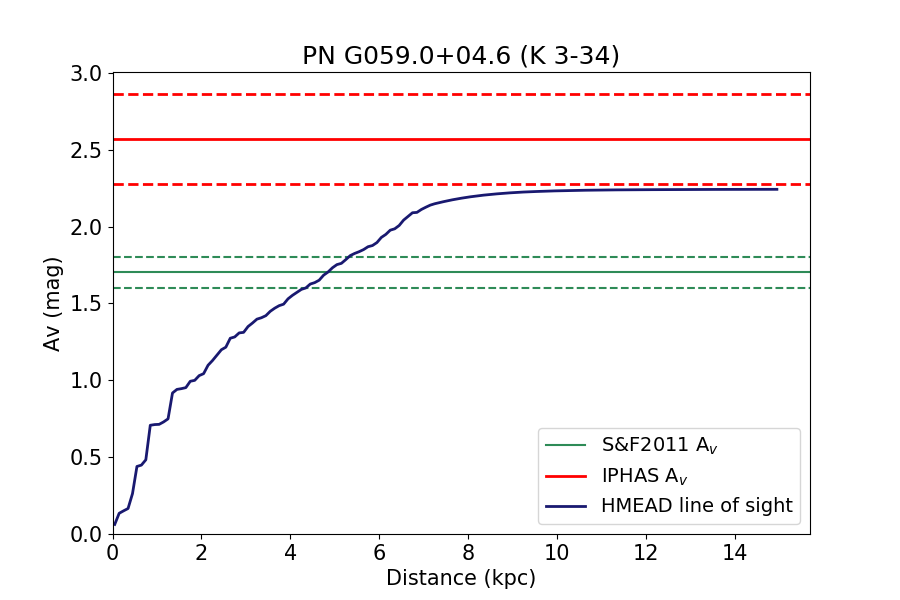}
  \subcaption{}
  \label{fig:K3-34}
  \end{subfigure}
\begin{subfigure}[b]{0.45\textwidth}
  \centering
  \includegraphics[width=\textwidth]{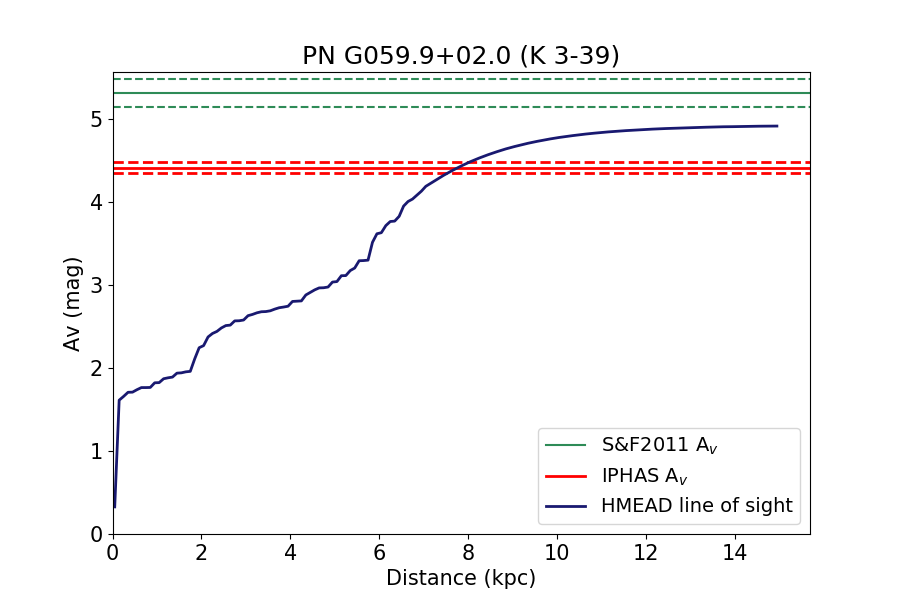}
  \subcaption{}
  \label{fig:K3-39}
  \end{subfigure}

  \caption{H-MEAD Extinction vs. Distance  plots for Hen 1-2, K 3-35, K 3-42, Hen 2-447, K 3-40, K 3-37, K 3-34 and K 3-39.}
  \label{fig:extCurves_App_6}
\end{figure*}

\begin{figure*}
\centering

\begin{subfigure}[b]{0.45\textwidth}
  \centering
  \includegraphics[width=\textwidth]{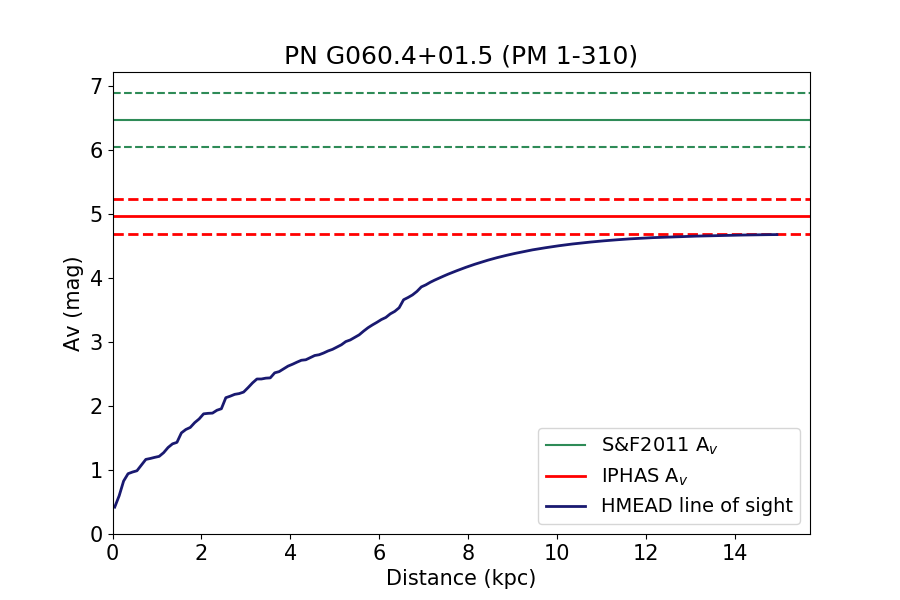}
  \subcaption{}
  \label{fig:PM1-310}
  \end{subfigure}
\begin{subfigure}[b]{0.45\textwidth}
  \centering
  \includegraphics[width=\textwidth]{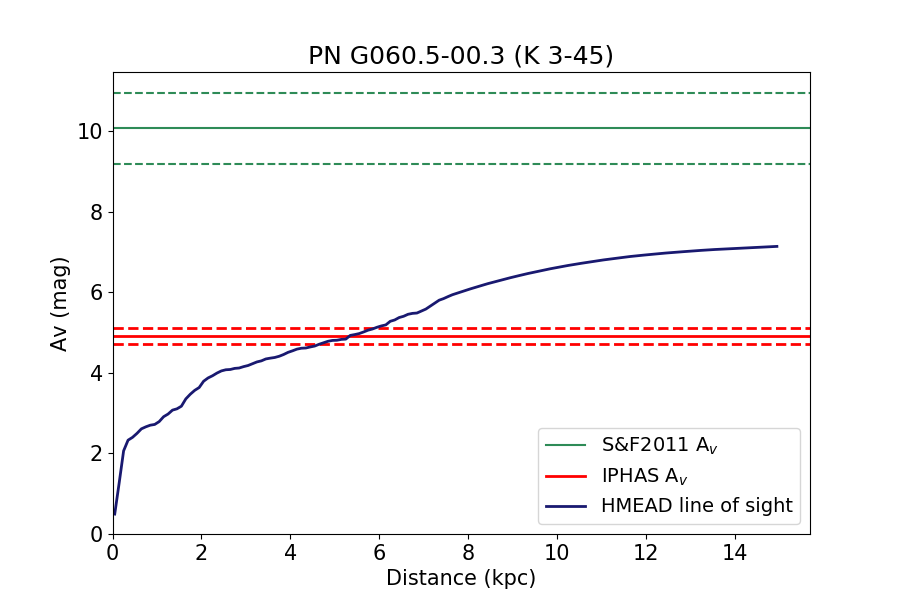}
  \subcaption{}
  \label{fig:K3-45}
  \end{subfigure}
  
\begin{subfigure}[b]{0.45\textwidth}
  \centering
  \includegraphics[width=\textwidth]{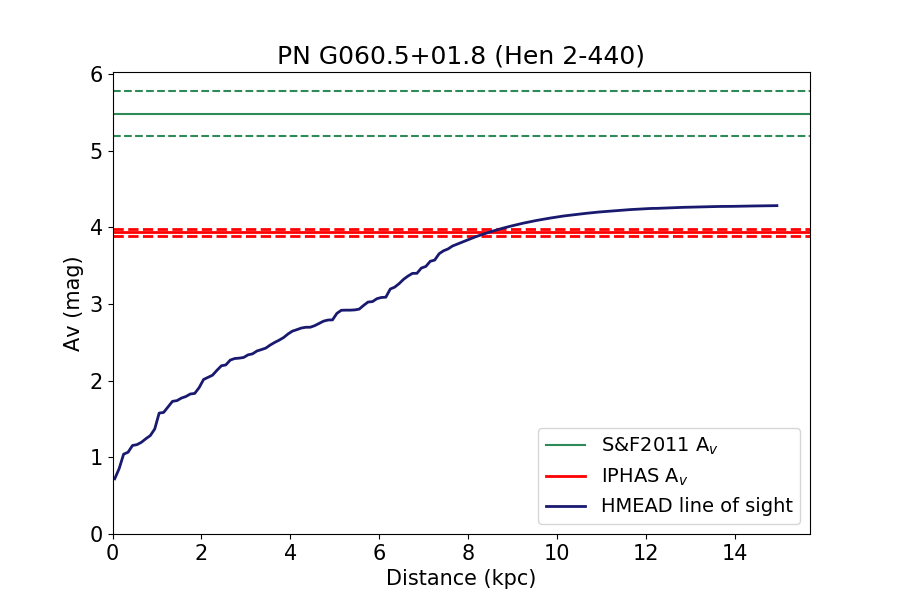}
  \subcaption{}
  \label{fig:Hen2-440}
  \end{subfigure}
\begin{subfigure}[b]{0.45\textwidth}
  \centering
  \includegraphics[width=\textwidth]{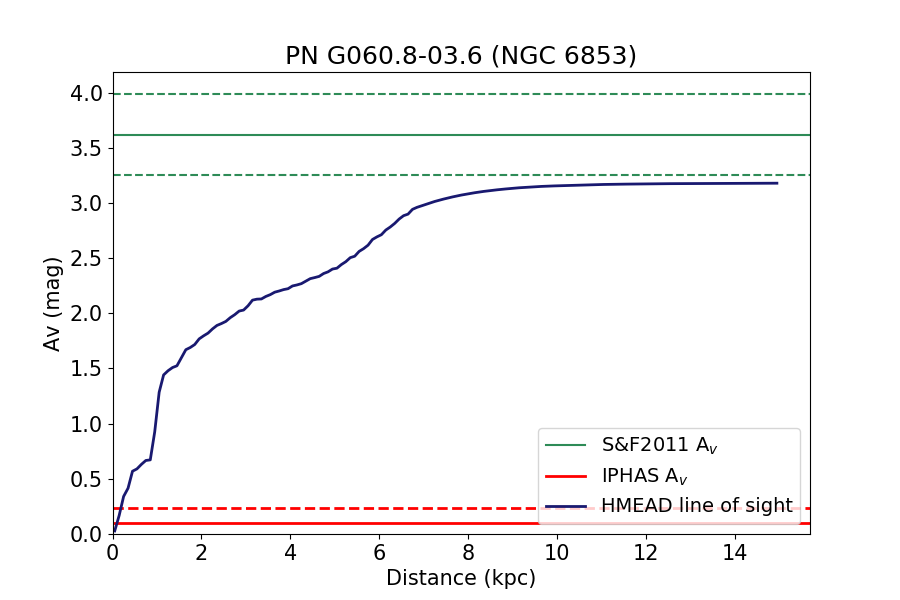}
 \subcaption{}
  \label{fig:NGC6853}
  \end{subfigure}

\begin{subfigure}[b]{0.45\textwidth}
  \centering
  \includegraphics[width=\textwidth]{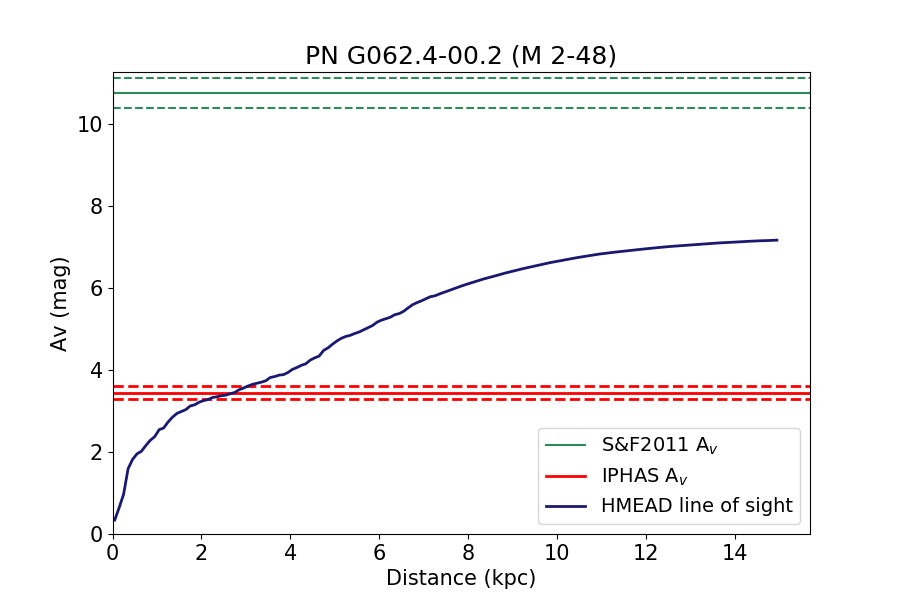}
 \subcaption{}
  \label{fig:M2-48}
  \end{subfigure}
\begin{subfigure}[b]{0.45\textwidth}
  \centering
  \includegraphics[width=\textwidth]{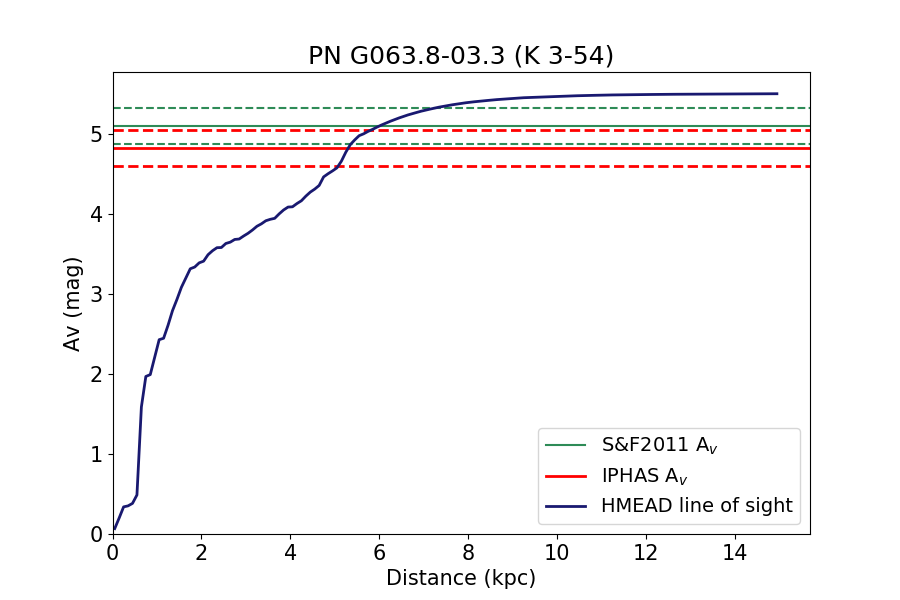}
  \subcaption{}
  \label{fig:K3-54}
  \end{subfigure}
  
 \begin{subfigure}[b]{0.45\textwidth}
  \centering
  \includegraphics[width=\textwidth]{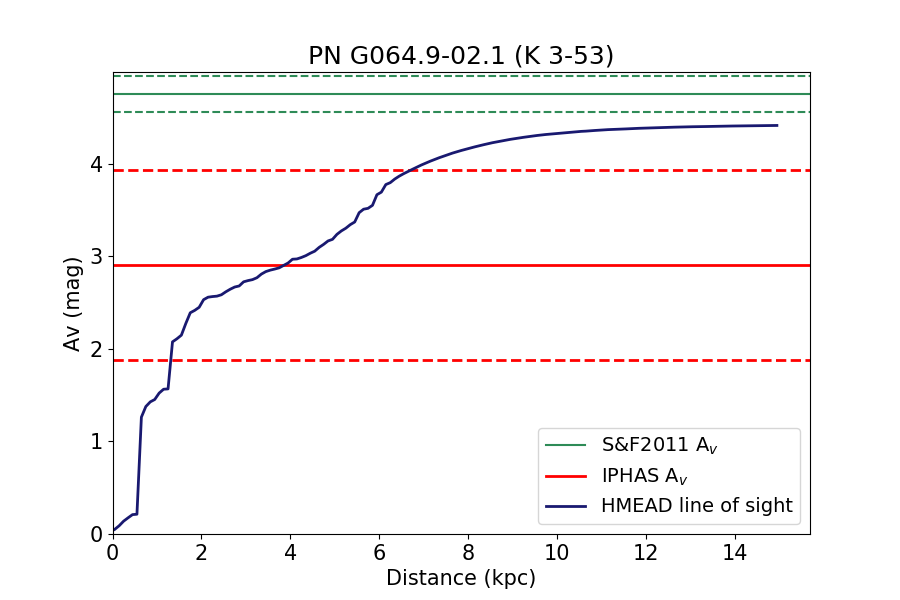}
  \subcaption{}
  \label{fig:K3-53}
  \end{subfigure}
\begin{subfigure}[b]{0.45\textwidth}
  \centering
  \includegraphics[width=\textwidth]{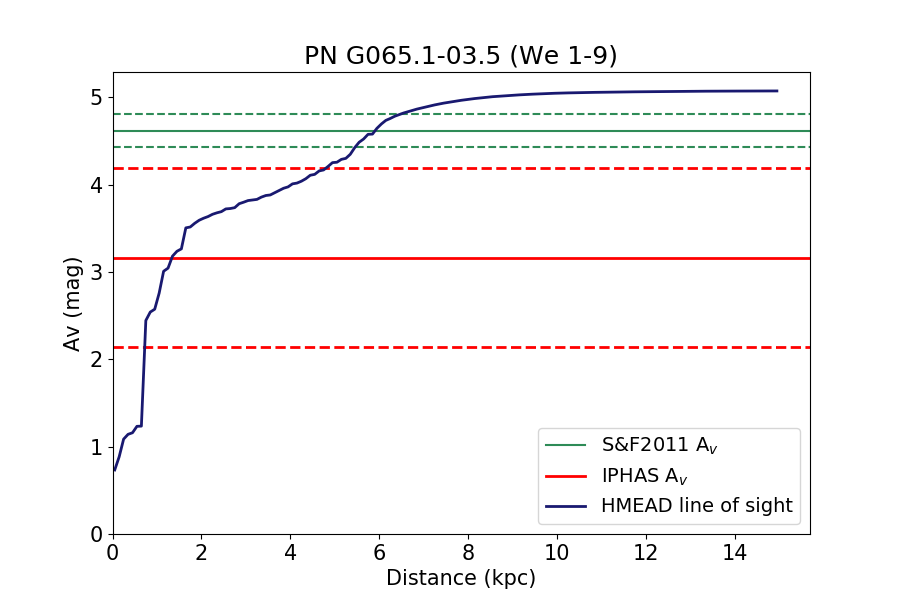}
  \subcaption{}
  \label{fig:We1-9}
  \end{subfigure}
  
  \caption{H-MEAD Extinction vs. Distance  plots for PM 1-310, K 3-45, Hen 2-440, NGC 6853, M 2-48, K 3-54, K 3-53 and We 1-9.}
  \label{fig:extCurves_App_7}
\end{figure*}

\begin{figure*}
\centering
  
\begin{subfigure}[b]{0.45\textwidth}
  \centering
  \includegraphics[width=\textwidth]{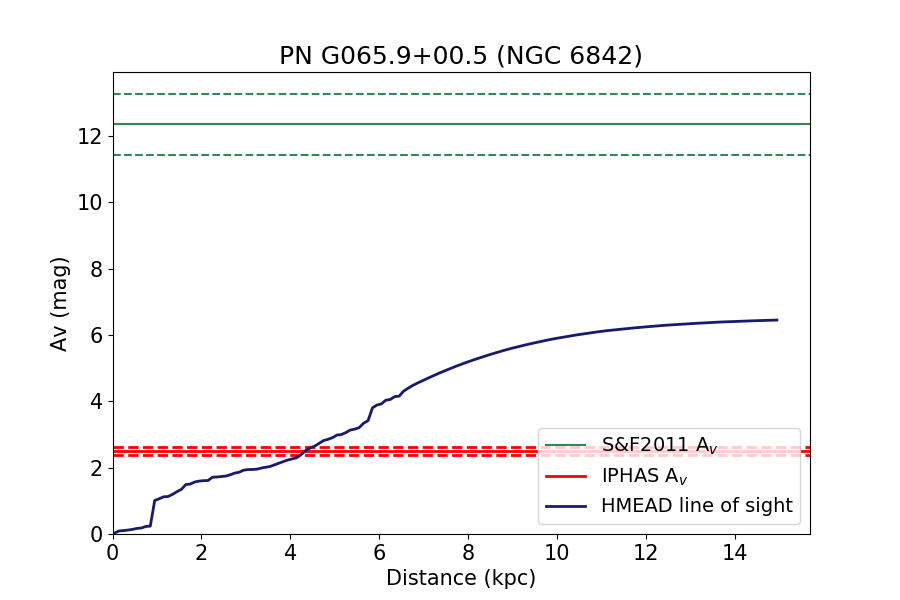}
  \subcaption{}
  \label{fig:NGC6842}
  \end{subfigure}
\begin{subfigure}[b]{0.45\textwidth}
  \centering
  \includegraphics[width=\textwidth]{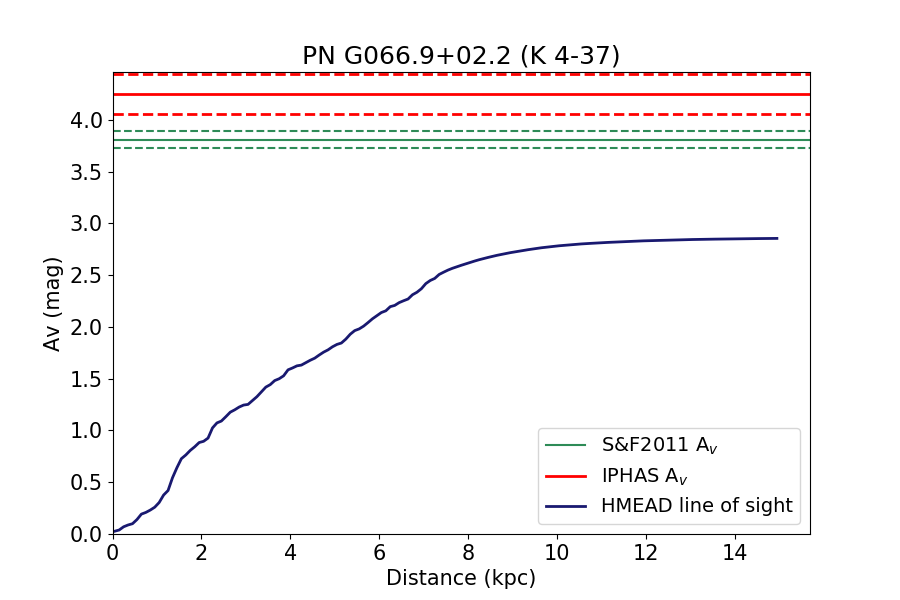}
  \subcaption{}
  \label{fig:K4-37}
  \end{subfigure}

\begin{subfigure}[b]{0.45\textwidth}
  \centering
  \includegraphics[width=\textwidth]{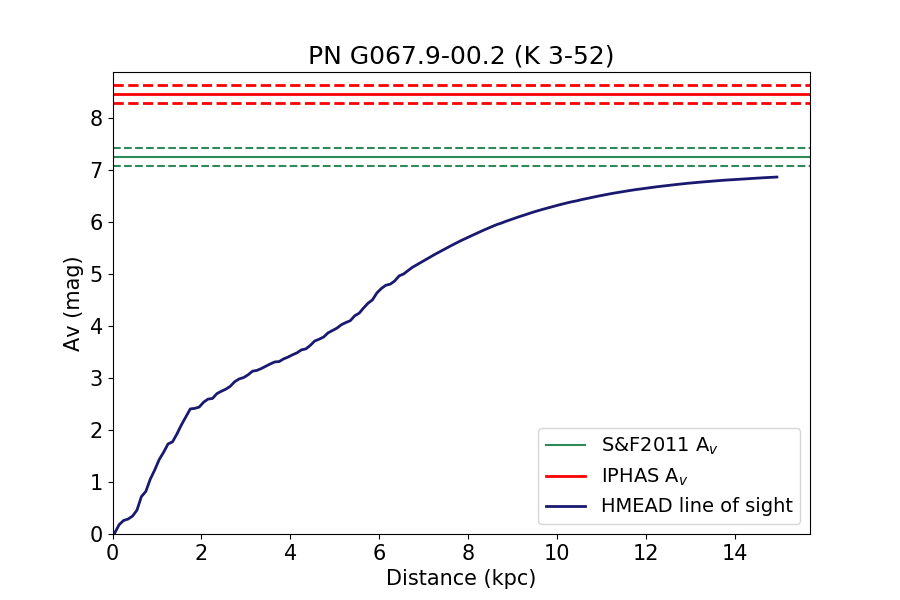}
 \subcaption{}
  \label{fig:K3-52}
  \end{subfigure}
\begin{subfigure}[b]{0.45\textwidth}
  \centering
  \includegraphics[width=\textwidth]{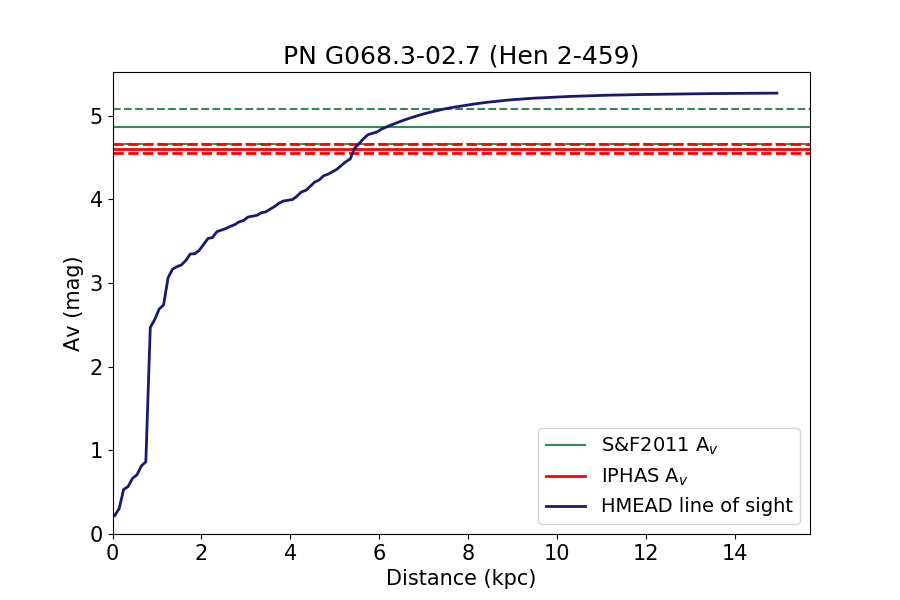}
  \subcaption{}
  \label{fig:Hen2-459}
  \end{subfigure}
  
\begin{subfigure}[b]{0.45\textwidth}
  \centering
  \includegraphics[width=\textwidth]{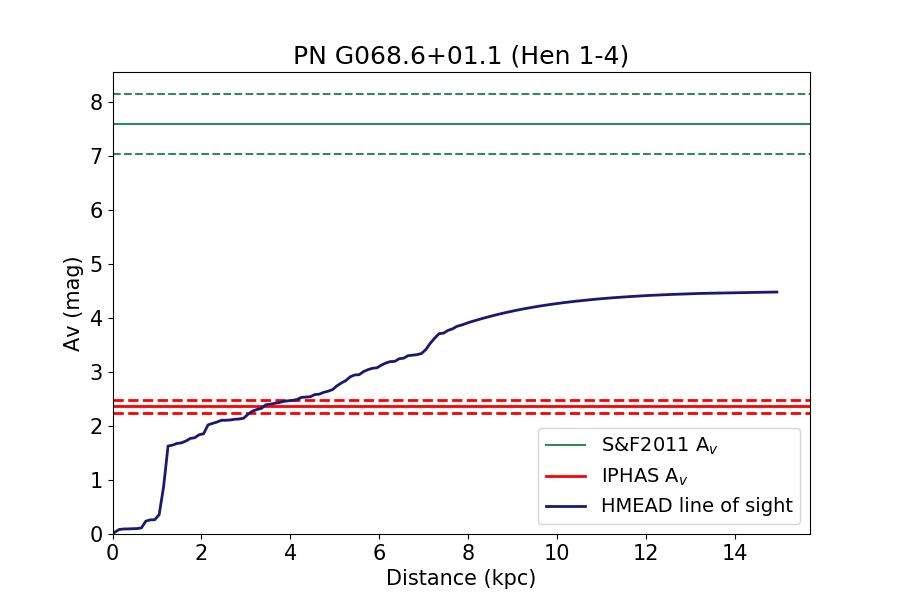}
 \subcaption{}
  \label{fig:Hen1-4}
  \end{subfigure}
\begin{subfigure}[b]{0.45\textwidth}
  \centering
  \includegraphics[width=\textwidth]{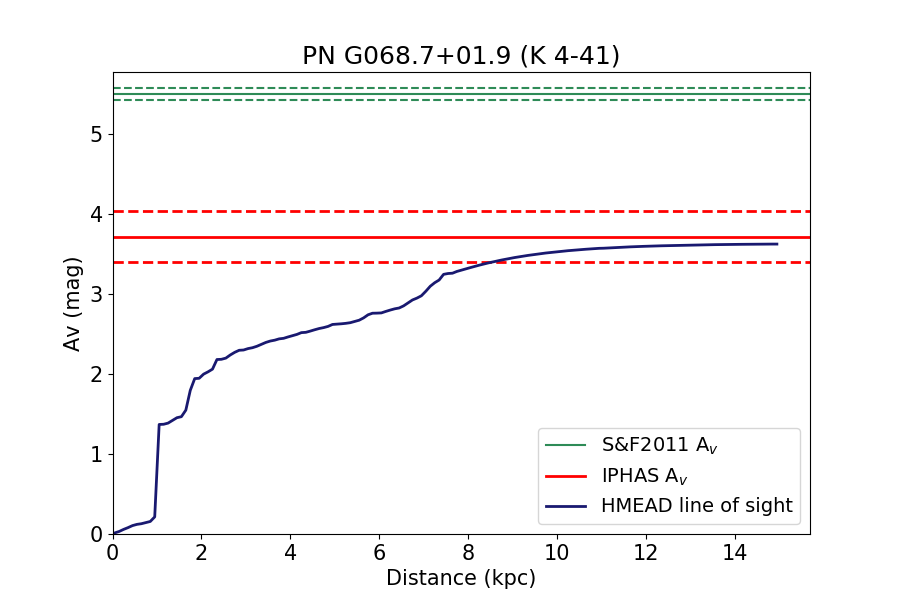}
  \subcaption{}
  \label{fig:K4-41}
  \end{subfigure}
  
\begin{subfigure}[b]{0.45\textwidth}
  \centering
  \includegraphics[width=\textwidth]{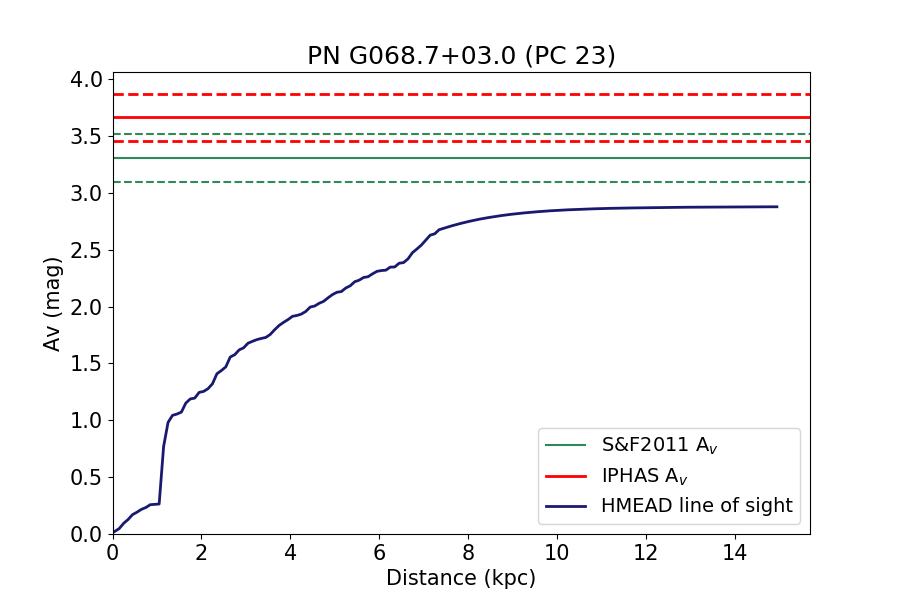}
  \subcaption{}
  \label{fig:PC23}
  \end{subfigure}
\begin{subfigure}[b]{0.45\textwidth}
  \centering
  \includegraphics[width=\textwidth]{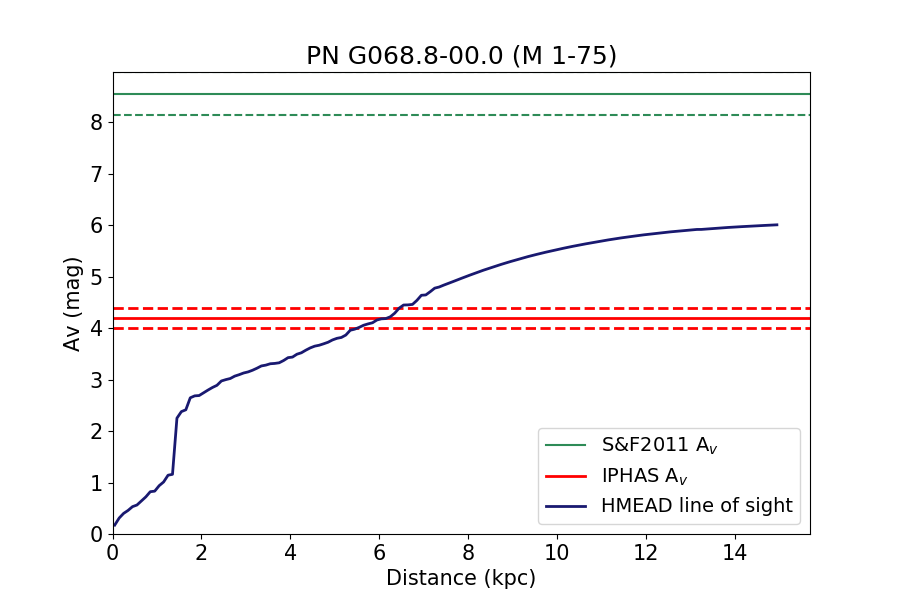}
  \subcaption{}
  \label{fig:M1-75}
  \end{subfigure}  
  
  \caption{H-MEAD Extinction vs. Distance plots for NGC 6842, K 4-37, K 3-52, Hen 2-459, Hen 1-4, K 4-41, PC 23 and M 1-75.}
  \label{fig:extCurves_App_8}
\end{figure*}

\begin{figure*}
\centering

\begin{subfigure}[b]{0.45\textwidth}
  \centering
  \includegraphics[width=\textwidth]{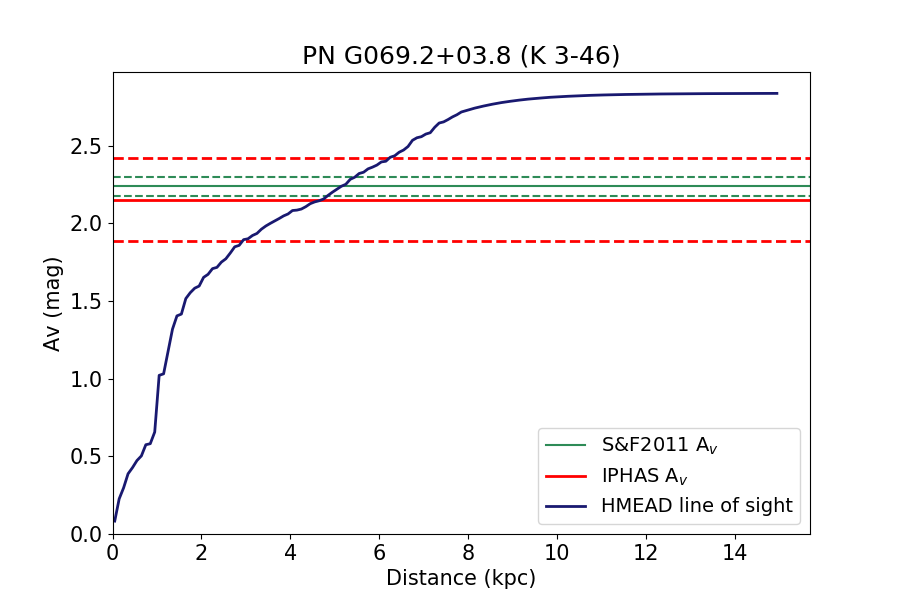}
  \subcaption{}
  \label{fig:K3-46}
  \end{subfigure}
\begin{subfigure}[b]{0.45\textwidth}
  \centering
  \includegraphics[width=\textwidth]{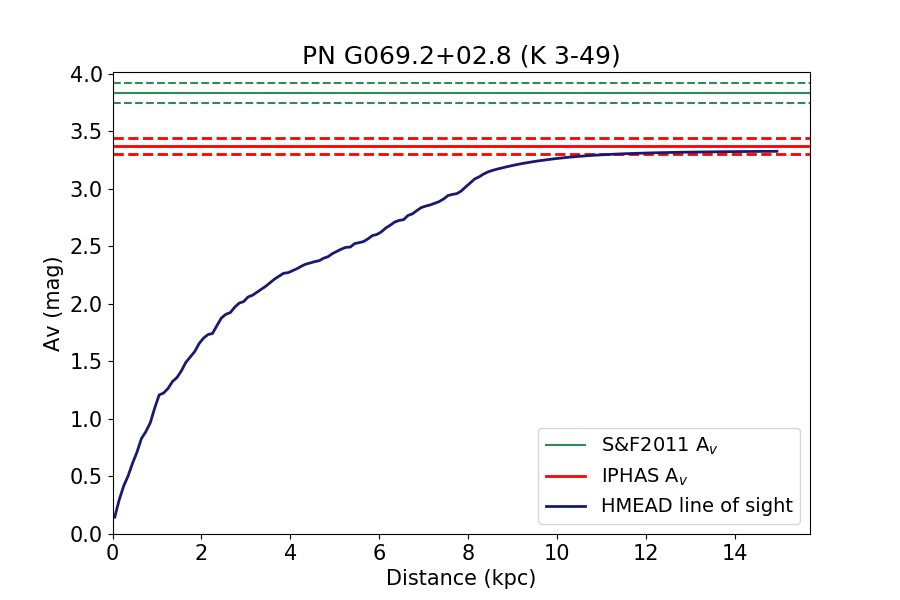}
  \subcaption{}
  \label{fig:K3-49}
  \end{subfigure}

\begin{subfigure}[b]{0.45\textwidth}
  \centering
  \includegraphics[width=\textwidth]{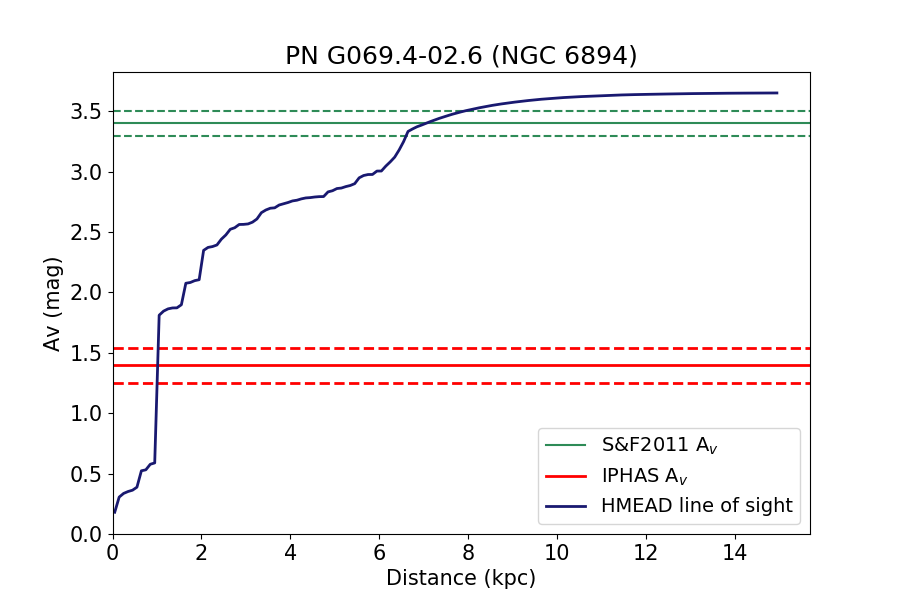}
 \subcaption{}
  \label{fig:NGC6894}
  \end{subfigure}
\begin{subfigure}[b]{0.45\textwidth}
  \centering
  \includegraphics[width=\textwidth]{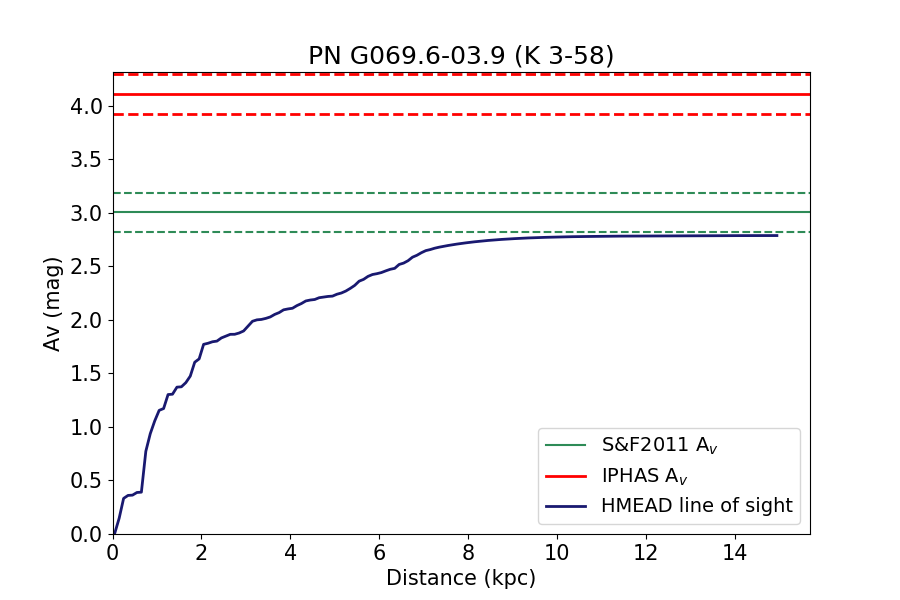}
  \subcaption{}
  \label{fig:K3-58}
  \end{subfigure}
  
\begin{subfigure}[b]{0.45\textwidth}
  \centering
  \includegraphics[width=\textwidth]{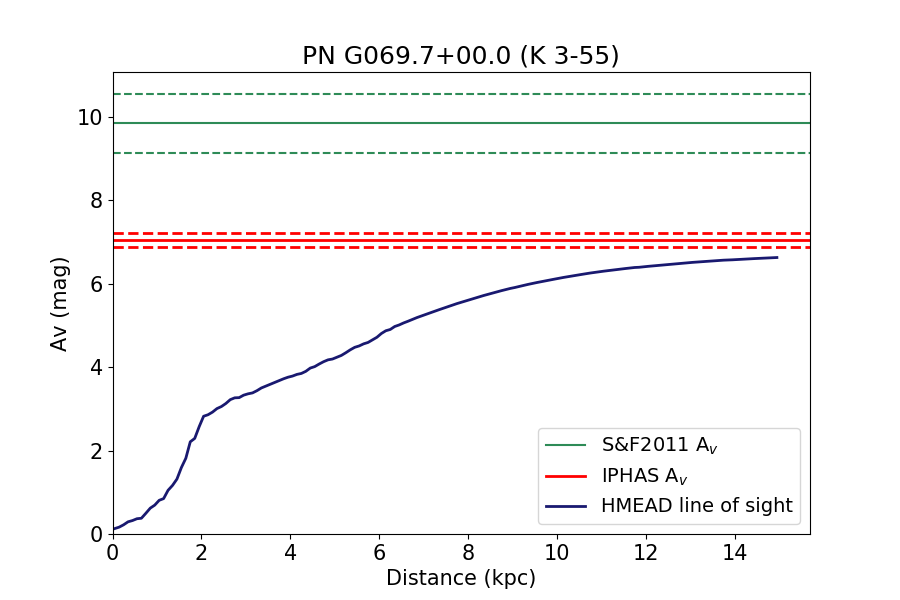}
 \subcaption{}
  \label{fig:K3-55}
  \end{subfigure}
\begin{subfigure}[b]{0.45\textwidth}
  \centering
  \includegraphics[width=\textwidth]{Distcurves/K3-35_LinAxes.png}
  \subcaption{}
  \label{fig:K3-35}
  \end{subfigure}
  
\begin{subfigure}[b]{0.45\textwidth}
  \centering
  \includegraphics[width=\textwidth]{Distcurves/K3-57_LinAxes.png}
  \subcaption{}
  \label{fig:K3-57}
  \end{subfigure}
\begin{subfigure}[b]{0.45\textwidth}
  \centering
  \includegraphics[width=\textwidth]{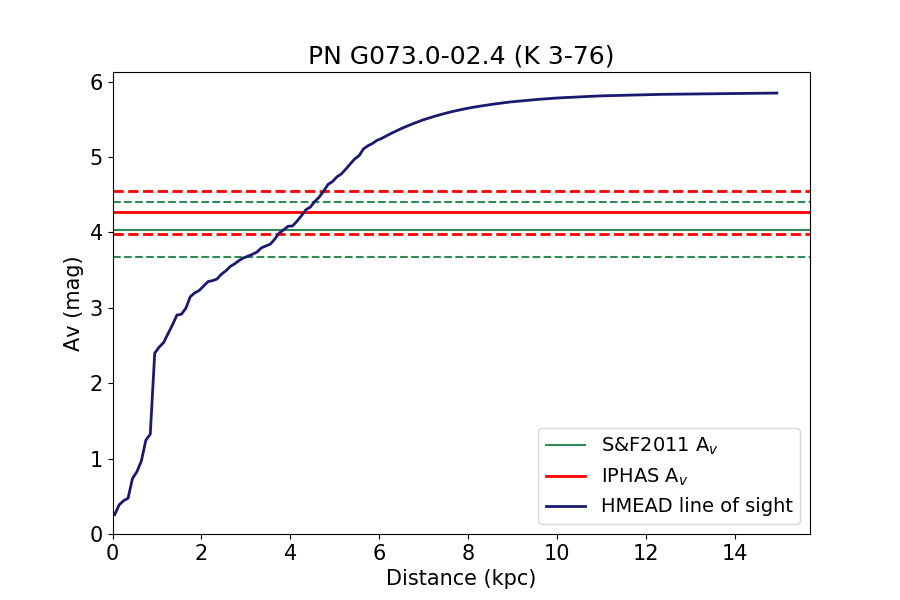}
  \subcaption{}
  \label{fig:K3-76}
  \end{subfigure}  
  
  \caption{H-MEAD Extinction vs. Distance plots for K 3-46, K 3-49, NGC 6894, K 3-58, K 3-55, K 3-35, K 3-57 and K 3-76.}
  \label{fig:extCurves_App_9}
\end{figure*}

\begin{figure*}
\centering

\begin{subfigure}[b]{0.45\textwidth}
  \centering
  \includegraphics[width=\textwidth]{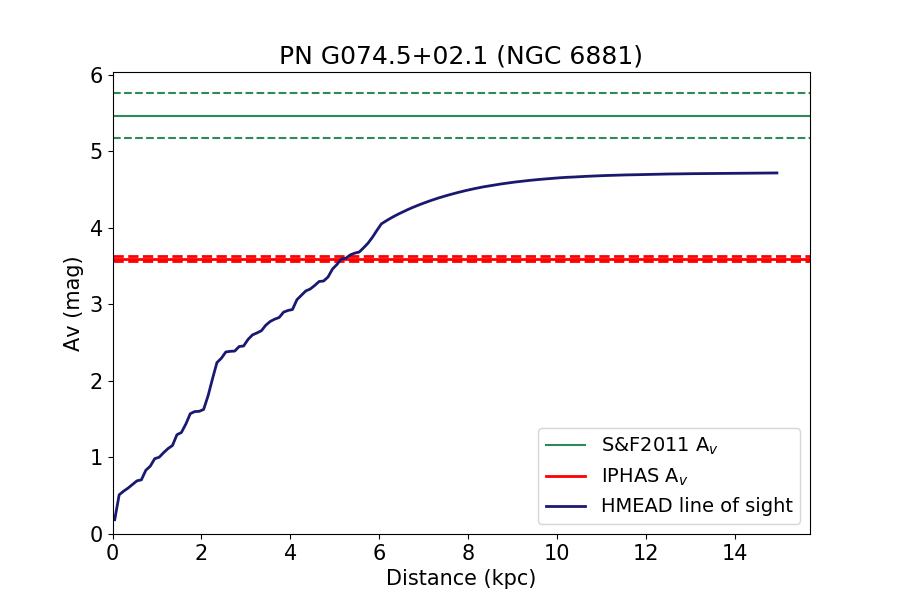}
  \subcaption{}
  \label{fig:NGC6881}
  \end{subfigure}
\begin{subfigure}[b]{0.45\textwidth}
  \centering
  \includegraphics[width=\textwidth]{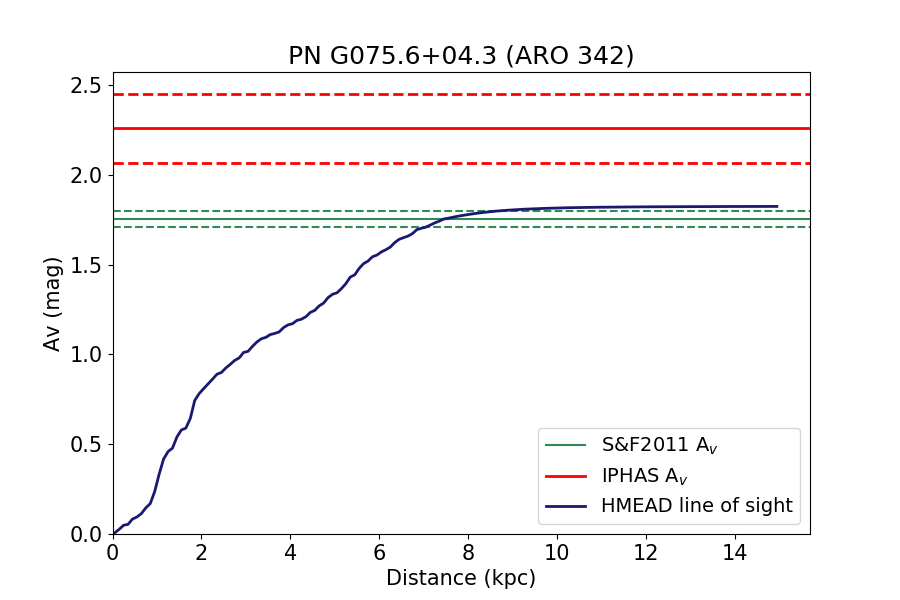}
  \subcaption{}
  \label{fig:ARO342}
  \end{subfigure}

\begin{subfigure}[b]{0.45\textwidth}
  \centering
  \includegraphics[width=\textwidth]{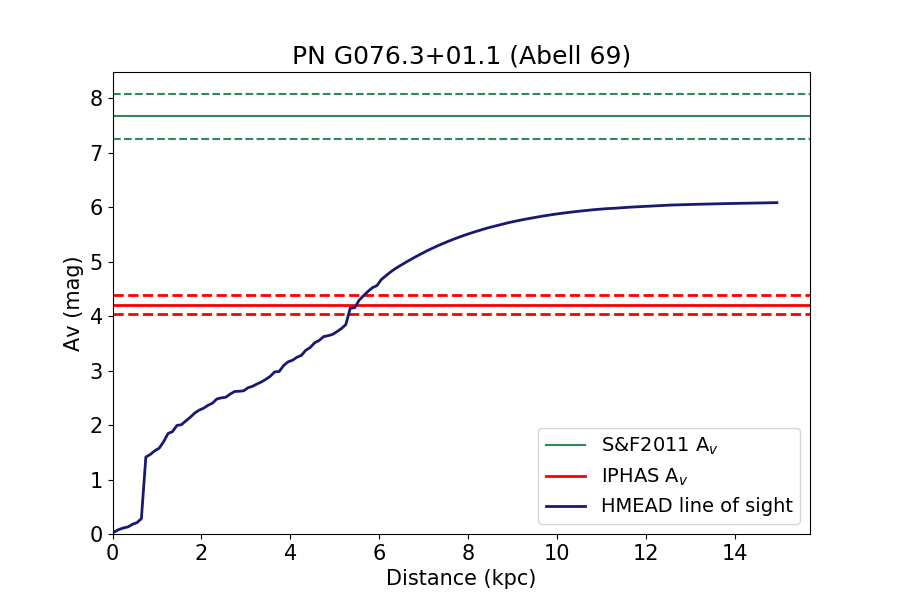}
 \subcaption{}
  \label{fig:Abell69}
  \end{subfigure}
\begin{subfigure}[b]{0.45\textwidth}
  \centering
  \includegraphics[width=\textwidth]{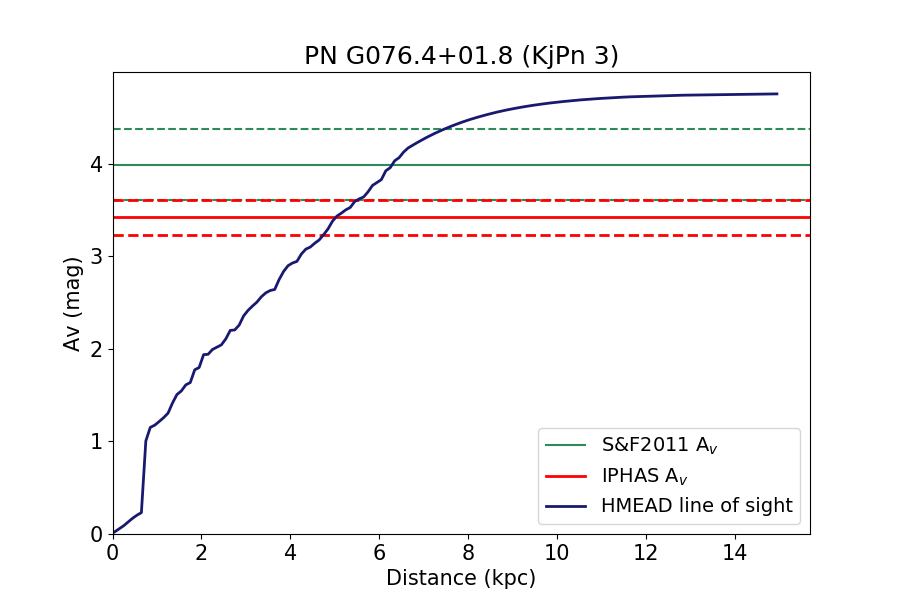}
  \subcaption{}
  \label{fig:KjPn3}
  \end{subfigure}
  
\begin{subfigure}[b]{0.45\textwidth}
  \centering
  \includegraphics[width=\textwidth]{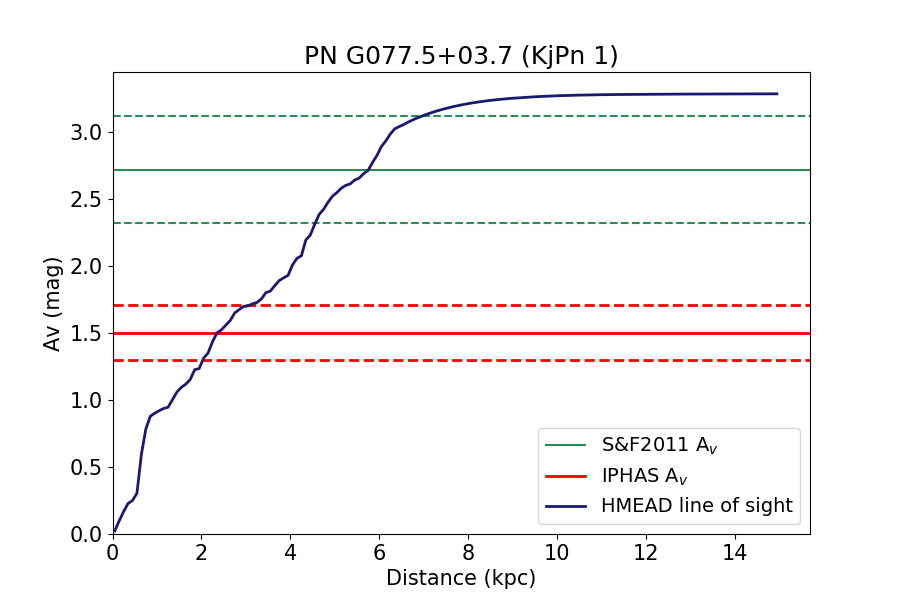}
 \subcaption{}
  \label{fig:KjPn1}
  \end{subfigure}
\begin{subfigure}[b]{0.45\textwidth}
  \centering
  \includegraphics[width=\textwidth]{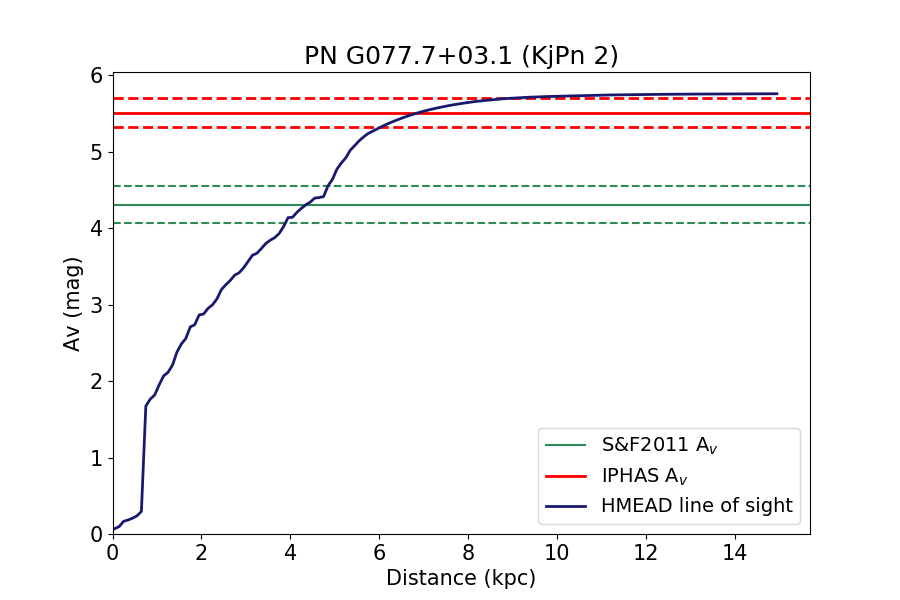}
  \subcaption{}
  \label{fig:KjPn2}
  \end{subfigure}
  
\begin{subfigure}[b]{0.45\textwidth}
  \centering
  \includegraphics[width=\textwidth]{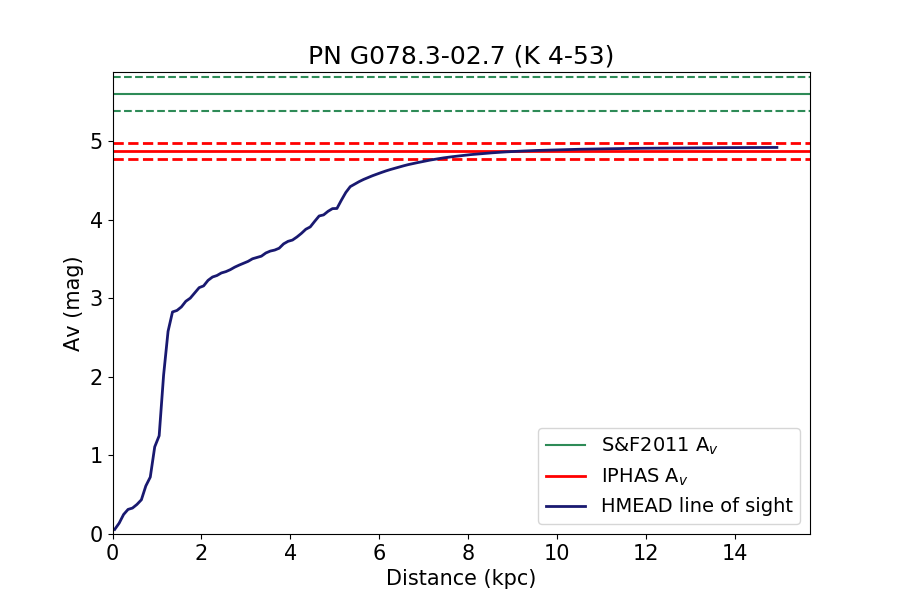}
  \subcaption{}
  \label{fig:K4-53}
  \end{subfigure}
\begin{subfigure}[b]{0.45\textwidth}
  \centering
  \includegraphics[width=\textwidth]{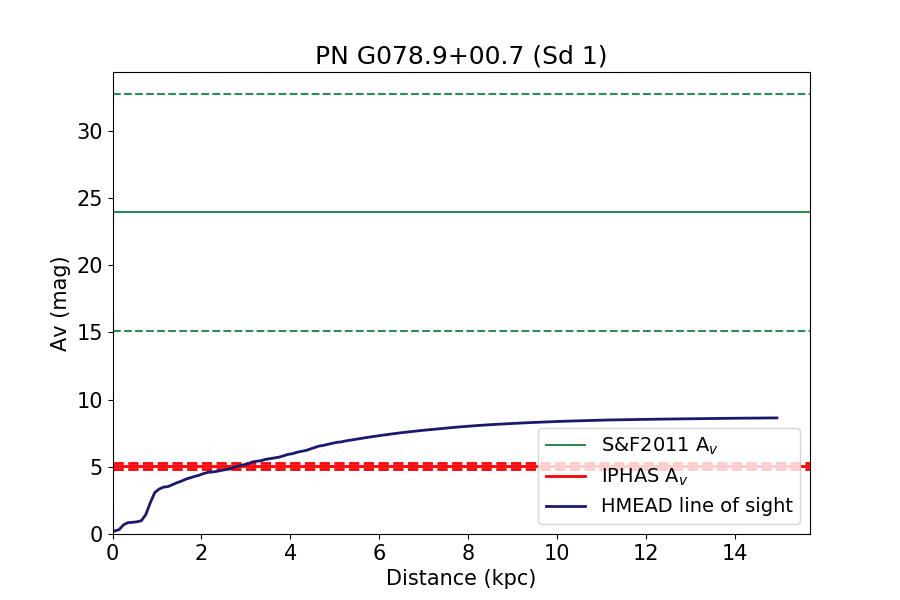}
  \subcaption{}
  \label{fig:Sd1}
  \end{subfigure}
  
  \caption{H-MEAD Extinction vs. Distance plots for NGC 6881, ARO 342, Abell 69, KjPn 3, KjPn 1, KjPn 2, K 4-53 and Sd 1.}
  \label{fig:extCurves_App_10}
\end{figure*}

\begin{figure*}
\centering

\begin{subfigure}[b]{0.45\textwidth}
  \centering
  \includegraphics[width=\textwidth]{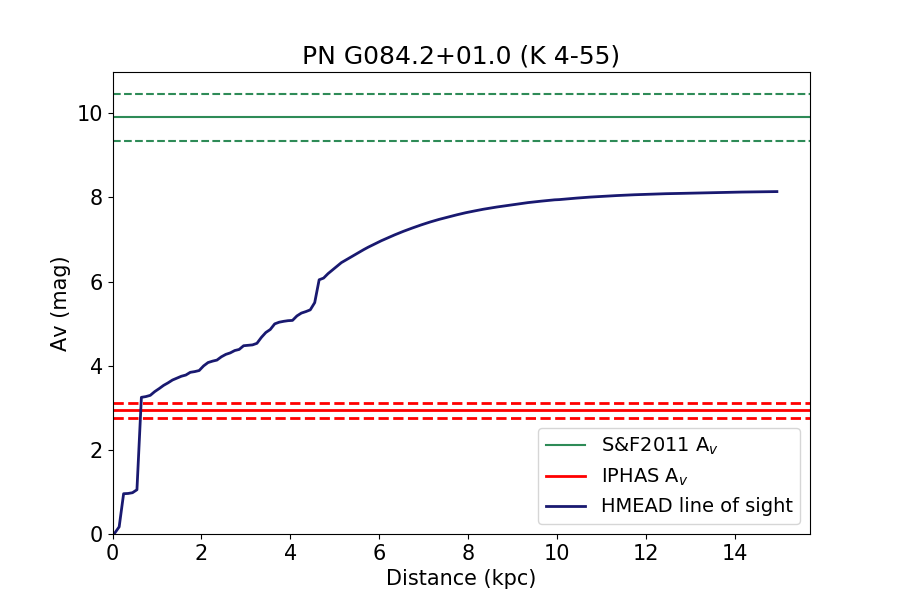}
  \subcaption{}
  \label{fig:K4-55}
  \end{subfigure}
\begin{subfigure}[b]{0.45\textwidth}
  \centering
  \includegraphics[width=\textwidth]{Distcurves/Abell71_LinAxes.png}
  \subcaption{}
  \label{fig:Abell71}
  \end{subfigure}

\begin{subfigure}[b]{0.45\textwidth}
  \centering
  \includegraphics[width=\textwidth]{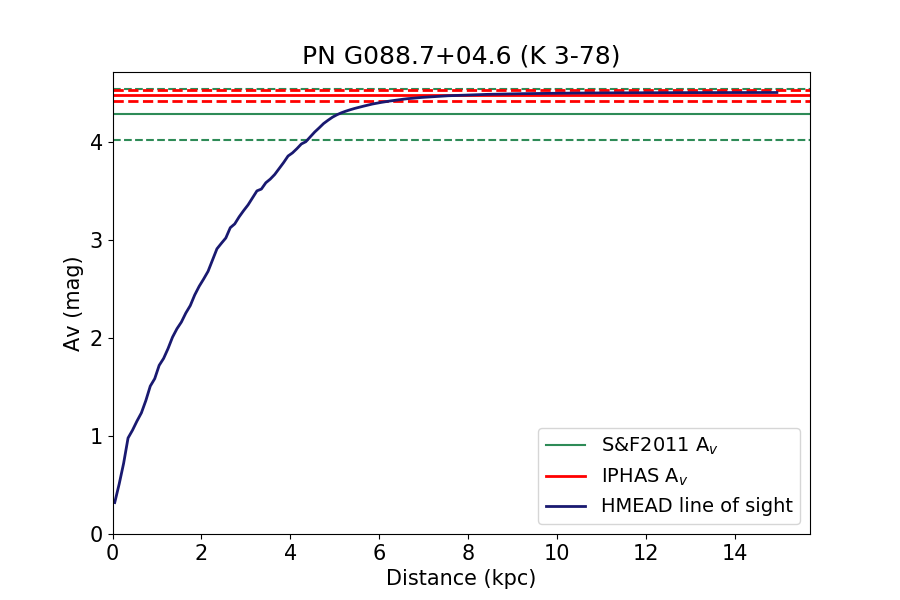}
 \subcaption{}
  \label{fig:K3-78}
  \end{subfigure}
\begin{subfigure}[b]{0.45\textwidth}
  \centering
  \includegraphics[width=\textwidth]{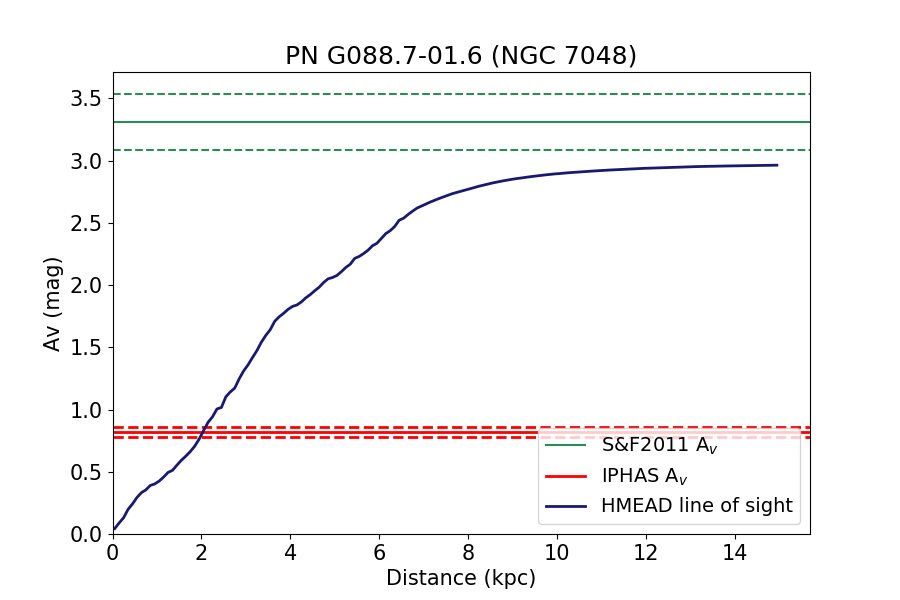}
  \subcaption{}
  \label{fig:NGC7048}
  \end{subfigure}
  
\begin{subfigure}[b]{0.45\textwidth}
  \centering
  \includegraphics[width=\textwidth]{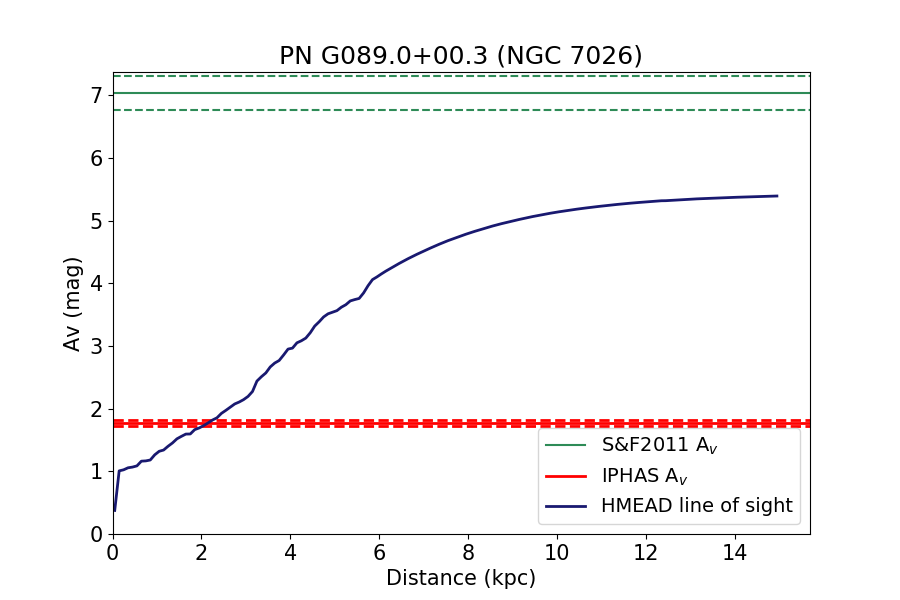}
 \subcaption{}
  \label{fig:NGC7026}
  \end{subfigure}
\begin{subfigure}[b]{0.45\textwidth}
  \centering
  \includegraphics[width=\textwidth]{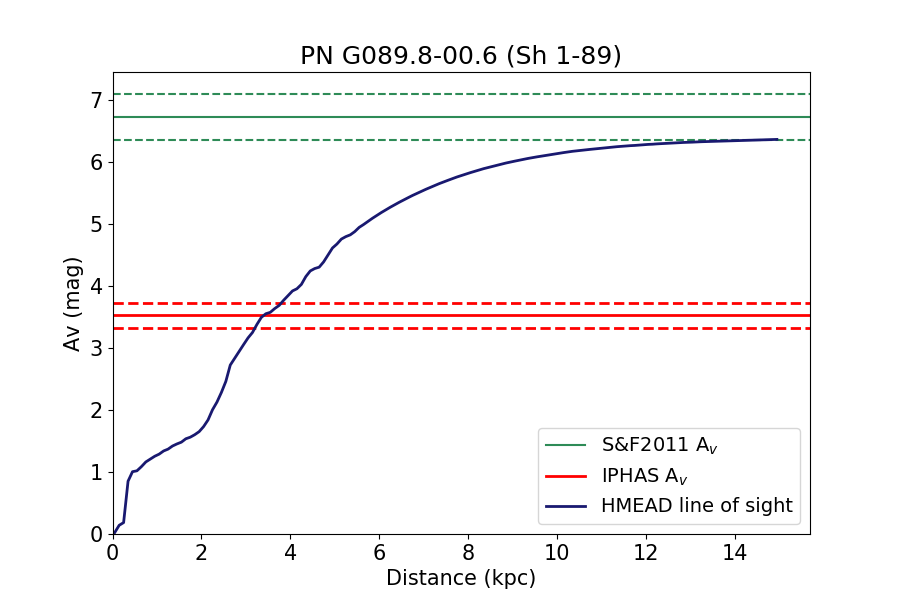}
  \subcaption{}
  \label{fig:Sh1-89}
  \end{subfigure}
  
 \begin{subfigure}[b]{0.45\textwidth}
  \centering
  \includegraphics[width=\textwidth]{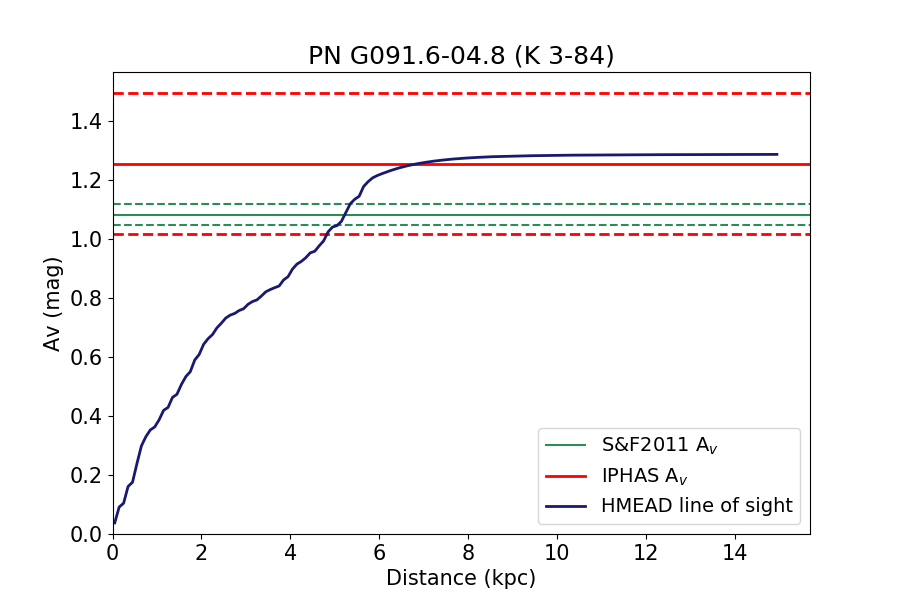}
  \subcaption{}
  \label{fig:K3-84}
  \end{subfigure}
\begin{subfigure}[b]{0.45\textwidth}
  \centering
  \includegraphics[width=\textwidth]{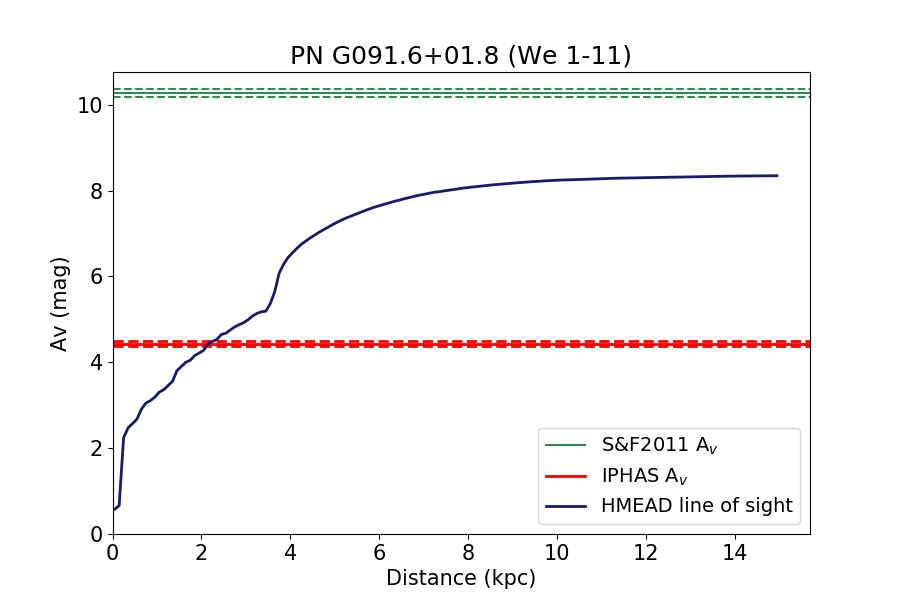}
  \subcaption{}
  \label{fig:We1-11}
  \end{subfigure}

  \caption{H-MEAD Extinctions vs. Distance plots for K 4-55, Abell 71, K 3-78, NGC 7048, NGC 7026, Sh 1-89, K 3-84 and We 1-11.}
  \label{fig:extCurves_App_11}
\end{figure*}

\begin{figure*}
\centering

\begin{subfigure}[b]{0.45\textwidth}
  \centering
  \includegraphics[width=\textwidth]{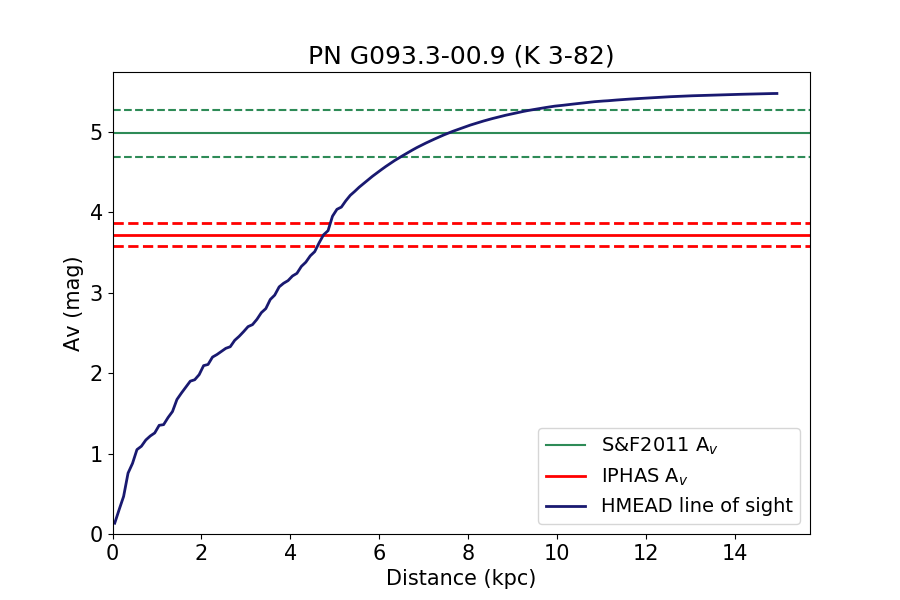}
  \subcaption{}
  \label{fig:K3-82}
  \end{subfigure}
\begin{subfigure}[b]{0.45\textwidth}
  \centering
  \includegraphics[width=\textwidth]{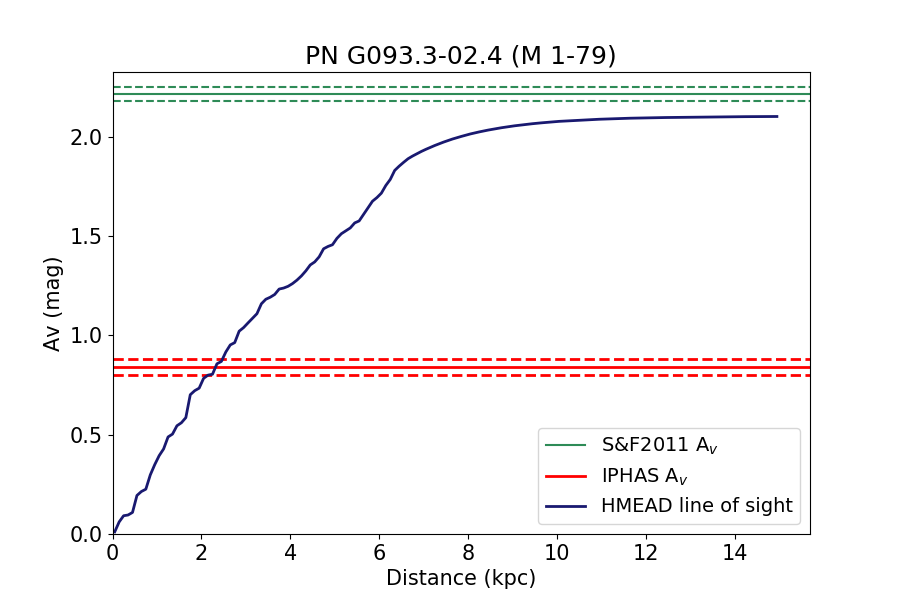}
  \subcaption{}
  \label{fig:M1-79}
  \end{subfigure}

\begin{subfigure}[b]{0.45\textwidth}
  \centering
  \includegraphics[width=\textwidth]{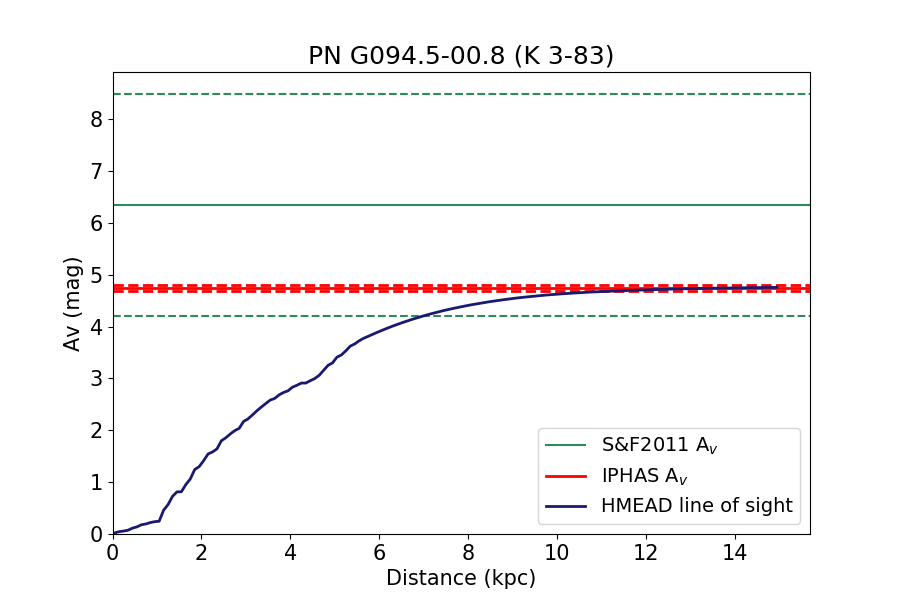}
 \subcaption{}
  \label{fig:K3-83}
  \end{subfigure}
\begin{subfigure}[b]{0.45\textwidth}
  \centering
  \includegraphics[width=\textwidth]{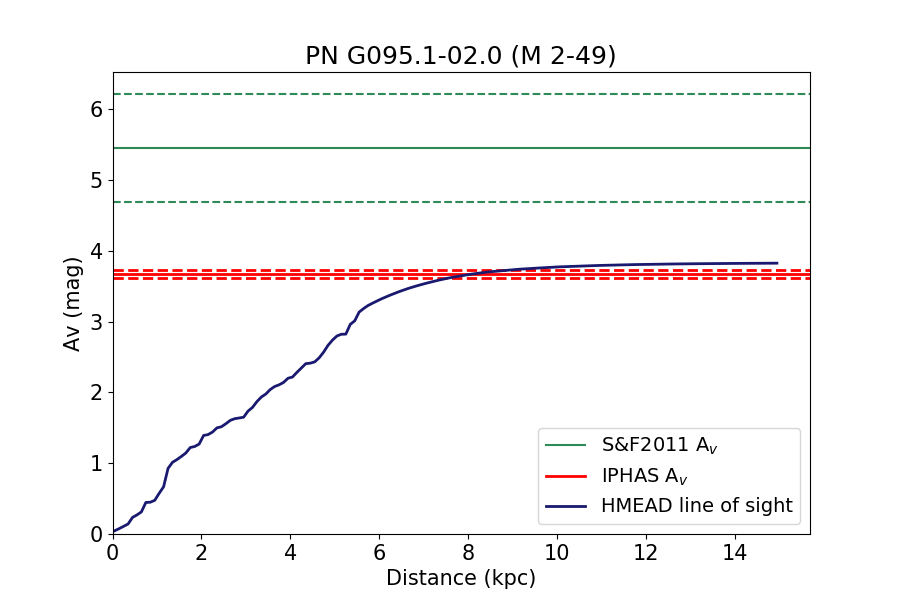}
  \subcaption{}
  \label{fig:M2-49}
  \end{subfigure}
  
\begin{subfigure}[b]{0.45\textwidth}
  \centering
  \includegraphics[width=\textwidth]{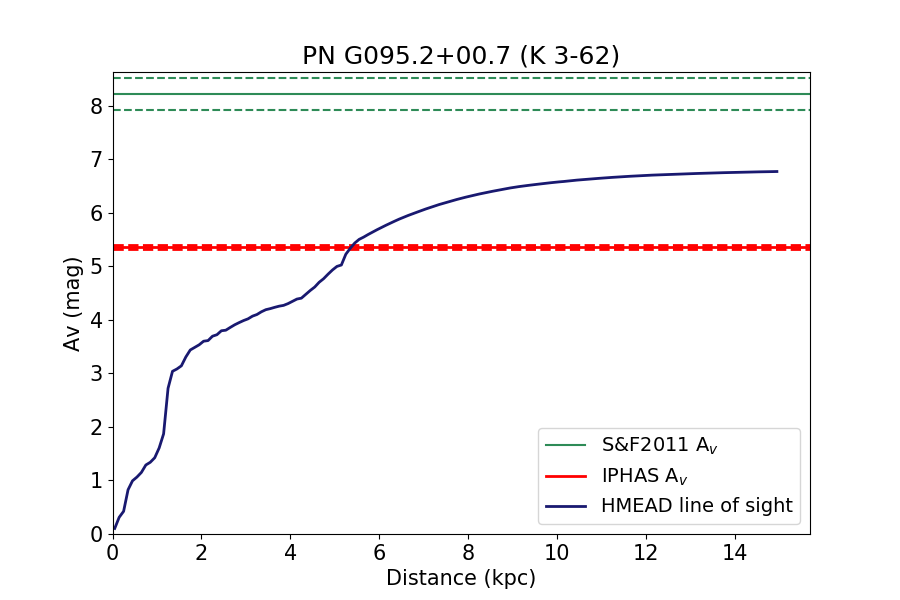}
 \subcaption{}
  \label{fig:K3-62}
  \end{subfigure}
\begin{subfigure}[b]{0.45\textwidth}
  \centering
  \includegraphics[width=\textwidth]{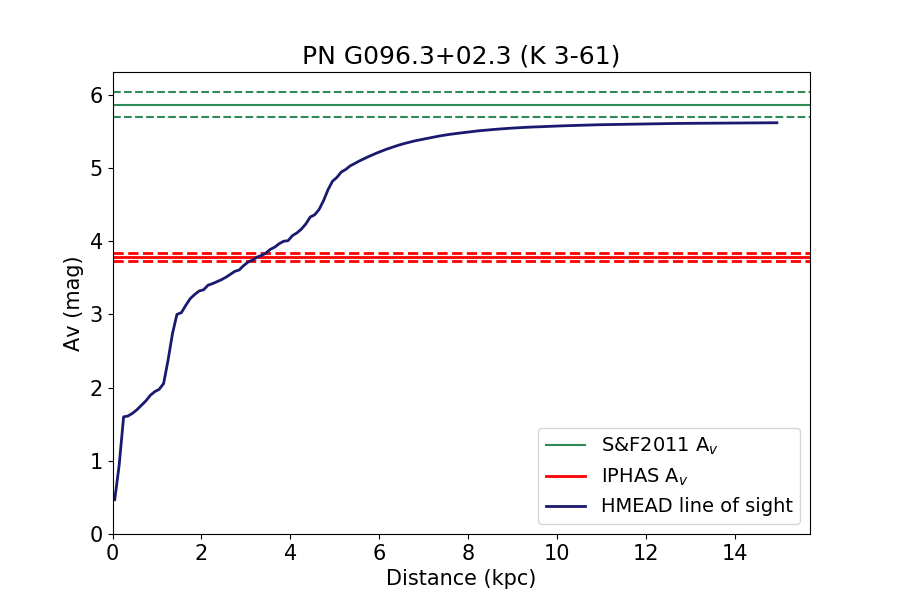}
  \subcaption{}
  \label{fig:K3-61}
  \end{subfigure}
  
\begin{subfigure}[b]{0.45\textwidth}
  \centering
  \includegraphics[width=\textwidth]{Distcurves/M2-50_LinAxes.png}
  \subcaption{}
  \label{fig:M2-50}
  \end{subfigure}
\begin{subfigure}[b]{0.45\textwidth}
  \centering
  \includegraphics[width=\textwidth]{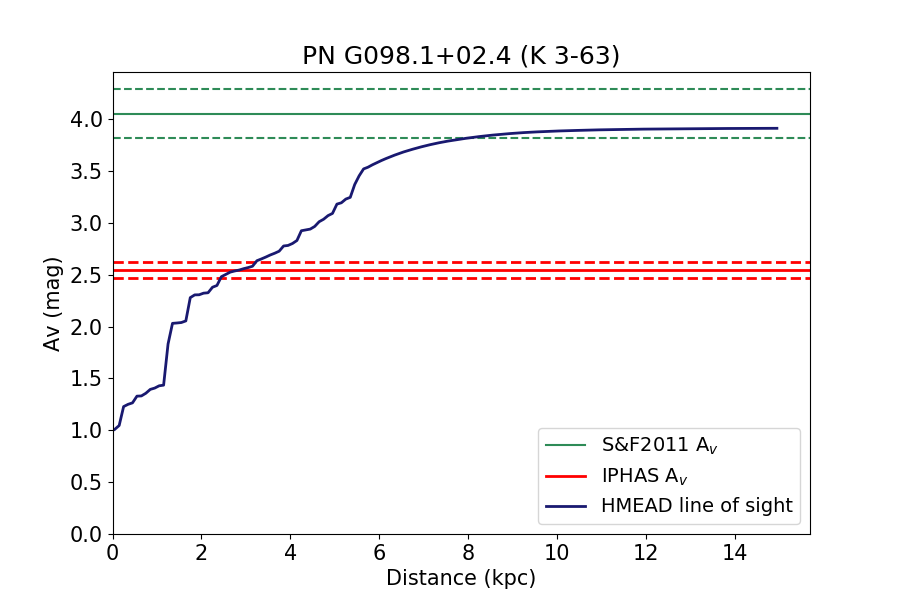}
  \subcaption{}
  \label{fig:K3-63}
  \end{subfigure}  
  
  \caption{H-MEAD Extinction vs. Distance plots for K 3-82, M 1-79, K 3-83, M 2-49, K 3-62, K 3-61, M 2-50 and K 3-63.}
  \label{fig:extCurves_App_12}
\end{figure*}

\begin{figure*}
\centering

\begin{subfigure}[b]{0.45\textwidth}
  \centering
  \includegraphics[width=\textwidth]{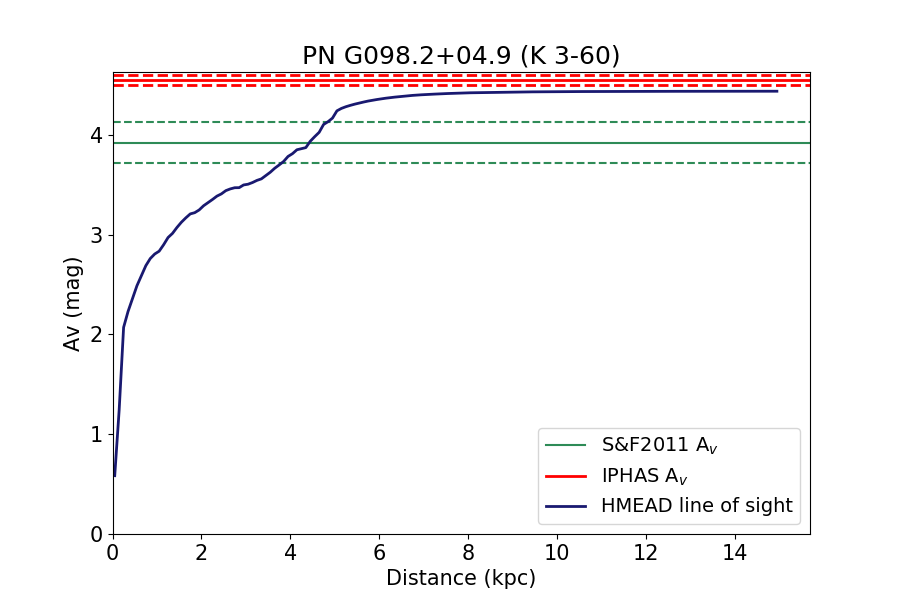}
  \subcaption{}
  \label{fig:K3-60}
  \end{subfigure}
\begin{subfigure}[b]{0.45\textwidth}
  \centering
  \includegraphics[width=\textwidth]{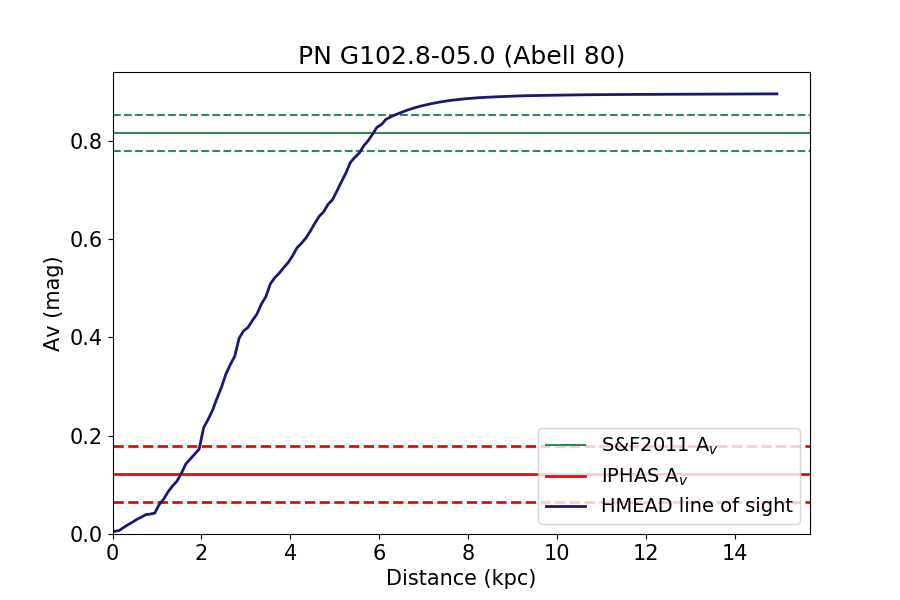}
  \subcaption{}
  \label{fig:Abell80}
  \end{subfigure}

\begin{subfigure}[b]{0.45\textwidth}
  \centering
  \includegraphics[width=\textwidth]{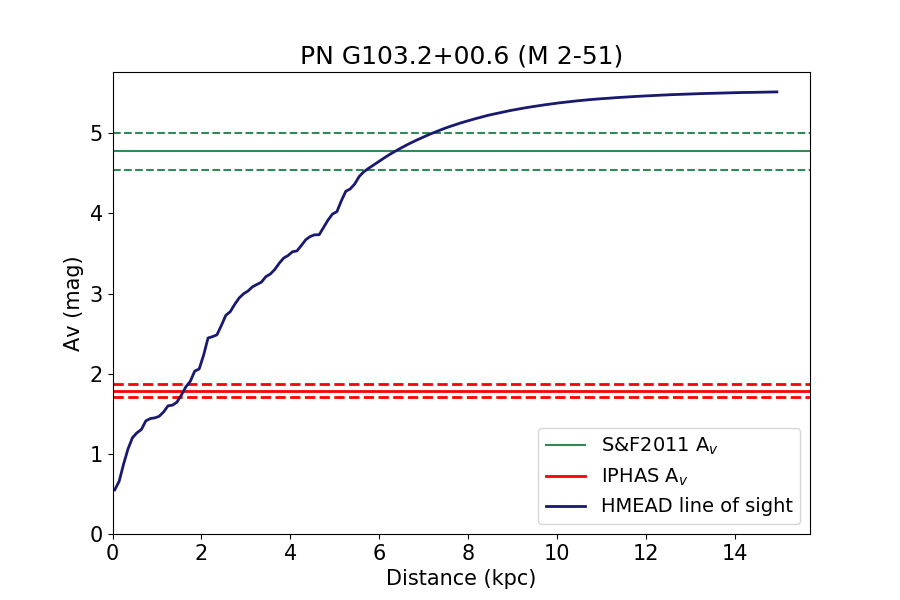}
 \subcaption{}
  \label{fig:KM2-51}
  \end{subfigure}
\begin{subfigure}[b]{0.45\textwidth}
  \centering
  \includegraphics[width=\textwidth]{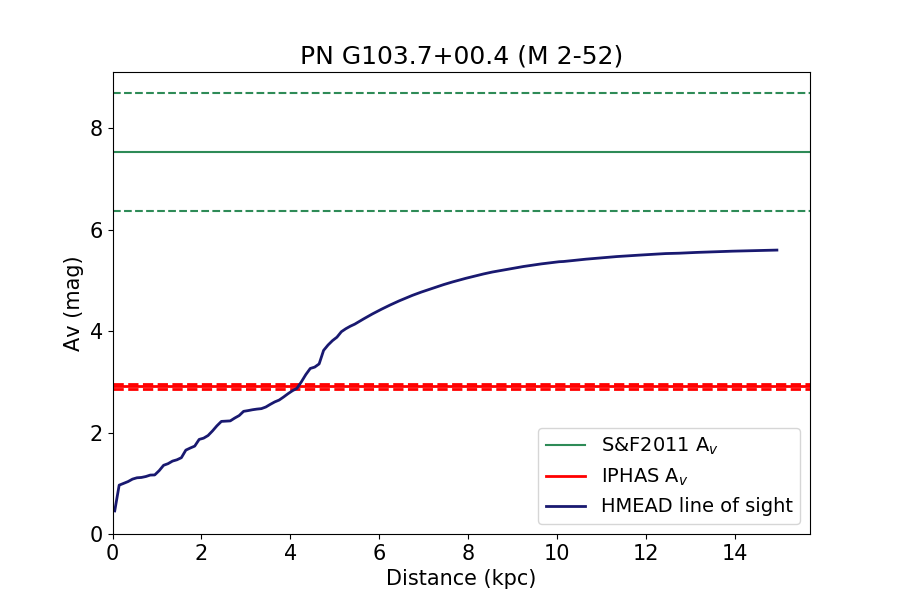}
  \subcaption{}
  \label{fig:M2-52}
  \end{subfigure}
  
\begin{subfigure}[b]{0.45\textwidth}
  \centering
  \includegraphics[width=\textwidth]{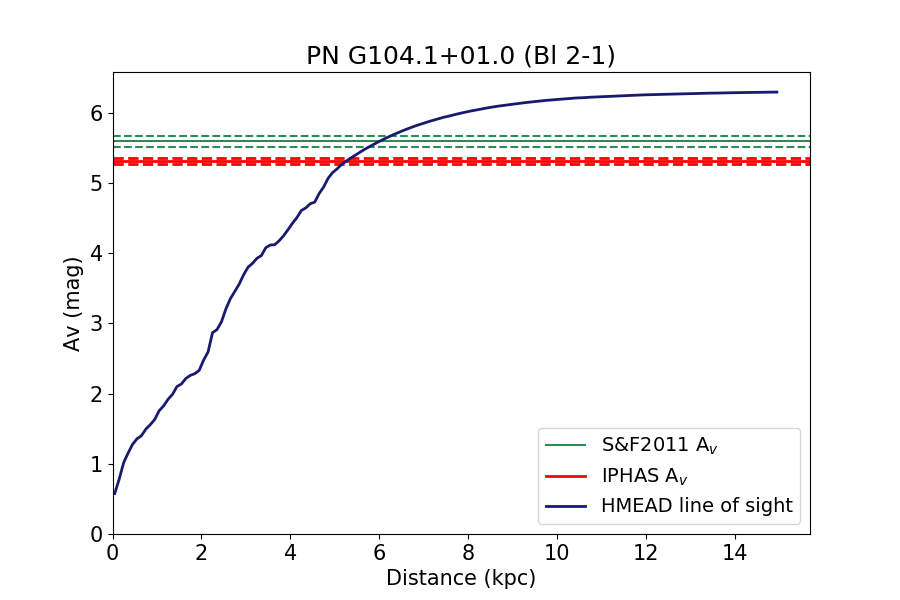}
 \subcaption{}
  \label{fig:Bl2-1}
  \end{subfigure}
\begin{subfigure}[b]{0.45\textwidth}
  \centering
  \includegraphics[width=\textwidth]{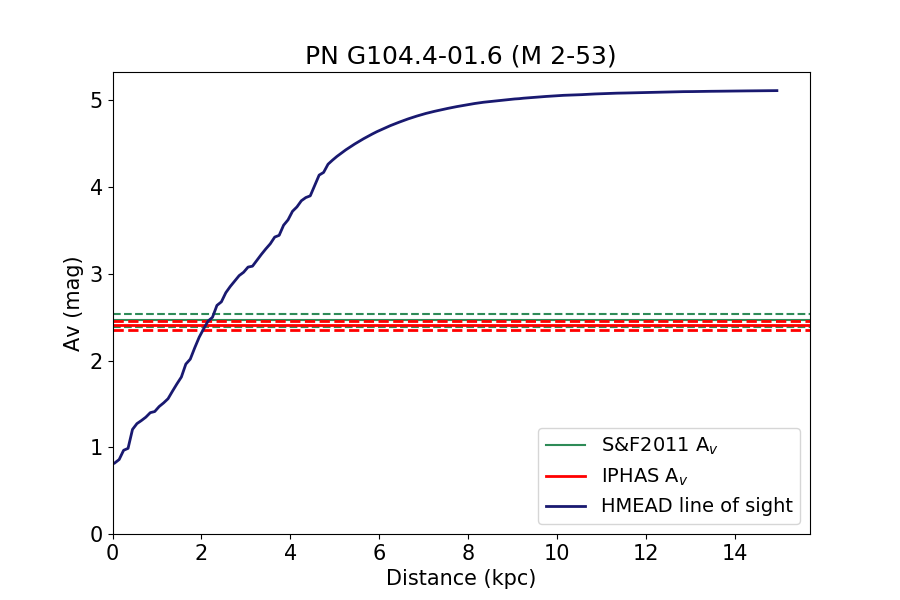}
  \subcaption{}
  \label{fig:M2-53}
  \end{subfigure}
  
\begin{subfigure}[b]{0.45\textwidth}
  \centering
  \includegraphics[width=\textwidth]{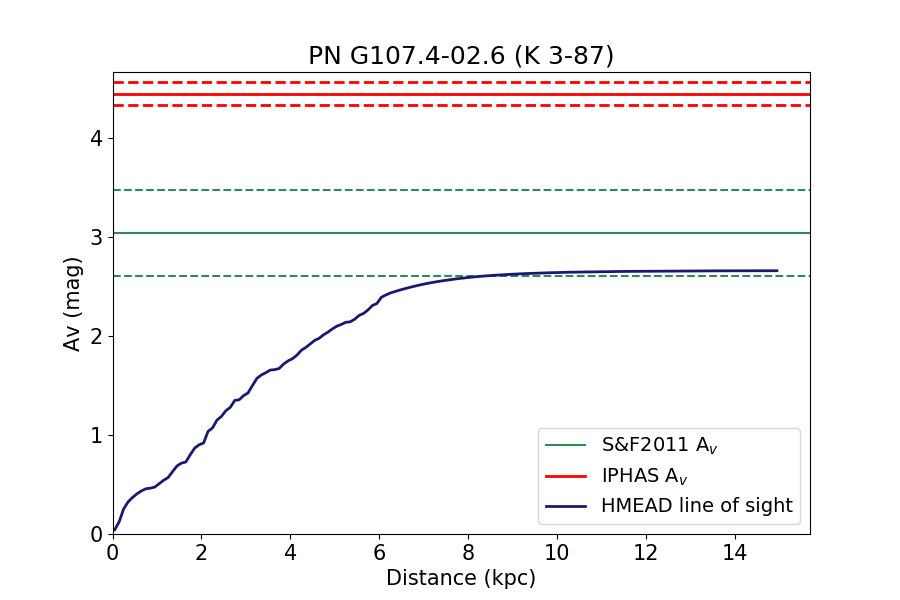}
  \subcaption{}
  \label{fig:K3-87}
  \end{subfigure}
\begin{subfigure}[b]{0.45\textwidth}
  \centering
  \includegraphics[width=\textwidth]{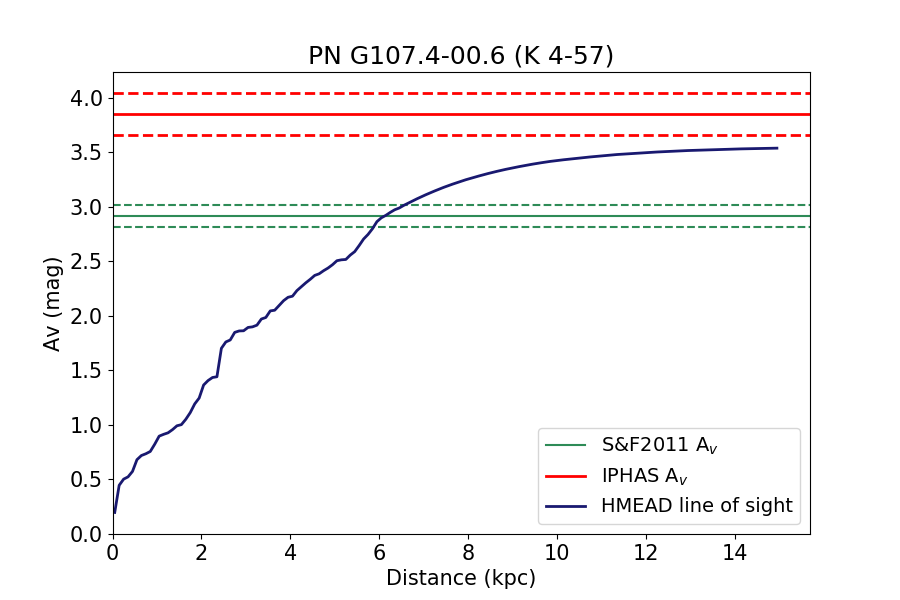}
  \subcaption{}
  \label{fig:K4-57}
  \end{subfigure}
  
  \caption{H-MEAD Extinction vs. Distance plots for K 3-60, Abell 80, M 2-51, M 2-52, Bl 2-1, M 2-53, K 3-87 and K 4-57.}
  \label{fig:extCurves_App_13}
\end{figure*}

\begin{figure*}
\centering

\begin{subfigure}[b]{0.45\textwidth}
  \centering
  \includegraphics[width=\textwidth]{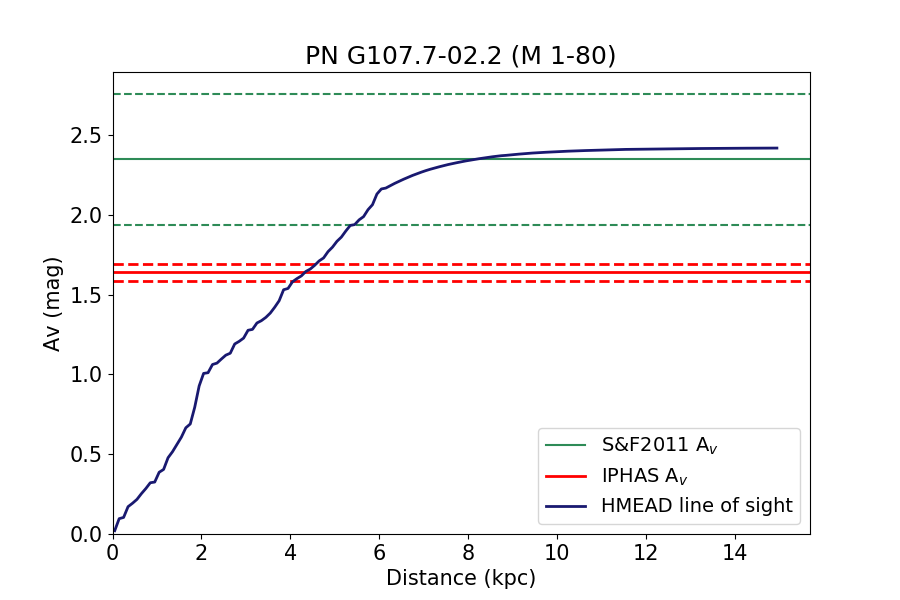}
  \subcaption{}
  \label{fig:M1-80}
  \end{subfigure}
\begin{subfigure}[b]{0.45\textwidth}
  \centering
  \includegraphics[width=\textwidth]{Distcurves/NGC7354_LinAxes.png}
  \subcaption{}
  \label{fig:NGC7354}
  \end{subfigure}

\begin{subfigure}[b]{0.45\textwidth}
  \centering
  \includegraphics[width=\textwidth]{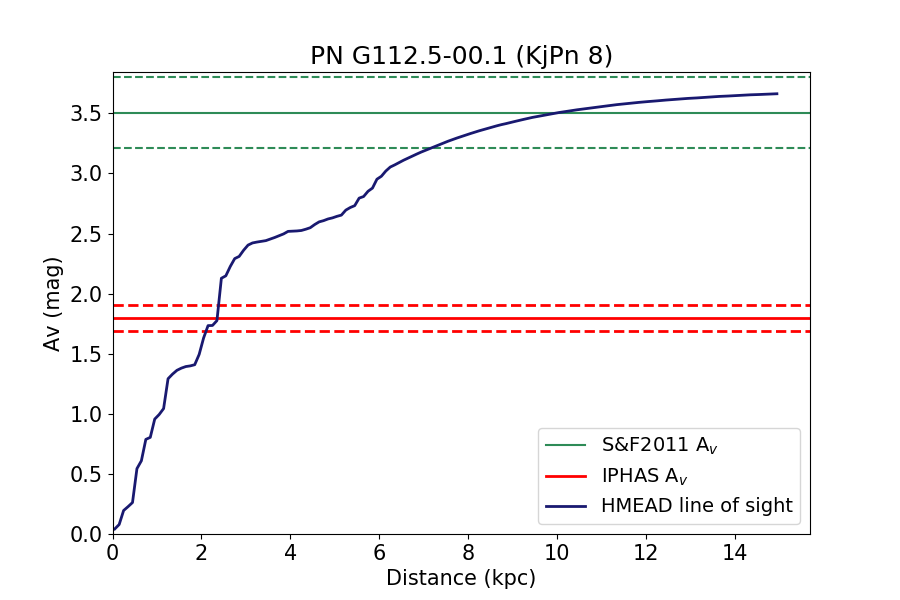}
 \subcaption{}
  \label{fig:KjPn8}
  \end{subfigure}
\begin{subfigure}[b]{0.45\textwidth}
  \centering
  \includegraphics[width=\textwidth]{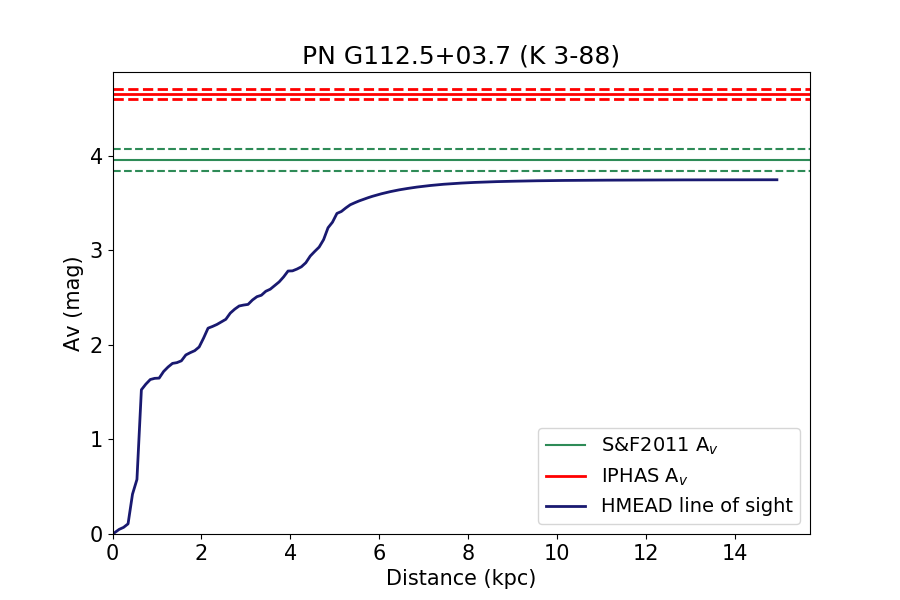}
  \subcaption{}
  \label{fig:K3-88}
  \end{subfigure}
  
\begin{subfigure}[b]{0.45\textwidth}
  \centering
  \includegraphics[width=\textwidth]{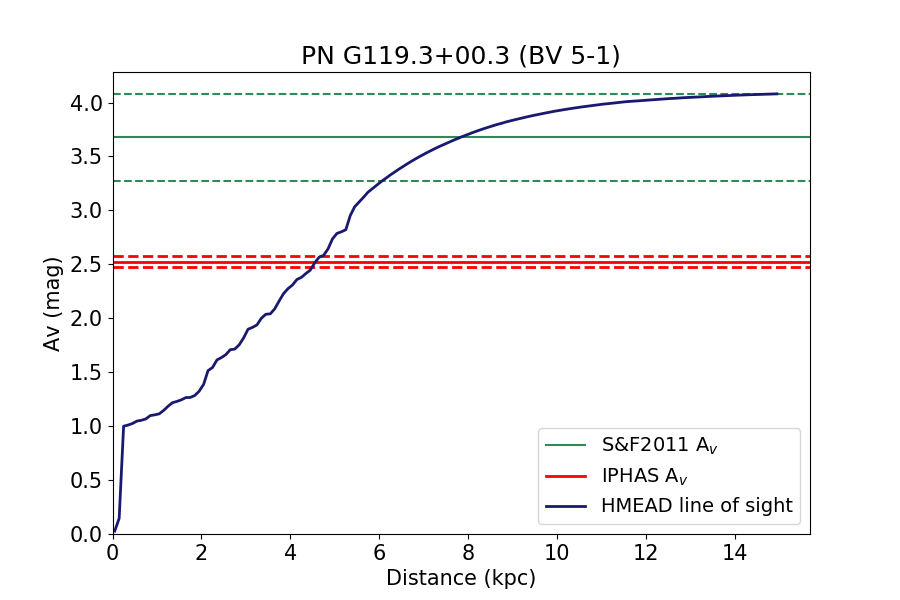}
 \subcaption{}
  \label{fig:BV5-1}
  \end{subfigure}
\begin{subfigure}[b]{0.45\textwidth}
  \centering
  \includegraphics[width=\textwidth]{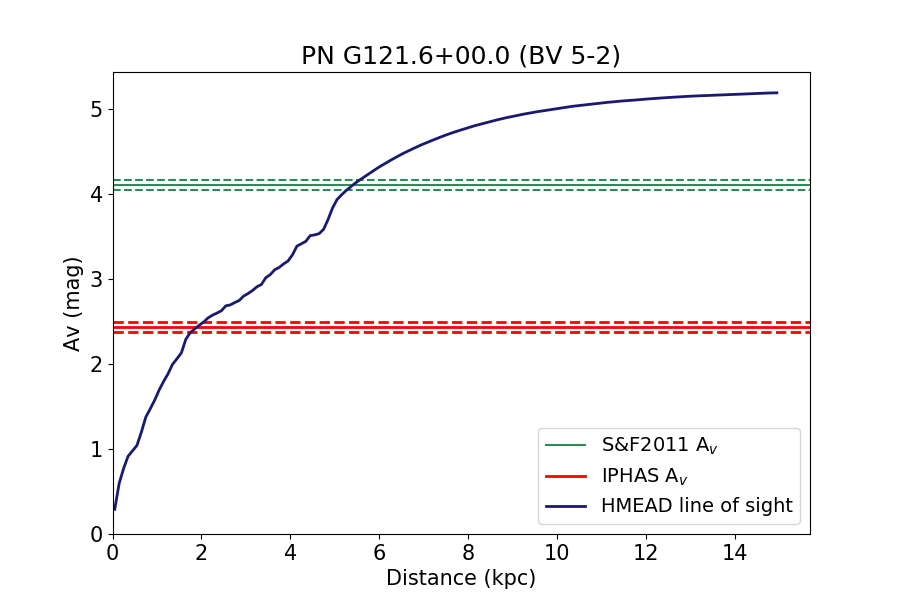}
  \subcaption{}
  \label{fig:BV5-2}
  \end{subfigure}
  
 \begin{subfigure}[b]{0.45\textwidth}
  \centering
  \includegraphics[width=\textwidth]{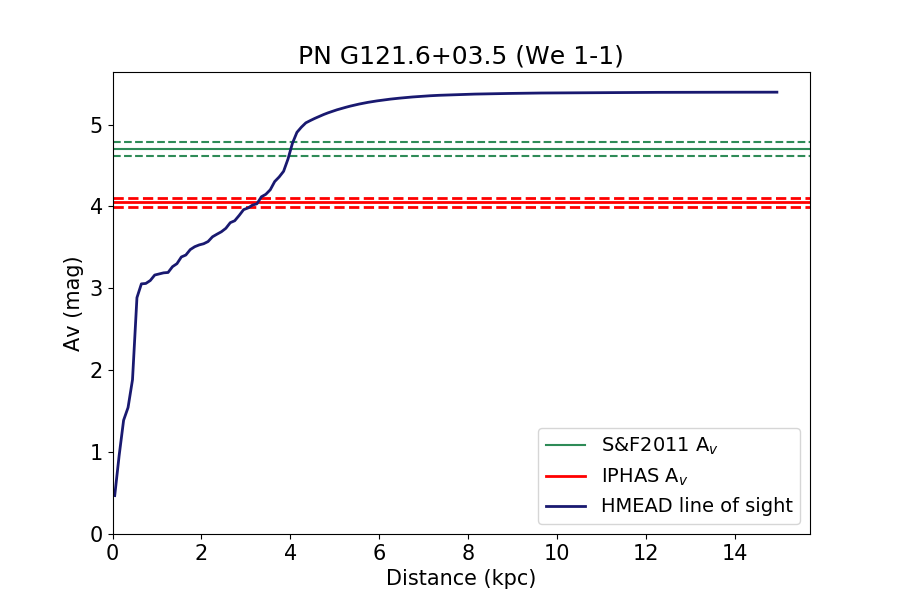}
  \subcaption{}
  \label{fig:We1-1}
  \end{subfigure}
\begin{subfigure}[b]{0.45\textwidth}
  \centering
  \includegraphics[width=\textwidth]{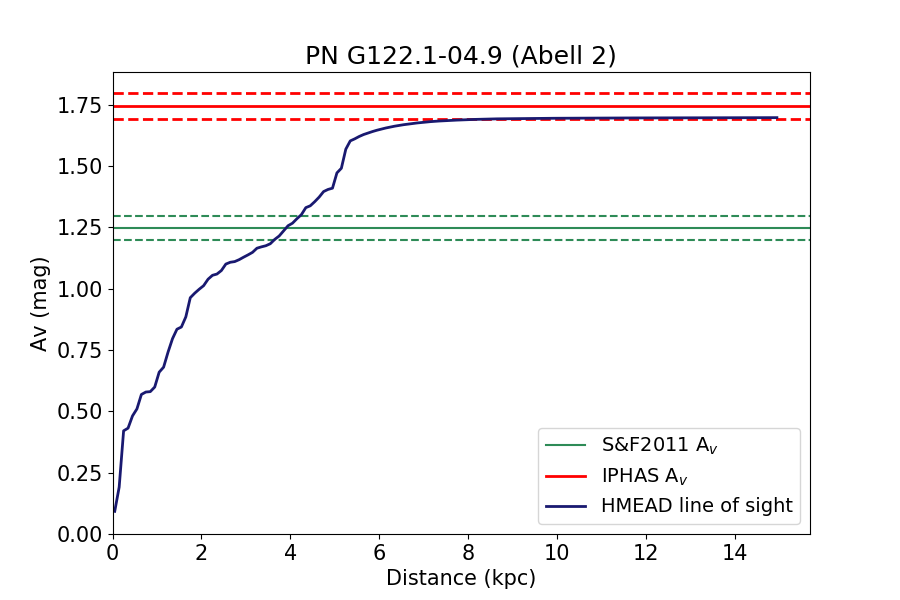}
  \subcaption{}
  \label{fig:Abell2}
  \end{subfigure}
  
 \caption{H-MEAD Extinction vs. Distance plots for M 1-80, NGC 7354, KjPn 8, K 3-88, BV 5-1, BV 5-2, We 1-1 and Abell 2.}
  \label{fig:extCurves_App_14}
\end{figure*}

\begin{figure*}
\centering

\begin{subfigure}[b]{0.45\textwidth}
  \centering
  \includegraphics[width=\textwidth]{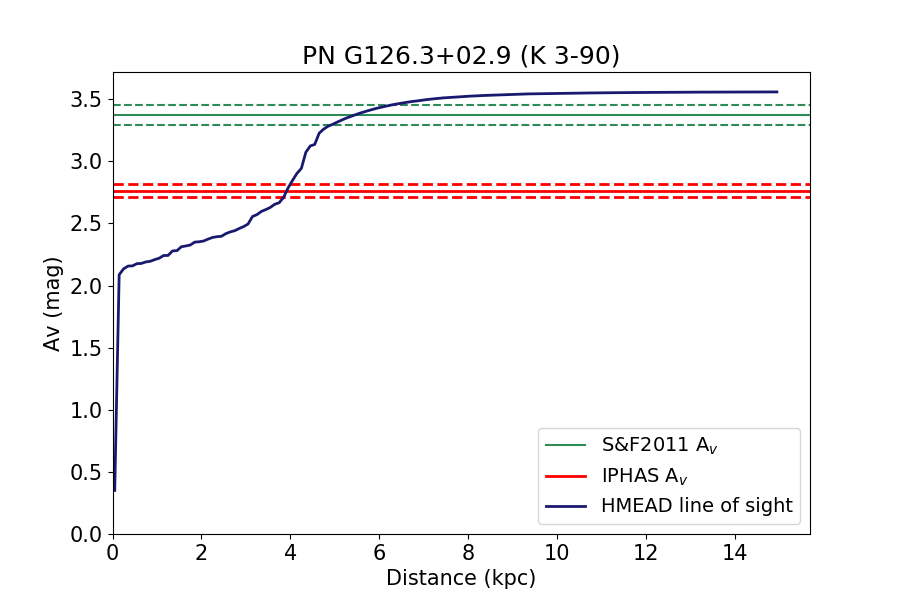}
  \subcaption{}
  \label{fig:K3-90}
  \end{subfigure}
\begin{subfigure}[b]{0.45\textwidth}
  \centering
  \includegraphics[width=\textwidth]{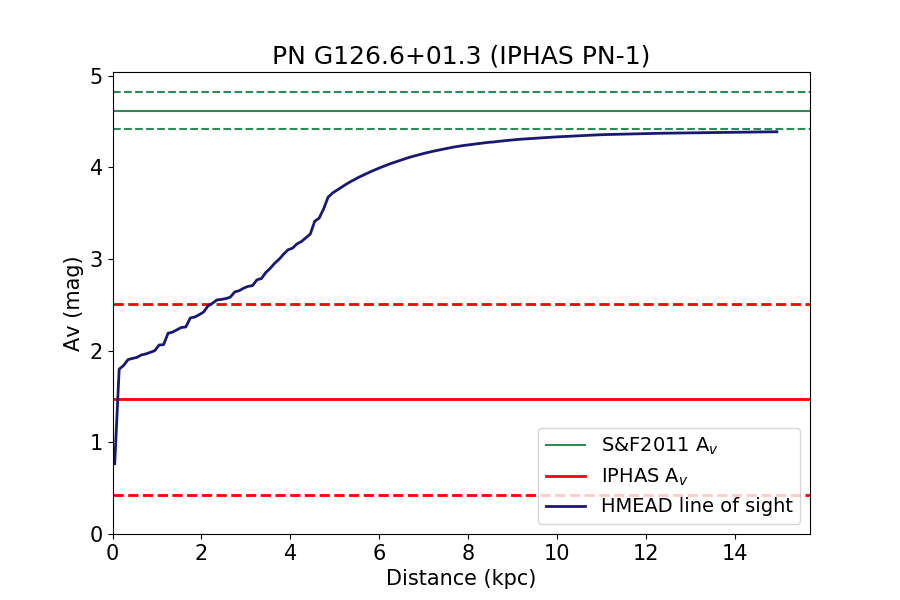}
  \subcaption{}
  \label{fig:IPHASPN-1}
  \end{subfigure}

\begin{subfigure}[b]{0.45\textwidth}
  \centering
  \includegraphics[width=\textwidth]{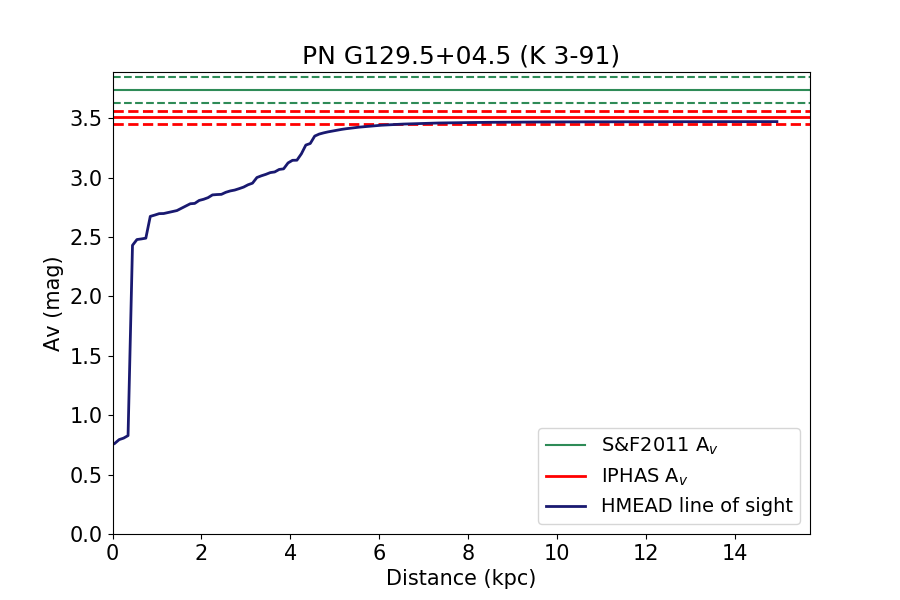}
 \subcaption{}
  \label{fig:K3-91}
  \end{subfigure}
\begin{subfigure}[b]{0.45\textwidth}
  \centering
  \includegraphics[width=\textwidth]{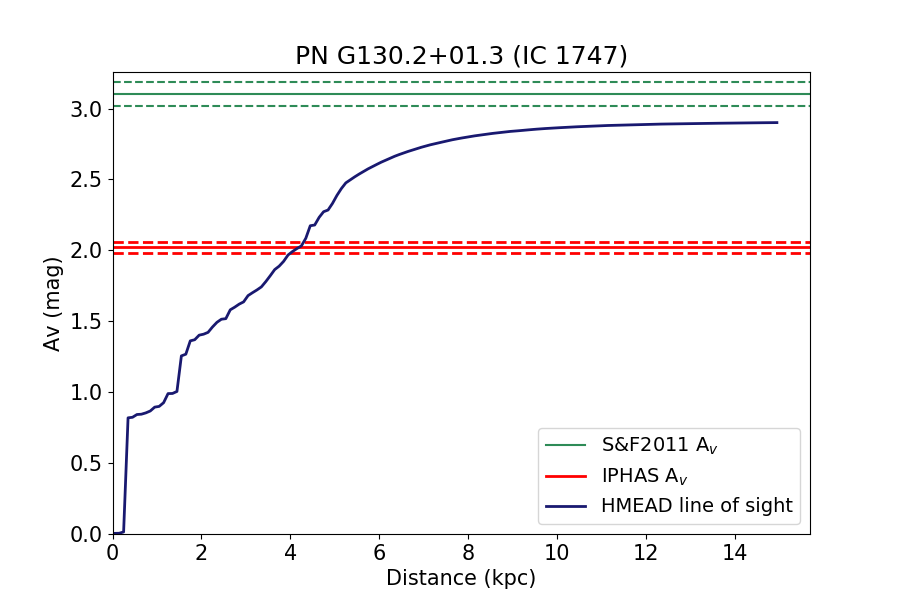}
  \subcaption{}
  \label{fig:IC1747}
  \end{subfigure}
  
\begin{subfigure}[b]{0.45\textwidth}
  \centering
  \includegraphics[width=\textwidth]{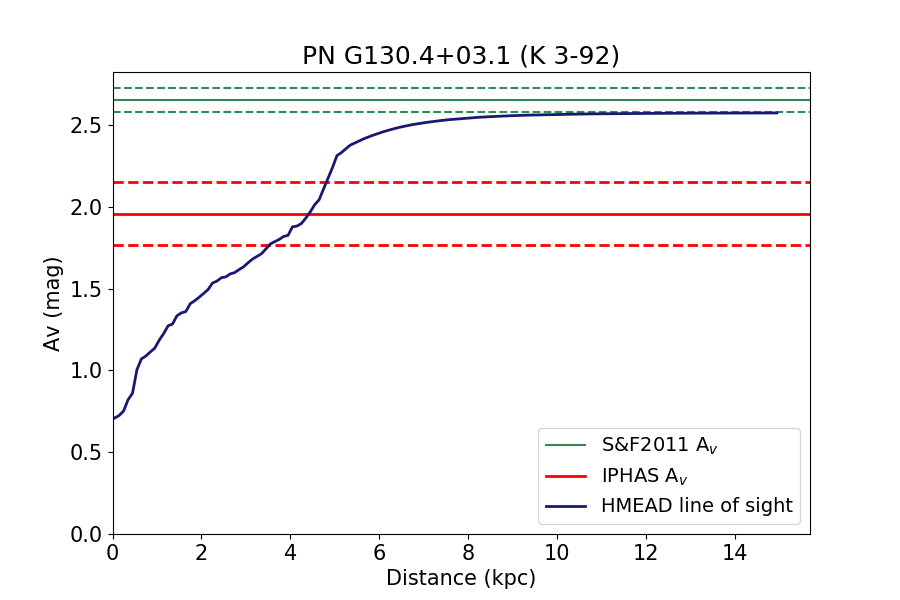}
 \subcaption{}
  \label{fig:K3-92}
  \end{subfigure}
\begin{subfigure}[b]{0.45\textwidth}
  \centering
  \includegraphics[width=\textwidth]{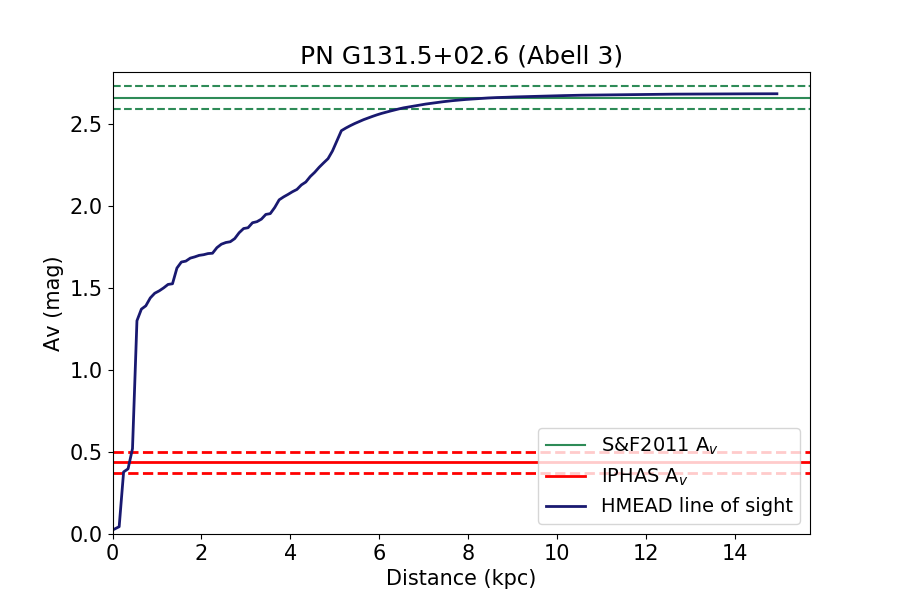}
  \subcaption{}
  \label{fig:Abell3}
  \end{subfigure}
  
\begin{subfigure}[b]{0.45\textwidth}
  \centering
  \includegraphics[width=\textwidth]{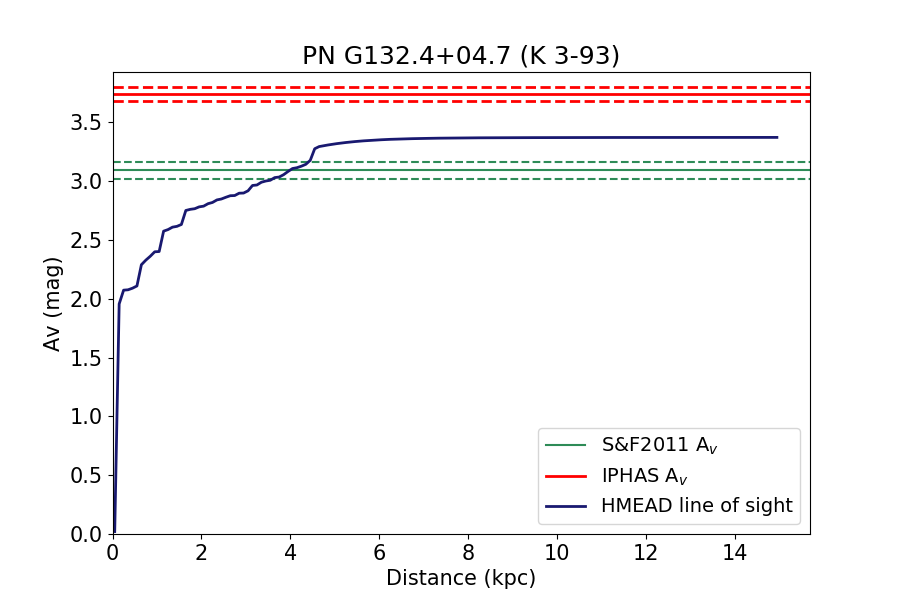}
  \subcaption{}
  \label{fig:K3-93}
  \end{subfigure}
\begin{subfigure}[b]{0.45\textwidth}
  \centering
  \includegraphics[width=\textwidth]{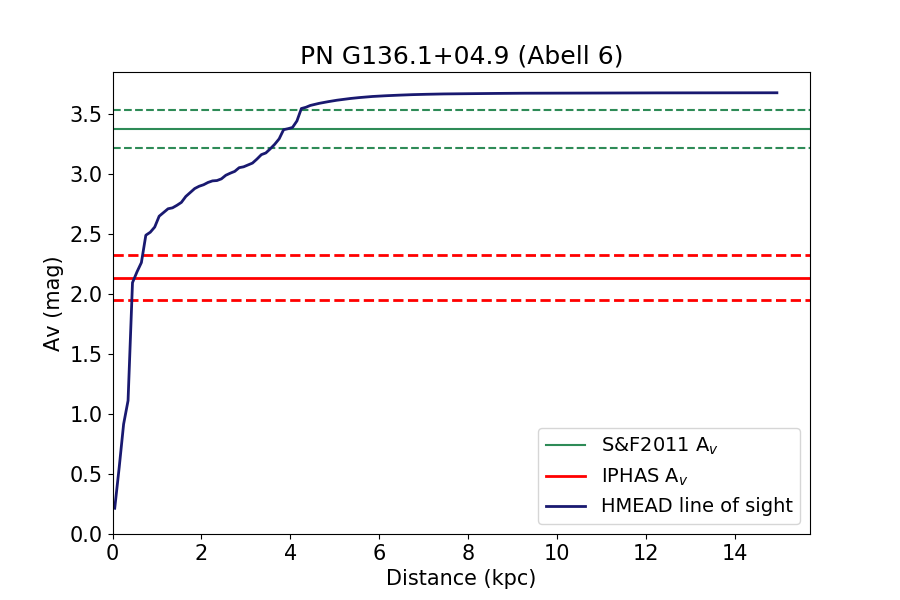}
  \subcaption{}
  \label{fig:Abell6}
  \end{subfigure}
  
  \caption{H-MEAD Extinction vs. Distance plots for K 3-90, IPHAS PN-1, K 3-91, IC 1747, K 3-92, Abell 3, K 3-93 and Abell 6.}
  \label{fig:extCurves_App_15}
\end{figure*}

\begin{figure*}
\centering

\begin{subfigure}[b]{0.45\textwidth}
  \centering
  \includegraphics[width=\textwidth]{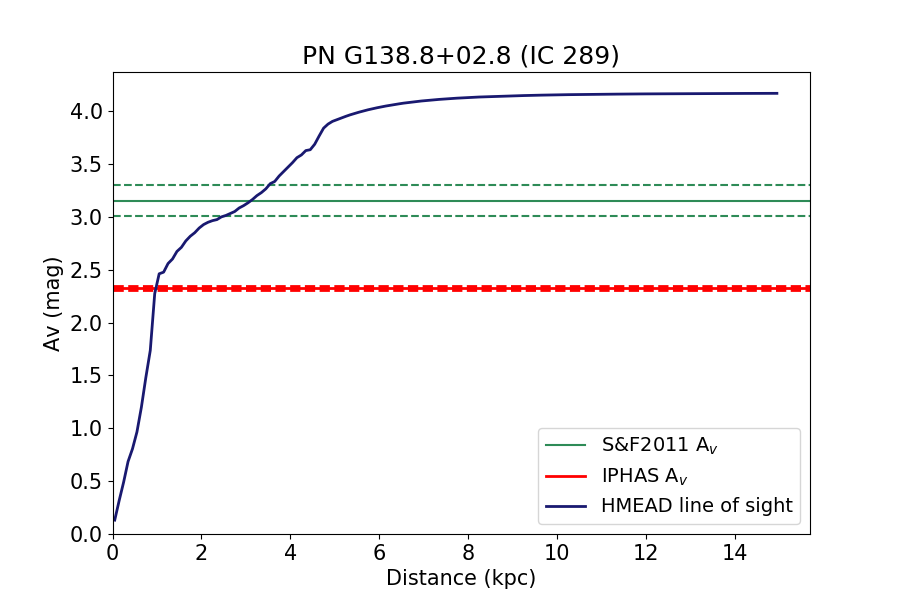}
  \subcaption{}
  \label{fig:IC289}
  \end{subfigure}
\begin{subfigure}[b]{0.45\textwidth}
  \centering
  \includegraphics[width=\textwidth]{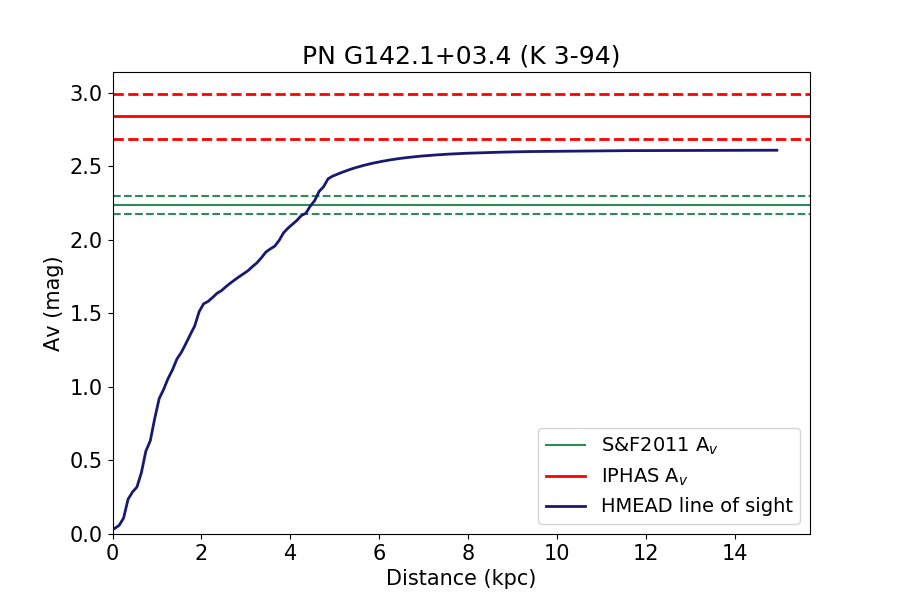}
  \subcaption{}
  \label{fig:K3-94}
  \end{subfigure}

\begin{subfigure}[b]{0.45\textwidth}
  \centering
  \includegraphics[width=\textwidth]{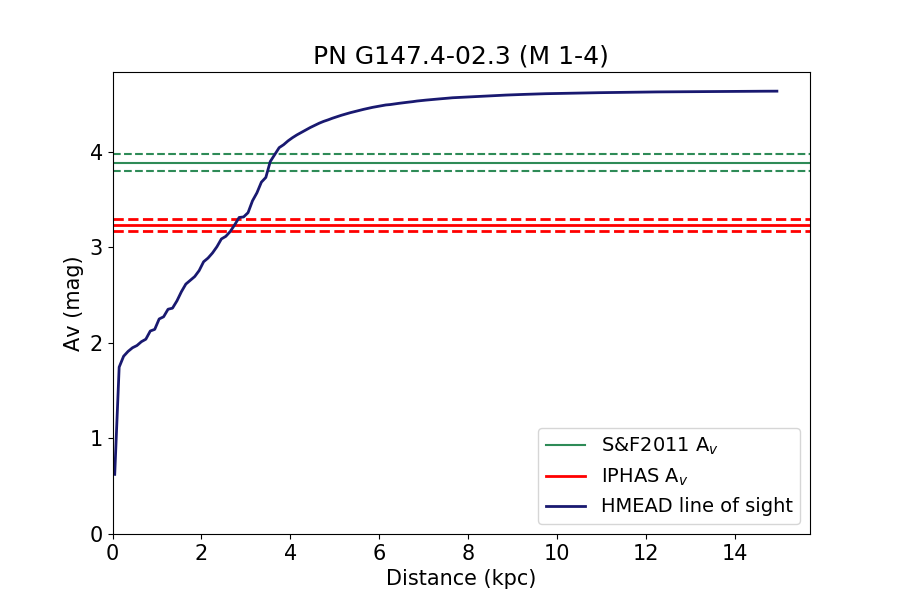}
 \subcaption{}
  \label{fig:M1-4}
  \end{subfigure}
\begin{subfigure}[b]{0.45\textwidth}
  \centering
  \includegraphics[width=\textwidth]{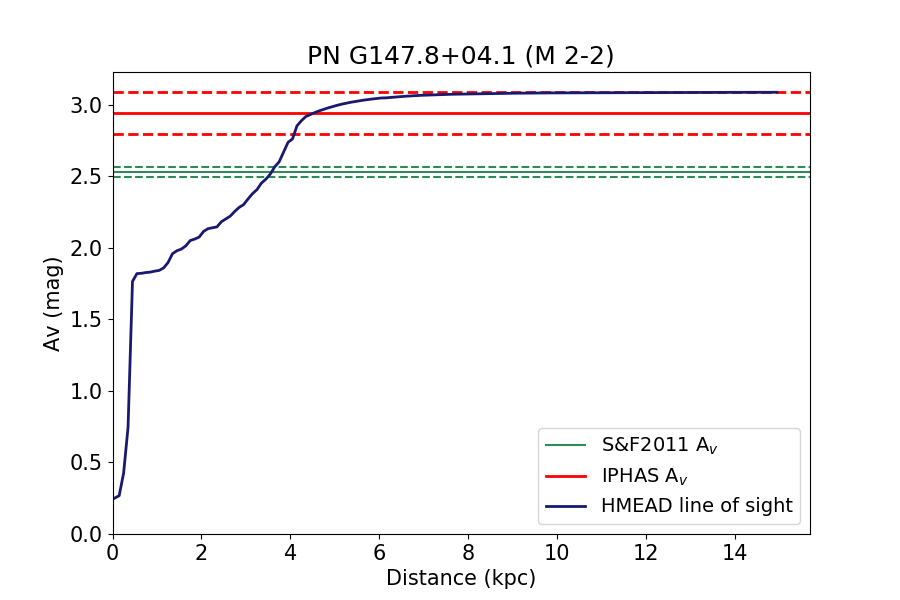}
  \subcaption{}
  \label{fig:M2-2}
  \end{subfigure}
  
\begin{subfigure}[b]{0.45\textwidth}
  \centering
  \includegraphics[width=\textwidth]{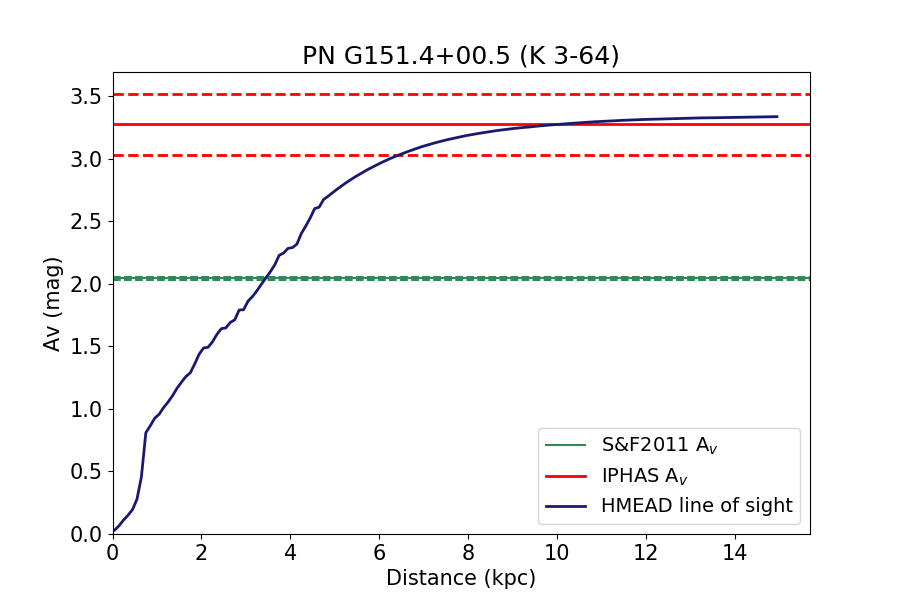}
 \subcaption{}
  \label{fig:K3-64}
  \end{subfigure}
\begin{subfigure}[b]{0.45\textwidth}
  \centering
  \includegraphics[width=\textwidth]{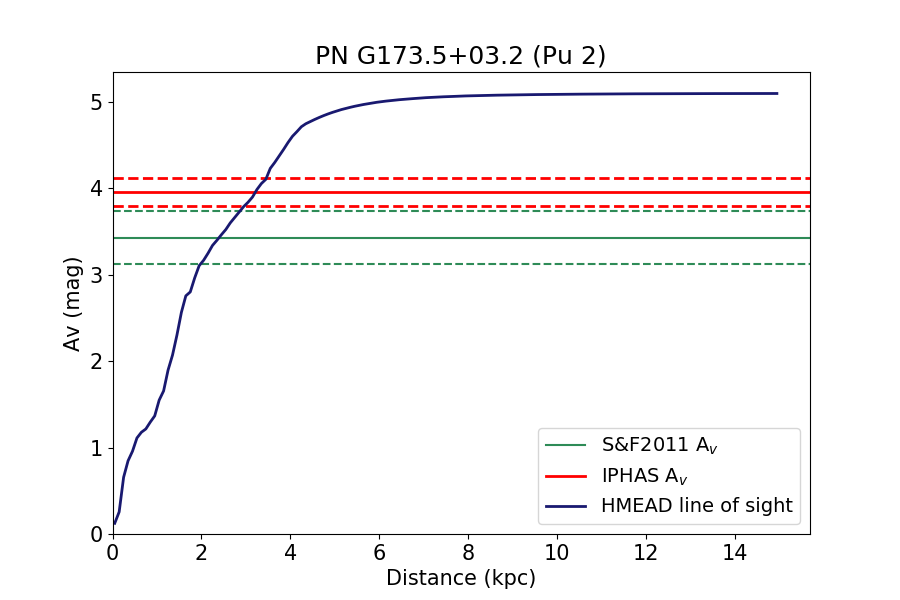}
  \subcaption{}
  \label{fig:Pu2}
  \end{subfigure}
  
\begin{subfigure}[b]{0.45\textwidth}
  \centering
  \includegraphics[width=\textwidth]{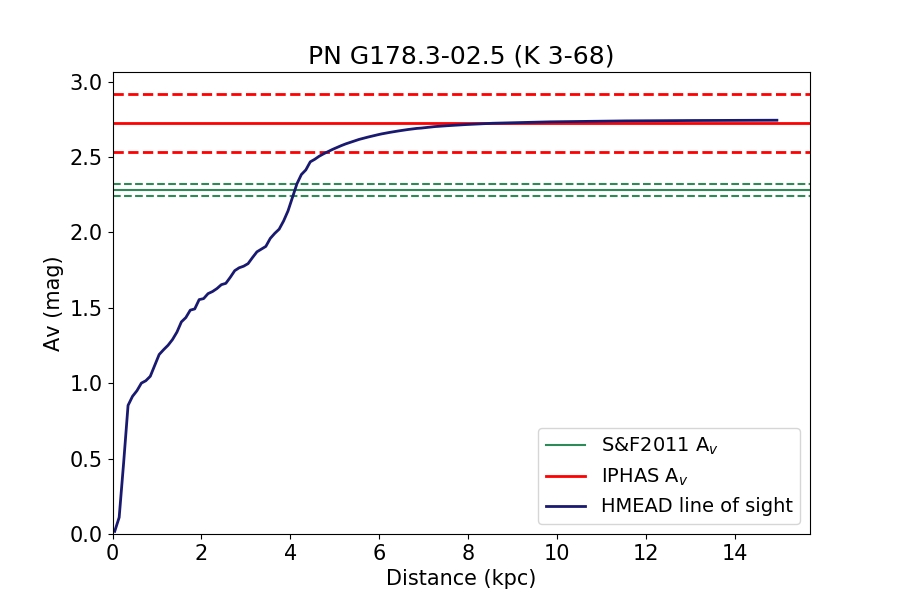}
  \subcaption{}
  \label{fig:K3-68}
  \end{subfigure}
\begin{subfigure}[b]{0.45\textwidth}
  \centering
  \includegraphics[width=\textwidth]{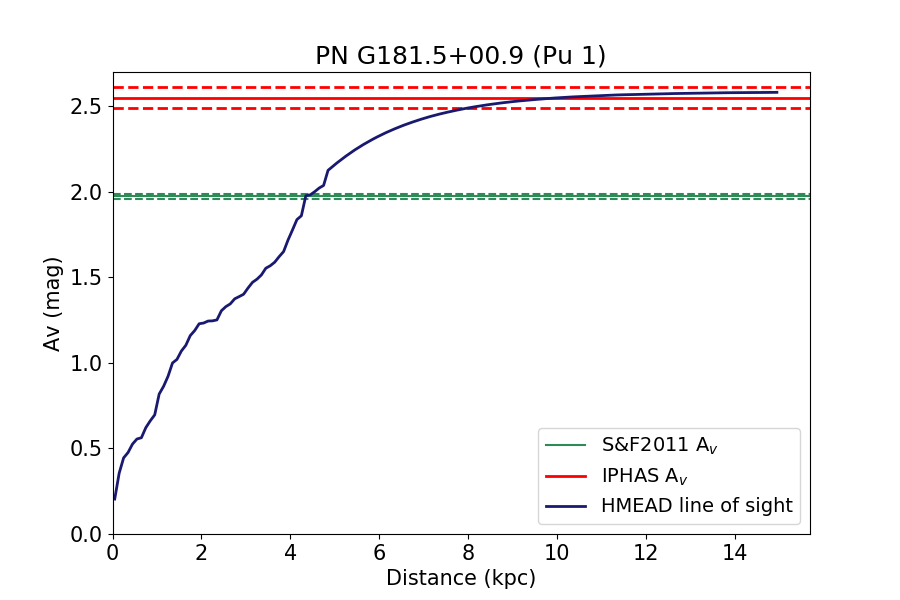}
  \subcaption{}
  \label{fig:Pu1}
  \end{subfigure}
  
  \caption{H-MEAD Extinction vs. Distance plots for IC 289, K 3-94, M 1-4, M 2-2, K 3-64, Pu 2, K 3-68 and Pu 1.}
  \label{fig:extCurves_App_16}
\end{figure*}

\begin{figure*}
\centering

\begin{subfigure}[b]{0.45\textwidth}
  \centering
  \includegraphics[width=\textwidth]{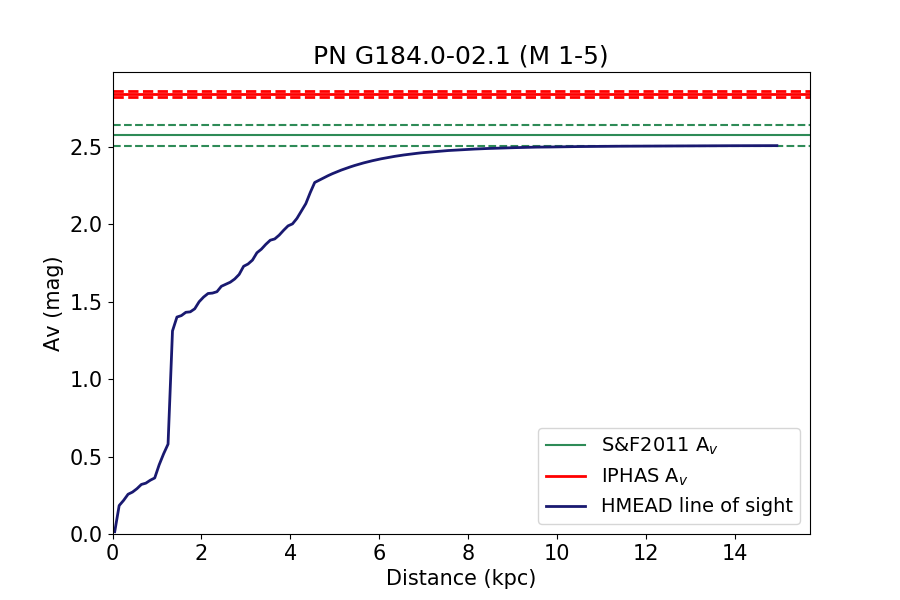}
  \subcaption{}
  \label{fig:M1-5}
  \end{subfigure}
\begin{subfigure}[b]{0.45\textwidth}
  \centering
  \includegraphics[width=\textwidth]{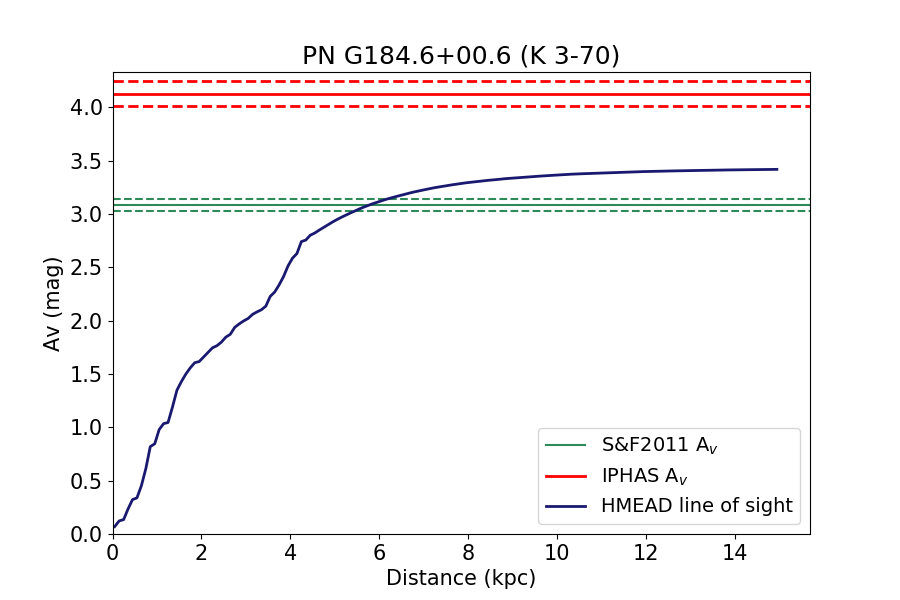}
  \subcaption{}
  \label{fig:K3-70}
  \end{subfigure}

\begin{subfigure}[b]{0.45\textwidth}
  \centering
  \includegraphics[width=\textwidth]{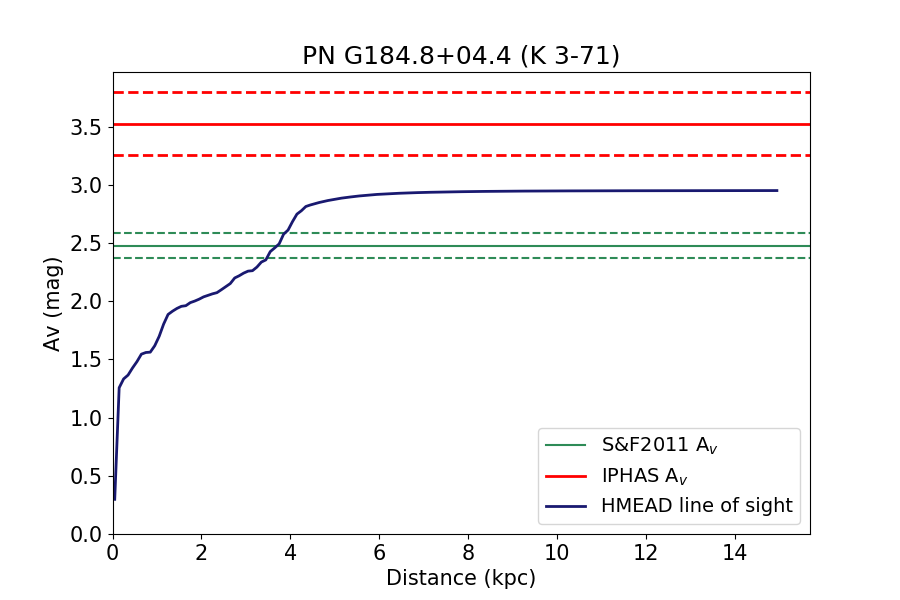}
 \subcaption{}
  \label{fig:K3-71}
  \end{subfigure}
\begin{subfigure}[b]{0.45\textwidth}
  \centering
  \includegraphics[width=\textwidth]{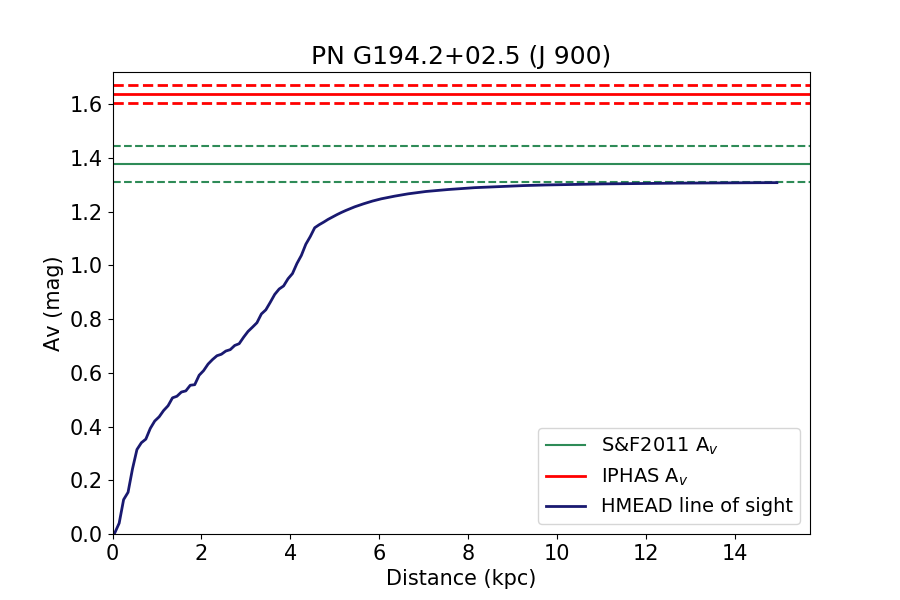}
  \subcaption{}
  \label{fig:J900}
  \end{subfigure}
\begin{subfigure}[b]{0.45\textwidth}
  \centering
  \includegraphics[width=\textwidth]{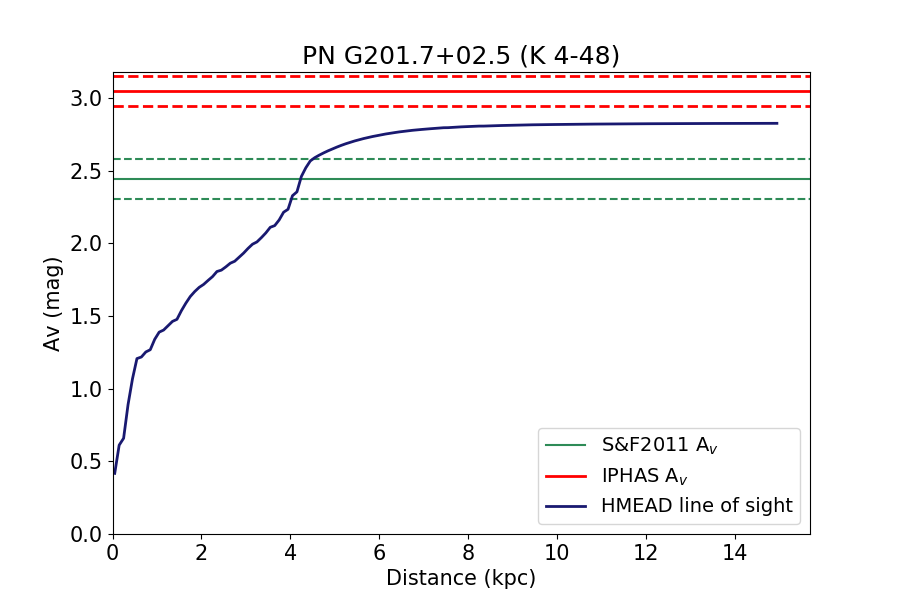}
  \subcaption{}
  \label{fig:K4-48}
  \end{subfigure}
\begin{subfigure}[b]{0.45\textwidth}
  \centering
  \includegraphics[width=\textwidth]{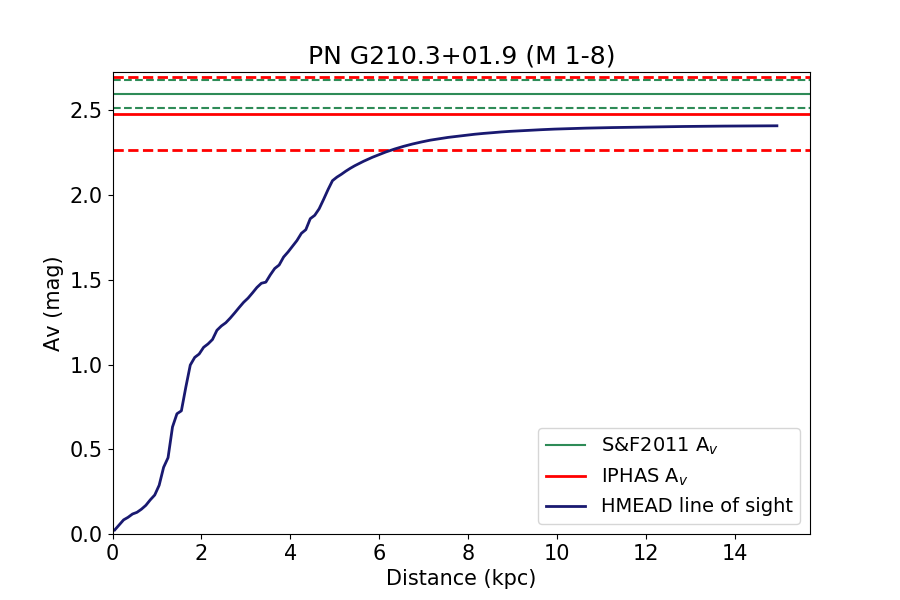}
  \subcaption{}
  \label{fig:M1-8}
  \end{subfigure}
  
\begin{subfigure}[b]{0.45\textwidth}

  \centering
  \includegraphics[width=\textwidth]{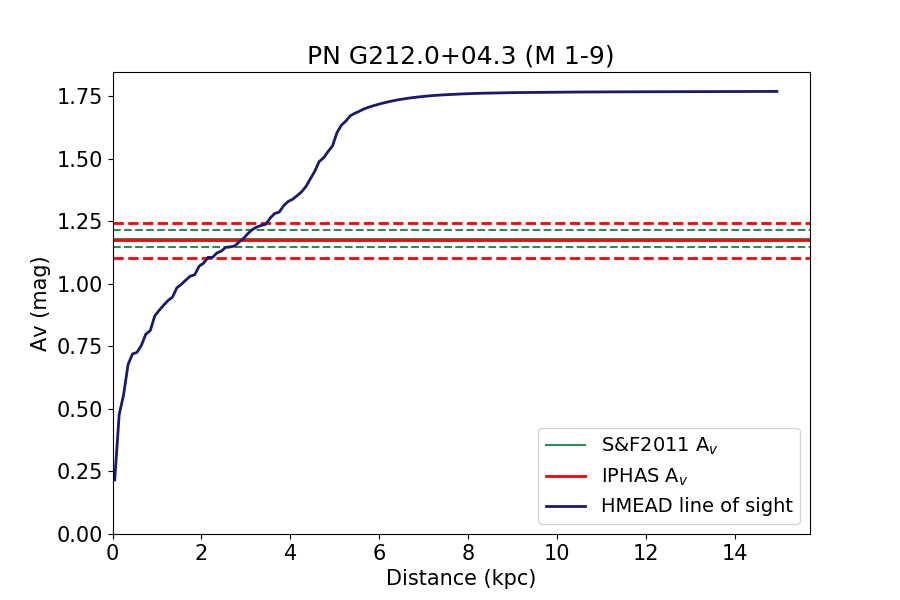}
  \subcaption{}
  \label{fig:M1-9}
  \end{subfigure}
 \caption{H-MEAD Extinction vs. Distance plots for M 1-5, K 3-70, K 3-71, J 900, K 4-48, M 1-8 and M 1-9.}
  \label{fig:extCurves_App_17}
\end{figure*}

